\documentclass[11pt,a4paper]{article}
\usepackage{placeins}
\usepackage{graphicx}
\usepackage{xcolor}
\usepackage{float}
\usepackage{afterpage}
\usepackage{amssymb,amsmath}
\usepackage{multirow,booktabs,multirow}
\usepackage{cite}
\usepackage[colorlinks=true, linkcolor=black!50!blue, urlcolor=blue, citecolor=blue, anchorcolor=blue]{hyperref}
\usepackage[font=small,labelfont=bf,margin=0mm,labelsep=period,tableposition=top]{caption}
\usepackage[a4paper,top=3cm,bottom=2.5cm,left=2.5cm,right=2.5cm,bindingoffset=0mm]{geometry}
\usepackage{tabularx}
\newcolumntype{C}[1]{>{\centering\arraybackslash}p{#1}}

\newcommand{\be}{\begin{equation}}
\newcommand{\ee}{\end{equation}}
\newcommand{\bea}{\begin{eqnarray}}
\newcommand{\eea}{\end{eqnarray}}
\newcommand{\bi}{\begin{itemize}}
\newcommand{\ei}{\end{itemize}}
\newcommand{\ben}{\begin{enumerate}}
\newcommand{\een}{\end{enumerate}}

\newcommand{\lp}{\left(}
\newcommand{\rp}{\right)}
\newcommand{\as}{\alpha_s}

\newcommand{\Ord}{\mathcal{O}}
\newcommand{\MSbar}{\overline{\text{MS}}}

\def\frac#1#2{{{#1}\over {#2}}}
\def\gsim{\gtrsim}
\def\lsim{\lesssim}

\newcommand{\draft}[1]{}

\newcommand{\muf}{\mu_\text{F}}
\newcommand{\mur}{\mu_\text{R}}

\def\beq{\begin{equation}}  
\def\eeq{\end{equation}}

\def\({\left(}
\def\){\right)}
\def\[{\left[}
\def\]{\right]}
\let\originalleft\left
\let\originalright\right
\renewcommand{\left}{\mathopen{}\mathclose\bgroup\originalleft}
\renewcommand{\right}{\aftergroup\egroup\originalright}

\numberwithin{equation}{section}
\numberwithin{figure}{section}
\numberwithin{table}{section}

\let\oldsubsection\subsection
\renewcommand\subsection[2][\subsectiontoc]{%
  \def\subsectiontoc{#2}%
  \oldsubsection[#1]{\boldmath #2}%
}

\let\oldsubsubsection\subsubsection
\renewcommand\subsubsection[2][\subsubsectiontoc]{%
  \def\subsubsectiontoc{#2}%
  \oldsubsubsection[#1]{\boldmath #2}%
}

\usepackage{xcolor}
\begin{document}
\newgeometry{top=1.5cm,bottom=1.5cm,left=2.5cm,right=2.5cm,bindingoffset=0mm}
\begin{titlepage}
\thispagestyle{empty}
\noindent
\vspace{-1.0cm}
\begin{flushright}
Edinburgh 2017/15\\
Nikhef/2017-027 \\
OUTP-17-12P
\end{flushright}
\vspace{.9cm}

\begin{center}
  {\LARGE \bf\boldmath
Parton distributions with small-$x$ resummation: \\[1.5ex]
    evidence for BFKL dynamics in HERA data}
\vspace{1.3cm}

{
  Richard~D.~Ball,$^1$
  Valerio~Bertone,$^2$
  Marco~Bonvini,$^3$
   Simone~Marzani,$^4$ \\
  Juan~Rojo,$^2$ and 
  Luca~Rottoli$^5$}

\vspace{0.8cm}
{\it \small ~$^1$ The Higgs Centre for Theoretical Physics, University of Edinburgh,\\
  JCMB, KB, Mayfield Rd, Edinburgh EH9 3JZ, Scotland
\\[1ex]
~$^2$ Department of Physics and Astronomy, VU University, NL-1081 HV Amsterdam,\\
and Nikhef Theory Group, Science Park 105, 1098 XG Amsterdam, The Netherlands
\\[1ex]
~$^3$ Dipartimento di Fisica, Sapienza Universit\`a di Roma\\
and INFN, Sezione di Roma, Piazzale Aldo Moro 5, 00185 Roma, Italy
\\[1ex]
~$^4$ Dipartimento di Fisica, Universit\`a di Genova\\
and INFN, Sezione di Genova, Via Dodecaneso 33, 16146, Italy
\\[1ex]
~$^5$ Rudolf Peierls Centre for Theoretical Physics, 1 Keble
Road,\\ University of Oxford, OX1 3NP Oxford, United Kingdom\\
}

\vspace{1.0cm}

{\bf \large Abstract}

\end{center}

We present a determination of the parton distribution functions of the proton
in which NLO and NNLO fixed-order calculations are supplemented by 
NLL$x$ small-$x$ resummation. 
Deep inelastic structure functions are computed consistently 
at NLO+NLL$x$ or NNLO+NLL$x$, 
while for hadronic processes small-$x$ resummation is included only 
in the PDF evolution, with kinematic cuts
introduced to ensure the fitted data lie in a region
where the fixed-order calculation of the hard cross-sections is reliable.
In all other respects, the fits use the same methodology and are based on the same
global dataset as the recent NNPDF3.1 analysis.
We demonstrate that the inclusion of small-$x$ resummation leads to a 
quantitative improvement in the perturbative description of the HERA
inclusive and charm-production
reduced cross-sections in the small $x$ region.
The impact of the resummation in our fits is greater at NNLO than at NLO, because
fixed-order calculations have a perturbative instability 
at small $x$ due to large logarithms that can be cured by resummation.
We explore the phenomenological implications of PDF sets
with small-$x$ resummation for the longitudinal structure function
$F_L$ at HERA, for parton luminosities and LHC benchmark cross-sections, 
for ultra-high energy neutrino-nucleus cross-sections, and for future
high-energy lepton-proton colliders such as the LHeC.

\vspace{0.8cm}

\end{titlepage}

\restoregeometry

\tableofcontents

\clearpage

\section{Introduction}
\label{sec:introduction}

The experiments at CERN's Large Hadron Collider (LHC) continue to explore particle physics both at the
high-energy and high-precision frontiers.
The outstanding quality of current and forthcoming LHC data
challenges the theory community to perform more precise calculations, so
that meaningful conclusions can be drawn when comparing these theoretical predictions to experimental measurements.
In this respect,
the tremendous effort put in place in order to arrive at precision calculations for hard-scattering
matrix elements and final-state parton evolution has to be accompanied by a comparable level of understanding of the
internal structure of the initial-state hadrons.

Global analyses of PDFs~\cite{Ball:2012cx,
  Alekhin:2017kpj,Accardi:2016qay,Dulat:2015mca,Harland-Lang:2014zoa,Jimenez-Delgado:2014twa}
(see~\cite{Gao:2017yyd,Ball:2015oha,Forte:2013wc,Butterworth:2015oua,
  Rojo:2015acz} for recent overviews)
are generally based on fixed-order perturbative calculations, at LO, NLO 
and NNLO.
However, it is well known that further logarithmic enhancements can affect partonic cross sections and DGLAP evolution kernels order by order in perturbation theory.
If we denote by $Q$ the hard scale of the process of interest and by $\sqrt{s}$ the center-of-mass energy of the colliding protons, we have logarithmic enhancements in two opposite limits, namely $Q^2 \sim s$ (the threshold region) and $Q^2 \ll s$ (the
high-energy region). Introducing the variable $x=Q^2/s$, the threshold
limit corresponds to large $x$, while the high-energy limit to small $x$. 

The LHC is exploring a vast kinematic range in $x$, potentially covering both extreme regions.
It is therefore crucially important to consistently assess the role of logarithmic corrections both at large and small $x$.
For instance, searches for new resonances at high mass are sensitive to PDFs in
the region between $0.1 \lsim x \lsim 0.7 $~\cite{Beenakker:2015rna}.
On the other hand, processes such as
forward production of  Drell-Yan lepton pairs
at small di-lepton
invariant masses~\cite{LHCb:2012fja}
and of $D$ mesons at small $p_T^D$\cite{Gauld:2015yia},
both measured by the LHCb collaboration,
probe values of $x$ at the other end of the spectrum, down to $x\sim 10^{-6}$.

Calculations that aim to describe these extreme regions of phase-space should in principle include resummation in the calculations of matrix elements and should make use of PDFs that were determined with a consistent theory. 
Threshold (large-$x$) resummation has already been included in PDF 
fits~\cite{Bonvini:2015ira} (see also Ref.~\cite{Corcella:2005us}) and 
dedicated studies which include threshold resummation in both the coefficient functions
and in the PDFs have been performed in the context of heavy supersymmetric 
particle production~\cite{Beenakker:2015rna}.
The inclusion of threshold resummation in PDF fits is straightforward 
because in the widely used $\MSbar$ scheme the DGLAP evolution kernels 
are not enhanced at large $x$~\cite{Korchemsky:1988si,Albino:2000cp}, so 
threshold resummation is only necessary for the coefficient functions, and 
can thus be included rather easily.

The situation is rather more intricate for small-$x$ resummation, 
because  both coefficient functions and splitting functions 
receive single-logarithmic contributions  to all orders in perturbation theory.
Small-$x$ resummation is based on the BFKL 
equation~\cite{Lipatov:1976zz,Fadin:1975cb,Kuraev:1976ge,Kuraev:1977fs,Balitsky:1978ic}. However, the naive application of the fixed coupling 
leading log $x$ (LL$x$) BFKL equation to small-$x$ deep-inelastic scattering 
(DIS) structure functions predicted
a much steeper growth than that actually observed by
the first HERA measurements~\cite{Abt:1993cb,Derrick:1993fta}, which
instead were
well reproduced by the predictions of LO and NLO running coupling 
DGLAP~\cite{Ball:1994du,Ball:1994kc,Forte:1995vs,Gluck:1994uf,Lai:1996mg,Martin:1994kn}.
This paradox was 
compounded by the computation of next-to-leading logarithmic (NLL$x$) 
corrections to the evolution
kernels~\cite{Fadin:1996nw,Fadin:1997hr,Fadin:1997zv,Camici:1997ij,Fadin:1998py},
which turned out to be large and negative, destabilizing the LL$x$
BFKL result. The correct implementation of small-$x$ resummation turns
out to require  the simultaneous resummation of
collinear and anti-collinear singularities in the small-$x$ evolution
kernels, together with a consistent resummation of running coupling
effects.    

This problem was tackled by several groups, see Refs.~\cite{Ball:1997vf,Ball:1999sh,Altarelli:1999vw,Altarelli:2000mh,Altarelli:2001ji,Altarelli:2003hk,Altarelli:2005ni,Ball:2007ra,Altarelli:2008xp,Altarelli:2008aj} (ABF), 
Refs.~\cite{Salam:1998tj,Ciafaloni:1998iv,Ciafaloni:1999yw,Ciafaloni:1999au,Ciafaloni:2000cb,Ciafaloni:2002xf,Ciafaloni:2003rd,Ciafaloni:2003kd,Ciafaloni:2005cg,Ciafaloni:2006yk,Ciafaloni:2007gf} 
(CCSS) 
and Refs.~\cite{Thorne:1999sg,Thorne:1999rb,Thorne:2001nr,White:2006yh}
(TW), which explored various theoretical and phenomenological aspects
of the problem, with the goal of achieving
consistent and phenomenologically viable frameworks that resum both collinear and 
high-energy logarithms simultaneously. Resummation corrections to
fixed-order evolution, when consistently implemented,  
were shown to be reasonably  small, thus 
explaining the success of the conventional unresummed description used
in standard PDF determinations.
More recently,
small-$x$ resummation based on the 
ABF formalism has been consistently matched to fixed NNLO for
perturbative evolution and deep-inelastic structure functions, and implemented in the public code {\tt HELL}~\cite{Bonvini:2016wki,Bonvini:2017ogt},
making small-$x$ resummation available for phenomenological applications.

On the other hand, while fixed-order DGLAP theory can provide a reasonable fit
to the inclusive HERA data, several groups have found
indications that the description of the  most precise
legacy datasets is not
optimal in the small-$x$ and small-$Q^2$ region, especially at
NNLO\footnote{It has been shown that the description of these data improves if either a higher-twist term or a phenomenological higher-order correction to $F_L$ is included~\cite{Harland-Lang:2016yfn}.}~\cite{Caola:2009iy,Caola:2010cy,Rojo:2015nxa,Abramowicz:2015mha,Harland-Lang:2016yfn,Abt:2016vjh,Abt:2017nkc}. 
Currently, the evidence that this tension is related to lack of small-$x$
resummation is inconclusive. The only way to show that it is
due to resummation would be to perform a complete global 
PDF analysis including small-$x$ resummation.
Since the effect of 
resummation is known to be small, at least in the kinematic region 
explored at HERA, it is necessary that these fits are
free of methodological bias.
The NNPDF framework~\cite{DelDebbio:2007ee,Ball:2008by,Ball:2009mk,Ball:2010de,Ball:2011uy,Ball:2013gsa,Ball:2014uwa,Ball:2016neh,Ball:2017nwa},
having been validated by a closure test, is thus ideal in this respect.

With these motivations, the goal of this paper is to present 
a state-of-the-art PDF determination in which  NLO and NNLO fixed-order 
perturbation theory is matched to NLL$x$ small-$x$ resummation.
This will be done by supplementing the
recent NNPDF3.1 PDF determination~\cite{Ball:2017nwa} with small-$x$ 
resummation of DGLAP evolution and DIS coefficient
functions using {\tt HELL}, thereby 
leading to resummed  PDF sets.
We will show that the 
inclusion of small-$x$
resummation significantly improves the quantitative description of the
small-$x$ and small-$Q^2$ HERA data, in particular at NNLO, both
for the inclusive and for the charm structure functions.
Our results fulfill a
program that was initiated more than 20 years ago,
when the first 
measurements of $F_2(x,Q^2)$ at
HERA
stimulated studies on the inclusion of small-$x$ resummation
in perturbative 
evolution~\cite{Ball:1995vc,Ball:1995qd,Ellis:1995gv,Forshaw:1995ga}.

The outline of this paper is as follows.
First, in Sect.~\ref{sec:th-review} we review the
implementation of small-$x$ resummation that we will use, and illustrate how
 resummation affects PDF evolution and DIS structure functions.
Then in Sect.~\ref{sec:fitsetup} we present the settings of our fits, which we dub NNPDF3.1sx,
and in particular we discuss the choice of kinematic cuts.
The results of the fits with small-$x$ resummation are discussed
in Sect.~\ref{sec:results}.
In Sect.~\ref{sec:diagnosis} we show the comparisons with the HERA
experimental data, and provide  detailed evidence for the onset of
resummation effects in the inclusive and charm-production structure functions.
We then perform a first exploration of the phenomenological
implications  of the NNPDF3.1sx fits at the LHC and beyond
in Sect.~\ref{sec:pheno}, and finally in Sect.~\ref{sec:summary} we
summarize and outline possible future developments.


\section{\boldmath Implementation of small-$x$ resummation}
\label{sec:th-review}

Here we briefly review the implementation of small-$x$ resummation
which will be adopted in the sequel.
First, we summarize the
general features of small-$x$ resummation theory, its main
ingredients, and available approaches to it.
We then discuss
separately the implementation and general phenomenology of small-$x$ resummation
of perturbative evolution, and of deep-inelastic structure functions.

\subsection{Basics of small-$x$ resummation}

In collinear factorization, the deep-inelastic scattering
structure functions can be expressed as
\beq\label{eq:xs}
\sigma(x,Q^2) = x
\sum_{i}\int_x^1 \frac{dz}{z}\,
\sigma_0\(Q^2,\as(\mur^2)\) \,C_{i}\(z,\as(\mur^2),\frac{Q^2}{\muf^2},\frac{Q^2}{\mur^2}\)\,f_{i}\left (\frac{x}{z},\muf^2\right) ,
\eeq
where $x=Q^2/s$, $\mur$ and $\muf$ are the renormalization and factorization scales, the sum runs over partons,
and we have factored out for convenience the Born cross-section $\sigma_0$. Similarly for hadronic processes 
\beq\label{eq:xsh}
\sigma(x,M^2) = x
\sum_{ij}\int_x^1 \frac{dz}{z}\,
\sigma_0\(M^2,\as(\mur^2)\) \,C_{ij}\(z,\as(\mur^2),\frac{M^2}{\muf^2},\frac{M^2}{\mur^2}\)\,{\cal L}_{ij}\left (\frac{x}{z},\muf^2\right) ,
\eeq
where $M^2$ is the invariant mass of the particles produced in the final state,
$x=M^2/s$, and the parton luminosities
\beq \label{eq:1D-lumi}
\mathcal{L}_{ij}(z,\mu^2) = \int_z^1 \frac{d w}w \, f_i\(\frac zw,\mu^2\) f_j(w,\mu^2).
\eeq
The scale dependence of the PDFs $f_{i}\left(x,\mu^2\right)$ is 
controlled by the DGLAP evolution equations
\beq \label{eq:dglap}
\mu^2 \frac{\partial}{\partial \mu^2} f_i(x,\mu^2) = \int_x^1 \frac{d z}{z} P_{ij}\(\frac{x}{z},\as(\mu^2) \) f_j(z,\mu^2)\, ,
\eeq
and knowledge of the splitting kernels $P_{ij}(x,\as)$ to $(k+1)$-loops allows 
for the resummation of collinear logarithms at N$^k$LO accuracy. 
The evolution kernels are currently known to 
NNLO (3 loops)~\cite{Moch:2004pa,Vogt:2004mw}, and 
partially even to N$^3$LO (4 loops)~\cite{Davies:2016jie,Moch:2017uml}. 

Single logarithms of $x$ affect higher order corrections to both
splitting functions and hard cross-sections. Specifically, the
generic all-order behaviour of the gluon-gluon splitting 
function is $P_{gg}\sim\frac{1}{x}\sum_n \as^{n} \ln^{n-1}\frac{1}{x}$. 
Small-$x$ logarithms are mostly relevant for PDFs in the singlet sector, 
{\it i.e.}\ the gluon and the quark singlet: small-$x$ (double) logarithms 
in nonsinglet PDFs are suppressed by an extra power of $x$. 
Partonic cross-sections (either
inclusive, or differential in rapidity or transverse momentum) can also 
contain small-$x$ logarithms, which depend on
the process and the observable. For gluon-induced
processes (such as Higgs or top production) resummation affects the
leading-order cross-section and it is thus a leading-log $x$ (LL$x$)
effect, while for quark-induced processes (such 
as Drell-Yan or deep-inelastic scattering) there must be a gluon-to-quark conversion, which
makes it a NLL$x$ effect. In either event at 
small $x$ and low scales the combination $\as \ln\frac{1}{x}$ can become large, 
spoiling fixed-order perturbation theory. In these circumstances it 
becomes necessary to resum the large logarithms in both splitting and 
coefficient functions in order to obtain reliable 
predictions. 

Small-$x$ resummation is based on the BFKL equation~\cite{Lipatov:1976zz,Fadin:1975cb,Kuraev:1976ge,Kuraev:1977fs,Balitsky:1978ic}, which can be written as 
an evolution equation in $x$ for off-shell gluons.
%
%
Knowledge of the BFKL kernel $K$ to $(k+1)$-loops allows for the resummation of small-$x$ logarithms to N$^k$LL$x$. The BFKL kernel is currently known to 2 loops~\cite{Fadin:1996nw,Fadin:1997hr,Fadin:1997zv,Camici:1997ij,Fadin:1998py}, and to 3 loops in the collinear approximation~\cite{Marzani:2007gk} (see Ref.~\cite{DelDuca:2008jg,Bret:2011xm,DelDuca:2011ae,DelDuca:2014cya,Caron-Huot:2016tzz,Caron-Huot:2017fxr} for other recent works on extending BFKL beyond NLL$x$). Thus, with current technology small-$x$ logarithms can be fully resummed to NLL$x$ accuracy.

A simultaneous resummation of collinear and high-energy logarithms can be obtained if one consistently combines the DGLAP and BFKL equations.
However, it turns out that this is far from trivial, particularly when 
the coupling runs, since the BFKL kernel also contains 
collinear (and anti-collinear) singularities which must be matched to those 
in DGLAP. 
This problem received great attention from several groups: Altarelli, Ball and Forte~\cite{Ball:1997vf,Ball:1999sh,Altarelli:1999vw,Altarelli:2000mh,Altarelli:2001ji,Altarelli:2003hk,Altarelli:2005ni,Ball:2007ra,Altarelli:2008xp,Altarelli:2008aj}, Ciafaloni, Colferai, Salam and Stasto~\cite{Salam:1998tj,Ciafaloni:1998iv,Ciafaloni:1999yw,Ciafaloni:1999au,Ciafaloni:2000cb,Ciafaloni:2002xf,Ciafaloni:2003rd,Ciafaloni:2003kd,Ciafaloni:2005cg,Ciafaloni:2006yk,Ciafaloni:2007gf} 
and Thorne and White~\cite{Thorne:1999sg,Thorne:1999rb,Thorne:2001nr,White:2006yh}, each of which produced resummed splitting functions for PDF evolution.
In the end, the theoretical ingredients used by the various groups were similar, thus leading to compatible results
(for a detailed comparison between the different approaches see~\cite{Dittmar:2005ed,Forte:2009wh}).
More recently, a public code named \texttt{HELL} (High-Energy Large Logarithms)~\cite{Bonvini:2016wki,Bonvini:2017ogt} has been produced to perform small-$x$ resummation to NLL$x$ of singlet splitting functions matched to NLO and NNLO fixed-order evolution.
\texttt{HELL} is largely based on the formalism developed by Altarelli, Ball and Forte (ABF)~\cite{Ball:1997vf,Ball:1999sh,Altarelli:1999vw,Altarelli:2000mh,Altarelli:2001ji,Altarelli:2003hk,Altarelli:2005ni,Ball:2007ra,Altarelli:2008xp,Altarelli:2008aj}.

In the ABF approach, one constructs perturbatively stable resummed results by combining three main ingredients: duality, {\it i.e.}\ consistency relations between the DGLAP and BFKL evolution kernels~\cite{Jaroszewicz:1982gr,Catani:1989sg,Ball:1997vf,Ball:1999sh}, which are used to construct a double-leading evolution kernel that simultaneously resums both collinear and small-$x$ logarithms;
symmetrization of the BFKL kernel in order to stabilize its perturbative expansion both in the collinear and anti-collinear regions of phase-space~\cite{Salam:1998tj,Altarelli:2005ni}, and thus in the region of asymptotically 
small $x$; and resummation of running coupling contributions, which despite being formally subleading are in fact dominant asymptotically, since they change
the nature of the small-$x$ singularity~\cite{Altarelli:2001ji,Ciafaloni:2002xf,Ciafaloni:2003rd,Altarelli:2003hk,Ball:2005mj,Thorne:1999sg}.
The resummation of gluon evolution with all the above ingredients consistently combined was originally achieved to NLO+NLL$x$ in Ref.~\cite{Ciafaloni:2003rd,Altarelli:2005ni}, while the inclusion of the quark contributions and the rotation to the physical basis of the singlet sector was completed in Ref.~\cite{Ciafaloni:2007gf,Altarelli:2008aj}. The matching to NNLO has been recently achieved in~\cite{Bonvini:2017ogt} and represents an important new development since it 
makes it possible to compare NNLO results with and without NLL$x$ 
small-$x$ resummation included.

Thanks to high-energy factorization~\cite{Catani:1990xk,Catani:1990eg,
Catani:1993ww,Catani:1994sq}
(generalized in Ref.~\cite{Caola:2010kv} to rapidity and in
Ref.~\cite{Forte:2015gve,Marzani:2015oyb} to transverse momentum 
distributions) it is
possible to also perform resummation of the leading small-$x$ logarithms in 
the coefficient functions both in deep inelastic cross-sections 
Eq.~(\ref{eq:xs}) and hadronic cross-sections Eq.~(\ref{eq:xsh}). 
The resummation relies on the resummation of
the splitting function, which must then be combined with a
computation of the hard cross-section with incoming off-shell
gluons. Such calculations have been made for a range of processes: 
heavy quark production~\cite{Catani:1990xk,Catani:1990eg,Collins:1991ty,Ball:2001pq}, DIS structure functions~\cite{Catani:1993rn,Catani:1994sq,Catani:1996sc}, Drell-Yan production~\cite{Marzani:2008uh,Marzani:2010ap}, direct photon production~\cite{Diana:2009xv,Diana:2010ef} and Higgs production~\cite{Hautmann:2002tu,Marzani:2008az,Caola:2011wq}. 
The use of these expressions to resum coefficient functions at fixed coupling is straightforward, but becomes more complicated when the coupling runs, due to the presence of anti-collinear singularities. This issue was resolved (both for photoproduction and hadroproduction processes) in Ref.~\cite{Ball:2007ra}, and used in Ref.~\cite{Altarelli:2008aj} to compute running coupling coefficient functions for DIS.

In order to discuss NLL$x$ resummation, we have to carefully specify the choice of factorization scheme. 
The so-called $Q_0\MSbar$ scheme is often introduced~\cite{Catani:1993ww,Catani:1994sq,Ciafaloni:2005cg,Marzani:2007gk}, and is preferred to the traditional $\MSbar$ because it gives more stable resummed results.
When expanded to fixed-order, the scheme-change factor between the two is $\mathcal{O}(\as^3)$, so NLL$x$ resummation in $Q_0\MSbar$ can be matched directly to the usual fixed order NNLO $\MSbar$ scheme calculation.

\subsection{Resummation of DGLAP evolution}

Resummed splitting functions take the generic form
\beq \label{eq:sf_match}
P_{ij}^\text{N$^k$LO+N$^h$LL$x$}(x)= P_{ij}^\text{N$^k$LO}(x)+ \Delta_k P_{ij}^\text{N$^h$LL$x$}(x),
\eeq
where the first contribution is the splitting function computed to fixed-order $k$ (so $k=0,1,2$ for LO, NLO and NNLO) and the second term is the resummed contribution, computed to either LL$x$ ($h=0$) or NLL$x$ ($h=1$), minus its expansion to the fixed order $k$ to avoid double counting.
We note that the splitting functions in the gluon sector ($P_{gg}$ and $P_{gq}$) contain LL$x$ and NLL$x$ contributions, while in the quark sector ($P_{qg}$ and $P_{qq}$) they only start at NLL$x$.
For this reason, there have been attempts to partially extend the resummation to the next logarithmic order (see \cite{Bartels:2002uz}) which however are not considered in this work.

\begin{figure}[t]
\centering
  \includegraphics[width=0.495\textwidth, page=1]{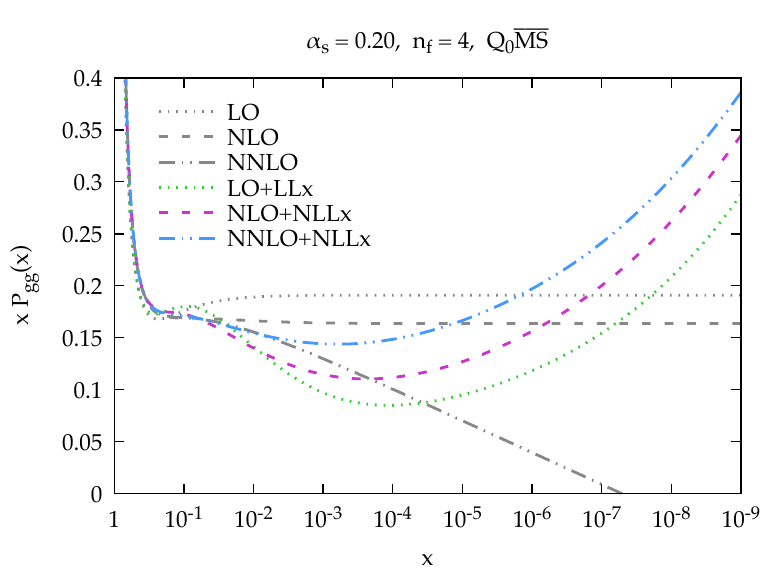}
  \includegraphics[width=0.495\textwidth, page=3]{plots/plot_P_nf4_paper_noband.pdf}
  \caption{\small Comparison of the fixed-order gluon-gluon
    $xP_{gg}(x,\as)$ (left) and the quark-gluon $xP_{qg}(x,\as)$ (right)
    splitting functions with the corresponding
    LO+LL$x$, NLO+NLL$x$ and NNLO+NLL$x$ results including small-$x$ resummation.
    The comparison is performed at a scale such that $\as=0.2$
    and in the $Q_0\MSbar$ scheme with $n_f=4$ active quark flavours.
  }
  \label{fig:splittingfunctions}
\end{figure}

In Fig.~\ref{fig:splittingfunctions} we show a 
comparison of the fixed-order gluon-gluon $xP_{gg}(x,\as)$ (left)
and the quark-gluon $xP_{qg}(x,\as)$ (right plot)
splitting functions with the resummed counterparts.
The comparison is performed in the $Q_0\MSbar$ factorization scheme, 
with $n_f=4$ active quark flavours and at a small scale such that $\as=0.2$.
We consider LL$x$ resummation matched to LO (for the gluon-gluon case), and NLL$x$ resummation matched to both NLO and NNLO. All calculations are performed using the \texttt{HELL} (version \texttt{2.0}) implementation of the ABF construction, and thus incorporate a number of technical improvements which makes the numerical implementation more robust, and allow the matching to NNLO fixed order as well as NLO: a detailed discussion and comparison is given in Refs.~\cite{Bonvini:2016wki,Bonvini:2017ogt}. 
The resummation of small-$x$ logarithms is more important 
at NNLO than at NLO, since at NNLO the fixed-order small-$x$ logarithms give rise to perturbative instabilities at small-$x$,
as visible from a comparison of the NLO and NNLO curves 
in Fig.~\ref{fig:splittingfunctions}.
Indeed, from the left hand plot, one can immediately see that for moderately small 
values of $x$ NLO gluon evolution is closer
to the all-orders result at small $x$ than NNLO evolution,
since for $10^{-6}\lesssim x\lesssim10^{-3}$ the NLO splitting kernels
are closer to the best prediction, NNLO+NLL$x$, than the NNLO ones.
Additionally, from the right plot, both resummed results for the gluon to quark spitting function are closer to NLO than to NNLO for $10^{-5}\lesssim x\lesssim10^{-1}$.
N$^3$LO evolution, when available~\cite{Davies:2016jie,Moch:2017uml}, will lead to even more significant instabilities at small $x$, due to the appearance of two extra powers of the small-$x$ logarithms (the leading NLO and NNLO logarithms are accidentally zero), and will make the inclusion of small-$x$ resummation even more crucial.

To facilitate the use of small-$x$ resummation,
the \texttt{HELL} code has been interfaced to the public
code \texttt{APFEL}~\cite{Bertone:2013vaa,Carrazza:2014gfa}.
Thanks to this \texttt{APFEL+HELL} interface,
it is straightforward to perform the PDF evolution (and the
computation of DIS structure functions)
with the inclusion of small-$x$ resummation effects.
Note that \texttt{APFEL+HELL} only implements the so-called ``exact'' solution of 
DGLAP evolution, rather than the 
``truncated'' solutions used in ABF (for example in Refs.~\cite{Ball:2007ra,Altarelli:2008xp,Altarelli:2008aj}), and nowadays 
routinely in NNPDF fits, in which subleading 
corrections are systematically expanded out~\cite{Ball:2008by}. For this 
reason we will use the exact solution throughout in this paper, to facilitate comparison between fixed-order and resummed results. 
Since the difference between the two 
solutions becomes smaller and smaller when
increasing the perturbative order, this choice does not affect
significantly our NNLO(+NLL$x$) results, but care should be taken when 
comparing the NLO PDFs from those of other NNPDF fits.

\begin{figure}[t]
\begin{center}
 \includegraphics[scale=0.91]{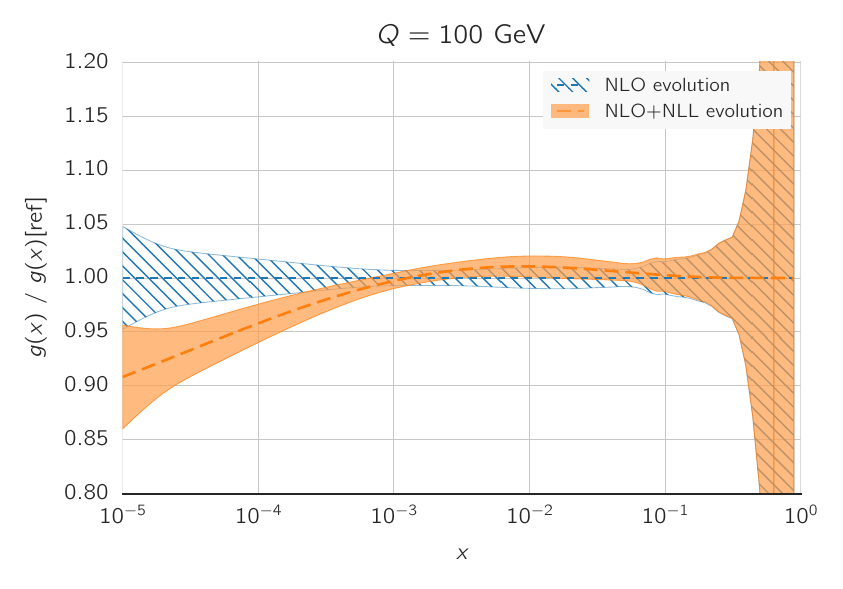}
 \includegraphics[scale=0.91]{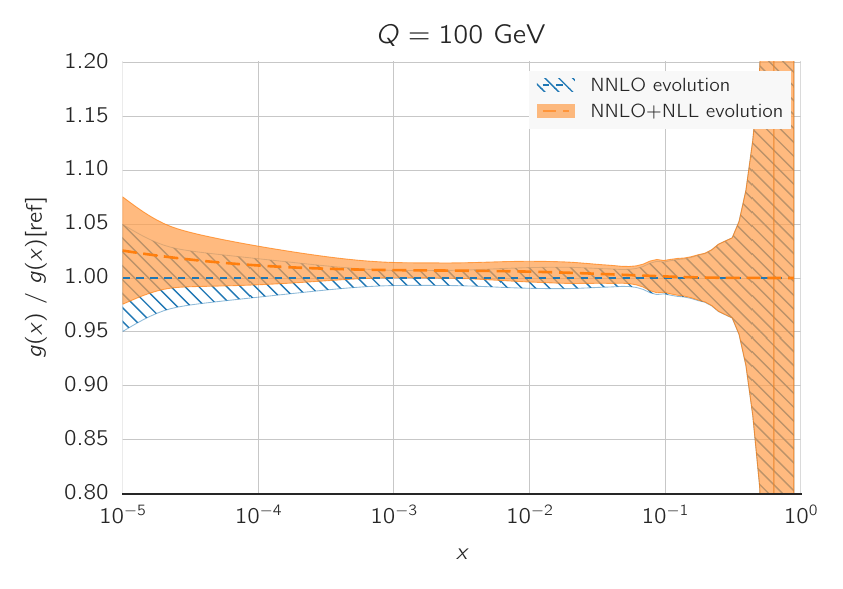}\\
 \includegraphics[scale=0.91]{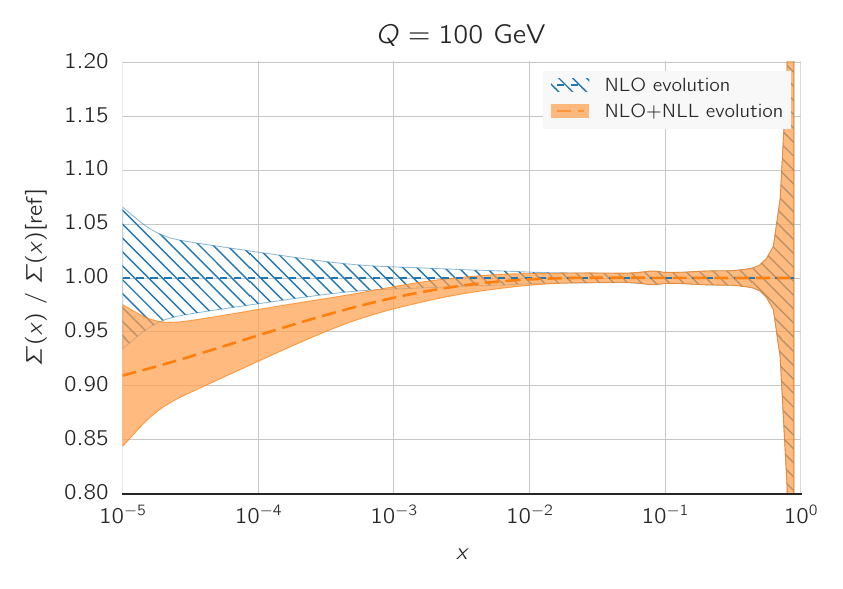}
 \includegraphics[scale=0.91]{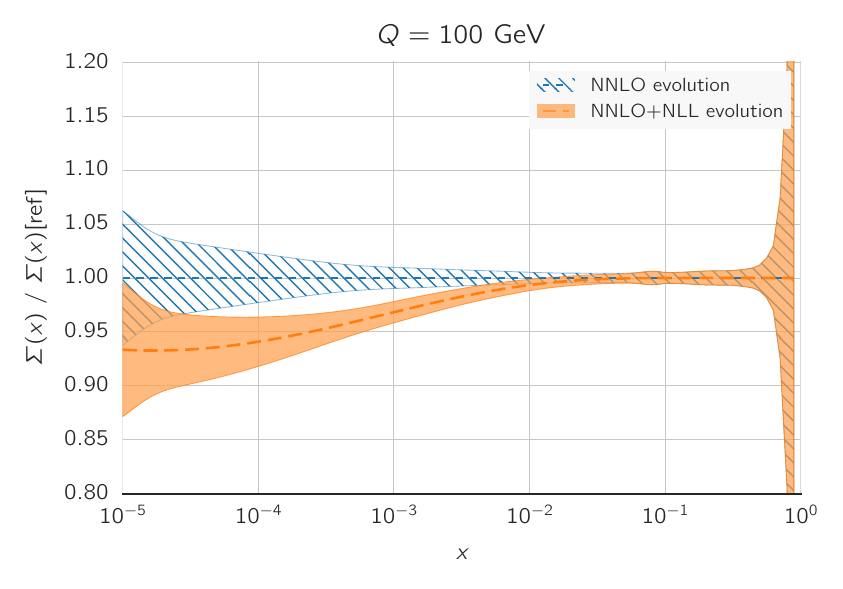}
 \caption{\small The ratio of the gluon (upper plots)
     and quark singlet (lower plots) for the evolution from
     a fixed boundary condition at $Q_0=1.65$ GeV up to $Q=100$ GeV
     using either fixed-order theory (NLO left, NNLO right) or 
     resummed theory (NLO+NLL$x$ left, NNLO+NLL$x$ right) for the DGLAP evolution.
     In this specific case, the input boundary condition
     has been chosen to be NNPDF3.1 NLO (NNLO).
  }
  \label{fig:resummedPDFs}
\end{center}
\end{figure}

We now investigate the effects induced
by evolving the PDFs
with resummed splitting kernels as compared to standard 
fixed-order DGLAP splitting functions.
In order to  illustrate these
effects, we take a given input PDF set
as fixed
at a low scale $Q_0$, that is,
a common boundary condition, and then evolve it
upwards using {\tt APFEL}+{\tt HELL}
with either fixed-order (NLO or NNLO) or resummed (NLO+NLL$x$ or NNLO+NLL$x$) theory.
In this way, we can determine what are the main differences induced at high scales by small-$x$
resummation in the PDF evolution; we stress however that the physical meaning of the resulting
comparison is limited,
as in a PDF fit with small-$x$ resummation the PDFs at low scales,
now taken to be equal to their fixed-order counterparts,
are likely to change significantly.

The results of this comparison
are collected in Fig.~\ref{fig:resummedPDFs}, where
we show the ratio of the gluon (upper plots)
and quark singlet (lower plots) as a function of $x$
for the evolution from
a fixed boundary condition at $Q_0=1.65$~GeV up to $Q=100$~GeV
using either (N)NLO fixed-order theory or (N)NLO+NLL$x$
resummed theory for the DGLAP evolution.
In this specific case, the input boundary condition
has been chosen to be NNPDF3.1 (N)NLO.
We observe that the effects of the different PDF evolution settings
are negligible at large and medium $x$, but can reach up to a few
percent at the smallest values of $x$ relevant for
the description of the data included in a PDF fit,
in particular the HERA structure functions.
Specifically,
we observe that resummation effects change the NLO evolution quite substantially for both the gluon and the quark singlet,
an effect which is reduced at NNLO for the gluon, while it remains of the same size (if not larger) for the quark singlet.
Although this study is purely illustrative and by no means predictive, it allows us to conclude that the effect of small-$x$ resummation in PDF evolution is in general sizeable and will certainly impact the determination of PDFs at small $x$.

\subsection{Resummation of DIS structure functions}\label{sec:resDIS}
     
Resummed results for DIS structure functions, including mass effects,
have been recently implemented in the public code \texttt{HELL},
version \texttt{2.0}~\cite{Bonvini:2017ogt}.
Analogously to Eq.~\eqref{eq:sf_match}, resummed and matched results can be written as
\beq \label{eq:cf_match}
C_{a,i}^\text{N$^k$LO+NLL$x$}(x)= C_{a,i}^\text{N$^k$LO}(x)+ \Delta_k C_{a,i}^\text{NLL$x$}(x),
\eeq
where the index $a$ denotes the type of structure function, $a=2, L, 3$, while the index $i$ refers to the incoming parton $i=q,g$. Note that in this paper we only consider NLL$x$ resummation of the partonic coefficient functions, since in DIS there are no LL$x$ contributions.
Consistently with the choice made for the evolution, we work in the $Q_0\MSbar$ scheme.

A consistent PDF fit which spans several orders of magnitude
in $Q^2$ further requires us to
consider a different number of active quark flavours at different energies, to
account for potentially large collinear logarithms due to massive quarks.
When crossing the threshold of a given heavy quark,
matching conditions which relate the PDFs above and below threshold are needed.
These matching conditions also contain small-$x$ logarithmic enhancements, which one can consistently resum.
As for DIS coefficient functions, the matching conditions are NLL$x$,
and their resummation, as well as the resummation of the massive coefficient 
functions~\cite{Catani:1996sc,Bonvini:2017ogt} is available 
in \texttt{HELL 2.0}.
These last ingredients make it straightforward to implement a resummation of 
the FONLL 
variable flavour number scheme~\cite{Forte:2010ta} 
used in the NNPDF fits. 

A careful treatment of charm is essential when addressing the impact of 
small-$x$ resummation on DIS structure functions, since the kinematic region where resummation is expected to be important (small $x$ and low $Q^2$) is rather close to the charm threshold. We thus fit the initial charm distribution, as in 
Ref.~\cite{Ball:2016neh}. The FONLL scheme can be readily extended to fitted charm,
in the process receiving an extra 
contribution~\cite{Ball:2015dpa}, denoted $\Delta_{\rm IC}$, 
which is currently known only at $\Ord(\as)$ \cite{Kretzer:1998ju,Ball:2015tna}.
When $\Delta_{\rm IC}$ is included, the phenomenological damping adopted
in the original FONLL formulation to smooth the transition to the regime
in which collinear logarithms are resummed does not have any effect~\cite{Ball:2015tna,Ball:2015dpa},
and is therefore omitted.
Since the $\Ord(\as)$ $\Delta_{\rm IC}$ contribution is then a small correction,
we expect the NNLO ($\Ord(\as^2)$) and small-$x$ resummation corrections to 
$\Delta_{\rm IC}$ to be practically insignificant
(see Ref.~\cite{Ball:2016neh} for a detailed discussion of this issue).

\begin{figure}[t]
\begin{center}
  \includegraphics[scale=0.95,page=2]{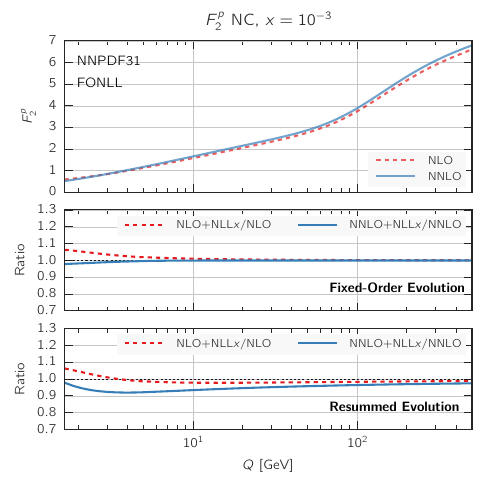}
  \includegraphics[scale=0.95,page=4]{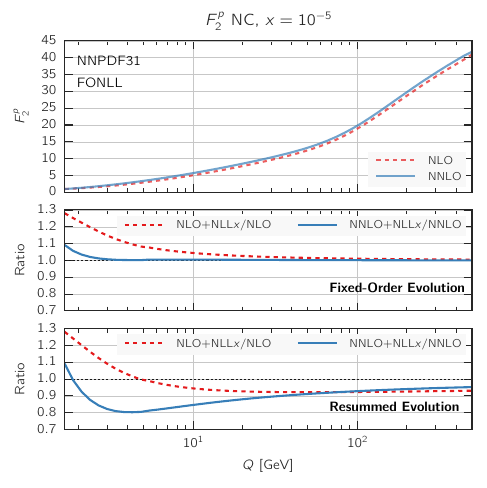}
  \caption{\small
   The proton neural-current (NC) structure function $F_2(x,Q)$ as a function of
    $Q$ for two different values of $x$ (left: $x=10^{-3}$; right: $x=10^{-5}$) and using different calculational schemes.
    In the top panels we show the structure function computed in fixed-order perturbation theory (NLO and NNLO).
    In the middle and bottom panels we show the ratio of resummed results (NLO+NLL$x$ and NNLO+NLL$x$) to their fixed-order 
    counterparts. In particular, in the middle panel the resummation is included in the coefficient function but not in the evolution,
    while in the bottom panel we resum both coefficient functions and 
parton evolution.
     The input boundary condition at $Q_0=1.65$ GeV
     has been chosen to be NNPDF3.1 NLO (NNLO), and all calculations 
are performed with $\as(m_Z)=0.118$, and a (pole) charm 
mass $m_c=1.51$ GeV.
  }
  \label{fig:resummedSFfixedPDF}
\end{center}
\end{figure}

To obtain a first qualitative estimate of the impact of small-$x$ resummation in the
DIS structure functions, we can compare theoretical predictions at (N)NLO with predictions
that include resummation. 
To disentangle the effect of resummation on PDF evolution from that in the coefficient functions in the
$Q_0\MSbar$ scheme, we take into account the effect of resummation in two steps. First, we compute structure functions
with the same (fixed-order) input PDFs, and include small-$x$ resummation in the coefficient functions only.
As a second step, we include resummation also in the DGLAP evolution, using a fixed input PDF boundary condition at a small scale $Q_0=1.65$~GeV, as previously done in Fig.~\ref{fig:resummedPDFs}.
Since, as already noticed, the use of a fixed boundary condition at a small scale
is not particularly physical, these results should be interpreted with care.

The proton structure function $F_2(x,Q)$ in neutral current (NC) DIS is shown in Fig.~\ref{fig:resummedSFfixedPDF} as a function of
$Q$ for two values of $x$, one moderate ($x=10^{-3}$, left plot) and one small ($x=10^{-5}$, right plot).
The upper panel of each plot shows the NLO and NNLO results.
The middle panel shows the ratio of resummed (N)NLO+NLL$x$ theory over the fixed-order (N)NLO results,
including resummation only in coefficient functions.
The lower panel, instead, shows the same ratio but with resummation included also in PDF evolution.
In all cases, we take the NNPDF3.1 boundary condition at (N)NLO at $Q_0=1.65$ GeV. 
As mentioned above,
heavy quark mass effects are included using the FONLL-B~(C) scheme~\cite{Forte:2010ta,Ball:2015tna,Ball:2015dpa} for the NLO~(NNLO) calculations,
supplemented with small-$x$ resummed contribution for the (N)NLO+NLL$x$ as described in Ref.~\cite{Bonvini:2017ogt}.

The comparison in Fig.~\ref{fig:resummedSFfixedPDF} is interesting from several points of view.
First of all, we observe that when resummation is included only in the coefficient functions its effect is rather mild,
almost negligible when matched to NNLO, even at rather small $x$ and 
at low scales.  
On the other hand, when including resummation in the PDF evolution, the situation changes.
In this case, we note that the differences between fixed-order and resummation are larger,
thus showing that in $F_2$ much of the impact of small-$x$ resummation arises from the
PDF evolution.
Moreover, the effects are always greater at NNLO than at NLO: 
at NNLO, effects of small-$x$ resummation can reach ten percent already for
$x\simeq 10^{-3}$, and twenty percent for $x\simeq 10^{-5}$.
This discussion suggests that at the level of PDF fits we expect 
little differences between fixed-order and resummed at NLO, but more 
significant differences at NNLO.

\begin{figure}[t]
\begin{center}
  \includegraphics[scale=0.95,page=2]{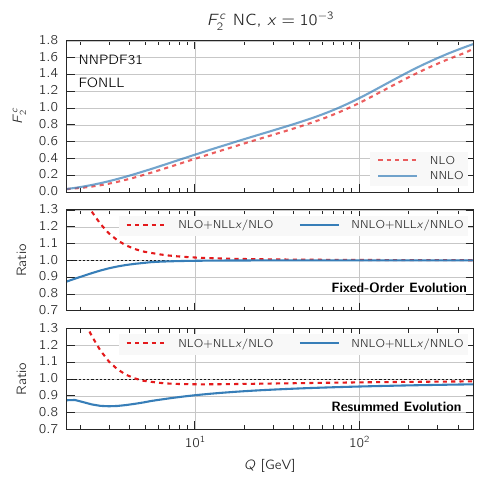}
  \includegraphics[scale=0.95,page=4]{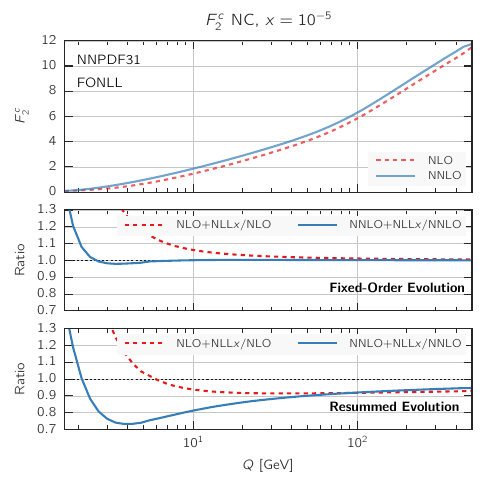}
  \caption{\small Same as Fig.~\ref{fig:resummedSFfixedPDF} for $F_2^c(x,Q)$,
    the charm
    component of the
 structure function $F_2(x,Q)$.
  }
  \label{fig:resummedSFfixedPDF_F2c}
\end{center}
\end{figure}

Next, in Fig.~\ref{fig:resummedSFfixedPDF_F2c} we show the same comparison
as in Fig.~\ref{fig:resummedSFfixedPDF} but now for $F_2^c(x,Q)$, the charm
component of the proton
structure function $F_2(x,Q)$.
By comparing Fig.~\ref{fig:resummedSFfixedPDF} and Fig.~\ref{fig:resummedSFfixedPDF_F2c}
we observe that the impact of small-$x$ resummation for
inclusive and charm structure functions is similar, 
except just above the charm threshold where the effects of the 
resummation in the charm coefficient function can be substantial. 
From this comparison, we see the importance of a careful treatment of 
mass effects close to the charm threshold, since these can change the 
size of the effect of small-$x$ resummation.

\begin{figure}[t]
\begin{center}
  \includegraphics[scale=0.95,page=2]{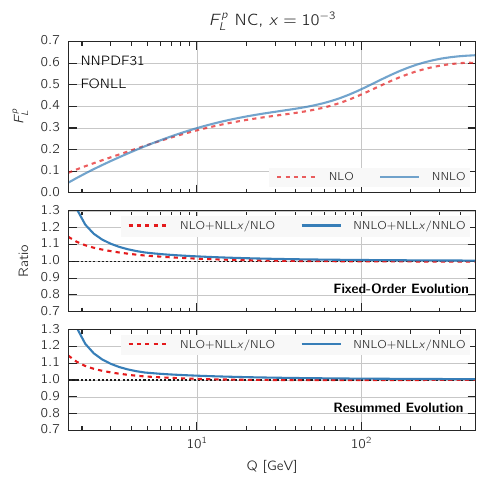}
  \includegraphics[scale=0.95,page=4]{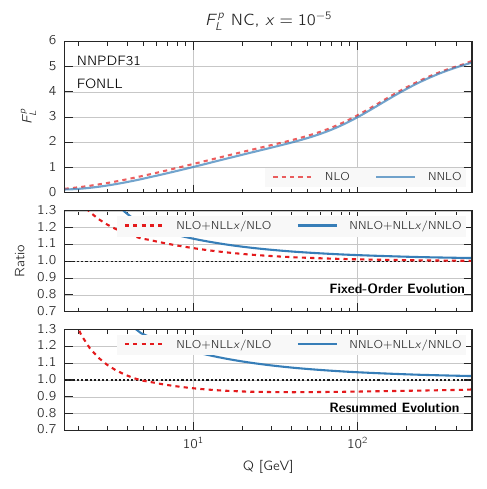}
  \caption{\small Same as Fig.~\ref{fig:resummedSFfixedPDF} for the proton
    longitudinal structure function $F_L(x,Q)$.
  }
  \label{fig:resummedSFfixedPDF_FL}
\end{center}
\end{figure}

Finally, in Fig.~\ref{fig:resummedSFfixedPDF_FL} we show the corresponding comparison
but this time for the longitudinal structure function $F_L(x,Q)$ in neutral current DIS.
Here we find that resummation effects in the coefficient functions only are 
substantially larger than in $F_2$,
and are now larger when matching resummation to NNLO than to NLO.
When resummation is included also in PDF evolution, the overall effect of resummation on $F_L$ is somewhat reduced at NLO,
thus showing some sort of compensation of the effects in PDF evolution and in partonic coefficient functions,
while it is enlarged at NNLO, which now reaches about a $30\%$ deviation at $x=10^{-5}$ at small $Q\sim5$~GeV.
The global pattern is similar to $F_2$, with differences smaller at NLO and more
significant at NNLO, though overall effect is somewhat bigger, consistently with the fact that $F_L$ is singlet dominated.
Given that $F_L$ contributes to the measured
reduced cross-sections $\sigma_{r,\rm NC}$ at high $y$,
which for the HERA kinematics corresponds to small $x$ and $Q^2$, this effect should be relevant for PDF fits.

%

\section{Fitting strategy}
\label{sec:fitsetup}

In this section we discuss the settings of the NNPDF3.1 fits with small-$x$ resummation, as well as of
their fixed-order counterparts, which are used as baseline 
comparisons.
In the following, we will denote these fits as NNPDF3.1sx, each
of them consisting of  $N_{\rm rep}=100$ Monte Carlo replicas.
We briefly present the input dataset, and review the theoretical treatment of the deep-inelastic and hadronic data used in the fit. 
We also discuss the strategy adopted for choosing appropriate kinematic cuts for both DIS
and hadronic processes.

\subsection{Fit settings}
The settings of the fits described in this work
follow closely those of the recent
NNPDF3.1 global analysis~\cite{Ball:2017nwa}.
In particular,
the same input dataset is used, which includes fixed-target~\cite{Arneodo:1996kd,Arneodo:1996qe,bcdms1,bcdms2,Whitlow:1991uw,Onengut:2005kv,
  Goncharov:2001qe,MasonPhD} and HERA~\cite{Abramowicz:2015mha}
DIS inclusive structure functions; charm and bottom cross-sections from HERA~\cite{Abramowicz:1900rp};
fixed-target Drell-Yan (DY)
production~\cite{Webb:2003ps,Webb:2003bj,Towell:2001nh,Moreno:1990sf};
gauge boson and inclusive jet production from the
Tevatron~\cite{Aaltonen:2010zza,Abazov:2007jy,Aaltonen:2008eq,Abazov:2013rja,D0:2014kma};
and electroweak boson production, inclusive jet, $Z$ $p_T$ distributions, and
$t\bar{t}$ total and differential
cross-sections from  ATLAS~\cite{Aad:2011dm,Aad:2013iua,Aad:2011fp,Aad:2011fc,Aad:2013lpa,
ATLAS:2012aa,ATLAS:2011xha,TheATLAScollaboration:2013dja,Aad:2015auj,Aaboud:2016btc,Aad:2014kva,Aaboud:2016pbd,Aad:2015mbv,Aad:2014qja,Aad:2014xaa},
CMS~\cite{Chatrchyan:2012xt,Chatrchyan:2013mza,Chatrchyan:2013tia,Chatrchyan:2013uja,Chatrchyan:2013faa,Chatrchyan:2012bra,Chatrchyan:2012ria,Khachatryan:2016pev,Khachatryan:2015luy,Khachatryan:2016mqs,Khachatryan:2015oqa,Khachatryan:2015oaa} and LHCb~\cite{Aaij:2012vn,Aaij:2012mda,Chatrchyan:2012bja,Aaij:2015gna,Aaij:2015zlq}
at $\sqrt{s}=7$ and 8 TeV.

As in the NNPDF3.1 analysis,
the charm PDF is fitted alongside the light quark PDFs~\cite{Ball:2016neh}, 
rather than being generated entirely from perturbative evolution off gluons and light 
quarks.
As usual in NNPDF, we use heavy quark pole masses~\cite{Ball:2016qeg}, 
and the charm quark pole mass is taken to be $m_c=1.51$~GeV. In all the results
presented here we take $\as(m_Z) = 0.118$.

The initial scale $Q_0$ at which PDFs are parametrized is chosen to be $Q_0=1.64$~GeV, {\it i.e.}\ $Q_0^2=2.69$~GeV$^2$, which is slightly smaller than the initial scale adopted in the NNPDF3.1 analysis, namely $Q_0=1.65$~GeV.
The main motivation for this choice of initial
scale is to be able to include the $Q^2=2.7$ GeV$^2$ bin in the HERA inclusive structure
function data~\cite{Abramowicz:2015mha},
which is expected to be particularly sensitive to the effects of small-$x$ resummation, and that was excluded from NNPDF3.1.
At the same time, the initial scale cannot be too low, to avoid entering a region in which
$\as$ is too large and the numerical reliability of the small-$x$ resummation implemented in the \texttt{HELL} code would be lost.\footnote{In its current implementation, \texttt{HELL 2.0} can only be used for values
  of $Q$ such that $\as(Q)\leq0.35$.}

In this work we have produced fits at fixed-order NLO and NNLO accuracy and corresponding 
resummed fits at NLO+NLL$x$ and NNLO+NLL$x$ accuracy. 
In the resummed fits, small-$x$ resummation is included both in the solution
of the evolution equations and in the deep-inelastic coefficient functions as 
discussed in Sect.~\ref{sec:th-review}. 
Heavy-quark mass effects are accounted for
using the FONLL-B and FONLL-C general-mass scheme~\cite{Forte:2010ta,Ball:2015dpa,Ball:2015tna}
for the NLO and NNLO fits respectively,  modified to include 
small-$x$ resummation effects when
NLO+NLL$x$ and NNLO+NLL$x$ theory is used as previously described. 

Theoretical predictions for the
Drell-Yan fixed-target and the hadron collider (Tevatron and LHC) cross-sections are obtained 
using fixed-order or resummed DGLAP evolution for (N)NLO and (N)NLO+NLL$x$ fits respectively,
but with their partonic cross-sections always evaluated at the corresponding fixed order. 
This approximation is due to the fact that the implementation of hadronic processes in \texttt{HELL} is still work in progress.
To account for this limitation, we cut all data in kinematic regions where small-$x$ corrections are expected to be significant, as explained in Sect.~\ref{sec:kincuts} below.

The settings for the evaluation of the hadronic hard-scattering
matrix elements are the same as in NNPDF3.1,
namely we use fast NLO calculations as generated by {\tt APPLgrid}~\cite{Carli:2010rw} and
{\tt FastNLO}~\cite{Wobisch:2011ij} tables,
which are combined before the fit with the DGLAP evolution kernels by means of
the {\tt APFELgrid} interface~\cite{Bertone:2016lga}.
For the NNLO fits, NNLO/NLO point-by-point $K$-factors are used~\cite{Ball:2017nwa}
using specific codes for each process: we use the code of~\cite{Czakon:2016dgf,Czakon:2015owf} for $t\bar{t}$ differential
distributions~\cite{Czakon:2016olj}; for the $Z$ $p_T$ distributions we use the calculation
of~\cite{Boughezal:2017nla,Boughezal:2015ded};
for Drell-Yan production we use {\tt FEWZ}~\cite{Gavin:2012sy}; while jet cross-sections
are treated using NLO matrix elements supplemented
by scale variation as additional theory systematics.

For comparison purposes, we have also produced DIS-only fits
for which small-$x$ resummation is included in both evolution
and coefficient functions for all data points included in the fit.
That is, in such fit, fully consistent small-$x$ resummed theory is used for the entire dataset.
Moreover, while PDF uncertainties are of course much larger due to the lack of hadronic data, the constraints
from the HERA structure functions are still the dominant ones in the small-$x$ region.
The comparison between the global and DIS-only NNPDF3.1sx fits is discussed in Sect.~\ref{sec:impactPDFs}.

\subsection{Kinematic cuts}
\label{sec:kincuts}

In the NNPDF3.1sx analysis, we apply the same experimental cuts as those of the
NNPDF3.1 fit~\cite{Ball:2017nwa} with two main
differences.
First, as discussed above, the lower $Q^2$ cut is reduced from $Q^2_{\rm min}=3.49$ GeV$^2$ in NNPDF3.1 to
$Q^2_{\rm min}=2.69$ GeV$^2$ here.
Thanks to this lower cut, we can now
include a further bin of the HERA inclusive cross-section data, specifically the one with $Q^2=2.7$ GeV$^2$.
In turn, this allows us to slightly extend the kinematic coverage of the small-$x$ region,
from $x_{\rm min} \simeq 4.6 \times 10^{-5}$ before, down to $x_{\rm min} \simeq 3 \times 10^{-5}$ now.
This lower cut also affects a handful of points at low $Q^2$ (although at larger values of $x$) of other fixed-target DIS experiments, which are therefore also included in the NNPDF3.1sx fits but not in NNPDF3.1.
The cut on $W^2\geq 12.5$ GeV$^2$ remains the same.

Moreover, no additional cuts are applied to the HERA charm
production cross-sections as compared
to the inclusive structure functions.
This was not the case in NNPDF3.1,
where some points at small-$x$ and $Q^2$ were excluded in the NNLO fit,
specifically those with $Q^2\le 8$ GeV$^2$.
We have explicitly
verified that the inclusion of these extra points does not
affect the resulting PDFs, though the $\chi^2$ of the $F_2^c$ data becomes
somewhat worse at NNLO.
Taking into account these two
differences,
from HERA we fit 1162 points for the inclusive structure
functions and 47 points for the $F_2^c$ data,
to be compared with 1145~(1145) and 47~(37) in NNPDF3.1 NLO~(NNLO) respectively.
The number of data points $N_{\rm dat}$ for
each of the DIS experiments included in NNPDF3.1sx
is collected in Table~\ref{tab:ndat1}.

\begin{table}[t]
  \centering
  \small
  \renewcommand{\arraystretch}{1.10}
\begin{tabular}{lc}
Experiment  & {$N_{\rm dat}$} \\
\toprule
NMC & 367 \\
SLAC & 80 \\
BCDMS & 581 \\
CHORUS & 886 \\
NuTeV dimuon & 79 \\
HERA I+II incl. NC & 1081  \\
HERA I+II incl. CC & 81  \\
HERA $\sigma_c^{\rm NC}$ & 47 \\
HERA $F_2^b$ & 29 \\
\bottomrule
Total       & {\bf 3231}    \\
\end{tabular}
\vspace{0.5cm}
\caption{\small The number of data points $N_{\rm dat}$ for
  each of the DIS experiments included in NNPDF3.1sx.}
\label{tab:ndat1}
\end{table}

The second main difference with respect to the NNPDF3.1 kinematic cuts is related to hadronic data.
As already discussed, for hadronic processes small-$x$ resummation
effects are included only in PDF evolution but not in the partonic cross-sections.
Therefore, in order to avoid biasing the fit
results, in the NNPDF3.1sx fits we include only those hadronic data
for which the effects of small-$x$ resummation on the coefficient function can be assumed to be
negligible.

Quantifying the impact of small-$x$ resummation on the
partonic coefficient functions would require the knowledge of such resummation.
Therefore, in order to estimate the region of sensitivity to small-$x$ logarithms, we resort to a more qualitative argument. The foundation of this argument is the observation that in a generic factorization scheme large logarithms appear both in the partonic coefficient functions and in the partonic evolution factors; in general, resummation corrections are thus expected to have a similar size both in the evolution and in the coefficient functions. This naive expectation is indeed confirmed by explicit calculations of 
hadronic resummed cross-sections~\cite{Ball:2007ra,Marzani:2009hu}, where it was found that the most common situation is a partial cancellation between the resummation corrections from evolution and those in the partonic cross-section. It follows that estimates based on the corrections due to resummed evolution alone will probably be conservative, in the sense that they will over-estimate the total resummation correction to the hadronic cross-section. 

In order to implement these cuts, we first introduce a parametrization of the resummation region in the $(x,Q^2)$ plane. Small-$x$ logarithmic corrections should in principle be resummed when $\as(Q^2) \ln 1/x$ approaches unity, since
the fixed-order perturbative expansion then breaks down.
We thus define our kinematic cut to the hadronic data in the NNPDF3.1sx fits such as to 
removes those data points for which
\beq
\label{eq:hcutdef}
\as(Q^2) \ln\frac1x  \geq H_{\rm cut} \, ,
\eeq
where $H_{\rm cut}\lesssim 1$ is a fixed parameter: the smaller $H_{\rm cut}$, the more data are removed. 
Assuming one-loop running for the strong coupling constant (which is enough for our purposes), Eq.~\eqref{eq:hcutdef} can instead be expressed as
\beq
\label{eq:kincutc}
\ln\frac{1}{x}\geq \beta_0 H_{\rm cut} \ln \frac{Q^2}{\Lambda^2} \, ,
\eeq
where $\Lambda\simeq88$~MeV is the QCD Landau pole for $n_f=5$, and $\beta_0\simeq0.61$. Thus the cut is a straight line in the plane of $\ln\frac{1}{x}$ and 
$\ln\frac{Q^2}{\Lambda^2}$, with gradient $\beta_0 H_{\rm cut}$. 

Note that the variable $x$ used in the definition of the cut, Eq.~\eqref{eq:hcutdef}, can in general only be related to the final-state
kinematic variables of hadronic observables by assuming leading-order kinematics. To see how this works in practice, consider for example weak gauge boson production: then $Q^2=M_V^2$, and for fixed $\sqrt{s}$ the cut translates into a maximum rapidity 
\be
y_{\rm max}= \ln\frac{M_V}{\sqrt{s}}+\beta_0H_{\rm cut}\ln\frac{M_V^2}{\Lambda^2}\, .
\ee
Thus in the case of $W$ boson production at $\sqrt{s}=7$ TeV, a cut of the form
of Eq.~(\ref{eq:kincutc}) with $H_{\rm cut}=0.5$ (0.7) would imply that
cross-sections with rapidities above $y_{\rm max} \simeq 0.3~(1.3)$ would be 
excluded from the fit.
In this case, the first (tighter) cut excludes all the LHC gauge boson production data except for a handful of points from the ATLAS and CMS measurements in the most central rapidity region.
The second (looser) cut instead allows to include most of the ATLAS and CMS gauge
boson production data.
However, the LHCb measurements are removed altogether for both values of the cut,
highlighting the sensitivity of forward $W,Z$ production data to the small-$x$ region.

\begin{figure}[t]
  \centering
  \includegraphics[width=0.49\textwidth]{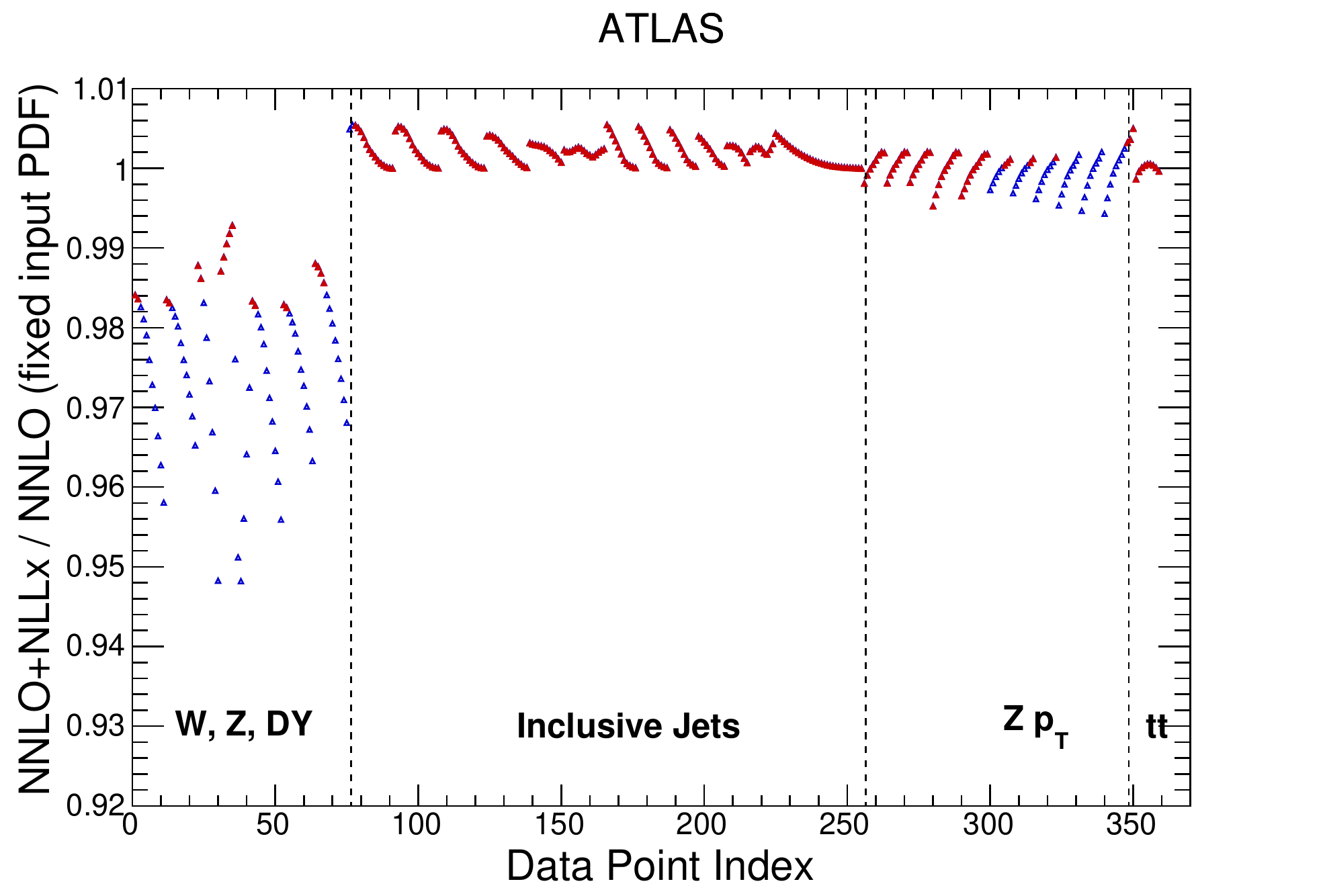}
  \includegraphics[width=0.49\textwidth]{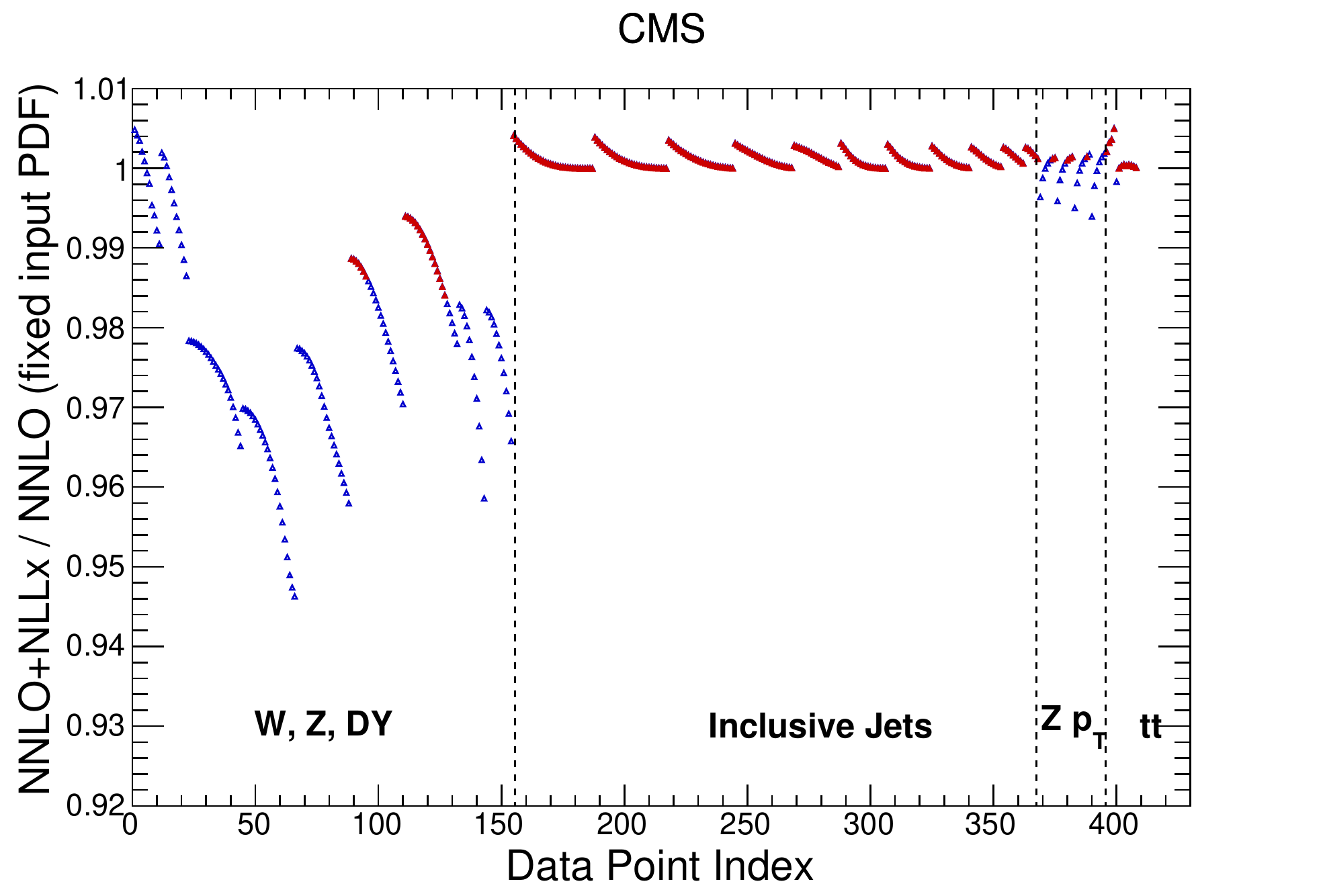}
  \includegraphics[width=0.49\textwidth]{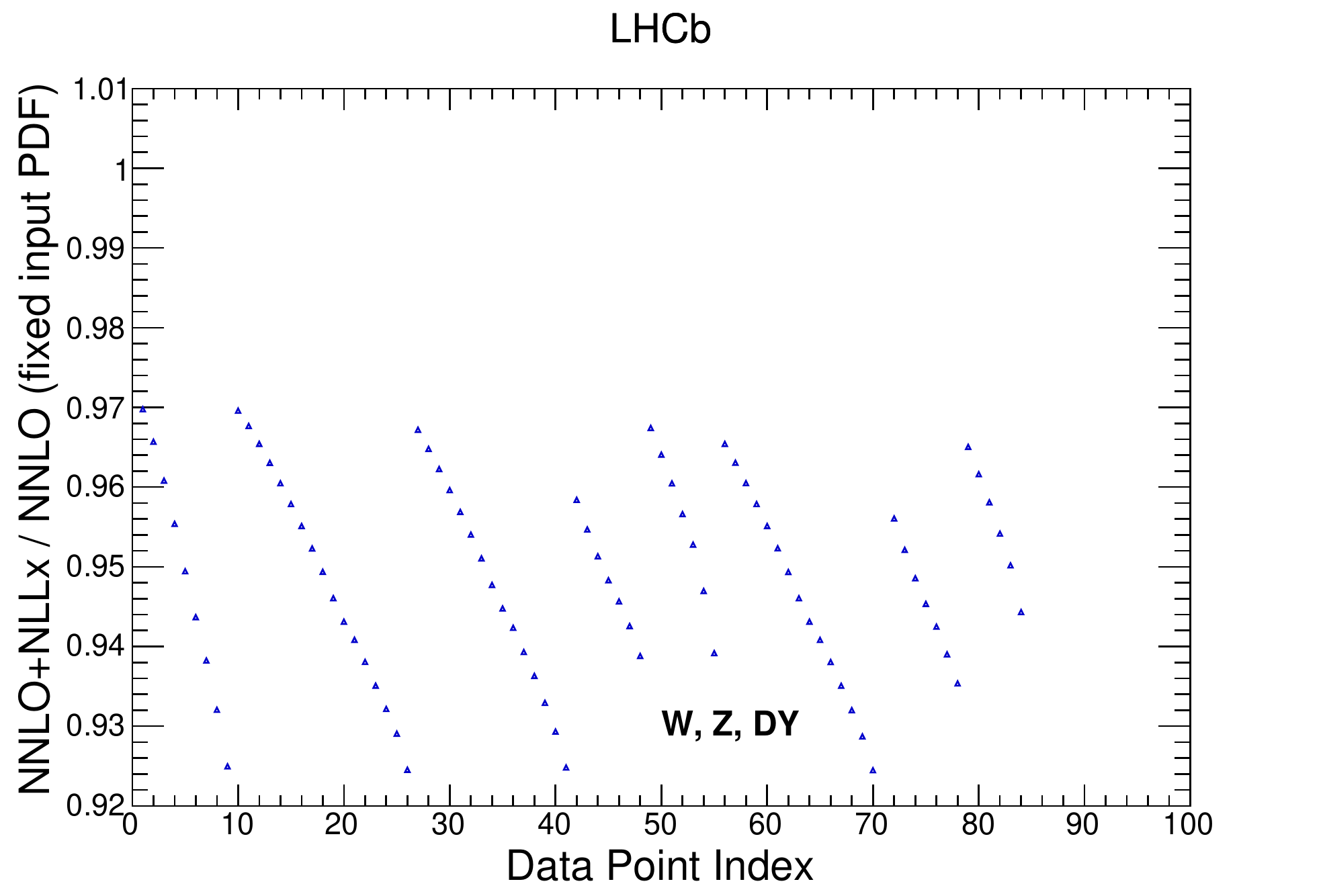}
  \includegraphics[width=0.49\textwidth]{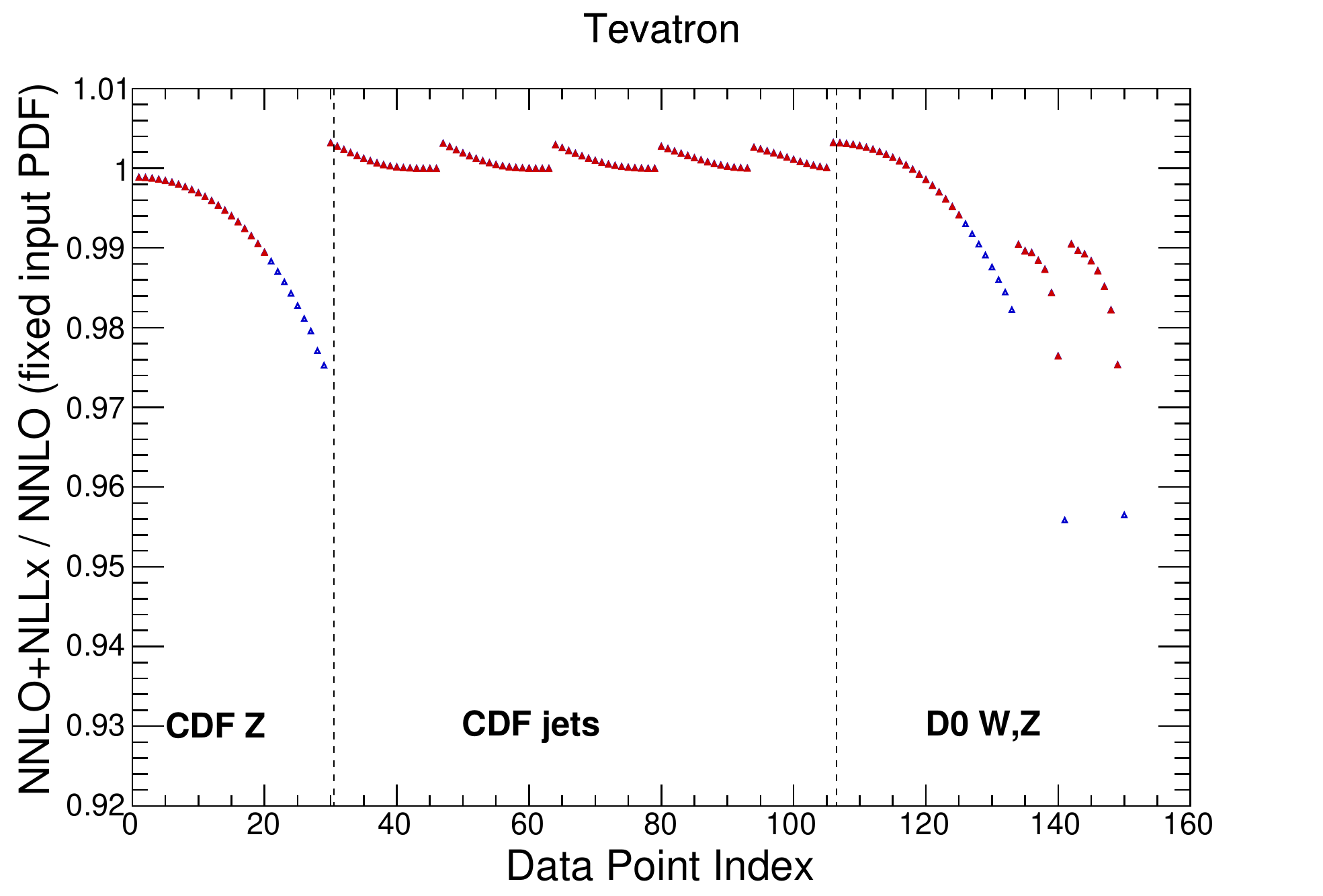}
  \caption{\small The ratio of hadronic cross-sections included
    NNPDF3.1 computed using a fixed input
    PDF at $Q_0=1.65$ GeV (in this case NNPDF3.1 NNLO) using either NNLO+NLL$x$ or
    NNLO theory for PDF evolution, always with NNLO partonic cross-sections.
    We show the results for ATLAS, CMS, LHCb, and the Tevatron, indicating the
    division of each experiment into families of processes.
    The empty blue triangles indicate those data points that
    are excluded from the NNPDF3.1sx fits with the default
    cut $H_{\rm cut}=0.6$, while the filled red ones indicate
    the points that satisfy the condition Eq.~(\ref{eq:kincutc}).
  }
  \label{fig:ratioXsec}
\end{figure}

It remains to determine the optimal value of $H_{\rm cut}$, 
in a way that minimizes at the same
time the amount of information lost from the dataset
reduction, but also the possible theoretical bias due to the missing
small-$x$ resummed coefficient functions.
In this work we will present results with three different values,
namely $H_{\rm cut}=0.5, 0.6$ and 0.7.
In Sect.~\ref{sec:results} we will motivate the choice of $H_{\rm cut}=0.6$ as our default value,
and show explicitly how the main findings on this work are independent of the specific
value of $H_{\rm cut}$ adopted.

Here we attempt to provide
an {\it a priori} argument to justify our choice by estimating the size of the resummation corrections through a comparison of the results obtained with fixed order and resummed parton evolution. Specifically, we take a fixed input 
PDF set (NNPDF3.1 NNLO) at $Q_0=1.65$~GeV
and evolve it using either NNLO or NNLO+NLL$x$ theory,
and then compute the convolution with fixed-order partonic coefficient 
functions.
The comparison is represented in Fig.~\ref{fig:ratioXsec},
where we show the ratio of hadronic cross-sections computed using NNLO+NLL$x$ evolution
over those computed using NNLO evolution.
We show the results for ATLAS, CMS, LHCb, and the Tevatron data
points included in NNPDF3.1, indicating the
division of each experiment into families of processes.
From this comparison, we see that 
the effects of small-$x$ resummation are likely to be significant only for the
$W$ and $Z$ Drell-Yan data, where they could be as large as up to $\sim 5\%$ for ATLAS and CMS,
and up to  $\sim 8\%$ for the forward LHCb measurements, while they are most likely negligible for
all other collider processes, such as jets, the $Z$ $p_T$,
and top quark pair production.
Given that the collider DY data have rather small experimental uncertainties, 
of the order of a few percent or even smaller, we should ensure that we cut
data where the effects
of small-$x$ resummation could be larger than $\sim 2\%$ (to be conservative).
We see from Fig.~\ref{fig:ratioXsec} that this is indeed
achieved with the default value of $H_{\rm cut}=0.6$:
for the included points, differences are always smaller than this threshold.

\begin{figure}[t]
\begin{center}
  \includegraphics[scale=0.63]{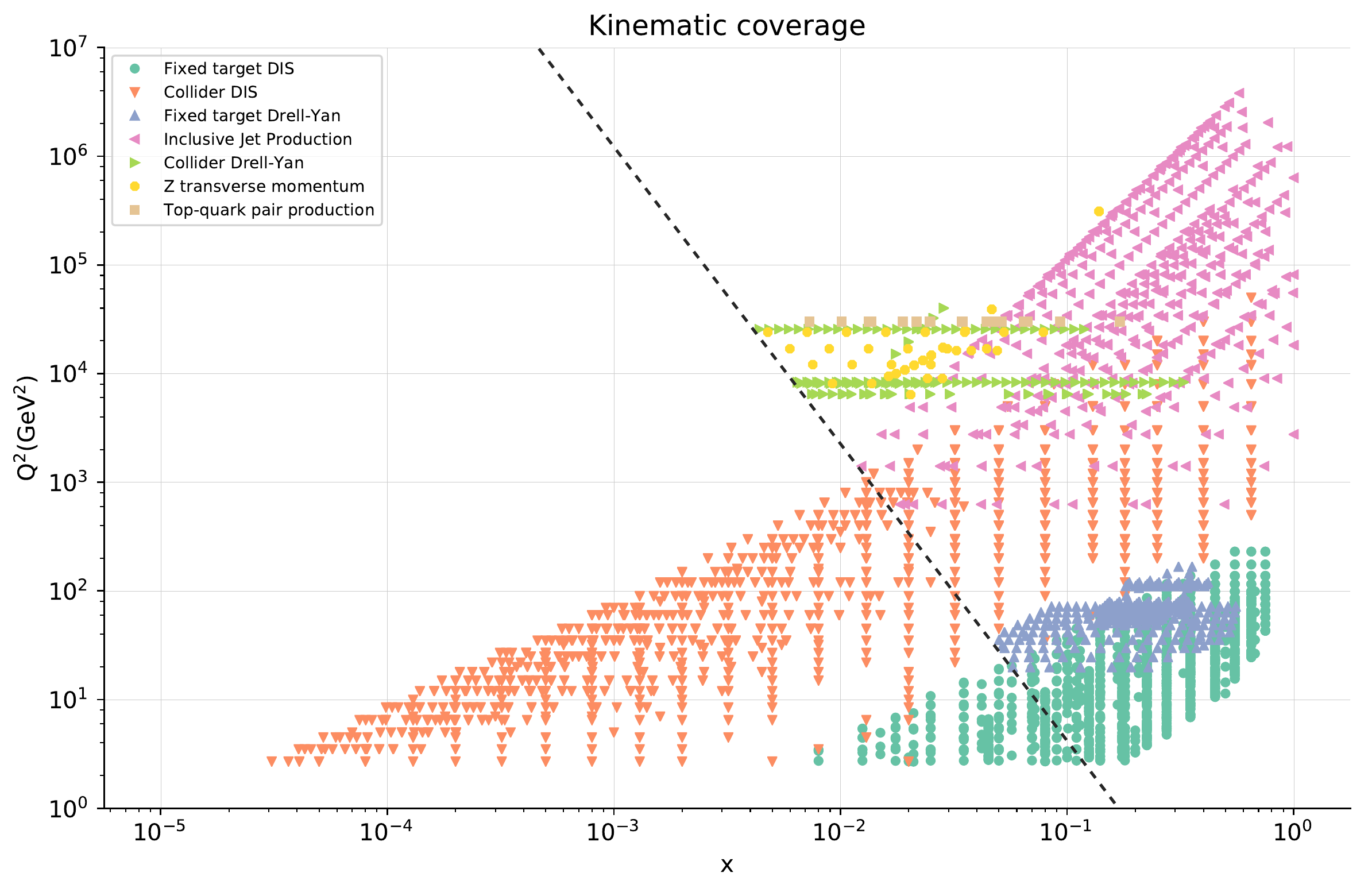}
  \caption{\small The kinematic coverage in the
    $(x,Q^2)$ plane of the data included in the NNPDF3.1sx fit
    with the default value of the kinematic cut to the hadronic data,
    $H_{\rm cut}=0.6$.
    The diagonal line indicates the value of the cut Eq.~\eqref{eq:kincutc},
    below which the hadronic data is excluded from the fit.
    For hadronic processes, the LO kinematics have been used to
    determine the  $(x,Q^2)$ values associated to each data bin.}
\label{fig:kinplotcuts}
\end{center}
\end{figure}

To summarize this discussion about the kinematic cuts
in the NNPDF3.1sx fits, we show in Fig.~\ref{fig:kinplotcuts}  the kinematic coverage in the
$(x,Q^2)$ plane of the data included in the present analysis,
for the default value $H_{\rm cut}=0.6$ of the cut to the hadronic data.
As mentioned above, for hadronic processes the LO kinematics have been used to
determine the values of $x$ and $Q^2$ associated to each data bin.
The diagonal line indicates the region below which the cut
defined in Eq.~(\ref{eq:kincutc}) removes hadronic data.
As a consequence of the kinematic cuts, the hadronic dataset is restricted to the large-$Q^2$
and medium- and large-$x$ region.

In Table~\ref{tab:ndat} we show the number of data points for the hadronic
data in the NNPDF3.1sx NNLO fits for with $H_{\rm cut}=0.5,0.6$ and 0.7.
The number in brackets corresponds to the values for the NLO fits, since
the kinematic cuts of the NNPDF3.1 fits~\cite{Ball:2017nwa} are slightly different
at NLO and at NNLO.
The main effect of the $H_{\rm cut}$ is on the Drell-Yan prediction
measurements from ATLAS and CMS, which in turn affects the quark and antiquark
flavour separation, and the $Z$ $p_T$ distributions, which provide information
on the gluon.
On the other hand, the inclusive jet and top-quark
pair production data, which are mostly sensitive to the
large-$x$ region, are essentially unaffected by the cut.
For completeness, we also provide the values
of $N_{\rm dat}$
when no cut is applied at all ($H_{\rm cut}=\infty$).
In the latter case, the fit also includes 85 (93) LHCb experimental points at NNLO (NLO).

\FloatBarrier
\begin{table}[htp]
\centering
  \footnotesize
  \renewcommand{\arraystretch}{1.10}
\begin{tabular}{l C{1.6cm}C{1.6cm}C{1.6cm}C{1.9cm}}
\multirow{2}{*}{Experiment}  & \multicolumn{4}{c}{$N_{\rm dat}$} \\
   &  $H_{\rm cut}=0.5$  & $H_{\rm cut}=0.6$  & $H_{\rm cut}=0.7$ & $H_{\rm cut}=\infty$ \\
\toprule
DY E866 $\sigma^d_{\rm DY}/\sigma^p_{\rm DY}$    &  11 & 13           & 14   &  15  \\
DY E886 $\sigma^p$        & 55 &     75        & 87 & 89 \\
DY E605  $\sigma^p$     &  85  &  85 &  85  & 85 \\
CDF $Z$ rap    &   12  &   20        &  29 & 29 \\
CDF Run II $k_t$ jets   &  76   &  76        & 76 & 76   \\
D0 $Z$ rap    &    12  &   20       &  28  &  28 \\
D0 $W\to e\nu$  asy    &  4    &  7 (8)        & 8 (12)  & 8~(13)\\
D0 $W\to \mu\nu$  asy    &  4   &   8 (9)      & 9 (10)  & 9~(10)\\
\midrule
ATLAS total   &  {\bf 230}   &   {\bf  258}      & {\bf 294}   & {\bf 354} \\
ATLAS $W,Z$ 7 TeV 2010    & 0   &   6         & 16 &30  \\
ATLAS HM DY 7 TeV    &   5   &   5       & 5  & 5\\
ATLAS $W,Z$ 7 TeV 2011    & 0   &  8        & 20 & 34 \\
ATLAS jets 2010 7 TeV     & 81    &  86        & 89 & 90  \\
ATLAS jets 2.76 TeV     &    56  &   59       &  59 & 59 \\
ATLAS jets 2011 7 TeV     &   31  &   31       &  31 & 31 \\
ATLAS $Z$ $p_T$ 8 TeV $(p_T^{ll},M_{ll})$   &  44   & 44          & 44& 44 \\
ATLAS $Z$ $p_T$ 8 TeV $(p_T^{ll},y_{ll})$ &    0  &   6        & 17 & 48 \\
ATLAS $\sigma_{tt}^{tot}$     &    3   &  3         &  3 & 3 \\
ATLAS $t\bar{t}$ rap    &  10    &    10       &  10  & 10 \\
\midrule
CMS total    &  {\bf   234}     &  {\bf 259}        &{\bf  316} & {\bf  409~(387)} \\
CMS $W$ asy 840 pb     & 0  & 0  & 7  & 11\\
CMS $W$ asy 4.7 fb     & 0  & 0  & 7  & 11 \\
CMS $W$ rap 8 TeV   &  0  & 0  & 12   & 22 \\
CMS Drell-Yan 2D 2011    &  8   &   24        & 44 & 110~(88) \\
CMS jets 7 TeV 2011     &   133   & 133          &  133& 133 \\
CMS jets 2.76 TeV    &   81      & 81         &   81 & 81 \\
CMS $Z$ $p_T$ 8 TeV $(p_T^{ll},y_{ll})$    & 3     &   10        & 19 & 28 \\
CMS $\sigma_{tt}^{tot}$     &    3    &  3        &  3 & 3\\
CMS $t\bar{t}$ rap    &    6   &   8        & 10  & 10\\
\midrule
LHCb total   &  {\bf   0}     &  {\bf 0}        &{\bf  0} & {\bf  85~(93)}\\ 
LHCb $Z$ rapidity 940 pb &    0     &   0       & 0 &  9\\ 
LHCb $Z\rightarrow ee$ rapidity 2 fb &     0     &   0       &   0 &   17\\ 
LHCb $W, Z \rightarrow \mu$ 7 TeV &    0     &   0       &   0 &   29 (33)\\ 
LHCb $ W, Z \rightarrow \mu$ 8 TeV &     0     &   0       &   0 &   30 (34)\\ 
\bottomrule
Total       & {\bf 723}   & {\bf  821 (823)}       & {\bf 946 (951)}  & {\bf 1187 (1179)}  \\
\end{tabular}
\vspace{0.5cm}
\caption{\small The number of data points $N_{\rm dat}$ for each of the hadronic
  experiments included in the NNLO NNPDF3.1sx global fits for different values of
  $H_{\rm cut}=0.5,0.6$ and 0.7, with the default value being $H_{\rm cut}=0.6$.
  The number in brackets corresponds to the values for the NLO fits, if different
  from the NNLO value.
  For completeness, we also show $N_{\rm dat}$ when the $H_{\rm cut}$ is not applied
  ($H_{\rm cut}=\infty$).
  The last row indicates the total number of hadronic data points included in the fit for each value of the cut.}
\label{tab:ndat}
\end{table}


\section{\boldmath Parton distributions with small-$x$ resummation}
\label{sec:results}

In this section we present the main results of this work, namely
the NNPDF3.1sx fits including the effects of
small-$x$ resummation.
We will present first the DIS-only fits and then the global fits, based on the dataset 
described in Sect.~\ref{sec:fitsetup}.
Unless otherwise specified, for the global fits we will use the default
cut $H_{\rm cut}=0.6$ for the hadronic data.

In the following, we will first discuss the DIS-only fits,
showing how small-$x$ resummation improves the fit quality and
affects the shape of the PDFs.
We then move to the global fits, and compare them to the DIS-only ones.
We find that the qualitative
results are similar, though PDF uncertainties are reduced.
We show the impact of resummation on the PDFs, and study the dependence on the cut used to remove the hadronic data potentially sensitive
to small-$x$ logarithms and for which we do not yet include resummation.
We show how our default choice for $H_{\rm cut}$ does not bias the fit,
and still allows us to determine PDFs whose uncertainties
are competitive with those of NNPDF3.1. 
We discuss in detail the role of the additional low-$Q^2$ HERA bin that we
include in this fit for the first time,
and how small-$x$ resummed theory is able to fit it satisfactorily.

We will further inspect the improved description of the HERA data in Sect.~\ref{sec:diagnosis}, where we will perform a number of diagnostic studies aimed at quantifying the onset of BFKL dynamics in the inclusive HERA structure functions.

\subsection{DIS-only fits}\label{sec:disonlyfits}

Let us start our discussion by considering the DIS-only fits, in which we include all the DIS data from
fixed-target and collider experiments described in Sect.~\ref{sec:fitsetup}.
For all these data, we have a complete theoretical description at resummed level,
thus allowing us to perform a fully consistent small-$x$
resummed fit.
First of all, in Table~\ref{tab:chi2tab_dis} we collect the $\chi^2/N_{\rm dat}$ values  
for the total and individual datasets computed with the PDFs fitted using NLO, NLO+NLL$x$,
NNLO and NNLO+NLL$x$ theory.
The $\chi^2$ values are computed using the experimental
definition of the covariance matrix, while
the $t_0$ definition~\cite{Ball:2009qv} was instead used during the fits,
as customary in the NNPDF analyses.
 In addition, we also show the difference in $\chi^2$ between the resummed
 and fixed-order results,
 \be
 \label{eq:deltachi2def}
 \Delta\chi^2_{\rm (N)NLO} \equiv \chi^2_{{\rm (N)NLO+NLL}x}-\chi^2_{\rm (N)NLO} \, ,
\ee
 which is useful to gauge how statistically significant are the differences
  between the fixed-order and resummed results for each experiment.

\begin{table}[t]
\centering
   \footnotesize
   \renewcommand{\arraystretch}{1.10}
   \begin{tabular}{l C{1.7cm}C{1.7cm}C{1.3cm}|C{1.7cm}C{2.2cm}C{1.3cm}}
     & \multicolumn{2}{c}{$\chi^2/N_{\rm dat}$}  & $\Delta\chi^2$  & \multicolumn{2}{c}{$\chi^2/N_{\rm dat}$}  & $\Delta\chi^2$ \\
 & NLO &  NLO+NLL$x$ &  &   NNLO & NNLO+NLL$x$   &  \\
 \toprule
NMC    &    1.31   &   1.32   &  +5  &   1.31   &   1.32   & +4 \\    
SLAC    &    1.25   &   1.28   &  +2  &  1.12   &   1.02    & $-8$ \\    
BCDMS    &    1.15   &   1.16   & +7   &  1.13   &   1.16     &  +14 \\    
CHORUS    &    1.00   &   1.01   &  +9  &   1.00   &   1.03    & +26  \\    
NuTeV dimuon    &    0.66   &   0.56      & $-8$ &      0.80   &   0.75   &   $-4$   \\
\midrule
HERA I+II incl. NC    &   1.13      &  1.13    &  $+6$   &  1.16    & 1.12       & $-47$    \\
HERA I+II incl. CC    &    1.11    &  1.09   &  $-1$   &  1.11    &   1.11      &  -    \\    
HERA $\sigma_c^{\rm NC}$    &    1.44   &   1.35   & $-5$   &   2.45   &   1.24   & $-57$ \\    
HERA $F_2^b$       &    1.06   &   1.14   & +2    &  1.12  &   1.17    & +2  \\    
\bottomrule
    {\bf Total}
    &  \bf  1.113   &\bf   1.119   & \bf +17 & \bf  1.139  & \bf  1.117  & \bf $\bf -70$ \\    
  \end{tabular}
  \vspace{0.5cm}
\caption{\small The values of $\chi^2/N_{\rm dat}$ for the total and the individual datasets included
  in the DIS-only NNPDF3.1sx NLO, NLO+NLL$x$, NNLO and NNLO+NLL$x$
  fits.
  The number of data points $N_{\rm dat}$ for each experiment is indicated in Table~\ref{tab:ndat1}.
  In addition, we also indicate the absolute difference $\Delta\chi^2$ between the resummed
  and fixed-order results, Eq.~\eqref{eq:deltachi2def}.
  We indicate with a dash the case $|\Delta\chi^2| < 0.5$.
}
\label{tab:chi2tab_dis}
\end{table}


We immediately observe that the NNLO+NLL$x$ fit has a
total $\chi^2/N_{\rm dat}$ that
improves markedly with respect to the NNLO result, which instead gives
the highest value of $\chi^2/N_{\rm dat}$.
The total $\chi^2/N_{\rm dat}$ is essentially the same in the NLO,
NLO+NLL$x$, and  NNLO+NLL$x$ fits.
As illustrated by the $\Delta\chi^2$ values of Table~\ref{tab:chi2tab_dis},
the bulk of the difference in the fit quality between the NNLO
and NNLO+NLL$x$ fits arises from the HERA inclusive neutral current 
and charm datasets,
which probe the smallest values of $x$,
and whose $\chi^2/N_{\rm dat}$ decrease from $1.16$ to $1.12$ ($\Delta\chi^2=-47$)
and from $2.45$ to $1.24$ ($\Delta\chi^2=-57$), respectively.

We note that the $\chi^2/N_{\rm dat}$ of the charm dataset is rather high at NNLO.
In fact, the description of the charm data can be rather sensitive
to the details of the heavy quark scheme.
For instance, we can set to zero the $\Delta_{\rm IC}$ term discussed in Sec.~\ref{sec:resDIS},
thus allowing the inclusion of a phenomenological-induced damping factor which has the role of suppressing
formally subleading terms numerically relevant at scales close to the charm threshold
(see~\cite{Forte:2010ta} and~\cite{Ball:2015tna,Ball:2015dpa}).\footnote
{Note that when the charm PDF is fitted, this manipulation is not really legitimate,
as contributions from an ``intrinsic'' component would be suppressed by the damping but may not be subleading.}
When the damping is included, we find that recomputing the $\chi^2/N_{\rm dat}$ of the charm dataset it becomes $1.10$ at NNLO.
On the other hand, the quality of resummed theory is very stable with respect to such a variation, and the $\chi^2/N_{\rm dat}$ of the charm data becomes $1.23$ ($\Delta\chi^2=+6$).
The rather high value of the charm data $\chi^2$ at NNLO with our default settings is mostly driven by a poor description of the low-$x$ and low-$Q^2$ bins.
Indeed, if we restrict our attention to the region which survives
the more conservative cut used in NNPDF3.1 ($Q^2 \ge 8$ GeV$^2$  for the HERA charm data),
we obtain $\chi^2/N_{\rm dat}=1.38$ at NNLO and $1.35$ at NNLO+NLL$x$ ($\Delta\chi^2=-1$) using our default settings.
The low-$Q^2$ region is somewhat affected by how the subleading terms are treated --- ultimately, this choice is driven by phenomenological reasons, and therefore it is possible that by tuning them one may achieve a satisfactory description of the data at NNLO, for instance by mimicking a perturbative behaviour\footnote{As observed in Ref.~\cite{Ball:2017nwa}, the fit quality to the charm data in NNLO global fits improves if the charm is perturbatively generated, but leads to an significant overall deterioration of the global $\chi^2$ with respect to a fit where the charm is independently parametrized.}; however, the same choice may be suboptimal at the resummed level.
Since at NLO(+NLL$x$) and with FONLL-B we achieve a satisfactory description of the charm data for all 47 points both at fixed order and at resummed level, here we shall use the same theory settings of the NNPDF3.1 paper, and interpret the more marked dependence on the subleading terms as a limitation of the fixed-order theory at NNLO.

We further observe that the description of the fixed-target DIS experiments, sensitive to the
medium and small-$x$ region, is not significantly affected by the inclusion of small-$x$ resummation,
giving us confidence that the resummed and matched
predictions reduce to their fixed-order counterpart where they should.
The only exception is the slight decrease in fit quality between the
NNLO and NNLO+NLL$x$ fits for BCDMS and CHORUS ($\Delta\chi^2=+14$ and $+26$, respectively).
As we will show in the next section, most of
these differences go away once the collider
dataset is included in the global fit, stabilizing the large $x$ PDFs.

Another interesting result from Table~\ref{tab:chi2tab_dis} is that
the effect of resummation is instead much less marked at NLO.
Indeed,
the NLO and NLO+NLL$x$ fits have very similar $\chi^2/N_{\rm dat}$: in particular
the $\chi^2$ change of the HERA inclusive (charm) dataset is rather small,
$\Delta\chi^2=+6\,(-5)$.
This is again not surprising, as the whole point of resummation is to cure instabilities in the fixed order perturbative expansion, by removing the large logarithms causing the instability and replacing them with all order results.
Thus the resummation is more important at NNLO than at NLO, and indeed would probably be yet
more important
at the next perturbative order (N$^3$LO).

\begin{figure}[t]
\centering
  \includegraphics[width=0.49\textwidth]{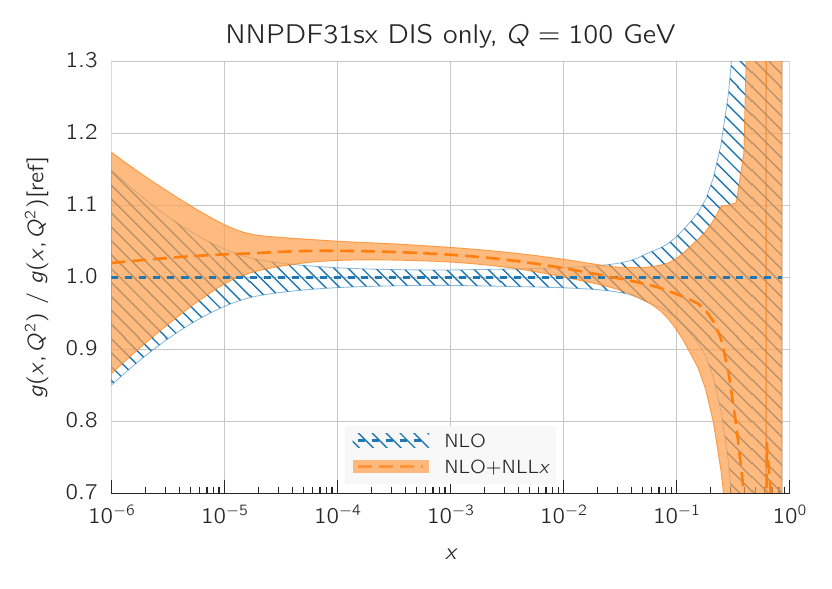}
  \includegraphics[width=0.49\textwidth]{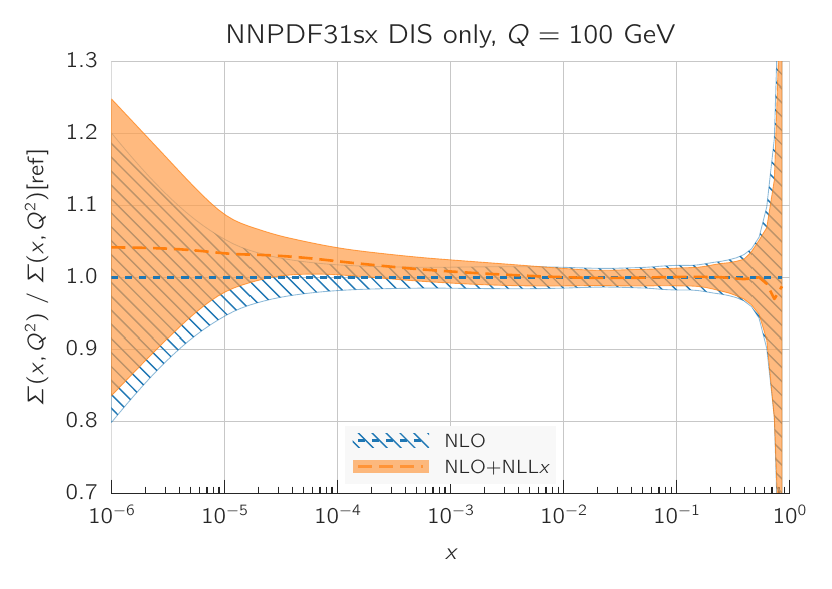}
  \includegraphics[width=0.49\textwidth]{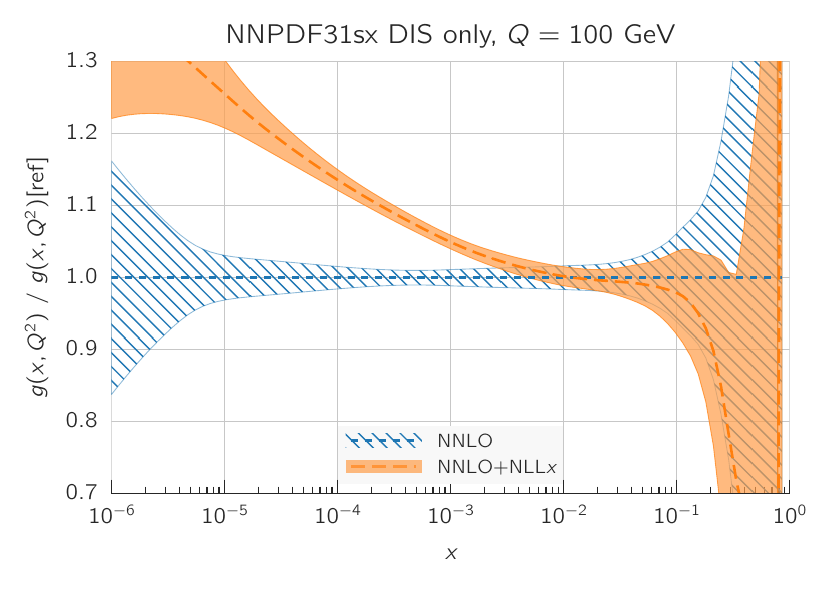}
  \includegraphics[width=0.49\textwidth]{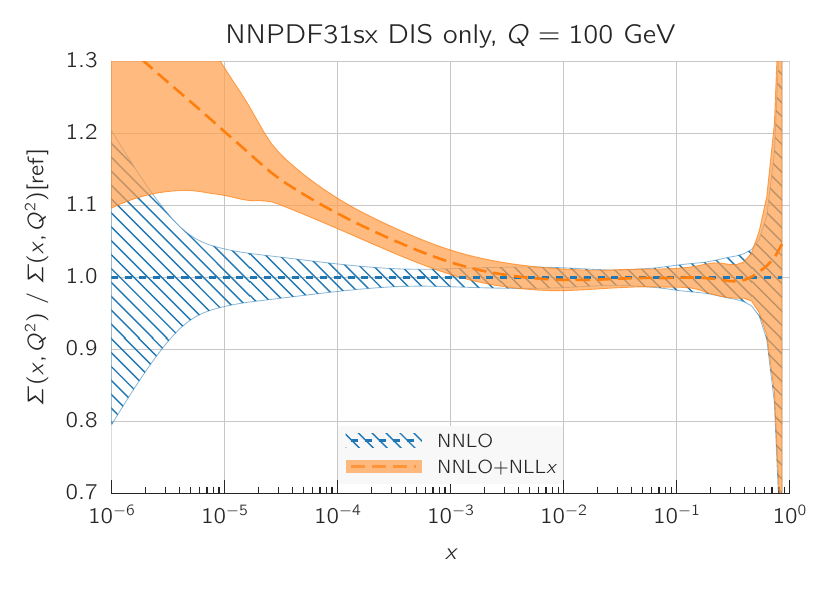}
  \caption{\small Comparison between the gluon (left) and the total quark singlet (right plots)
    from the NLO and NLO+NLL$x$
    (upper plots) and from the NNLO and NNLO+NLL$x$
    DIS-only fits
    (lower plots).
    The comparison is performed at $Q=100$~GeV, normalized to the central value
    of the corresponding fixed-order fit, and the bands indicate
    the 68\% confidence level PDF uncertainties.}
  \label{fig:PDFfit_results_singlet_gluon}
\end{figure}

We can see this result more clearly by considering the resulting 
fitted PDFs and their uncertainties.
In Fig.~\ref{fig:PDFfit_results_singlet_gluon} we show
the ratio between the gluon (left) and the total quark singlet (right)
at $Q=100$~GeV in the NLO+NLL$x$ fit as compared
to the NLO baseline (upper plots) and in the NNLO+NLL$x$ fit
as compared to the NNLO baseline (lower plots).
In this comparison, as well as in subsequent PDF plots,
the bands represent the 68\% confidence level PDF uncertainty.
Consider first the NLO+NLL$x$ fit. Here 
the resummation has a moderate effect:
the resummed gluon PDF is somewhat enhanced between $x=10^{-5}$ and $x=10^{-2}$,
with the PDF uncertainty bands only partially overlapping, whilst 
the shift in central values for the singlet is well within the 
PDF uncertainties.
This remains true down to the smallest values of $x$: even for values
as small as $x\simeq 10^{-6}$ the shifts of the central value of
the singlet and the gluon PDF due to the resummation are less than 10\%.
This is a consequence of the fact that, as discussed in 
Sect.~\ref{sec:th-review}, NLO theory is a reasonably good approximation to the fully resummed result at small-$x$, and any differences are such that can be reabsorbed into small changes in the gluon PDF.

The situation is rather different at NNLO+NLL$x$.
In this case, we see that starting from $x\lsim 10^{-3}$ the
resummed gluons and quarks are systematically higher
than in the baseline NNLO fit, by an amount which ranges
from 10\% for $x\sim 10^{-4}$ up to 20\% for
$x\sim 10^{-5}$ (though note that in this analysis there are
no experimental constraints for $x\lsim 3\times 10^{-5}$). The shifts 
outside central values are significantly outside the PDF uncertainty bands, 
yet result in an improvement in the quality of the fit. 

Note that we are performing these comparisons at the electroweak scale 
$Q\sim100$~GeV, where there are no DIS data and where the effect of resummed evolution is combined with the change of the fitted PDFs at low scales.
This has the advantage of showing that several 
observables at the LHC characterized by electroweak scales
are likely to be sensitive to small-$x$ resummation through the PDFs, particularly when measurements can be performed at high rapidities.
Therefore, for such observables, the use of small-$x$ resummed PDFs (and 
coefficient functions) is probably going to be necessary in order to obtain reliable theoretical predictions.

\begin{figure}[t]
\centering
  \includegraphics[width=0.49\textwidth]{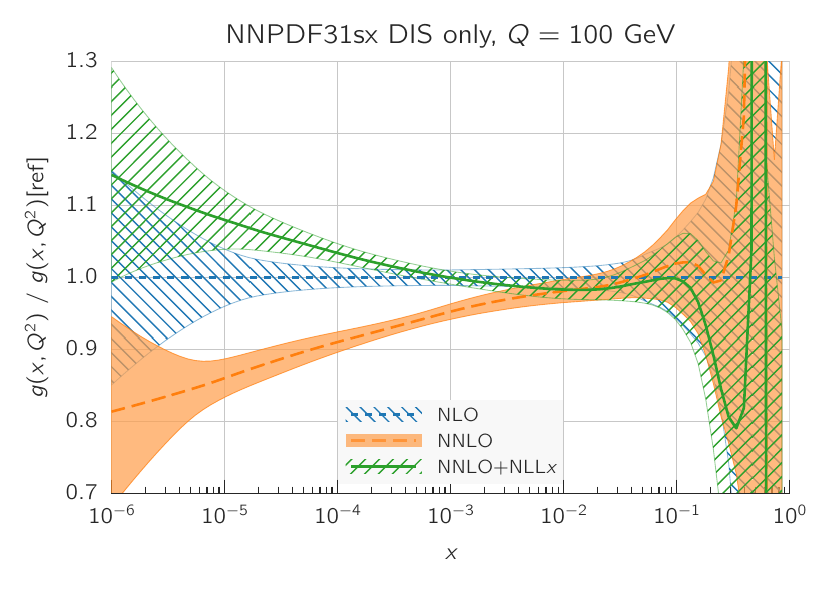}
  \includegraphics[width=0.49\textwidth]{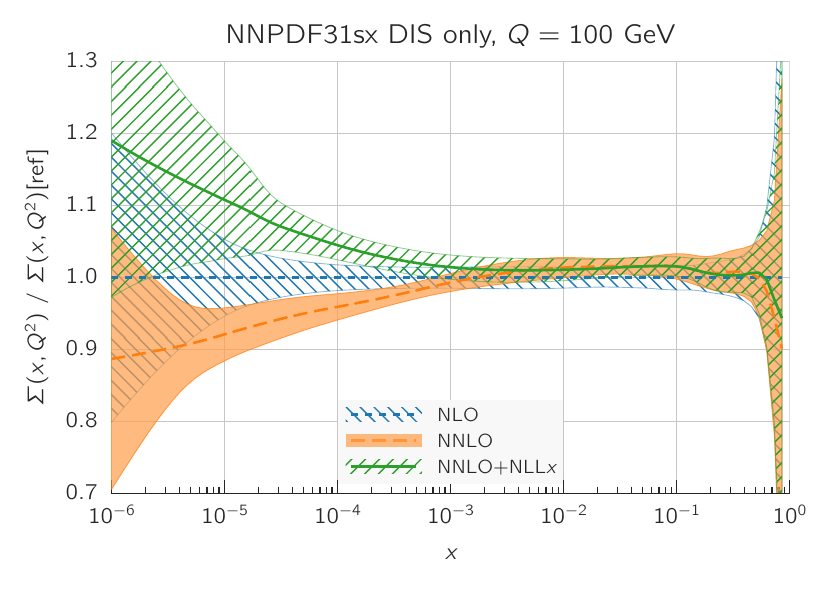}
  \caption{\small Comparison between the gluon (left) and
    quark
    singlet (right plot) PDFs in the NNPDF3.1sx DIS-only fits using
    NLO, NNLO, and  NNLO+NLL$x$ theory at $Q=100$~GeV, normalized
    to the central value of the former.}
  \label{fig:gluon_PDF_small_x_comparison_fo}
\end{figure}

In Fig.~\ref{fig:PDFfit_results_singlet_gluon}
we observed that including resummation leads to a significantly larger
shift in the small-$x$ quark singlet and gluon PDFs
at NNLO than at NLO.
This is so despite the fact that from
the point of view of small-$x$ resummation the information
added is the same in both cases, and that the resummed splitting
and coefficient functions at small $x$ are quite similar
whichever fixed-order calculation they are matched to.
The explanation of this paradoxical result is that fixed-order perturbation theory is unstable at small $x$ due to the small-$x$ logarithms, and while this instability is quite small at NLO, due to accidental zeros in some of the coefficients, it is significant at NNLO, and would probably become very substantial at N$^3$LO. 
To better illustrate
this effect, and the way it is cured by resummation, in 
Fig.~\ref{fig:gluon_PDF_small_x_comparison_fo}
we compare the NLO, NNLO and NNLO+NLL$x$ results for the gluon and
singlet PDFs in the baseline fits at $Q=100$~GeV,
normalized to the NLO prediction.
We find that the NNLO results are systematically below
the NLO ones for $x\leq 10^{-2}$,
and that the net effect of adding NLL$x$ resummation
to the NNLO fit is to bring it more in line with the
NLO (and thus as well with the NLO+NLL$x$) result.
This provides an explanation of our previous observation that NNLO theory 
fits small-$x$ DIS data worse than NLO, while NNLO+NLL$x$ provides the best description of all.

\begin{figure}[t]
\centering
   \includegraphics[width=\textwidth]{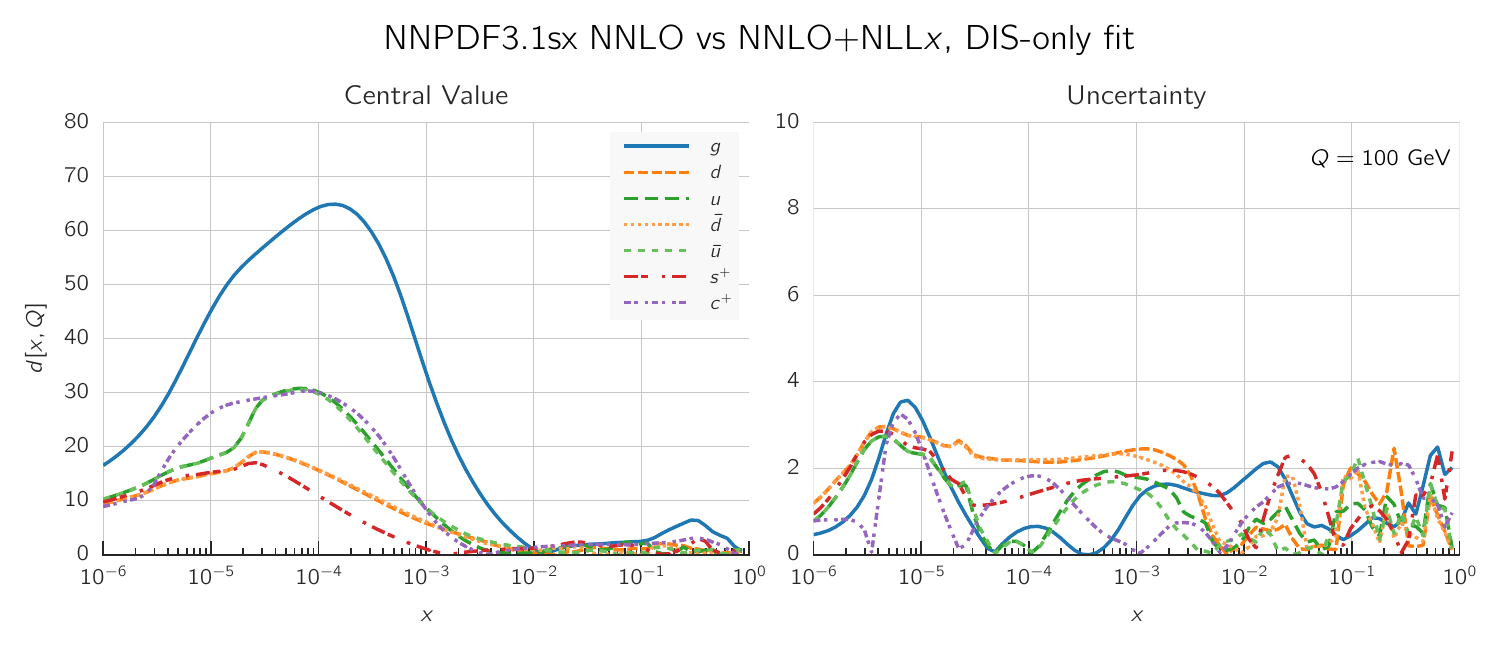}
  \caption{\small The statistical distances between the
    central values (left) and the PDF uncertainties (right plot)
    of the NNPDF3.1sx NNLO and NNLO+NLL$x$ fits at $Q=100$~GeV
    in the flavour basis.}
  \label{fig:distance-sx-dis}
\end{figure}

So far we focused on the gluon and quark singlet, as small-$x$ resummation affects PDFs in the singlet sector.
To quantify the effect of resummation on the PDFs in the physical basis
it is convenient to use a distance estimator, as defined
in Refs.~\cite{Ball:2010de,Ball:2014uwa}. This allows us to represent
in a concise way how two PDF fits differ among themselves, both at the level
of central values and of PDF uncertainties.
In Fig.~\ref{fig:distance-sx-dis} we show these distances between the
central values (left) and the PDF uncertainties (right)
of the NNPDF3.1sx NNLO and NNLO+NLL$x$ fits at $Q=100$~GeV.
Since these fits are based on $N_{\rm rep}=100$ replicas
each, a distance of $d\sim 10$ corresponds to a variation of one-sigma
of the central values or the PDF uncertainties in units of the corresponding
standard deviation.

From the comparison in Fig.~\ref{fig:distance-sx-dis} we see that
the impact of using NNLO+NLL$x$ theory peaks between $x\simeq 10^{-3}$
and $x\simeq 10^{-5}$, where $d\gsim 30$, meaning that the central value
shifts by more than three times the corresponding PDF uncertainty.
The gluon is the most affected PDF, followed by the
charm and then by the light quark PDFs.
Note that the differences are not restricted
to the region of very small-$x$, since for gluons $d\sim 10$ already at $x\simeq 5\cdot 10^{-3}$,
relevant for the production of electroweak scale particles such as $W$ and $Z$ bosons
at the LHC.
On the other hand, the impact of using NNLO+NLL$x$ theory is as expected
small for the PDF uncertainties, since from the experimental
point of view very little new information is being added into the fit.
However, as we will discuss in greater detail in Sect.~\ref{sec:impactlowQ2},
adding small-$x$ resummation has allowed us to lower the minimum value 
of $Q^2$ for the HERA data included in the fits --- which in turn extends to
smaller $x$ the PDF kinematic coverage, thus reducing PDF uncertainties in the very small-$x$ region.

\begin{figure}[t]
\centering
  \includegraphics[width=0.49\textwidth]{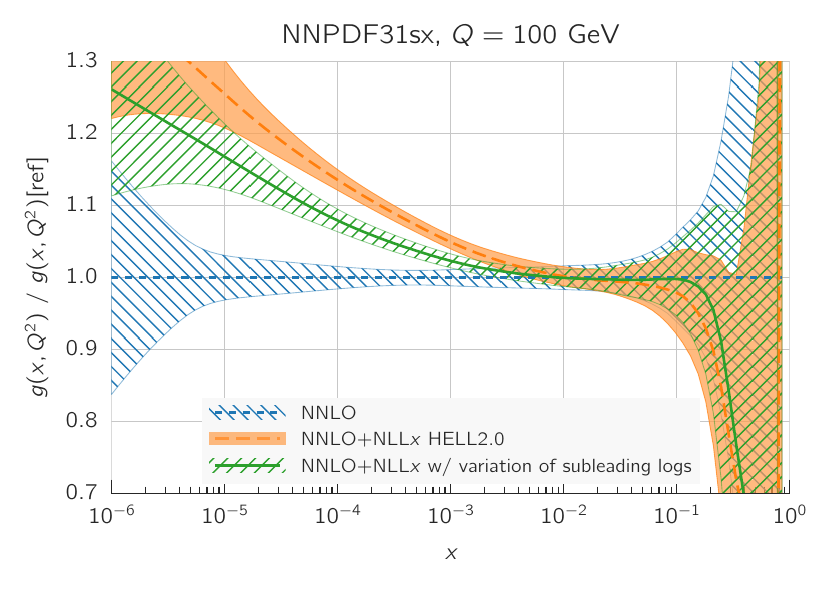}
  \includegraphics[width=0.49\textwidth]{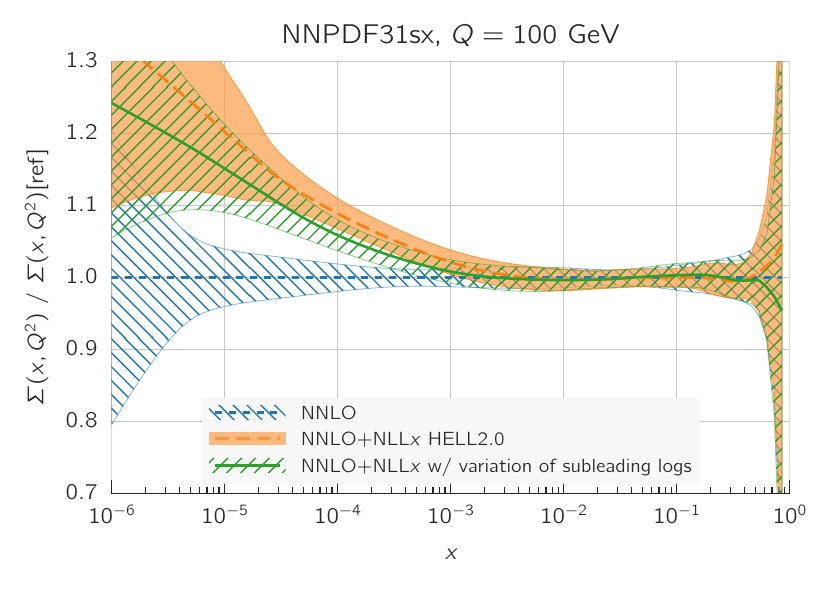}
  \caption{\small Comparison between the gluon (left) and the total quark singlet (right plots)
    from the NNLO and NNLO+NLL$x$ DIS-only fits, including the variant of the resummation which differs by subleading terms,
    as discussed in the text.}
  \label{fig:PDFfit_HELLvar}
\end{figure}
Before moving to the global fits, we want to briefly investigate how our results
are sensitive to unknown subleading logarithmic contributions.
Indeed, the results of Ref.~\cite{Bonvini:2017ogt}
are provided with an uncertainty band aimed at estimating the impact of
subleading (NNLL$x$) contributions not predicted by NLL$x$ resummation.
Ideally, the uncertainty band should be included as a theory uncertainty in the fit procedure;
however, at the moment the inclusion of theory uncertainties in PDF fits is still under study.
Nevertheless, we can investigate the effects of such uncertainties by performing another fit in which we change the resummation
by subleading terms. A simple way to do it in a consistent manner is to vary by subleading terms
the anomalous dimension used for the resummation of coefficient functions and of $P_{qg}$.
As the resummed gluon splitting function depends on the resummed $P_{qg}$,
all splitting functions and coefficient functions are affected by this change.
More specifically, the so-called LL$^\prime$ anomalous dimension used in \texttt{HELL 2.0} (and hence in this work)
is replaced with the full NLL$x$ anomalous dimension, as proposed originally in Refs.~\cite{Altarelli:2008aj}.
The effect of this variation is contained within the uncertainty bands of Ref.~\cite{Bonvini:2017ogt}.

The result of this fit, based on the same DIS-only dataset considered so far and performed at NNLO+NLL$x$ accuracy,
is fully consistent with that obtained with the baseline theory settings.
The fit quality is essentially unaffected, and the $\chi^2$ variations with respect to the numbers in Table~\ref{tab:chi2tab_dis}
are compatible with statistical fluctuations.
Most PDFs are not sensitive to this variation, except the gluon and the quark singlet, which do change a little,
to accommodate the different subleading terms in the splitting functions and coefficient functions.
These PDFs are shown in Fig.~\ref{fig:PDFfit_HELLvar} and compared with the default \texttt{HELL 2.0} result.
In both cases the new PDFs are smaller than our default ones, i.e.\ closer to the NNLO results.
This is mostly due to a harder resummed $P_{qg}$ in the varied resummation, which is therefore closer to its NNLO counterpart, at intermediate values of $x$, than our default resummation. 
For the gluon in particular, the new results are not compatible within the uncertainty bands with our default fit,
highlighting that the PDF uncertainty does not cover the theory uncertainty from missing higher orders.
However, all the qualitative conclusions remain unchanged.

\subsection{Global fits}

We now turn to consider the global fits, based
on the complete
dataset described in Sect.~\ref{sec:kincuts}.
We first show the results of the fits, obtained with the default
cut parameter $H_{\rm cut}=0.6$,
highlighting similarities and differences with respect to the DIS-only fits,
and we discuss the impact of resummation on the PDFs.
We then study the dependence of our results upon variation of the
value of $H_{\rm cut}$.
Finally, we discuss in some detail the description of the low-$Q^2$ HERA
bin which we include in the NNPDF31sx fits.

\subsubsection{Fit results and comparison to the DIS-only fits}
\label{sec:impactPDFs}

\begin{table}[p]
\centering
  \scriptsize
  \renewcommand{\arraystretch}{1.10}
   \begin{tabular}{l C{1.5cm}C{1.5cm}C{1.3cm}|C{1.5cm}C{1.7cm}C{1.3cm}}
     & \multicolumn{2}{c}{$\chi^2/N_{\rm dat}$}  & $\Delta\chi^2$  & \multicolumn{2}{c}{$\chi^2/N_{\rm dat}$}  & $\Delta\chi^2$ \\
     & NLO &  NLO+NLL$x$ &  &   NNLO & NNLO+NLL$x$   &  \\
\toprule
NMC     & 1.35   &   1.35   & +1    & 1.30   &   1.33  &  +9  \\    
SLAC      &  1.16  &   1.14   & $-$1  &   0.92   &   0.95  & +2 \\    
BCDMS      & 1.13   &   1.15   & +12  &   1.18   &   1.18   & +3  \\    
CHORUS     & 1.07   &   1.10   & +20   &  1.07   &   1.07   &  $-$2 \\    
NuTeV dimuon     &  0.90   &   0.84   & $-$5   & 0.97   &   0.88  & $-$7  \\
\midrule
HERA I+II incl. NC    &  1.12      &  1.12   & -2    &   1.17   &   1.11    &  $-62$  \\
HERA I+II incl. CC    &    1.24   &   1.24   & -    &   1.25   &  1.24      &  $-1$   \\
HERA $\sigma_c^{\rm NC}$     &  1.21   &   1.19   & $-$1   &   2.33   &   1.14  & $-$56 \\    
HERA $F_2^b$                &  1.07   &   1.16   & +3   &  1.11   &   1.17    & +2  \\
\midrule
DY E866 $\sigma^d_{\rm DY}/\sigma^p_{\rm DY}$     &  0.37   &   0.37   & -    &  0.32   &   0.30    & -  \\    
DY E886 $\sigma^p$                              &  1.06   &   1.10   & +3    &  1.31   &   1.32  & -  \\    
DY E605  $\sigma^p$                             &  0.89   &   0.92   &  +3   &  1.10   &   1.10  & -  \\    
CDF $Z$ rap                                     &  1.28   &   1.30   &  -   &  1.24   &   1.23   &  -  \\    
CDF Run II $k_t$ jets                           &  0.89   &   0.87   &  $-$2   &  0.85   &   0.80  & $-$4  \\    
D0 $Z$ rap                                      &  0.54   &   0.53   &  -   &  0.54   &   0.53  & -  \\    
D0 $W\to e\nu$  asy                             &  1.45   &   1.47   & -    &  3.00   &   3.10    & +1  \\    
D0 $W\to \mu\nu$  asy                           &  1.46   &   1.42   &  -   & 1.59   &   1.56    &  -  \\
\midrule
ATLAS total                                 &  1.18   &   1.16   &  $-$7    &   0.99   &   0.98    &  $-2$   \\    
ATLAS $W,Z$ 7 TeV 2010                      &  1.52   &   1.47   &  -    &  1.36   &   1.21     & $-1$   \\    
ATLAS HM DY 7 TeV                           &  2.02   &   1.99   &   -   &   1.70   &   1.70     & - \\    
ATLAS $W,Z$ 7 TeV 2011                      &  3.80   &   3.73   &   $-$1   &   1.43   &   1.29  & $-1$  \\    
ATLAS jets 2010 7 TeV                       &  0.92   &   0.87   &  $-$4   &  0.86   &   0.83  &  $-2$   \\    
ATLAS jets 2.76 TeV                         &  1.07   &   0.96   &  $-$6  &  0.96   &   0.96    &  -  \\    
ATLAS jets 2011 7 TeV                       &  1.17   &   1.18   & -  &  1.10   &   1.09  & $-1$   \\    
ATLAS $Z$ $p_T$ 8 TeV $(p_T^{ll},M_{ll})$    &  1.21   &   1.24   & +2    &  0.94   &   0.98  &  +2  \\    
ATLAS $Z$ $p_T$ 8 TeV $(p_T^{ll},y_{ll})$    &  3.89   &   4.26   & +2   &  0.79   &   1.07    & +2  \\    
ATLAS $\sigma_{tt}^{tot}$                    &  2.11   &   2.79   & +2      &  0.85   &   1.15  & +1 \\    
ATLAS $t\bar{t}$ rap                        &  1.48   &   1.49   & -    &  1.61   &   1.64    & -  \\
\midrule
CMS total                               &  0.97   &   0.92   &  $-$13  &  0.86   &   0.85    & $-3$ \\    
CMS Drell-Yan 2D 2011                   &  0.77   &   0.77   & -    & 0.58   &   0.57  &  - \\    
CMS jets 7 TeV 2011                     &  0.88   &   0.82   &  $-$9  & 0.84   &   0.81   & $-3$ \\    
CMS jets 2.76 TeV                       &  1.07   &   0.98   &  $-$7 &  1.00   &   1.00    &  - \\    
CMS $Z$ $p_T$ 8 TeV $(p_T^{ll},y_{ll})$  &  1.49   &  1.57    &  +1   & 0.73   &   0.77   &  - \\    
CMS $\sigma_{tt}^{tot}$                  &  0.74   &   1.28   & +2    & 0.23   &   0.24   &  - \\    
CMS $t\bar{t}$ rap                      &  1.16   &   1.19   &  -   &   1.08   &   1.10   &  - \\    
\bottomrule
{\bf Total }    &    \bf  1.117   &\bf   1.120   & \bf +11   &   \bf  1.130   & \bf  1.100   &  \bf $\bf-121$   \\    
\end{tabular}
\vspace{0.5cm}
\caption{Same as Table~\ref{tab:chi2tab_dis}, now for the global NNPDF3.1sx NLO, NLO+NLL$x$,
  NNLO and NNLO+NLL$x$ fits, corresponding to
  the baseline value of $H_{\rm cut}=0.6$ for the cut to the hadronic data.
  %
}
\label{tab:chi2tab_pertorder}
\end{table}


We start by considering the quality of the global NNPDF3.1sx fits
at NLO, NLO+NLL$x$, NNLO and NNLO+NLL$x$, using the default value of $H_{\rm cut}=0.6$
for the hadronic data cut discussed in Sect.~\ref{sec:kincuts}.
The values of the $\chi^2/N_{\rm dat}$ for the total and the individual datasets
are shown in Table~\ref{tab:chi2tab_pertorder}.
As in the DIS-only case, in this table we also include
the absolute $\chi^2$ difference between the resummed and
fixed-order results, $\Delta\chi^2$  Eq.~(\ref{eq:deltachi2def}).
We observe that the NNPDF3.1sx fit
based on NNLO+NLL$x$ theory leads to
the best overall fit quality, $\chi^2/N_{\rm dat}= 1.100$.
The NNLO fit, on the other hand, has again the highest 
$\chi^2/N_{\rm dat}=1.130$, so that the overall improvement
is $\Delta\chi^2=-121$.
Whilst resummation proves particularly beneficial 
at NNLO, the effect at NLO is very mild; the $\chi^2/N_{\rm dat}\simeq 1.120$
at NLO+NLL$x$ is compatible, within statistical fluctuations,
with the $1.117$ obtained with fixed-order theory, that is, $\Delta\chi^2=+11$.
Note that in the NNPDF3.1 fits the NNLO $\chi^2$ was markedly better than the
NLO one~\cite{Ball:2017nwa}: this is no longer the case here, since the 
high-precision Drell-Yan and $Z$ $p_T$ data points, which are poorly described by NLO theory, are now partly removed by the $H_{\rm cut}$ cut.

The improvement of the $\chi^2$ at NNLO+NLL$x$
is essentially due to the HERA charm and neutral current structure function
data.
On one hand, as we already noticed in the 
DIS-only fits, by using NNLO+NLL$x$ theory one achieves
an improved description of the
precise HERA NC inclusive structure function measurements, whose 
 $\chi^2/N_{\rm dat}$ decreases from $1.17$ in the NNLO fit
to $1.11$ in the NNLO+NLL$x$ fit, $\Delta\chi^2=-62$.
A marked improvement is also achieved for the HERA
charm cross-sections, whose $\chi^2/N_{\rm dat}$ goes down
from 2.33 to 1.14, $\Delta\chi^2=-56$. These two datasets are thus sufficient to explain the overall improvement in the total $\chi^2$.

We also find that NNLO theory describes better 
than the corresponding NLO theory
the ATLAS and CMS measurements,
particularly the recent high-precision data
such as the ATLAS $W,Z$ 2011 rapidity distributions, and the
ATLAS and CMS 8 TeV $Z$ $p_T$ distributions.
Specifically, the $\chi^2/N_{\rm dat}$ total values for ATLAS and CMS
is $1.18\, (1.16)$ and $0.97\, (0.92)$ in the NLO(+NLL$x$) fits respectively, decreasing
to $0.99\, (0.98)$ and $0.86\, (0.85)$ when using NNLO (+NLL$x$) theory.
It is interesting that in all cases the resummed fits are slightly better than their
fixed order counterparts.

Despite the improved description of the large-$Q^2$ collider data with respect to 
the NLO theory, the NNLO fit turns out to have the highest $\chi^2$ of the four
theories, as in the DIS-only case.
The main reason is the poor description of the HERA inclusive
and charm dataset, which 
contain almost one third ($N_{\rm dat}=1209$) of the number of data points
included in the fit ($N_{\rm dat}=3930$).
Moreover, we
observe that the effects of small-$x$ resummation at NNLO are confined to the HERA data;
the differences between the $\chi^2$ values of the (N)NLO and (N)NLO+NLL$x$ fits
for the other datasets are being all rather small.
This is in agreement with the findings of the DIS-only fits,
and with the fact that hadronic data potentially sensitive to small-$x$
effects have been cut.
Specifically, in the NNLO fits there is no other dataset besides 
the HERA inclusive and charm data with $|\Delta\chi^2|\ge 10$.

Comparing the values of the $\chi^2/N_{\rm dat}$ for
the DIS experiments in the global and DIS-only fits,
we notice that once resummation is accounted for, the global fit is if anything slightly better than the DIS-only fit.
In particular for the inclusive HERA data, where
$\chi^2/N_{\rm dat}$ is $1.16~(1.12)$ at NNLO(+NLL$x$) in the DIS-only fits,
we have $\chi^2/N_{\rm dat} = 1.17~(1.11)$ in the global fits, so that $\Delta\chi^2$ decreases from $-47$ to $-62$ in the global fit.
The other significant difference between the global and DIS-only fits
appears in the NuTeV dimuon data, which is fit somewhat less well in the global fit 
(irrespective of resummation) due to the tension
with the LHC data relative to the proton strangeness,
especially with the ATLAS $W,Z$ 2011 rapidity distributions~\cite{Ball:2017nwa}.

\begin{figure}[ht]
\centering
  \includegraphics[width=\textwidth]{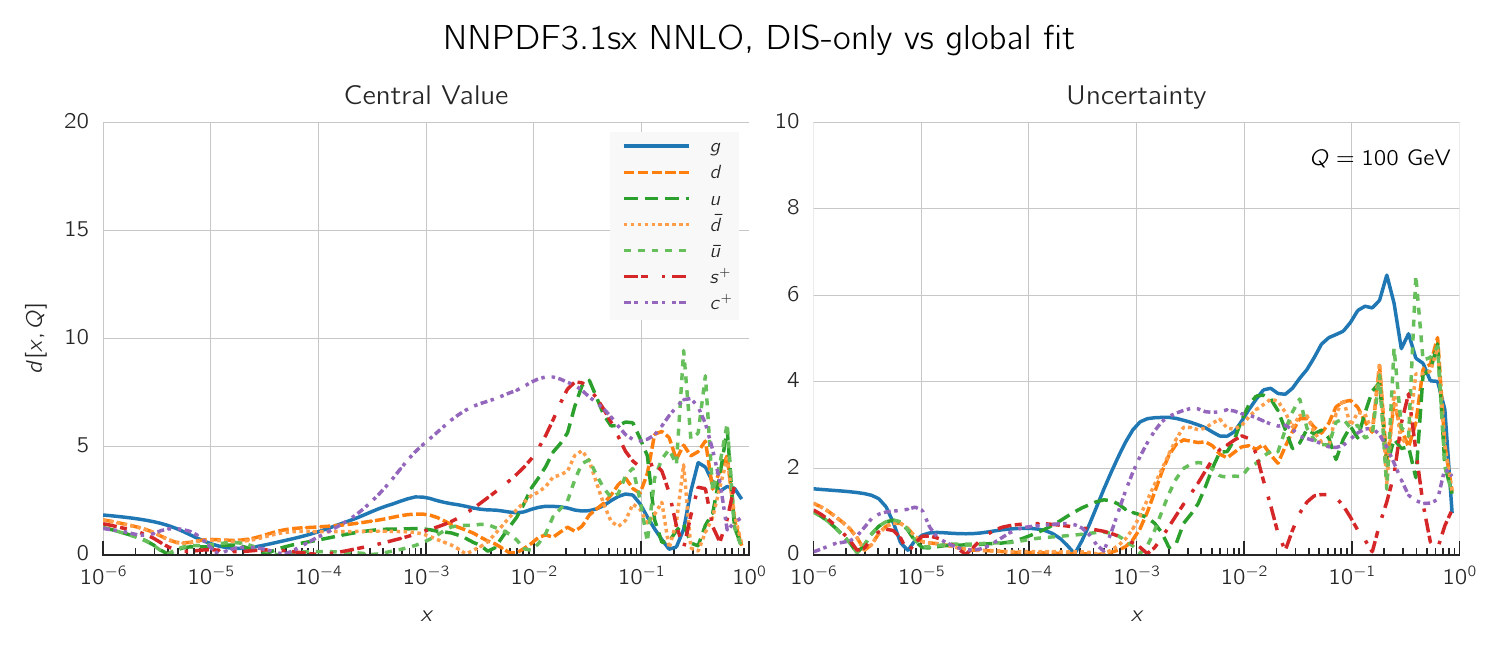}
  \caption{\small Same as Fig.~\ref{fig:distance-sx-dis}
    for the comparison between the fixed-order NNLO NNPDF3.1sx
    DIS-only and global fits.
    Note that the range of the $y$ axis on the left plot
    has been reduced.}
  \label{fig:distance-sx-global-dis}
\end{figure}

We now move to the impact of small-$x$ resummation on the global dataset PDFs.
First, we quantify the differences between the global and the DIS-only fits,
taking as a representative the baseline fixed-order NNLO fit.
We start by showing the distance estimator in Fig.~\ref{fig:distance-sx-global-dis},
both for the central value (left) and the PDF uncertainty (right), at $Q=100$~GeV.
Due to the conservative kinematic cut imposed on the collider observables,
the distances between the global and DIS-only fits are moderate
and localized to the medium and large-$x$ region, while the small-$x$ region is pretty much unchanged.
The PDF flavour which is most affected is the charm PDF, whose distance is
about $10$ for $x\sim10^{-2}$. 
The decrease in PDF uncertainties in the global dataset at medium and 
large-$x$ is clearly visible,
especially for the gluon PDF which is only constrained
in an indirect way by the DIS structure function data.

\begin{figure}[ht]
\centering
   \includegraphics[width=0.49\textwidth]{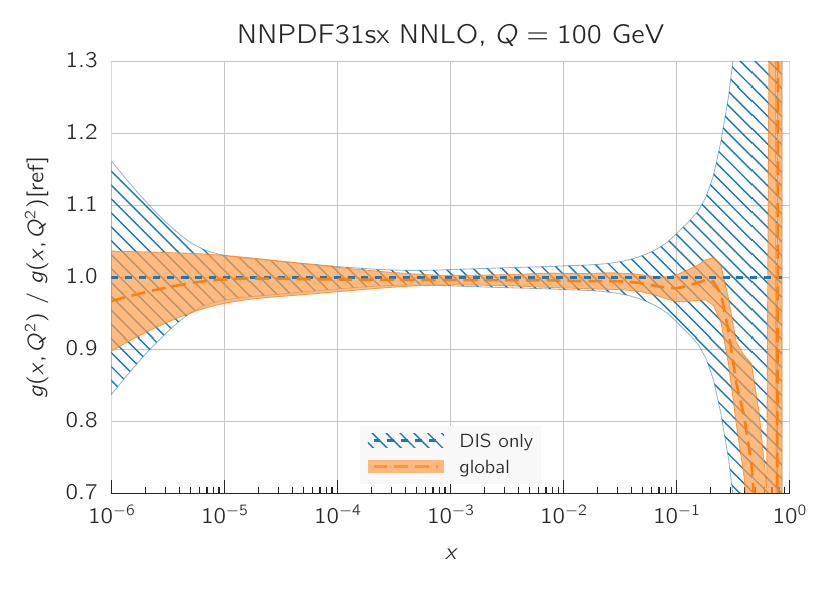}
  \includegraphics[width=0.49\textwidth]{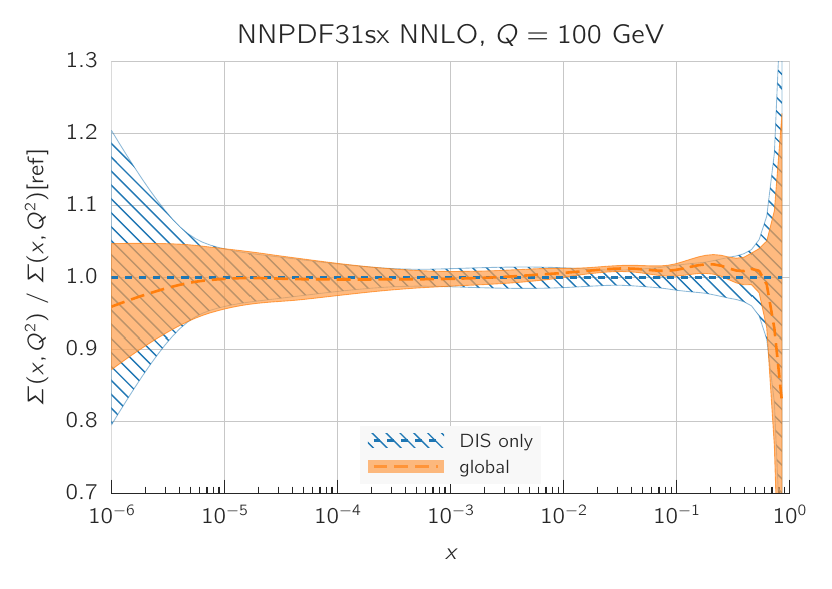}
  \includegraphics[width=0.49\textwidth]{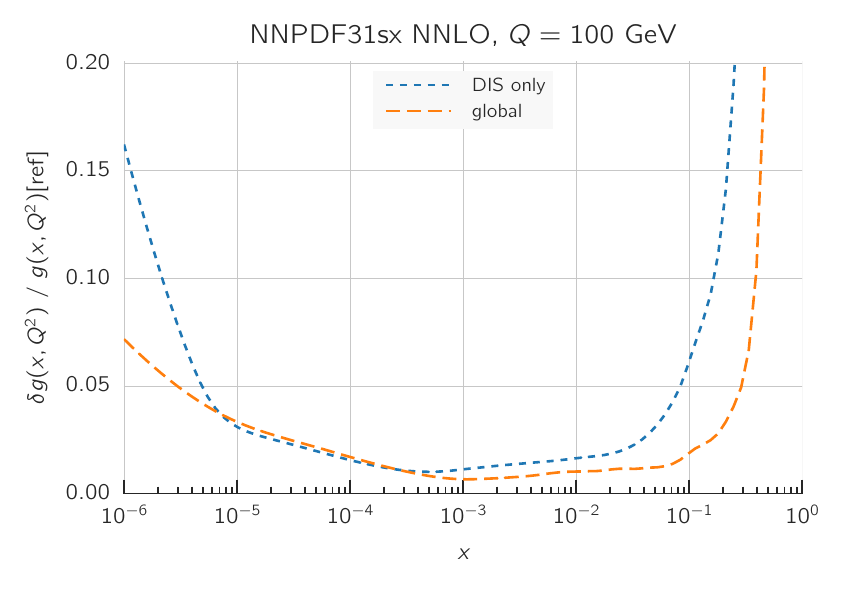}
  \includegraphics[width=0.49\textwidth]{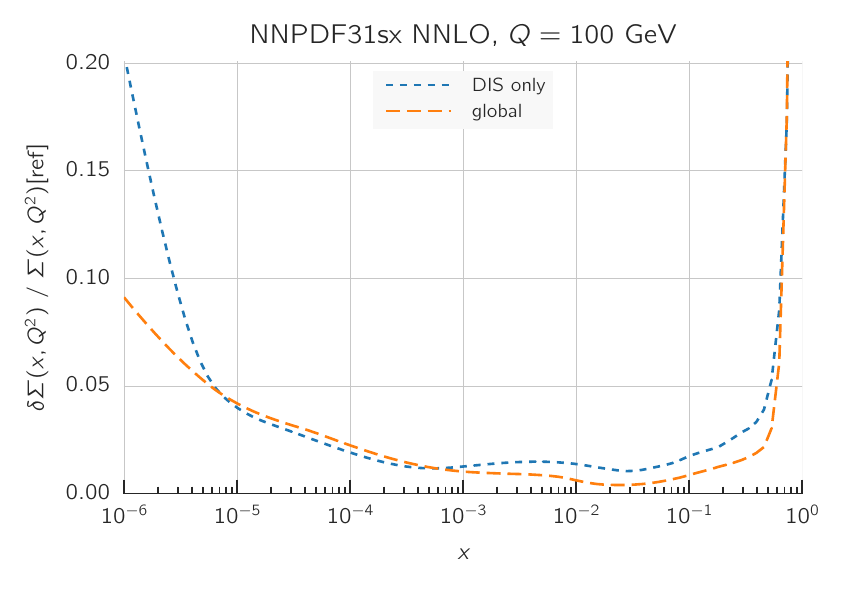}
  \caption{\small Comparison between the gluon (left)
    and the total quark singlet (right) at  $Q=100$~GeV between the
    NNPDF3.1sx NNLO DIS-only and global fits.
    The upper plots show the ratio of global fit results over the DIS-only fit results,
    while the bottom plots compare the relative PDF uncertainty between
    the two fits.}
  \label{fig:pdfcomp-smallx-nnlo-global-vs-dis}
\end{figure}

In Fig.~\ref{fig:pdfcomp-smallx-nnlo-global-vs-dis} we show a direct comparison between the gluon (left)
and the total quark singlet (right) at $Q=100$~GeV between the
NNPDF3.1sx NNLO DIS-only and global fits.
The upper plots show the ratio of global fit results over the DIS-only fit results,
while the bottom plots compare the relative PDF uncertainty between the two fits.
At the level of central values, there is good consistency at the one-sigma level; for $x\lsim 0.1$, the central values of the DIS-only
and global fits are very close to each other.
Concerning PDF uncertainties, the improvement in going from
DIS-only to global is very clear, especially in the large-$x$ region
for the gluon where the DIS-only fit exhibits much larger uncertainties.
The global fit also exhibits somewhat smaller uncertainties in the extrapolation
region for $x \lesssim 10^{-5}$, even if at small-$x$ the direct constraints are essentially the same in the two cases. However, given the large size of PDF uncertainties in this region, the observed differences are consistent with statistical fluctuations.

\subsubsection{Features of the small-$x$ resummed PDFs from the global fit}
The comparison done so far demonstrates that the use of the global dataset
is very beneficial from the point of view of the PDF uncertainties,
while it does not affect the qualitative and quantitative results at small $x$.
Therefore, the global fits will be considered from now on the baseline
NNPDF3.1sx fits, and we will focus on these results for subsequent
applications and studies.
Therefore, before moving forward, it is interesting to analyse the features
of these 
fits in more detail.

\begin{figure}[ht]
\centering
  \includegraphics[width=\textwidth]{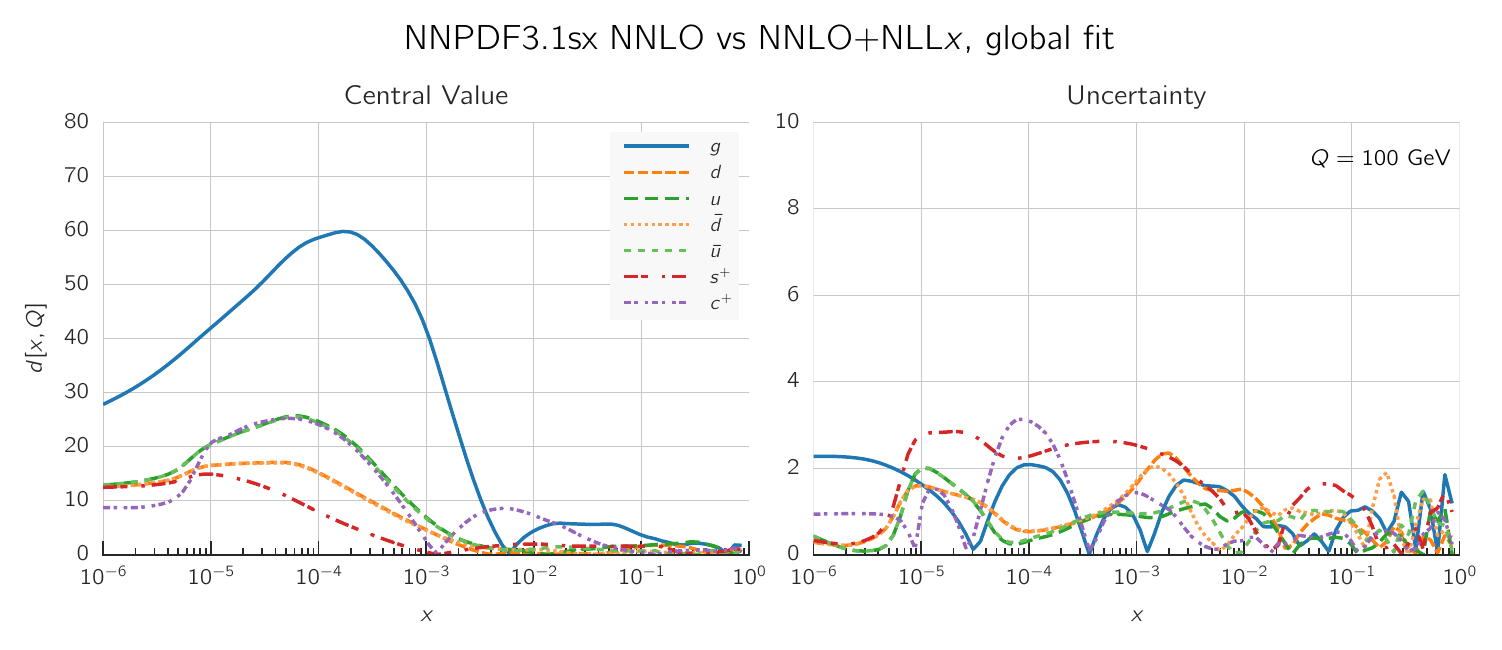}
   \caption{\small Same as Fig.~\ref{fig:distance-sx-dis}
    for the NNPDF3.1sx global fits.}
    \label{fig:distance-sx-global}
\end{figure}

We focus on the results at NNLO and NNLO+NLL$x$, as at NLO
the impact of resummation is less significant (just as in the DIS-only fits)
and also less important from the point of view of applications to the LHC and future high-energy collider physics.
In Fig.~\ref{fig:distance-sx-global} we show the same
distance comparison as in Fig.~\ref{fig:distance-sx-dis} but
now for the NNPDF3.1sx global fits.
By comparing this figure with the corresponding DIS-only case,
we see that in the global fits the qualitative features are the same.
The increased significance of the distances at large $x$ observed in the
global fit as compared to the DIS-only is a direct consequence of the reduced 
PDF uncertainties in the global fit, rather than to a shift in the 
central values. 

\begin{figure}[ht]
\centering
  \includegraphics[width=0.49\textwidth]{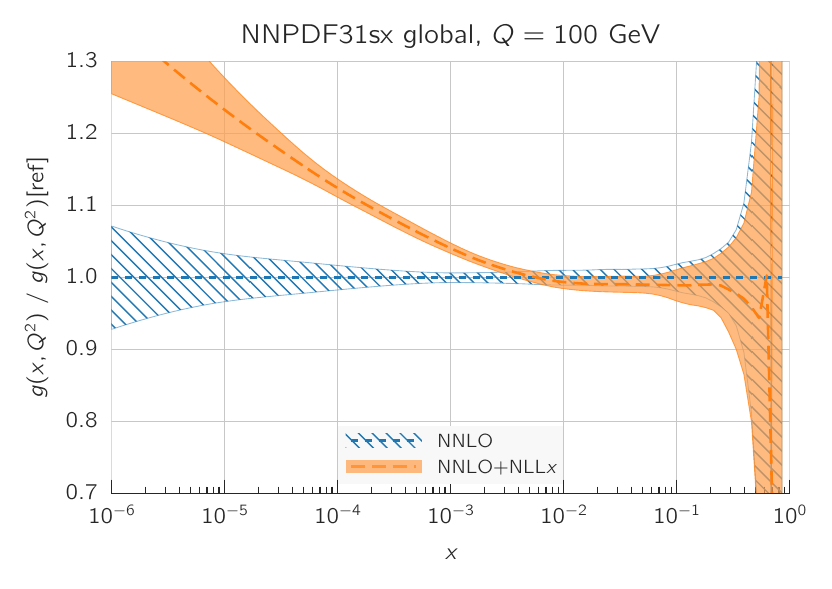}
  \includegraphics[width=0.49\textwidth]{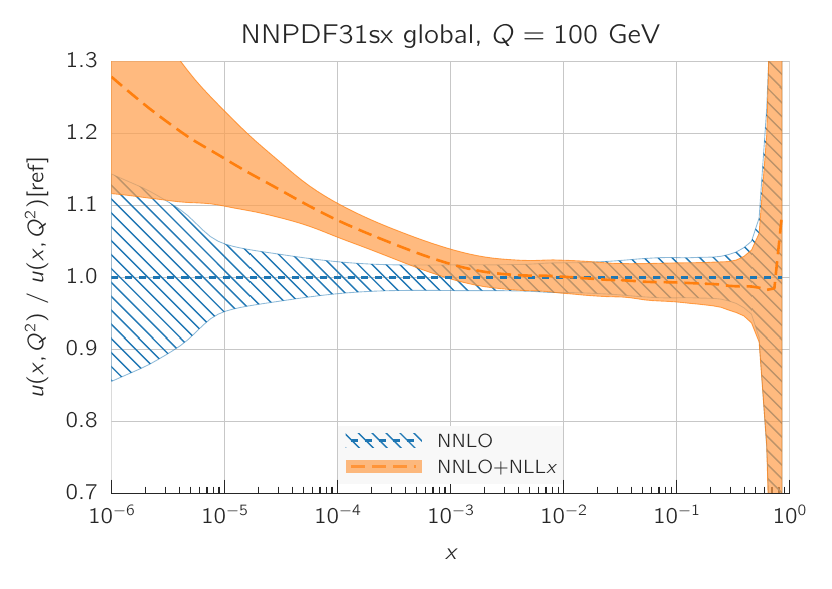}
  \includegraphics[width=0.49\textwidth]{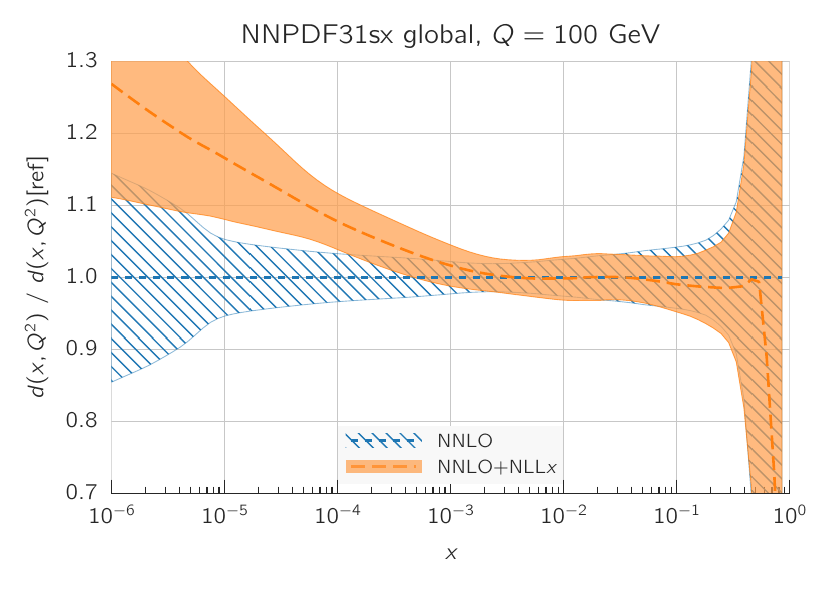}
  \includegraphics[width=0.49\textwidth]{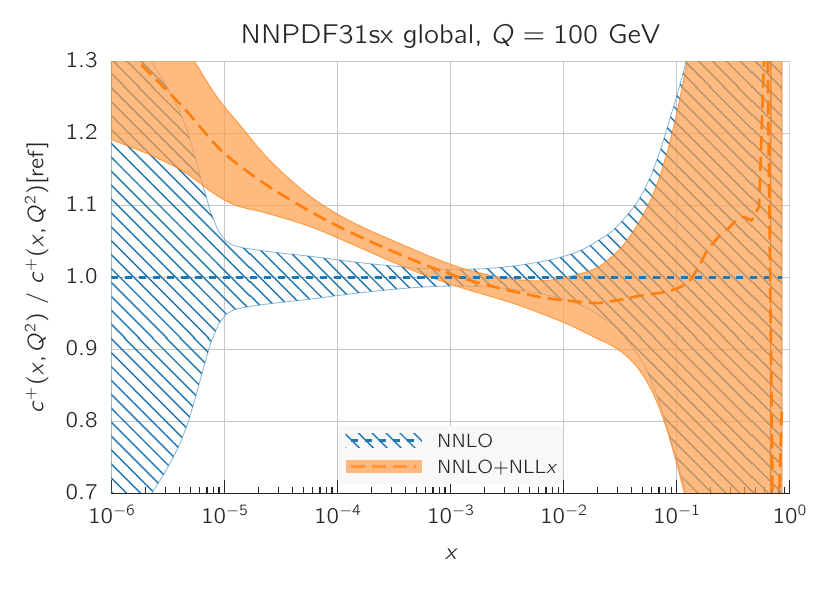}
  \caption{\small Comparison of the NNPDF3.1sx  NNLO and NNLO+NLL$x$ global fits
     at $Q=100$~GeV.
    We show the  gluon PDF and the charm, up, and down quark PDFs,
    normalized to the central value of the baseline NNLO fit.}
  \label{fig:smallx-nnlo-global}
\end{figure}

To visualize these effects, in Fig.~\ref{fig:smallx-nnlo-global} we show the flavour
combinations most affected by resummation
(as indicated in the distance plot of Fig.~\ref{fig:distance-sx-global}),
namely the gluon, charm, up and down PDFs, at a typical electroweak scale of $Q=100$~GeV.
The impact of NLL$x$ resummation is very similar for all the quark 
combinations: the effect is mild for $x\gtrsim 10^{-3}$, whilst it
increases at small-$x$, by an amount which
is however mostly consistent with the one or two sigma PDF uncertainties.
The effect is rather more marked for the gluon, where
the NNLO+NLL$x$ fit can be up to 30\% bigger at $x\simeq 10^{-6}$,
well outside the uncertainty band.
Thus the main impact of high-energy resummation
is to strongly enhance the gluon and mildly enhance the quarks at small-$x$.

\begin{figure}[htb]
\centering
   \includegraphics[width=0.49\textwidth]{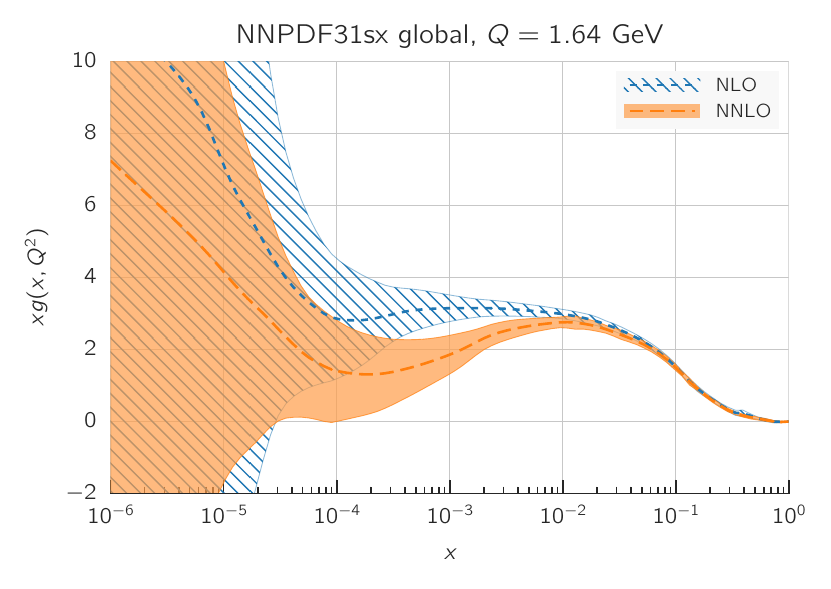}
 \includegraphics[width=0.49\textwidth]{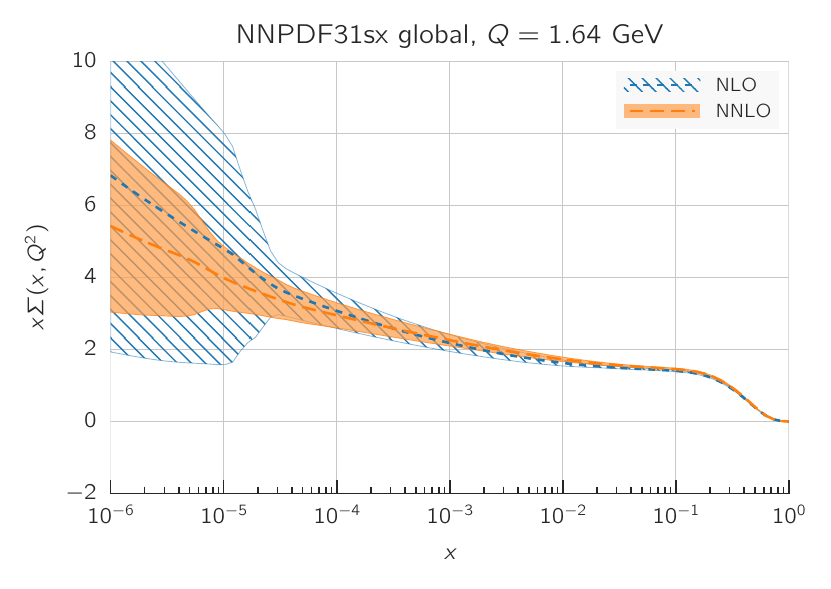}
   \includegraphics[width=0.49\textwidth]{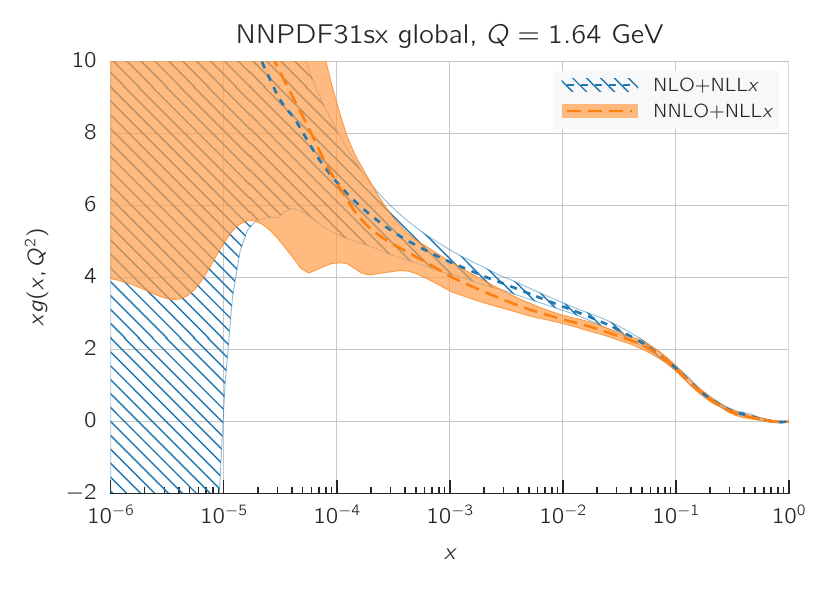}
 \includegraphics[width=0.49\textwidth]{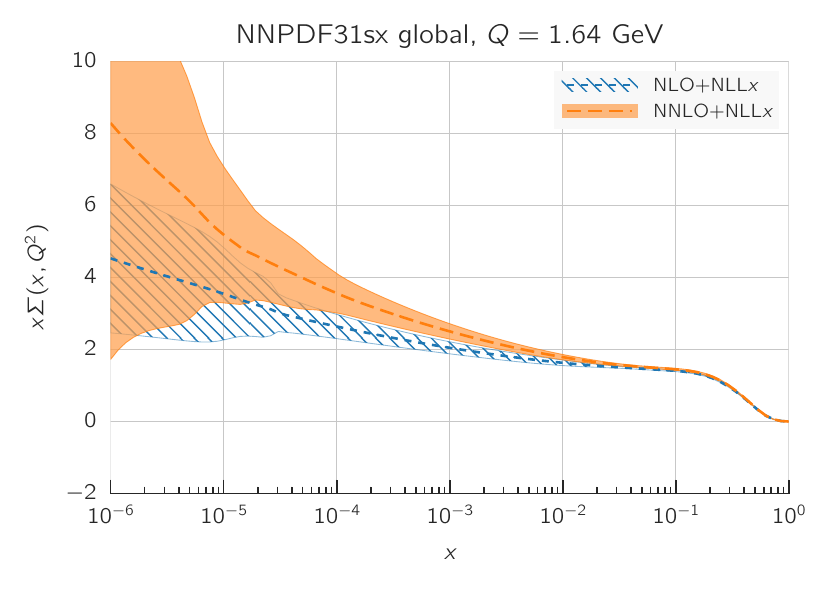} 
 \caption{\small Comparison of NLO and NNLO fit results at the input parametrization scale of $Q=1.64$~GeV (upper plots), and of NLO+NLL$x$ and NNLO+NLL$x$ fit results at the input parametrization scale of $Q=1.64$~GeV (lower plots). Left plots: gluon; right plots: quark singlet.}
 \label{fig:smallx-nnlo-global-lowQ}
\end{figure}

To conclude the discussion on the results of the global NNPDF3.1sx fits,
we move away from the electroweak scale and consider the PDFs at the
input parametrization scale $Q_0$.
This comparison
is interesting because it disentangles the effects of small-$x$ resummation
on the fitted PDFs from those due to the evolution
from low to high scales.
With this motivation, we
show in Fig.~\ref{fig:smallx-nnlo-global-lowQ} the gluon and the quark singlet at the fit scale $Q_0=1.64$ GeV.
In the case of the total quark singlet, we see that the impact of resummation is
moderate, with a one sigma increase at small $x$ in the NNLO+NLL$x$ fit which 
helps to improve the fit to the low $Q^2$ HERA data. The slightly larger effects seen at higher scales are thus mostly driven by the evolution,
that mixes the singlet with the gluon.
On the other hand, the effects of resummation are more marked for the fitted 
gluon, where we see explicitly a drop in the NNLO gluon at small $x$ driven 
by perturbative instability, which disappears on resummation in such a way 
that the NNLO+NLL$x$ gluon is rather flat, and indeed very close to the NLO and 
NLO+NLL$x$ gluon.
Note that the resummation thus extends the perturbative 
region at small $x$: even at $Q_0=1.64$ GeV the fitted gluon remains stable, 
and it seems likely that one would have to go to even lower scales (below the 
charm threshold) before the kind of instability seen in NNLO fixed order 
perturbation theory sets in.
Note that we would not expect the same to be 
true of N$^3$LO perturbation theory: the unresummed logarithms at N$^3$LO are 
considerably larger than those at NNLO, and thus the need for resummation 
at N$^3$LO would be even more pressing than at NNLO.

\FloatBarrier

\subsubsection{Dependence on the value of $H_{\rm cut}$}
\label{sec:dephcut}

Thus far we have only discussed the results of the global fit obtained using
the default cut to the hadronic data, identified as $H_{\rm cut}=0.6$.
We now discuss the dependence of the fit results with respect to variations 
of this
choice, both from the point of view of the fit quality and of the impact at 
the PDF level.
In doing so, we provide further
motivation for the choice of $H_{\rm cut}=0.6$
for our default global fits.

\begin{table}[tp]
  \centering
     \scriptsize
     \renewcommand{\arraystretch}{1.10}
     \begin{tabular}{lC{1.3cm}C{1.75cm}C{1.3cm}C{1.75cm}C{1.3cm}C{1.75cm}}
   & \multicolumn{2}{c}{$H_{\rm cut}=0.5$}  & \multicolumn{2}{c}{$H_{\rm cut}=0.6$}  & \multicolumn{2}{c}{$H_{\rm cut}=0.7$} \\
 & NNLO & NNLO+NLL$x$   & NNLO & NNLO+NLL$x$   & NNLO & NNLO+NLL$x$ \\
\toprule
NMC    &    1.31   &   1.31   &   1.30   &   1.33  &  1.31   &   1.36  \\    
SLAC    &    1.03   &   0.96   &   0.92   &   0.95 &  0.92   &   0.88  \\    
BCDMS    &    1.18   &   1.18   &   1.18   &   1.18 &  1.18   &   1.14  \\    
CHORUS    &    1.04   &   1.03   &   1.07   &   1.07 &  1.10   &   1.10  \\    
NuTeV dimuon    &    0.68   &   0.82   &   0.97   &   0.88 &   0.91   &   1.06  \\    
\midrule
HERA I+II incl. NC    &    1.17   &   1.11   &   1.17   &   1.11 &  1.17   &   1.12  \\    
HERA I+II incl. CC    & 1.23      &  1.23    & 1.23     &  1.24   &  1.26   & 1.26     \\    
HERA $\sigma_c^{\rm NC}$    &    2.34   &   1.17   &   2.33   &   1.14 &   2.43   &   1.17  \\    
HERA $F_2^b$     &    1.10   &   1.16   &   1.11  &   1.17 &   1.11   &   1.17  \\
\midrule
DY E866 $\sigma^d_{\rm DY}/\sigma^p_{\rm DY}$    &    0.34   &   0.35   &   0.32   &   0.30 &   0.38   &   0.36  \\    
DY E886 $\sigma^p$    &    0.99   &   0.96   &   1.31   &   1.32 &  1.33   &   1.28  \\    
DY E605  $\sigma^p$    &    1.05   &   1.03   &   1.10   &   1.10  &   1.17   &   1.10  \\    
CDF $Z$ rap    &    1.49   &   1.47   &   1.24   &   1.23  &   1.55   &   1.46  \\    
CDF Run II $k_t$ jets    &    0.83   &   0.80   &   0.85   &   0.80 &  0.85   &   0.86  \\    
D0 $Z$ rap    &    0.71   &   0.72   &   0.54   &   0.53  & 0.65   &   0.64  \\    
D0 $W\to e\nu$  asy     &    4.16   &   4.18   &   3.00   &   3.10 & 2.85   &   2.90  \\    
D0 $W\to \mu\nu$  asy    &    1.78   &   1.81   &   1.59   &   1.56  & 1.41   &   1.50  \\
\midrule
ATLAS  total  &    1.00   &   0.97   &   0.99   &   0.98 & 1.05   &   1.01  \\    
ATLAS $W,Z$ 7 TeV 2010    &    -   &   -   &   1.36   &   1.21  & 1.07   &   0.95  \\    
ATLAS HM DY 7 TeV    &    1.55   &   1.61   &   1.70   &   1.70 & 1.62   &   1.72  \\    
ATLAS $W,Z$ 7 TeV 2011    &    -   &   -   &   1.43   &   1.29 & 2.11   &   1.75  \\    
ATLAS jets 2010 7 TeV     &    0.88   &   0.82   &   0.86   &   0.83  & 0.92   &   0.89  \\    
ATLAS jets 2.76 TeV     &    0.94   &   0.87   &   0.96   &   0.96 & 0.98   &   0.93  \\    
ATLAS jets 2011 7 TeV     &    1.09   &   1.08   &   1.10   &   1.09 & 1.11   &   1.08  \\    
ATLAS $Z$ $p_T$ 8 TeV $(y_{ll},M_{ll})$       &    0.99   &   1.04   &   0.94   &   0.98 & 0.94   &   0.98  \\    
ATLAS $Z$ $p_T$ 8 TeV $(p_T^{ll},M_{ll})$    &    -   &   -   &   0.79   &   1.07 & 0.61   &   0.73  \\    
ATLAS $\sigma_{tt}^{tot}$     &    0.91   &   1.22   &   0.85   &   1.15 & 0.84    &   1.12  \\    
ATLAS $t\bar{t}$ rap    &    1.76   &   1.73   &   1.61   &   1.64 & 1.55   &   1.56  \\
\midrule
CMS  total  &    0.88   &   0.84   &   0.86   &   0.85 & 0.90   &   0.88  \\    
CMS $W$ asy 840 pb     &    -   &   -   &   -   &   -  & 0.41   &   0.39  \\    
CMS $W$ asy 4.7 fb     &    -   &   -   &   -   &   - & 1.25   &   1.23  \\    
CMS Drell-Yan 2D 2011    &    0.57   &   0.84   &   0.58   &   0.51 & 0.95   &   1.01  \\    
CMS $W$ rap 8 TeV    &    -   &   -   &   -   &   - &  0.85   &   0.64  \\    
CMS jets 7 TeV 2011     &    0.83   &   0.76   &   0.84   &   0.81 & 0.84   &   0.81  \\    
CMS jets 2.76 TeV    &    1.00   &   0.95   &   1.00   &   1.00 & 1.00   &   0.98  \\    
CMS $Z$ $p_T$ 8 TeV $(p_T^{ll},M_{ll})$    &    1.20   &   1.55   &   0.73   &   0.77 & 0.74   &   0.77  \\    
CMS $\sigma_{tt}^{tot}$     &    0.24   &   0.28   &   0.23   &   0.24 & 0.23   &   0.23  \\    
CMS $t\bar{t}$ rap    &    0.78   &   0.78   &   1.08   &   1.10   & 0.91   &   0.92  \\    
\bottomrule
{\bf Total }    &  \bf  1.120   &\bf   1.085   & \bf  1.130   & \bf  1.100 & \bf 1.142   & \bf  1.112  \\    
     \end{tabular}
     \vspace{0.5cm}
\caption{\small Same as Table~\ref{tab:chi2tab_pertorder},
  now comparing the values of the  $\chi^2/N_{\rm dat}$ for the
  global NNLO and NNLO+NLL$x$ fits obtained
  with different values of the hadronic data cut, $H_{\rm cut}=0.5, 0.6$
  and 0.7.
  Note that fits with different values of $H_{\rm cut}$ have in general
  a different number of data points in the hadronic experiments,
  as indicated in Table~\ref{tab:ndat}.
  Columns 4 and 5 of this table correspond to the same numbers
  as those in columns 5 and 6 of Table~\ref{tab:chi2tab_pertorder}.
  For ease of comparison, the $\delta\chi^2$ variations
  among fits with different cuts, Eq.~(\ref{eq:deltachi2variationcuts}),
  are collected in Table~\ref{tab:chi2tab_ccut_deltachi2}.
}
\label{tab:chi2tab_ccut}
\end{table}


\begin{table}[tp]
  \centering
     \scriptsize
     \renewcommand{\arraystretch}{1.10}
     \begin{tabular}{lC{1.4cm}C{1.5cm}C{0.9cm}|C{1.5cm}C{1.7cm}C{0.9cm}}
       & \multicolumn{2}{c}{$\chi^2(0.6)-\chi^2(0.5)$}
       &   \multirow{2}{*}{$\delta N_{\rm dat}$}
       & \multicolumn{2}{c}{$\chi^2(0.7)-\chi^2(0.6)$} &
       \multirow{2}{*}{$\delta N_{\rm dat}$} \\
 & NNLO & NNLO+NLL$x$&  & NNLO & NNLO+NLL$x$   &   \\
\toprule  
NMC &  $-4$  & $+6$  &    -   & $+2$     & $+12$    & - \\
SLAC &  $-9$ &  $-1$ &   -     & -    &   $-5$   &-\\
BCDMS &  $-2$&$-1$&     -   &    -    &  $-21$     &-\\
CHORUS & $+32$  & $+38$  &   -     & $+20$     & $+28$   &- \\
NuTeV dimuon&$+23$&  $+5$ &  -       &   $-5$  & $+14$ &- \\
\midrule
HERA I+II incl NC & $-5$ & $-7$ & -   & $+2$ & $+11$   &- \\
HERA I+II incl CC & $+2$  & $+1$ &  - & $+1$ & $+2$ &- \\
HERA $\sigma_c^{\rm NC}$ &  $-1$  &$-1$&    -     &   $+4$    &  $+1$   & -\\
HERA $F_2^b$  & -  &-&   -      &  -     &  $+1$   &  -\\
\midrule
DY E866 $\sigma^d_{\rm DY}/\sigma^p_{\rm DY}$ & - & -       & $+2$      & -      & $+1$   &  $+1$ \\
DY E886 $\sigma^p$ & $+44$  &  $+47$  & $+20$         &  $+17$     &  $+12$   & $+12$ \\
DY E605  $\sigma^p$ &  $+5$  &$+7$&      -   &  $+5$      & -    & - \\
CDF $Z$ rap &  $+7$ &$+7$&  $+8$  &  $+20$             &  $+17$    & $+9$\\
CDF Run II $k_t$ jets & $+1$&-&   -       &  -    & $+5$    & -\\
D0 $Z$ rap &  $+2$&$+2$&  $+8$  &  $+8$          &  $+7$  &  $+8$\\
D0 $W\to e\nu$  asy  &  $+4$&  $+5$ &  $+3$       &    $+2$   & $+1$   & $+1$ \\
D0 $W\to \mu\nu$  asy & $+6$ & $+5$  &  $+4$      &   -     &  $+1$  & $+1$ \\
\midrule
ATLAS  total  & $+25$  &$+29$&  $+27$       &  $+53$     &  $+43$   & $+36$\\
ATLAS $W,Z$ 7 TeV 2010 &$+8$  &  $+7$ &  $+6$      &    $+8$    &  $+8$   & $+10$\\
ATLAS HM DY 7 TeV &  $+1$ & -         & -  &    -   &   -    & $+1$ \\
ATLAS $W,Z$ 7 TeV 2011 &$+11$  &  $+10$ & $+8$        &  $+30$     &  $+25$   & $+12$ \\
ATLAS jets 2010 7 TeV  &$+3$  &$+5$&   $+5$      &  $+7$     &  $+7$   & $+3$  \\
ATLAS jets 2.76 TeV  & $+4$ & $+7$  &  $+3$       & $+1$      &   $-2$   & -\\
ATLAS jets 2011 7 TeV  &  -&-&   -       &  -     &-      & - \\
ATLAS $Z$ $p_T$ 8 TeV $(y_{ll},M_{ll}) $ &$+5$  & $+6$  & $+6$        & $+6$      &    $+6$ & $+11$ \\
ATLAS $Z$ $p_T$ 8 TeV $(p_T^{ll},M_{ll})$ & $-2$ &  $-3$ & -       &    -    &   -  & - \\
ATLAS $\sigma_{tt}^{tot}$  &- & -  &  -        &  -    &  -   & -\\
ATLAS $t\bar{t}$ rap &  $-2$ &  $-1$ &  -      &  $-1$       & $-1$    & - \\
\midrule
CMS  total  & $+17$&$+24$&   $+25$       & $+60$     &  $+57$   & $+57$ \\
CMS $W$ asy 840 pb  &-  &-&  -      &   $+8$     & $+8$  & $+7$  \\
CMS $W$ asy 4.7 fb  &-&-&     -    &   $+10$    & $+9$    & $+7$ \\
CMS Drell-Yan 2D 2011 &  $+9$  &$+7$&  -       &    $+28$   & $+31$   & $+20$  \\
CMS $W$ rap 8 TeV &  - & -&     $+16$     &  $+11$    &  $+11$    & $+12$ \\
CMS jets 7 TeV 2011  &  $+1$&  $+7$ &  -       &    -   & -   &  -\\
CMS jets 2.76 TeV & -  &  $+4$ & -    &     -     &  $-2$    & - \\
CMS $Z$ $p_T$ 8 TeV $(p_T^{ll},M_{ll})$ &  $+4$ &  $+3$ &  $+7$       &  $+6$     &  $+7$  &  $+9$ \\
CMS $\sigma_{tt}^{tot}$  & - & -  &     -    &   -    &  -   & -\\
CMS $t\bar{t}$ rap &$+4$ & $+4$& $+2$        &   -    &  -   & $+2$\\
\bottomrule
  \end{tabular}
  \vspace{0.5cm}
  \caption{The differences
    $\delta\chi^2\equiv \chi^2\lp H_{\rm cut}^{(1)}\rp -\chi^2\lp H_{\rm cut}^{(2)}\rp$,
    Eq.~(\ref{eq:deltachi2variationcuts}),
  for the global fits reported in Table~\ref{tab:chi2tab_ccut}.
  Since the fits with different values of $H_{\rm cut}$ have in general
  a different number of data points for the hadronic experiments,
  we also indicate in each case the difference
  $\delta N_{\rm dat}=N_{\rm dat}\lp H_{\rm cut}^{(1)}\rp -N_{\rm dat}\lp H_{\rm cut}^{(2)}\rp$.}
\label{tab:chi2tab_ccut_deltachi2}
\end{table}


To begin with, we study the
dependence of the quality of the NNPDF3.1sx fits
as a function of the value of the cut parameter $H_{\rm cut}$
applied to the hadronic data.
In Table~\ref{tab:chi2tab_ccut} we show
a comparison of the NNLO and NNLO+NLL$x$ values of
the $\chi^2/N_{\rm dat}$ for the fits with $H_{\rm cut}=0.5,0.6$
and $0.7$.
In addition, to better appreciate the variations for $\chi^2$
for the fits with different $H_{\rm cut}$ cuts,
in Table~\ref{tab:chi2tab_ccut_deltachi2} we also show
the differences
\be
\label{eq:deltachi2variationcuts}
\delta\chi^2\equiv \chi^2 \lp H_{\rm cut}^{(1)}\rp-\chi^2\lp H_{\rm cut}^{(2)}\rp\,  ,
\ee
for the global fits obtained using NNLO and NNLO+NLL$x$ theory.
To highlight that in general fits varying $H_{\rm cut}$ have different
number of data points, we also indicate in the same table the difference
 $\delta N_{\rm dat}=N_{\rm dat}\lp H_{\rm cut}^{(1)}\rp -N_{\rm dat}\lp H_{\rm cut}^{(2)}\rp$
 for each experiment.

 The main general feature that we note from the comparisons in Tables~\ref{tab:chi2tab_ccut}
 and~\ref{tab:chi2tab_ccut_deltachi2} is that the $\chi^2/N_{\rm dat}$
 values  exhibit a rather
 moderate dependence on the specific value of the kinematic cut to the hadronic data.
Concerning the total dataset, the $\chi^2/N_{\rm dat}$ values
slightly increase as $H_{\rm cut}$ is raised and the dataset is enlarged: in particular,
for the NNLO~(NNLO+NLL$x$) fits, the values of $\chi^2/N_{\rm dat}$
for the total dataset are $1.120$, $1.130$, and $1.142$ ($1.085$, $1.100$, and $1.112$)
for $H_{\rm cut}=0.5, 0.6$ and $0.7$ respectively.
The fact that the fit quality of
both the fixed-order and resummed fits is
slightly better for $H_{\rm cut}=0.5$
is a direct consequence of the more restrictive dataset.

In the case of the NNLO+NLL$x$ fits,
the difference between the  $\chi^2/N_{\rm dat}$
of the fit with $H_{\rm cut}=0.6$ and the fit with $H_{\rm cut}=0.7$ is larger than a
statistical fluctuation.
This might be an indication that the deterioration of the fit with $H_{\rm cut}=0.7$
could be related to non-negligible effects of unresummed small-$x$ logarithms in
the extra hadronic data that are included in this fit.
This conjecture is supported by the fact that, while with $H_{\rm cut}=0.6$
the resummation improves 
the total $\chi^2$  over the fixed order by around 120 points, for 
$H_{\rm cut}=0.7$ the improvement is reduced to less than 100 points. 
On the other hand, the same trend is also visible in the NNLO fits,
and there it can be partly explained by the contributions from some collider
 points that are in tension between the DIS data, for instance,
the ATLAS $W,Z$ 2011 rapidity distributions and the neutrino data.
We also find that
the more conservative fit with $H_{\rm cut}=0.5$ also 
improves with resummation by even more than the
$H_{\rm cut}=0.6$ fit (around 140 points),
thus 
suggesting that our default cut value  is safe, in the sense that 
it is not affected by large unresummed logarithms in the hadronic processes.

\begin{figure}[t]
  \centering
    \includegraphics[width=\textwidth]{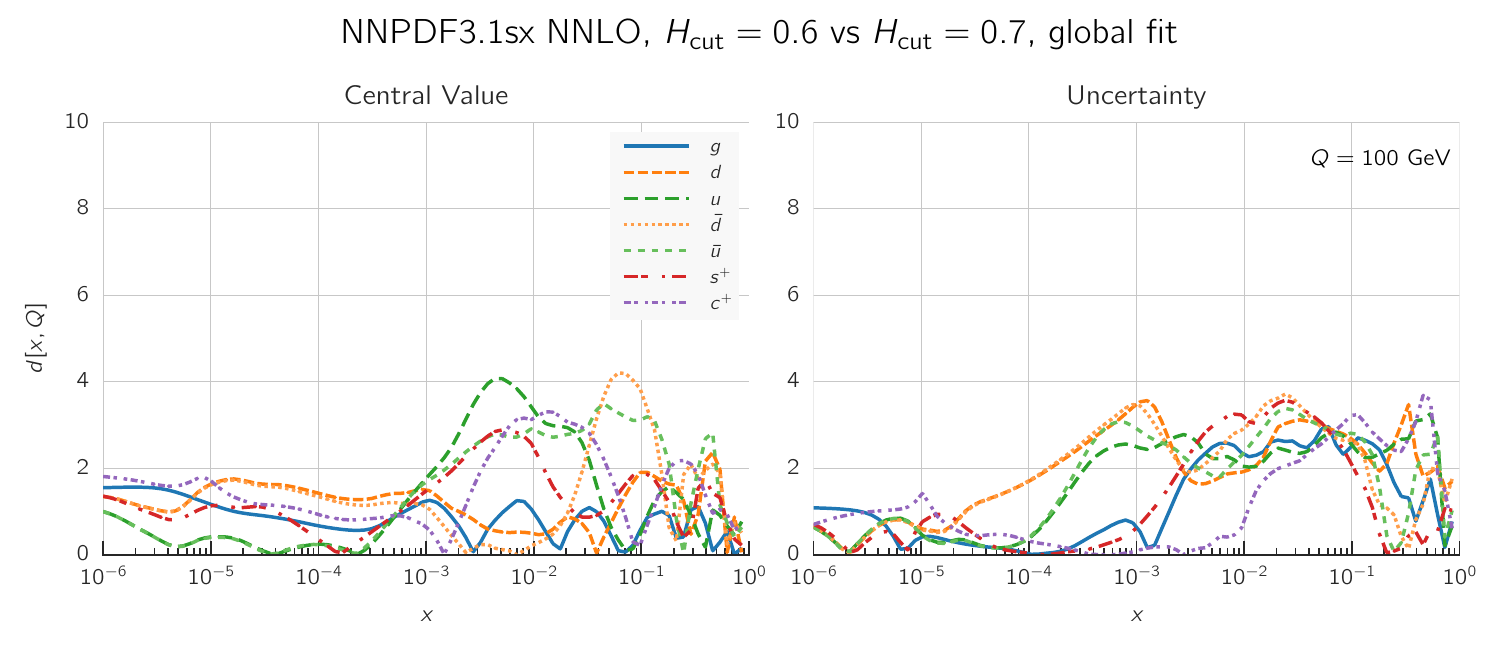}
    \vspace{0.3cm}
    \includegraphics[width=\textwidth]{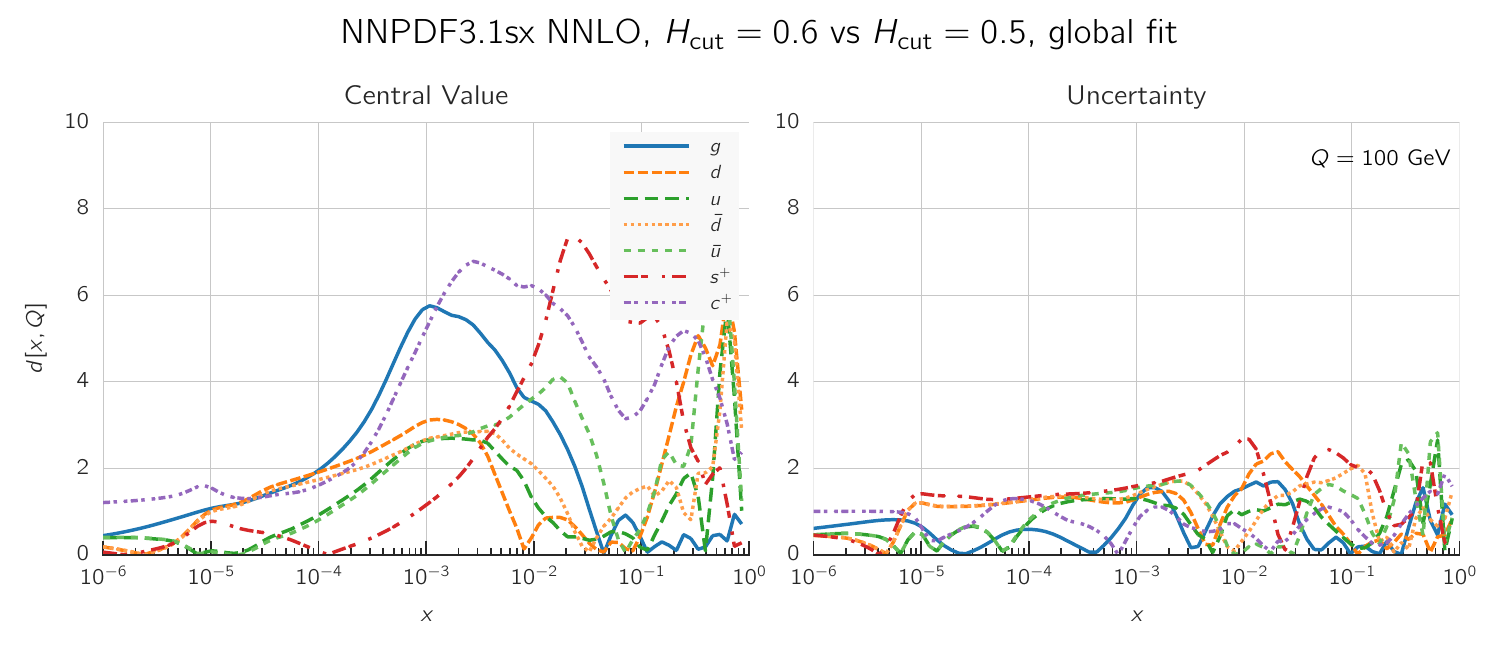}
    \caption{\small Same as Fig.~\ref{fig:distance-sx-dis}
      for the comparison of the baseline NNPDF3.1sx NNLO+NLL global
      fit with $H_{\rm cut}=0.6$ with the corresponding fits
      with $H_{\rm cut}=0.7$ (upper) and $H_{\rm cut}=0.5$ (lower plots).}
    \label{fig:distances-HcutDep}
\end{figure}
 
We further investigate the impact on the PDFs of the various choices of $H_{\rm cut}$.
We show in Fig.~\ref{fig:distances-HcutDep} the distance estimator
to compare the default NNPDF3.1sx NNLO+NLL$x$ global
fit with $H_{\rm cut}=0.6$ with the corresponding fits with the $H_{\rm cut}=0.7$  and $H_{\rm cut}=0.5$ fits.
In terms of central values, we see that differences are well below PDF
uncertainties (which corresponds to $d\simeq 10$) 
when comparing $H_{\rm cut}=0.7$ to $H_{\rm cut}=0.6$.
On the contrary, the distances between the $H_{\rm cut}=0.5$ fit and 
the $H_{\rm cut}=0.6$ fit are larger, especially for charm and strangeness at 
$x \gsim 10^{-3}$. 
This comparison indicates that there is no real benefit in loosening
the cut from $H_{\rm cut}=0.6$ to $H_{\rm cut}=0.7$ (since differences at the
PDF level are small, and the possibility of biasing the fit higher)
whilst it is indeed advantageous to use $H_{\rm cut}=0.6$ rather
than the tighter cut $H_{\rm cut}=0.5$, thanks to the increase in
PDF constraints provided by the additional data.

\begin{figure}[t]
  \centering
    \includegraphics[width=0.49\textwidth]{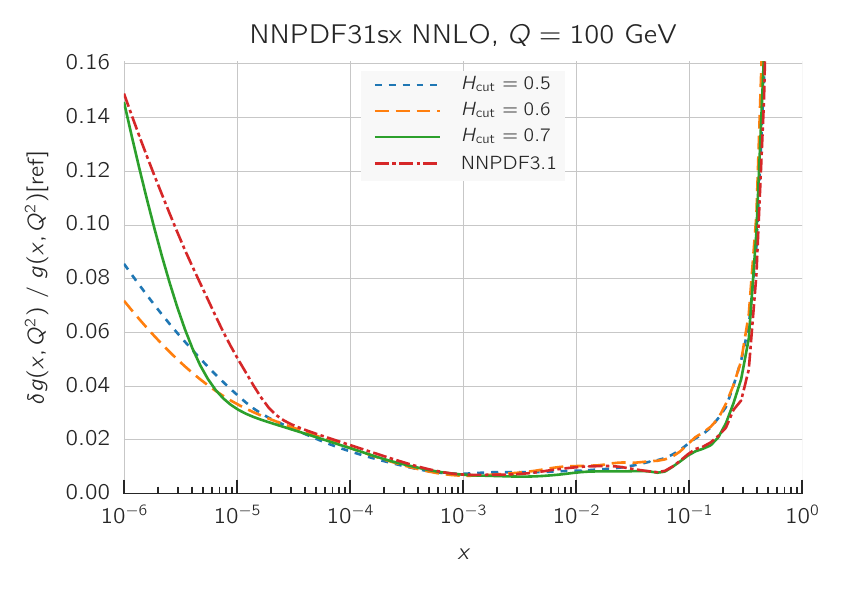}
    \includegraphics[width=0.49\textwidth]{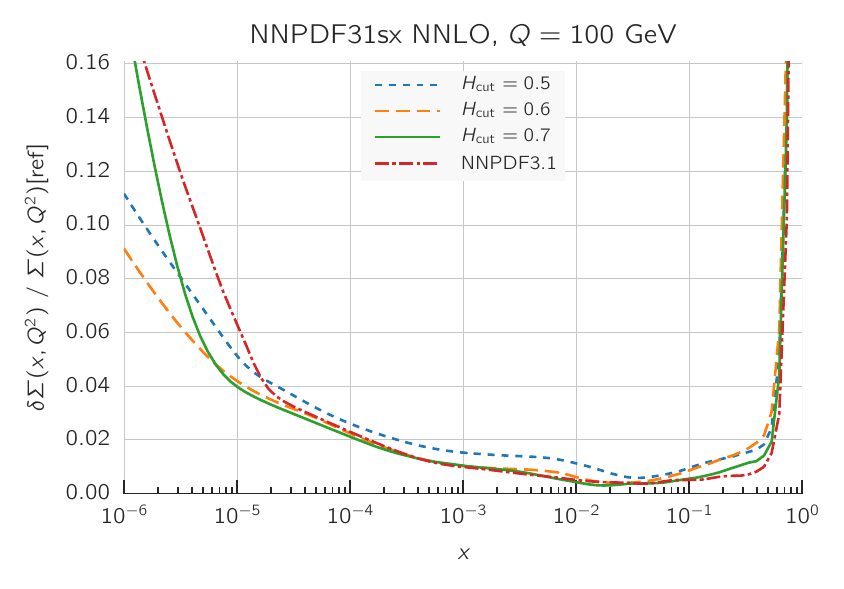}
    \includegraphics[width=0.49\textwidth]{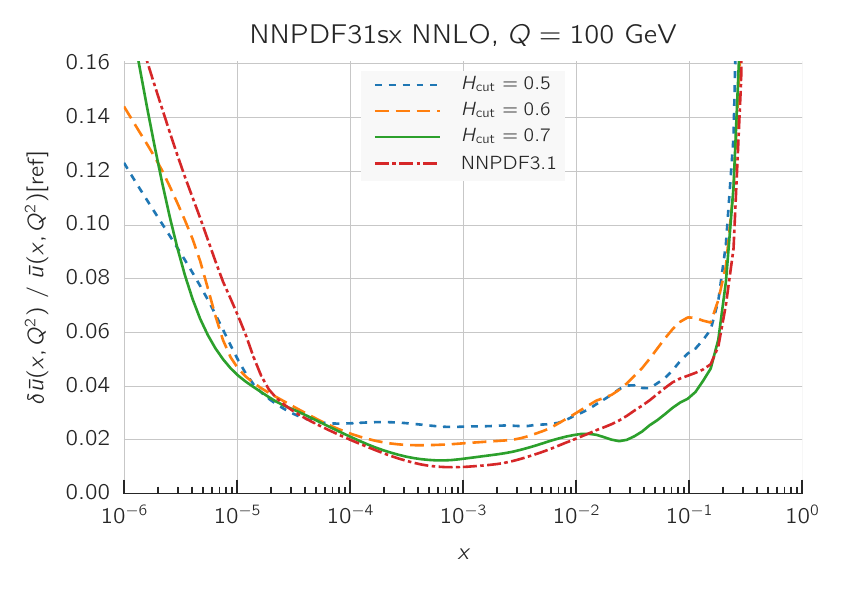}
    \includegraphics[width=0.49\textwidth]{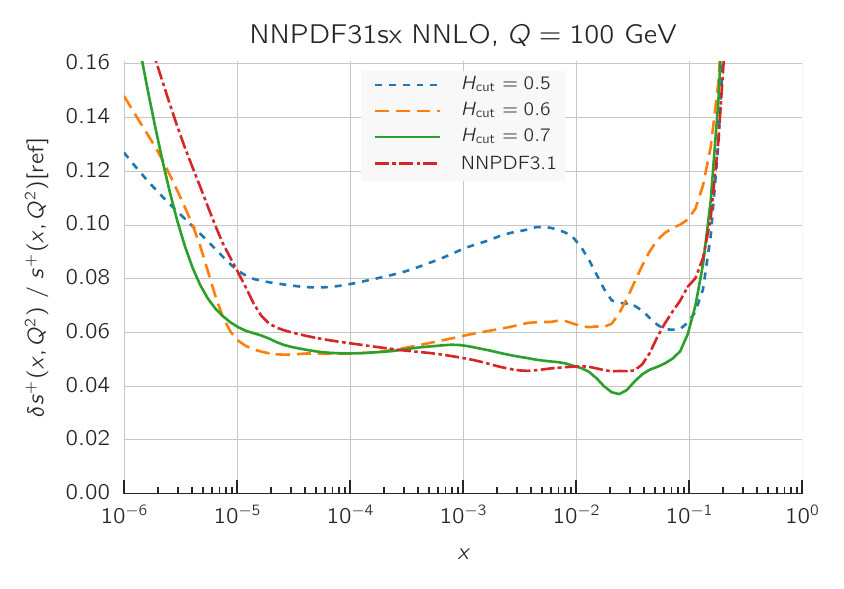}
    \caption{\small The relative PDF uncertainties in the NNPDF3.1sx NNLO fits
      with the three different values of the $H_{\rm cut}$ cut on the hadronic
      data, compared with those from NNPDF3.1.
      We show the gluon, the quark singlet, the anti-up quark, and the total
      strangeness, at $Q=100$~GeV.}
    \label{fig:ERR-cutdep}
\end{figure}

Finally, in Fig.~\ref{fig:ERR-cutdep} we show the relative
PDF uncertainties in the NNPDF3.1sx NNLO fits
with the three different values of the $H_{\rm cut}$ cut on the hadronic
data.
For completeness, we also include in this comparison
the results of the NNPDF3.1 fit.
Specifically, we show the gluon, the quark singlet, the anti-up quark, and the total
strangeness, at $Q=100$~GeV.
From the comparison we see that as expected the smaller
the value of $H_{\rm cut}$, the more marked the increase in PDF uncertainties.
On the other hand, we see that for $H_{\rm cut}=0.6$ the results
are already competitive with those of NNPDF3.1.
We also find that in the small-$x$ region, PDF uncertainties are smaller in the NNPDF3.1sx
fits than in the NNPDF3.1 ones, especially for our default value of $H_{\rm cut}=0.6$, due
to the lowering of the $Q_{\rm min}^2$ kinematic cut (see also the discussion
in Sect.~\ref{sec:impactlowQ2}).
 
Summarizing, we have provided here
a number of indications that the NNPDF3.1sx fit with $H_{\rm cut}=0.6$
is not biased by hadronic data sensitive to small-$x$ resummation,
and at the same time we have demonstrated that the resulting
PDF uncertainties are competitive, though still larger, with those of NNPDF3.1.
These considerations provide further weight
for our default choice of the $H_{\rm cut}$ cut to the hadronic data.

\subsubsection{The role of the $Q^2=2.7$~GeV$^2$ bin}
\label{sec:impactlowQ2}

We have stressed in Sect.~\ref{sec:fitsetup} that,
as opposed to NNPDF3.1,  we include in the NNPDF3.1sx
fits an additional low $Q^2$ bin of the inclusive HERA dataset,
specifically the one with $Q^2=2.7$~GeV$^2$.
This choice has the important advantage of extending the kinematic coverage of the fits
from $x_{\rm min} \simeq 5 \times 10^{-5}$ down to $x_{\rm min} \simeq 3 \times 10^{-5}$.
The main reason why this bin was excluded from previous NNPDF fits (as well
as in most other global PDF fits) is its low value of
$Q^2$, which lies at the boundary between perturbative and non-perturbative dynamics, and
where fixed-order perturbation theory might not be fully appropriate.
Here we show
that this failure is not actually due to non-perturbative dynamics, but rather
it represents a limitation of the fixed-order expansion in the small-$x$ region
enhanced by the larger value of $\as$.
Indeed, we find that
once NNLO fixed-order perturbation theory is supplemented by NLL$x$ resummation,
this bin can be described with similar quality as the rest of the HERA data.
 
\begin{figure}[t]
  \centering
    \includegraphics[width=0.49\textwidth]{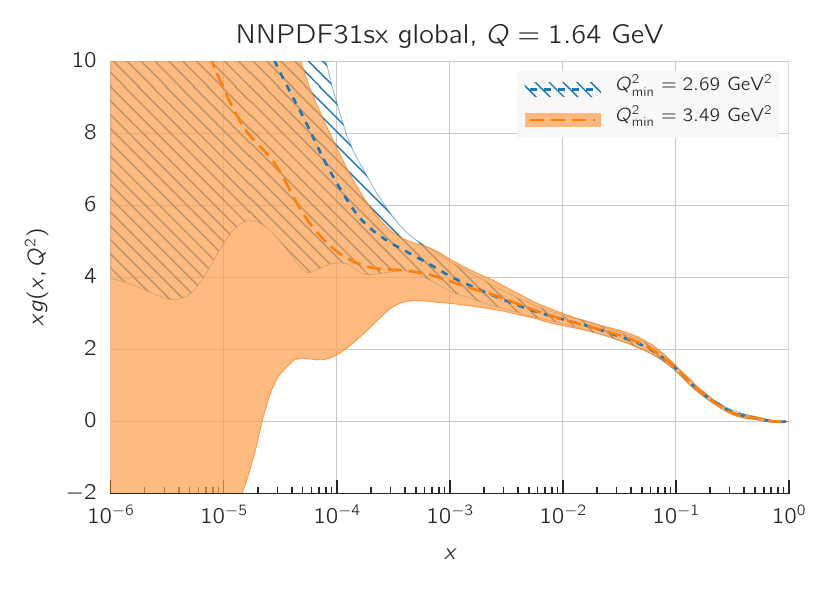}
    \includegraphics[width=0.49\textwidth]{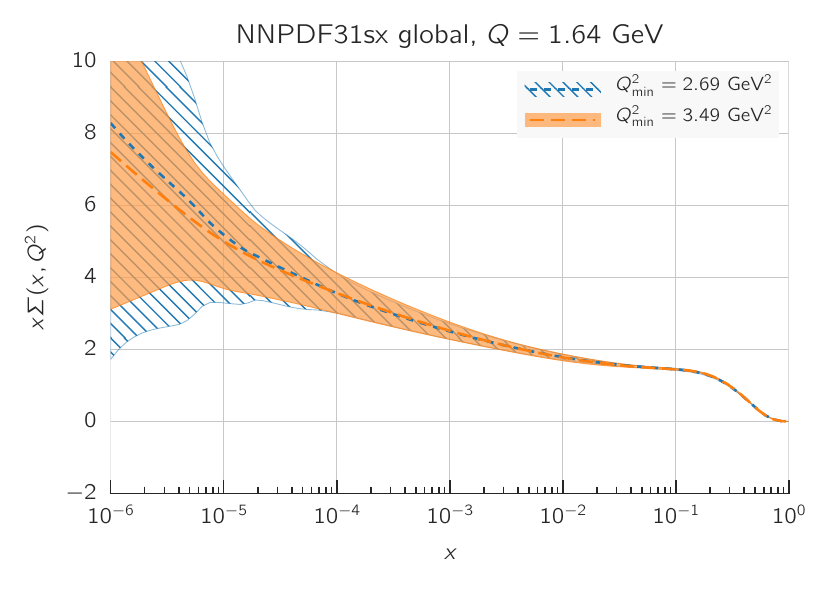}
    \includegraphics[width=0.49\textwidth]{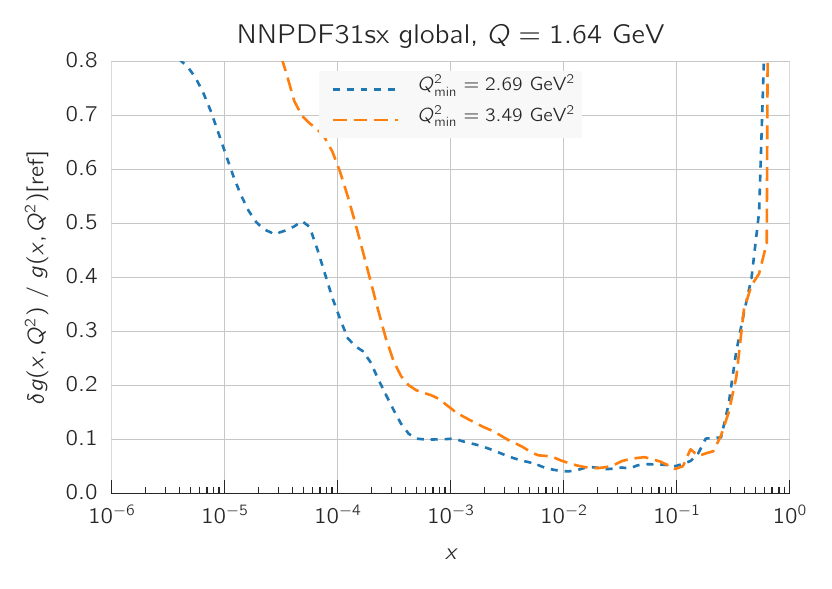}
    \includegraphics[width=0.49\textwidth]{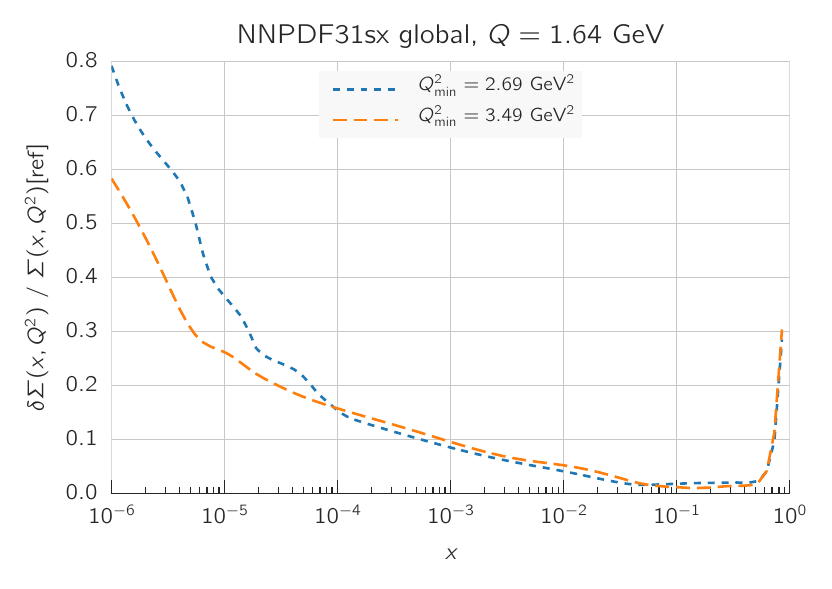}
    \caption{\small Comparison of the gluon (left) and quark singlet (right)
      at  $Q_0=1.64$~GeV
      between the  NNPDF3.1sx NNLO+NLL$x$ fits with
      the $Q_{\rm min}^2=2.69$~GeV$^2$ (baseline) and $Q_{\rm min}^2=3.49$~GeV$^2$
      kinematic cuts (upper plots)
      together with the corresponding relative PDF uncertainties (lower plots)}
    \label{fig:ERR-smallx-nnlo-global-lowQ}
\end{figure}

To illustrate this point, we have computed the  values of $\chi^2/N_{\rm dat}$
for the $N_{\rm dat}=17$ data points that constitute the $Q^2=2.7$ GeV$^2$ bin
of the inclusive HERA structure function dataset.
We find that the  values of $\chi^2/N_{\rm dat}$ for this bin
are 1.64 and 1.34 in the NNPDF3.1sx NLO+NLL$x$ and NNLO+NLL$x$ fits.
These results can be compared with the corresponding values
in the NLO and NNLO fits, which turn out to be 2.04 and 3.04, respectively.
The trend is the same as that for the total NC HERA inclusive dataset
(see Table~\ref{tab:chi2tab_pertorder}), namely with the NNLO+NLL$x$ (NNLO) fit
leading to the best (worst) overall description, and with the NLO and NLO+NLL$x$
values in between.
Interestingly, we also see that for this specific fit NNLO+NLL$x$ theory
leads to a rather better $\chi^2$ than the NLO+NLL$x$ one, although the small
number of data points prevents drawing any strong conclusion
from this observation.

Once we have established that the fit quality to the $Q^2=2.7$~GeV$^2$ HERA bin
is satisfactory when NLL$x$ resummation is included,
 we can next turn to study the constraints that this bin has on the small-$x$ PDFs.
With this motivation,
we have performed a global fit at NNLO+NLL$x$ with the same settings as the NNPDF3.1sx
baseline 
but raising the low $Q^2$ kinematic
cut from $Q_{\rm min}^2=2.69$~GeV$^2$ to $Q_{\rm min}^2=3.49$~GeV$^2$, as in NNPDF3.1,
so that the HERA bin with $Q^2=2.7$~GeV$^2$  is excluded.
In the latter case, the lowest HERA bin included is the one with $Q^2=3.5$ GeV$^2$.

The results are shown in Fig.~\ref{fig:ERR-smallx-nnlo-global-lowQ},
where the gluon and the quark singlet PDFs obtained in the 
NNPDF3.1sx NNLO+NLL$x$ fits with and without this additional bin
are compared at the input
parametrization scale of $Q_0=1.64$~GeV, together with their 
relative PDF uncertainties.
We find that the inclusion of this extra $Q^2$ bin leads to a significant reduction of
small-$x$ uncertainty of the gluon
in the region which is constrained by the data ($x\gtrsim10^{-5}$),
while the quark singlet is essentially unaffected.
These results illustrate how the use of an improved theory, NNLO+NLL$x$
in this case, can lead indirectly to a decrease of the  PDF uncertainties
due to the possibility of including more data in the fits from a wider kinematic range.

\section{\boldmath Small-$x$ resummation and HERA structure functions}
\label{sec:diagnosis}

The results of the previous section provided two main pieces of information.
First of all, the inclusion of small-$x$ resummation 
improves the description of those datasets which represent the best probe of
the small-$x$ region, namely the inclusive and charm HERA structure functions.
Second, the impact of resummation at the level of PDFs can be sizable.
In this section, we focus on the HERA data in the small-$x$ and small-$Q^2$ region,
in order to further quantify the improvement in its description when
fixed-order theory is supplemented by NLL$x$ resummation.

We first compare the HERA structure functions at low $x$ with
fixed-order and resummed theoretical
predictions, both for the inclusive
and charm  reduced  cross-sections  as well as
for the longitudinal structure
function $F_L$.
In all cases, 
we highlight the improved description that
is achieved once NNLO+NLL$x$ theory is used in all cases.
To quantitatively investigate the evidence for the onset of small-$x$ resummation 
in the HERA data, we  introduce several estimators
building upon the set of diagnostic tools
first presented in Refs.~\cite{Caola:2010cy,Caola:2009iy}.
We finally study how removing HERA data at low-$x$ and low-$Q^2$
affects global NNLO fits, 
and we discuss how the resulting PDFs are modified at medium- and large-$x$.
This way, it is possible to assess whether the inclusion of data poorly described
in a fixed-order analysis might be a source of bias for high-$Q^2$ phenomenology.

\subsection{The HERA data in the small-$x$ region}

\begin{figure}[p]
\centering
  \includegraphics[width=0.49\textwidth,page=1]{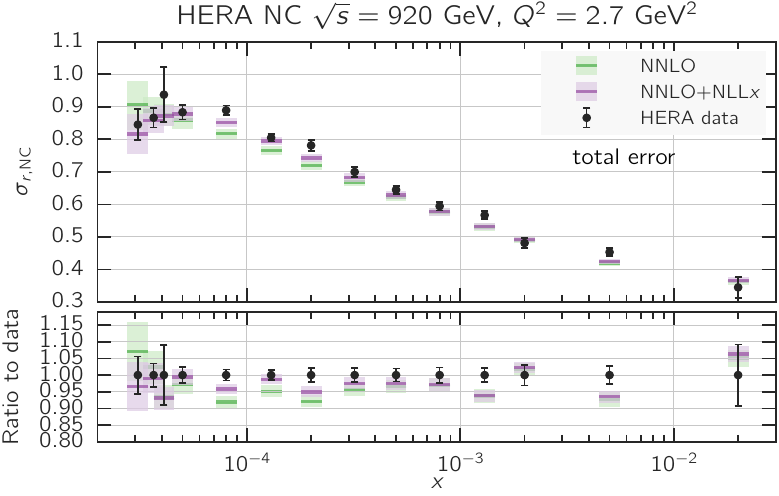}
  \includegraphics[width=0.49\textwidth,page=1]{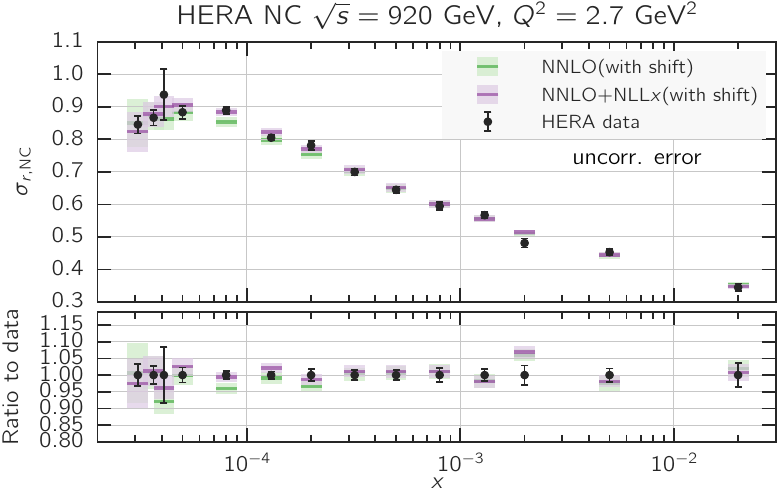}\\
  \includegraphics[width=0.49\textwidth,page=2]{plots/HERA920_samerange.pdf}
  \includegraphics[width=0.49\textwidth,page=2]{plots/HERA920_samerange_shift.pdf}\\
  \includegraphics[width=0.49\textwidth,page=3]{plots/HERA920_samerange.pdf}
  \includegraphics[width=0.49\textwidth,page=3]{plots/HERA920_samerange_shift.pdf}\\
  \includegraphics[width=0.49\textwidth,page=4]{plots/HERA920_samerange.pdf}
  \includegraphics[width=0.49\textwidth,page=4]{plots/HERA920_samerange_shift.pdf}
  \caption{\small Comparison between the HERA NC
   reduced cross-section from the $\sqrt{s}=920$~GeV
   dataset and the results of the NNLO and NNLO+NLL$x$ fits with the
   corresponding PDF uncertainties.    
    We show the results for the first four bins in $Q^2$ above
    the $Q^2_{\rm min}$ kinematic cut.
    For each bin we also show in the bottom panel the ratio of
    the theory predictions to the experimental data.
    The plots on the right show the theoretical prediction including the
    shifts as discussed in the text.}
  \label{fig:datathcomp}
\end{figure}

In order to investigate in greater detail how well resummed theory describes the low-$Q^2$ HERA
cross-sections,
we first perform a comparison of the theoretical predictions obtained using the results of the
NNPDF3.1sx NNLO and NNLO+NLL$x$ global fits to the experimental data.
To begin with, in Fig.~\ref{fig:datathcomp} we show the neutral-current (NC) reduced cross-section,
defined as
\be
\label{eq:sigmaRed2}
\sigma_{r,\rm NC}(x,Q^2,y)\equiv \frac{d^2\sigma_{\rm NC}}{dx dQ^2}\cdot
\frac{Q^4x}{2\pi\alpha Y_+}=F_2(x,Q^2)-\frac{y^2}{Y_+}F_L(x,Q^2)\, ,
\ee
where $Y_+=1+(1-y)^2$ and $y=\frac{Q^2}{sx}$ is the inelasticity.
This comparison is performed
for the first four bins in $Q^2$ above 
our $Q_{\rm min}^2$ kinematic cut of the $\sqrt{s}=920$~GeV dataset,
corresponding to $Q^2=2.7, 3.5, 4.5$ and 6.5~GeV$^2$ respectively.
In the left plots, the uncertainty of the experimental data points is given by the sum in quadrature of 
the various sources of uncorrelated and correlated uncertainties, whereas the theoretical predictions
include the associated PDF uncertainty.
In the right plots, instead, only the uncorrelated uncertainties are shown in the data, and the correlations
are taken into account via shifts which modify the theoretical prediction~\cite{Pumplin:2002vw}
and facilitate the graphical comparison.
Note that these correlations are included in the $\chi^2$ definition.
However, unlike in a Hessian approach, in a Monte Carlo method one does
not determine the best-fit systematic shifts.
Rather, here we have computed them a posteriori, under the assumption that the uncertainties are gaussian,
which is not necessarily true in a Monte Carlo fit. Therefore, this comparison must be interpreted with care.
%
%
%

From this comparison, we see that for $x\gsim 5\times 10^{-4}$ the results
of the NNLO and NNLO+NLL$x$ fits are essentially identical; 
in both cases, the theoretical predictions undershoot the data.
The trend changes for values of $x$ smaller than $5\times 10^{-4}$, where the NNLO and the NNLO+NLL$x$ predictions start to differ.
Around this value, we observe that the reduced cross-section exhibits
a slope change too: the data stop rising and, after a turnover, the reduced cross-section starts decreasing.
As a result, the NNLO prediction starts to overshoot the data, whereas the NNLO+NLL$x$ prediction is in reasonable agreement with the
data for $x \lesssim 10^{-4}$.
It is worth observing that the differences between the NNLO and NNLO+NLL$x$  
predictions are relatively small and concern only a limited number of points.
By looking at the bottom panels in Fig.~\ref{fig:datathcomp}, where we show the ratio 
to the experimental data, we see that
the two predictions differ by at most $10\%$ and only for the smallest values of $x$.  
Yet the combined HERA dataset is so precise that the improvement in the 
description provided by small-$x$ resummation is clearly visible at the $\chi^2$
level, as was shown in Tab.~\ref{tab:chi2tab_pertorder}, and will be discussed further
below in Sect.~\ref{subsec:heramap}.

The improved description of the inclusive reduced cross-section data at small-$x$
can be in part traced back to the role of the longitudinal structure function $F_L(x,Q)$. 
As reviewed in Sect.~\ref{sec:th-review}, $F_L$
is particularly sensitive to the effects of small-$x$
resummation, and in particular to deviations from
the DGLAP framework.
The reason is that it vanishes at the Born level, and therefore
it receives gluon-initiated contributions already
at its first non-trivial order.
As shown in Fig.~\ref{fig:resummedSFfixedPDF_FL}, the
differences between the NNLO and NNLO+NLL$x$ can be as large
as $\sim 30\%$ at the lowest $x$ and $Q^2$ bins for which there
are data available.
As a consequence, at small-$x$ and small-$Q^2$ the contribution of $F_L$ to
$\sigma_{r,\rm NC}$ can be significant, see Eq.~(\ref{eq:sigmaRed2}),
thus partly explaining the
differences between the NNLO and NNLO+NLL$x$ predictions observed
in Fig.~\ref{fig:datathcomp}.
Therefore, it is useful to compare the predictions also
for the longitudinal structure function $F_L$ in the NNLO
and NNLO+NLL$x$ fits.

\begin{figure}[t]
\centering
  \includegraphics[page=1,width=0.79\textwidth]{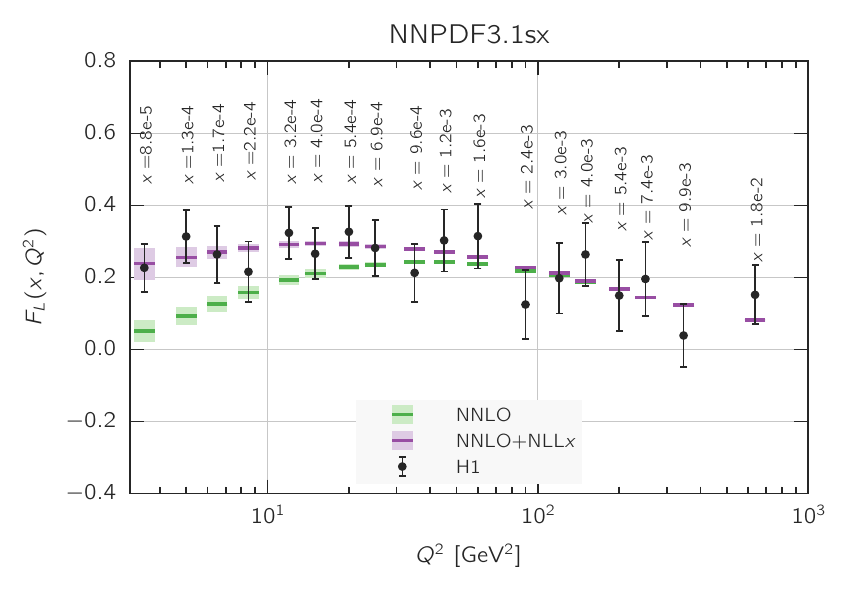}
  \caption{\small The longitudinal structure function
    $F_L(x,Q^2)$ as a function of $Q^2$ for different $x$ bins
    for the most recent H1 measurement~\cite{Andreev:2013vha},
    comparing the results of the NNLO and NNLO+NLL$x$ fits.}
  \label{fig:FLres}
\end{figure}

In Fig.~\ref{fig:FLres} we compare the latest measurements of
$F_L$ from the H1 collaboration~\cite{Andreev:2013vha}\footnote
{The $F_L$ structure function has also been measured by the ZEUS collaboration~\cite{Abramowicz:2014jak},
  but with a reduced kinematic coverage of the small-$x$ region.
  The ZEUS measurement is in mild tension with the H1 measurement, 
though it is affected by larger experimental uncertainties.}
with the predictions from the NNPDF3.1sx NNLO and NNLO+NLL$x$ fits.
Note that our fits already include the constraints
from $F_L$, not directly but rather via
its contribution to the NC reduced cross-section,
Eq.~(\ref{eq:sigmaRed2}).
In this comparison, the experimental uncertainties have been added
in quadrature, and each value of $Q^2$
corresponds to a different $x$ bin as indicated in the plot.
The NNPDF3.1sx results are shown down to
the smallest scale for which one can reliably compute a 
prediction,\footnote{The H1 measurement includes
three further bins at small-$Q^2$, reaching down to $Q^2=1.5$ GeV$^2$.} which is set
by the initial parametrization scale $Q_0^2=2.69$~GeV$^2$.

We see that for $Q^2 \lsim 100$~GeV$^2$ there are significant differences
between the NNLO+NLL$x$ and the NNLO predictions, which 
can be traced back to a combination of the corresponding differences for 
the input small-$x$ gluon
and those in the splitting and coefficient functions 
(see Fig.~\ref{fig:resummedSFfixedPDF_FL}).
The NNLO+NLL$x$ result is larger than the NNLO result by a significant amount: 
at $Q^2 \simeq 10$~GeV$^2$, the resummed calculation is more than a factor 2 larger
than the NNLO result.
Moreover, while at NNLO $F_L$ starts becoming negative
at small $x$ and $Q^2$ (below the scale where the positivity constraints are
imposed in the NNPDF fits) the NNLO+NLL$x$ result instead exhibits
a flat behavior even for the smallest values of $Q^2$.
The larger value of $F_L$ with the NNLO+NLL$x$ theory leads to a lower reduced cross 
section at high $y$, with a more pronounced turnover, thus giving a better description of $\sigma_{r,\rm NC}$ at small-$x$, as shown in Fig.~\ref{fig:datathcomp}.

\begin{figure}[t]
\centering
  \includegraphics[width=0.495\textwidth,page=1]{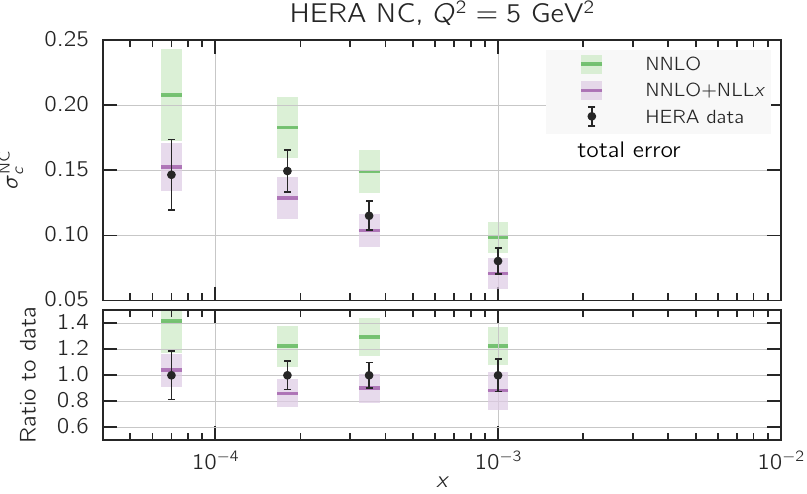}
  \includegraphics[width=0.495\textwidth,page=1]{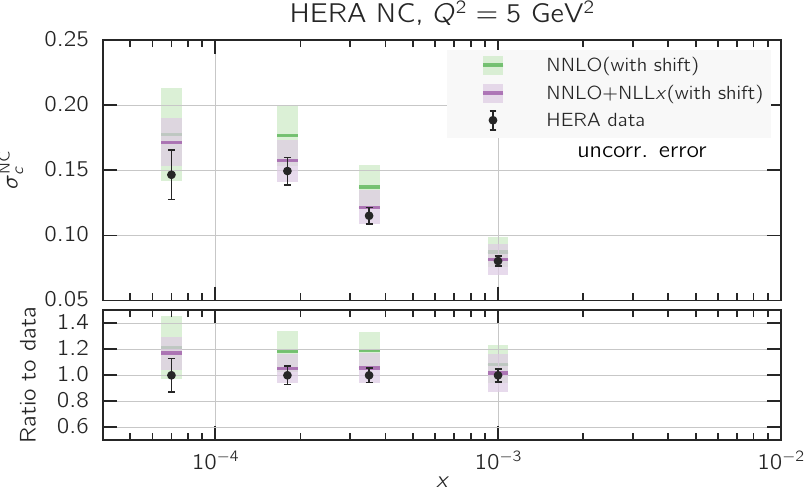}
  \includegraphics[width=0.49\textwidth,page=2]{plots/HERAF2C_samerange_NNLO.pdf}
  \includegraphics[width=0.49\textwidth,page=2]{plots/HERAF2C_samerange_NNLO_shift.pdf}
  \caption{\small Same as Fig~\ref{fig:datathcomp} for the
  HERA charm production cross-sections.}
  \label{fig:datathcompF2c}
\end{figure}

Finally, in Fig.~\ref{fig:datathcompF2c} we show a similar
comparison to that of  Fig~\ref{fig:datathcomp}, this time
for the HERA charm production reduced cross-sections.
Here we also show the two $Q^2$ bins about the
lower $Q^2_{\rm min}$ cut, which in this case correspond to
the $Q^2=5$ and 7 GeV$^2$ bins.
We find that especially for the bin with $Q^2=5$ GeV$^2$,
the NNLO+NNL$x$ prediction agrees well with the HERA data while
the NNLO one overshoots it.
We remind again the reader that these
graphical comparisons do not
take into account the correlations between systematic 
uncertainties.
The large difference between the $\chi^2$ at NNLO and at NNLO+NLL$x$
is therefore only partially captured by Fig.~~\ref{fig:datathcompF2c}.
As we shall see in greater detail in Sect.~\ref{subsec:heramap}, 
also in this case the deterioration of the NNLO $\chi^2$
with respect to the NNLO+NNL$x$ result
shown in Table~\ref{tab:chi2tab_pertorder} stems mostly
from the low-$Q^2$, low-$x$ bins.

Note that the HERA charm cross-sections are extracted
from the experimentally measured
fiducial cross-section~\cite{Abramowicz:1900rp}
by extrapolation
to the full phase space using the fixed-order $\mathcal{O}( \as^2)$
calculation of the {\tt HVQDIS} program~\cite{Harris:1997zq}, based  on the fixed-flavour number scheme.
This should be contrasted with the inclusive neutral current structure
function measurements, which are determined from the outgoing
lepton kinematics and therefore do not assume any theory input.
Given that we have shown that fixed-order and resummed predictions
for $F_2^c$ can exhibit important differences at small-$x$,
such theory-based extrapolation based on the 
$\mathcal{O}(\as^2)$ fixed-order
calculation might introduce a bias whose size is difficult to quantify.
It is quite possible that a more consistent analysis of the raw data based instead on an extrapolation using resummed theoretical predictions might further improve the already good
agreement of the extracted charm cross-section with the NNLO+NLL$x$ fit. 

\subsection{Quantifying the onset of small-$x$ resummation in the HERA data}\label{subsec:heramap}
In this section we resort to a number of
statistical estimators to identify more precisely  
the onset of small-$x$ resummation in the inclusive
and charm HERA measurements.
First, we perform 
a detailed $\chi^2$ analysis,
which we then complement by a study of the pulls
between theory and HERA data.

 \subsubsection{$\chi^2$ analysis}\label{sec:chi2}

The $\chi^2/N_{\rm dat}$ values summarized in Table~\ref{tab:chi2tab_pertorder}
indicate that the fit quality of the inclusive HERA structure functions
improves when resummation effects are included: 
this is particularly true at NNLO, where
the total $\chi^2$ drops by $\Delta\chi^2=-121$ units in the NNLO+NLL$x$ fit.
We now want to identify the origin of this improvement,
and investigate to what extent it arises from 
a better description of the data
in the small-$x$ and small-$Q^2$ region where the
effects of small-$x$ resummation are expected to be most important.

To achieve this goal, we have recomputed the $\chi^2/N_{\rm dat}$ values of the HERA inclusive
and charm cross-sections
using the NNPDF3.1sx NLO, NNLO, NLO+NLL$x$, and NNLO+NLL$x$ global fits with the default
choice $H_{\rm cut}=0.6$, excluding those data points for which 
\beq
\label{eq:dcutdef}
\as(Q^2) \ln\frac1x  \geq D_{\rm cut} \, .
\eeq
The condition Eq.~\eqref{eq:dcutdef} is designed to exclude data 
for which the small-$x$ logarithmic terms are expected to be of the same size at all orders in
the coupling $\as$,
thus potentially spoiling the perturbative behaviour of the theoretical predictions at fixed order.

\begin{figure}[t]
  \centering
    \includegraphics[width=0.49\textwidth]{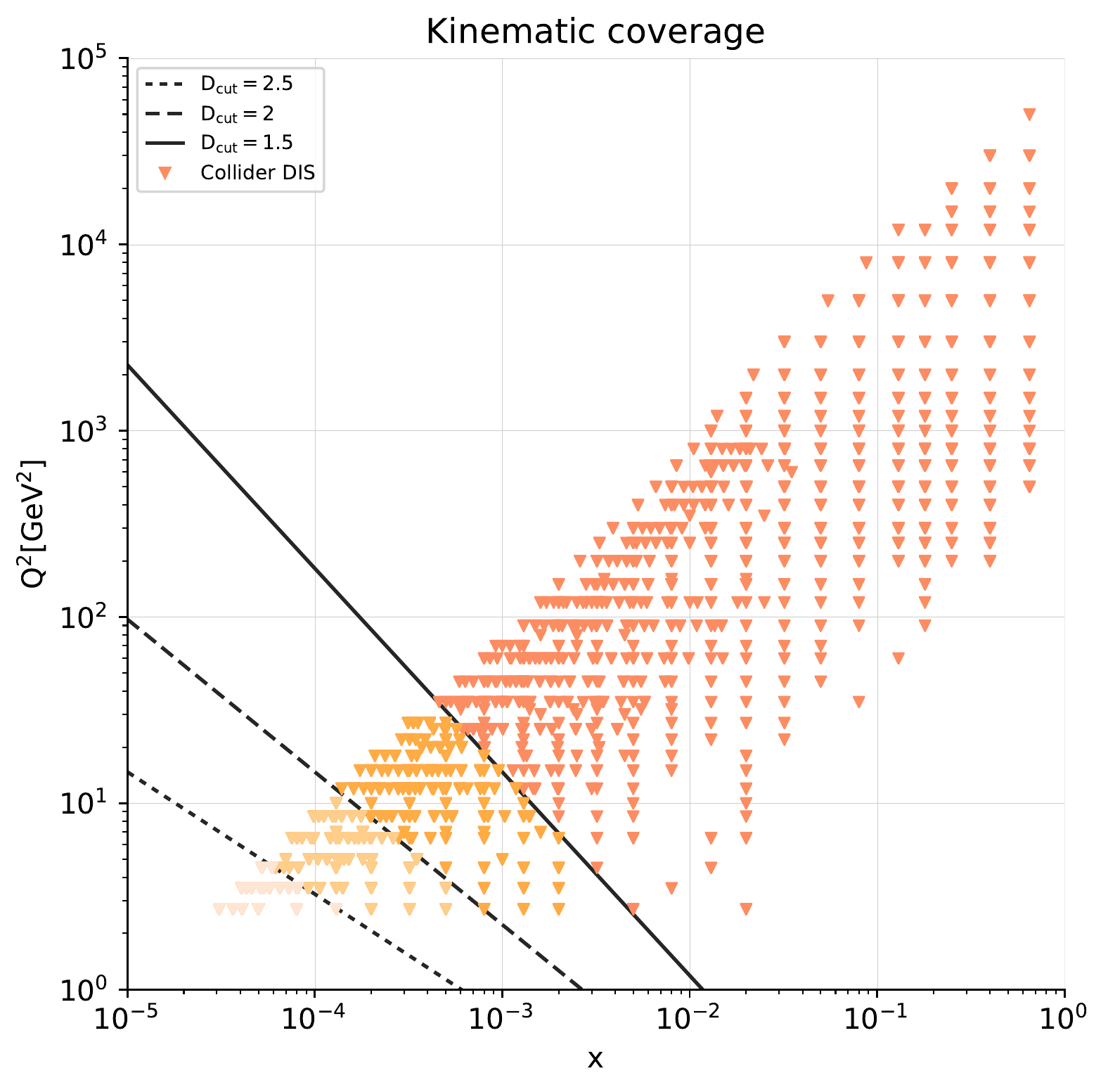}
    \includegraphics[width=0.49\textwidth]{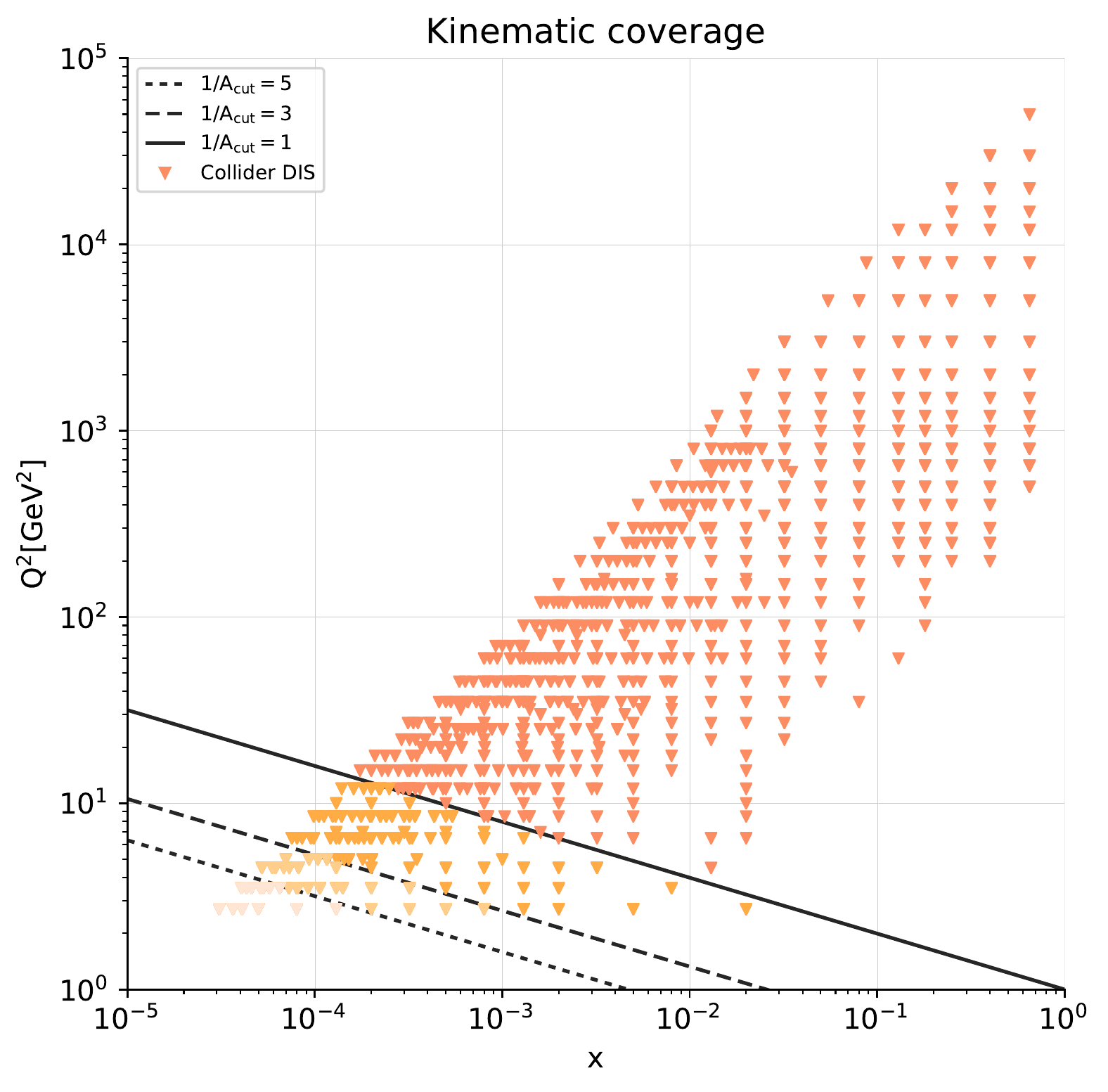}
    \caption{\small The kinematic coverage of the
      HERA inclusive structure function data that enters
      the NNPDF3.1sx fits. 
      The tilted lines represent representative values of the
      cut to DIS data applied after the fit to study evidence for
      BFKL effects at small-$x$ and small-$Q^2$.
   Left plot: perturbative-inspired cut Eq.~(\ref{eq:dcutdef});
   right plot: saturation-inspired cut
   Eq.~(\ref{eq:saturationcut}).
   Note that the data points affected by the various cuts are plotted
   with different shades.
    }
  \label{fig:kincutDIS}
\end{figure}

From basic considerations (see also Sect.~\ref{sec:kincuts}), one would expect fixed-order perturbation
theory to break down for $\as(Q^2) \ln\frac1x$ of order $1$.
The parameter $D_{\rm cut}$ should thus be of order $1$ as well.
By varying the value of $D_{\rm cut}$, 
we can vary the number of data points excluded from the computation of the $\chi^2/N_{\rm dat}$.
For sufficiently small values of $D_{\rm cut}$, all contributions
which potentially spoil perturbation theory should be cut away,
and we should thus find that small-$x$ resummation does not improve the 
quality of the fit.
Then as we increase $D_{\rm cut}$, more data points at small $x$ and $Q^2$ will be included, and the effects of the resummation should become apparent.
A kinematic plot showing the HERA structure function
data which are cut for various values of $D_{\rm cut}$ 
is shown in the left panel of Fig.~\ref{fig:kincutDIS}.
We emphasize that this cut should not be confused with the
$H_{\rm cut}$ cut defined in Eq.~\eqref{eq:hcutdef}, which was used to determine which hadronic data enter in the fit;
here the parameter $D_{\rm cut}$ applies only to DIS structure functions and is used as an
\emph{a posteriori} diagnosis tool after the fit has been performed.

\begin{figure}[t]
  \centering
    \includegraphics[width=0.49\textwidth]{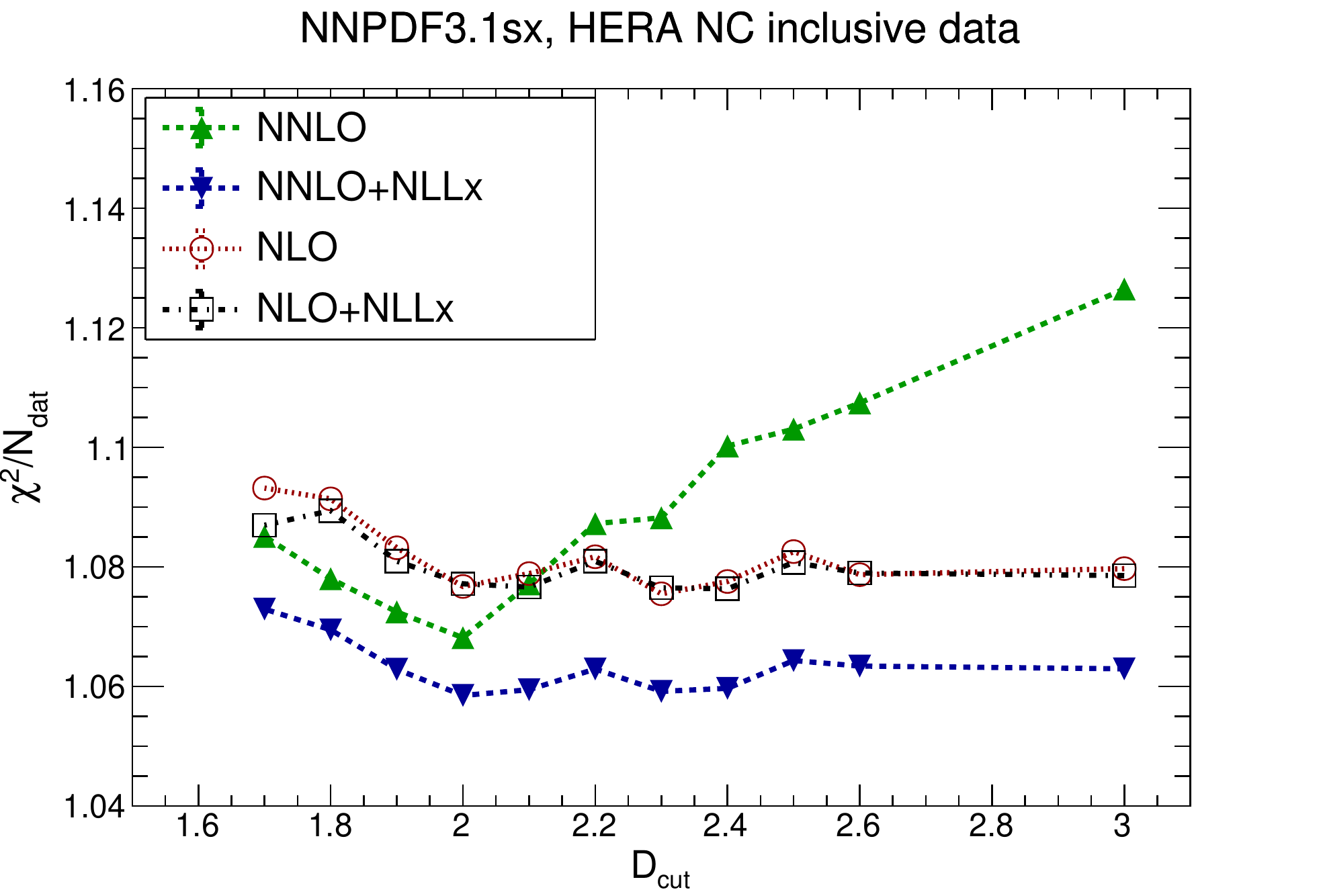}
    \includegraphics[width=0.49\textwidth]{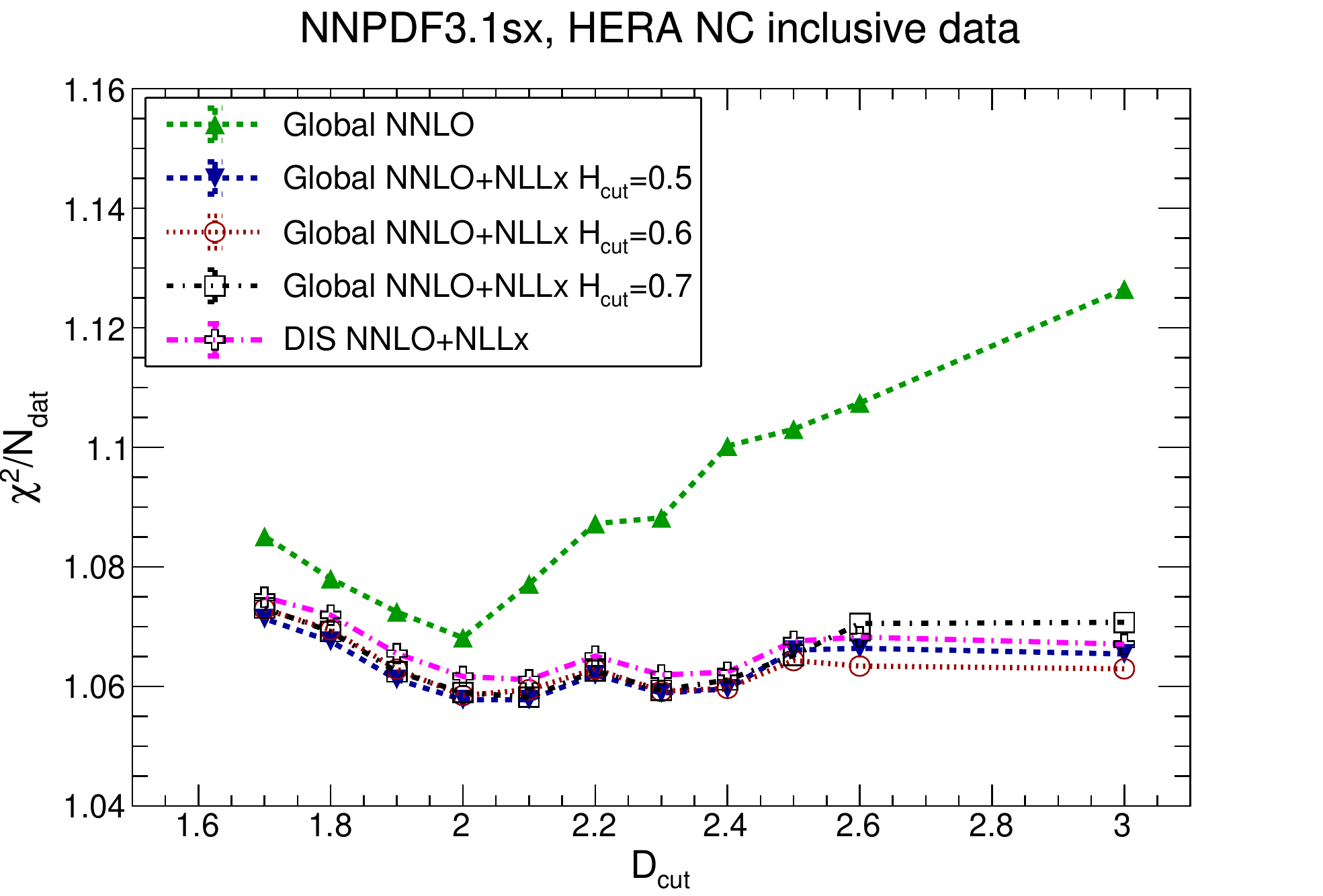}
    \includegraphics[width=0.49\textwidth]{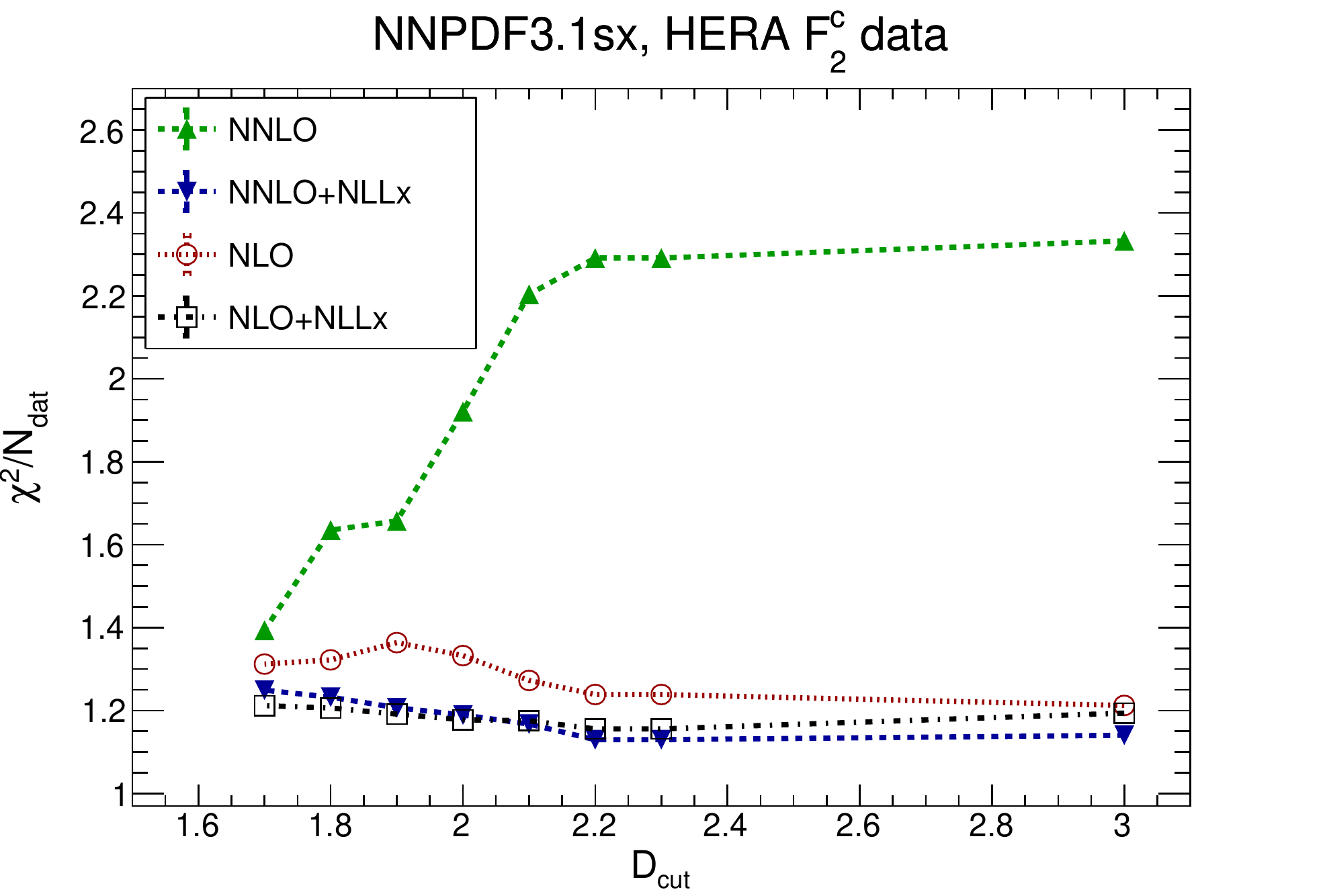}
    \includegraphics[width=0.49\textwidth]{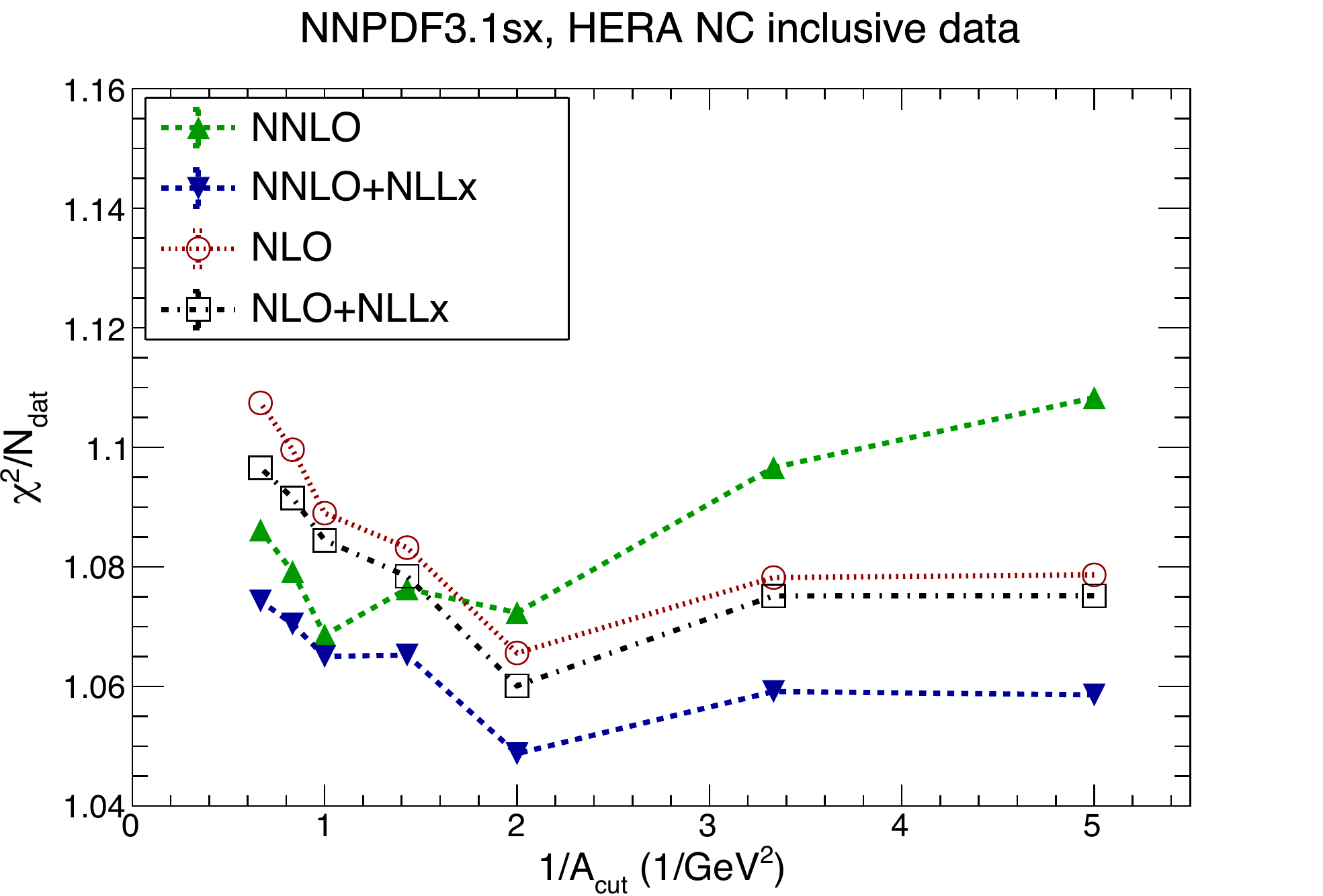}
   \caption{\small Upper left: the values of $\chi^2/N_{\rm dat}$
     in the NNPDF3.1sx global fits
    for the HERA NC inclusive structure function data for different
    values of the cut $D_{\rm cut}$ Eq.~(\ref{eq:dcutdef}), comparing
     the results of the NLO, NLO+NLL$x$, NNLO,
    and NNLO+NLL$x$ fits.
    Upper right: same comparison, now between the global NNLO
    and NNLO+NLL$x$ baseline fits with the
    NNLO+NLL$x$ global fits with $H_{\rm cut}=0.5$
    and 0.7 and with the DIS-only fit.
    Bottom left: same as above for the HERA
    charm production data.
    Bottom right: same as upper left,   now
     with the saturation-inspired
     cut Eq.~(\ref{eq:saturationcut}).
       \label{fig:chi2-profile-hera-smallx}
   }
\end{figure}

In Fig.~\ref{fig:chi2-profile-hera-smallx} we display the values of $\chi^2/N_{\rm dat}$
for the HERA neutral current inclusive (top left) and charm (bottom left)
reduced cross-sections as a function of $D_{\rm cut}$.
First of all, we observe that at NNLO the $\chi^2/N_{\rm dat}$ increases
sharply for $D_{\rm cut}\gsim 2$, or, equivalently, as more
data from the small-$x$ and small-$Q^2$ region are included, both for
the inclusive and the charm data.
On the other hand, this trend disappears for the NNLO+NLL$x$ fits:
in this case the value of $\chi^2/N_{\rm dat}$ is flat
for all $D_{\rm cut}$ values in the studied range.

Another interesting feature of these plots
is that the stability with respect to the value of $D_{\rm cut}$ is also present
for the NLO and NLO+NLL$x$ fits.
Indeed, the $\chi^2/N_{\rm dat}$ values for the NLO, NLO+NLL$x$, and NNLO+NLL$x$ fits all exhibit a rather similar
shape.
This is of course a consequence of the fact that, as shown in Sect.~\ref{sec:results},
the PDFs obtained from the fits using these
three theories are rather close to each other, whereas the NNLO PDFs are very different at small $x$.
Remarkably, for the inclusive data especially the NNLO+NLL$x$  fits lead to a 
better $\chi^2/N_{\rm dat}$ than the NLO and NLO+NLL$x$ ones, presumably due to 
the additional NNLO corrections included in the NNLO+NLL$x$ matched 
calculations.
This result highlights the importance of the NNLO corrections for 
the optimal description of the medium and large-$x$ HERA data.

The results of Fig.~\ref{fig:chi2-profile-hera-smallx}
demonstrate that fixed-order NNLO theory does not provide a satisfactory 
description of either the inclusive or charm DIS data at 
small $x$ and small $Q^2$.
The better description is instead achieved by including NLL$x$
effects, providing direct evidence of the need for small-$x$ resummation  
at small-$x$.
Moreover, we observe that the rise in the $\chi^2/N_{\rm dat}$
values of the NNLO fits becomes very significant for $D_{\rm cut}\gsim 2$.
This means that BFKL effects at NNLO approximately start to become important when
\be
\ln \frac{1}{x} \gsim 1.2 \ln \frac{Q^2}{\Lambda^2} \, ,
\ee
see Eq.~(\ref{eq:kincutc}),
which implies, for instance, that the effects of small-$x$ resummation become phenomenologically relevant
around
$x\simeq 8\times 10^{-4}$ ($2.7\times 10^{-4}$) for $Q^2=2.7$ GeV$^2$ (6.5 GeV$^2$).
This estimate is consistent with the results presented in Sects.~\ref{sec:th-review} and~\ref{sec:results}.

To study whether the treatment of the hadronic data in the PDF fits can modify this conclusion,
in the upper right panel of Fig.~\ref{fig:chi2-profile-hera-smallx} we
also compare the $\chi^2/N_{\rm dat}$
values as a function of $D_{\rm cut}$ for the NNPDF3.1sx NNLO+NLL$x$ global
fits with the three $H_{\rm cut}$ values
discussed in Sect.~\ref{sec:dephcut}, namely
$H_{\rm cut}=0.5$, $0.6$ and $0.7$, 
as well as with the global $H_{\rm cut}=0.6$ NNLO fit
and the NNLO+NLL$x$ DIS-only fit.
These comparison illustrate that our quantitative conclusions are to a very good
approximation independent of the specific cut applied to 
the hadronic data: very similar NNLO+NLL$x$ results are found
in the global fit irrespective of the value of $H_{\rm cut}$, as well
as for the corresponding DIS-only fit.
We have also verified that the same conclusion holds for the NLO and NLO+NLL$x$ fits.

In Refs.~\cite{Caola:2010cy,Caola:2009iy}, a similar cutting 
exercise was performed, but in that case the specific form of the
cut to the small-$x$ and small-$Q^2$ data was inspired by
saturation arguments. 
Specifically, the condition used to exclude data points was
\be
\label{eq:saturationcut}
Q^2 x^{\lambda} \ge A_{\rm cut} \, ,
\ee
with $\lambda=0.3$.
The value of $A_{\rm cut}$ determines how stringent is the cut:
the larger its value, the more data points excluded (so $1/A_{\rm cut}$ behaves
qualitatively in the same way as $D_{\rm cut}$).
While the inspiration for the cut Eq.~(\ref{eq:saturationcut}) is different
from that of Eq.~(\ref{eq:dcutdef}) (which is based instead
on perturbative considerations), the practical result is the same,
with only some differences on the exact shape of the cut
in the $(x,Q^2)$ plane (see the right panel of Fig.~\ref{fig:kincutDIS}).
The results for the $\chi^2/N_{\rm dat}$ as a function of $1/A_{\rm cut}$
are shown in the bottom right panel of Fig.~\ref{fig:chi2-profile-hera-smallx},
and indeed confirm that the trend is essentially the same, irrespectively of the
specific details of how the small-$x$ and $Q^2$ data are cut.

In summary, the results collected in Fig.~\ref{fig:chi2-profile-hera-smallx}
clearly demonstrate the
onset of BFKL dynamics in the small-$x$ and $Q^2$ region for both
the inclusive and charm HERA data.
Specifically, we find that the use of NNLO+NLL$x$ theory gives
the best description of the HERA data in
the small-$x$ region, while NNLO theory gives a significantly worse description.
Moreover, our results also allow us to determine 
the kinematic region where small-$x$ resummation effects
start to become phenomenologically relevant, thus providing useful
guidance to estimate their reach at the LHC as well as for future colliders.
  
\subsubsection{Pull analysis}
\label{sec:pull}

A complementary approach to further investigate the onset 
of BFKL dynamics in the low-$x$ region, and
to make connection with the analysis of Refs.~\cite{Caola:2010cy,Caola:2009iy},
is provided by 
the calculation of the relative pull between experimental data
and theory.
This relative pull is defined as
\be
\label{eq:relativedistances}
P_i^{\rm rel}(x,Q^2)\equiv  \frac{\big|\sigma_{{\rm data},i}-\sigma_{{\rm th},i}\big|}{
(\sigma_{{\rm data},i}+\sigma_{{\rm th},i})/2 
} \, ,
\ee
where the normalization is given by the average of central values.
  This estimator allows us to  quantify the absolute
 size of the differences between data and theory in units
 of the cross-section itself.
Here we focus on the results computed with NNLO and
NNLO+NLL$x$  theory, using the  NNPDF3.1sx sets obtained in the 
respective global fits with the default cut $H_{\rm cut}=0.6$.

\begin{figure}[t]
  \begin{center}
    \includegraphics[width=0.49\textwidth]{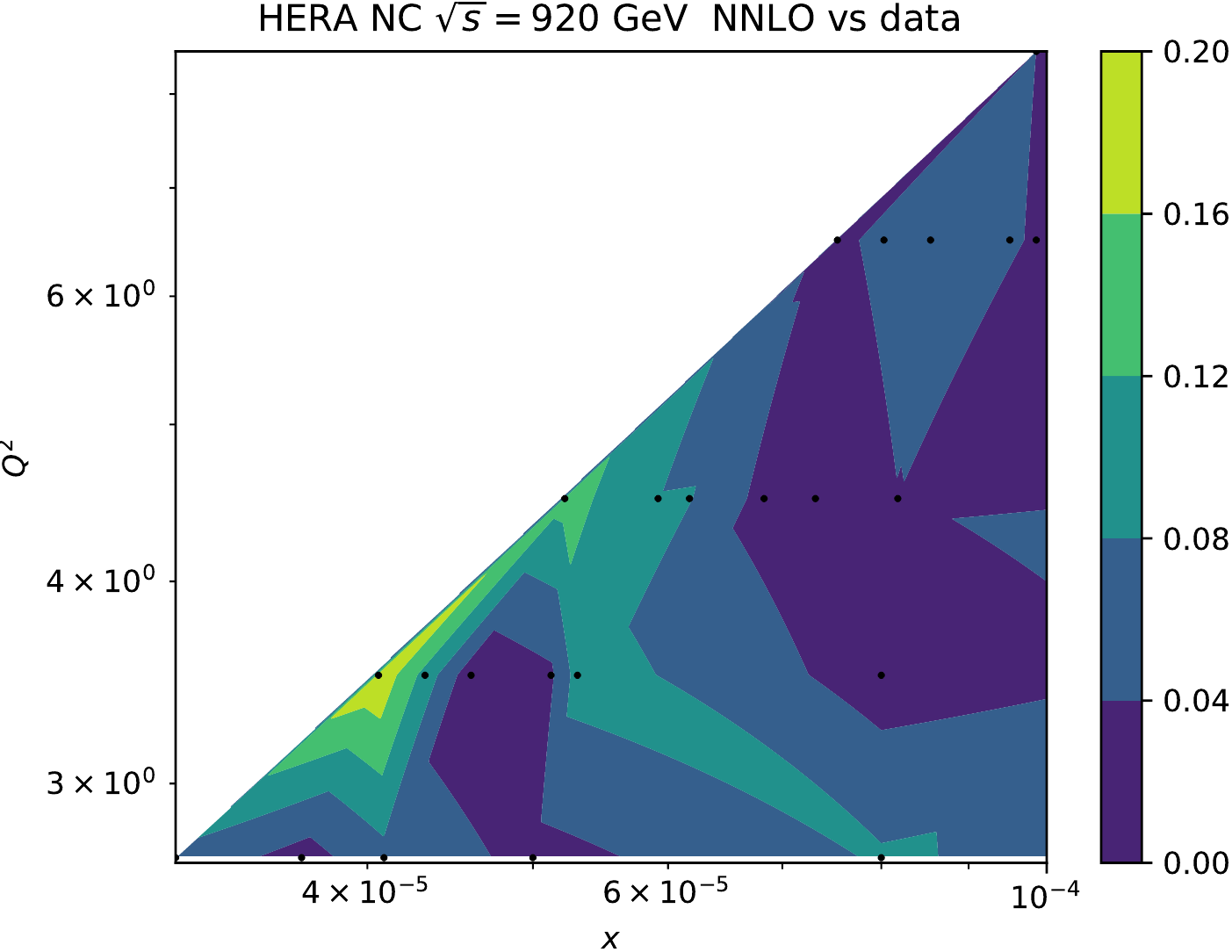}
    \includegraphics[width=0.49\textwidth]{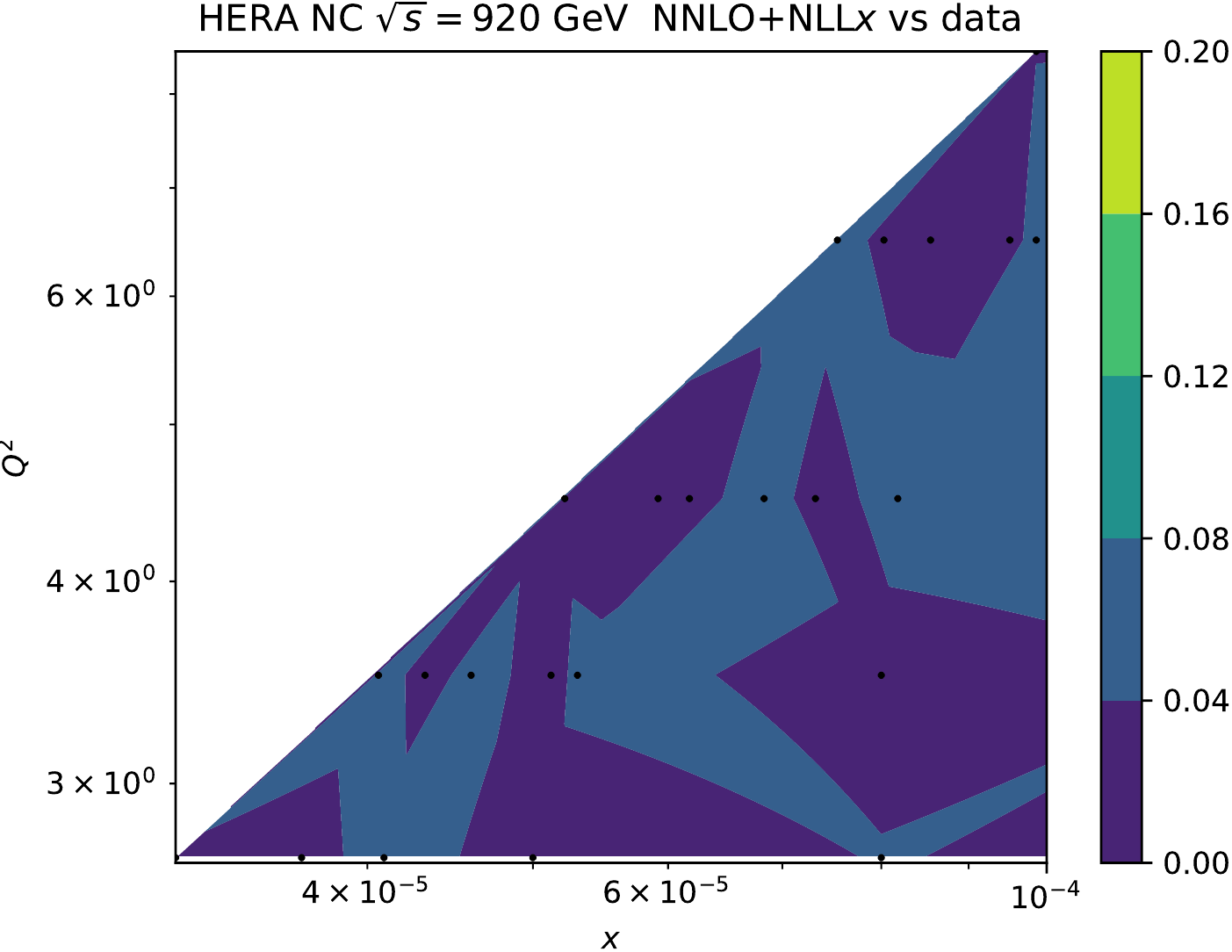}
    \caption{Left panel: interpolated representation of the relative pull
      Eq.~(\ref{eq:relativedistances})
      between the HERA NC reduced cross-section data at $\sqrt{s}=920$ GeV and the
      NNLO fit, in the small-$x$
      and small-$Q^2$ region.
      Right panel: same as the left panel now
      for the NNLO+NLL$x$ fit.
  }
  \label{fig:reldiff-data}
\end{center}
\end{figure}

To visualize the differences between data and theory
in the small-$x$ and small-$Q^2$ region, we can represent  
the relative pull $P_i^{\rm rel}(x,Q^2)$, Eq.~(\ref{eq:relativedistances}),
as a function of ($x$,$Q^2$) in the relevant region of the kinematic
plane. 
In Fig.~\ref{fig:reldiff-data} we  show an interpolated
representation of $P^{\rm rel}(x,Q^2)$
for the HERA neutral-current dataset
at $\sqrt{s}=920$ GeV and the NNLO
and NNLO+NLL$x$ fits.
In the case of the NNLO fit, the relative differences between
theory and data can be up to $\sim 20\%$ at small-$x$ and $Q^2$, and reduce
to less than a few percent at larger $x$ or $Q^2$.
On the other hand, the agreement between data and theory is markedly
improved in the case of the NNLO+NLL$x$ fit: the quality
of the data description is essentially the same everywhere in the region
considered,
and the relative differences between data and theory
are a everywhere below the $8\%$ level.
These plots show that by using NNLO+NLL$x$ theory, one can
achieve a satisfactory description of the inclusive HERA measurements
in the entire region spanned by the available data.

\begin{figure}[t]
  \centering
  \includegraphics[scale=0.59]{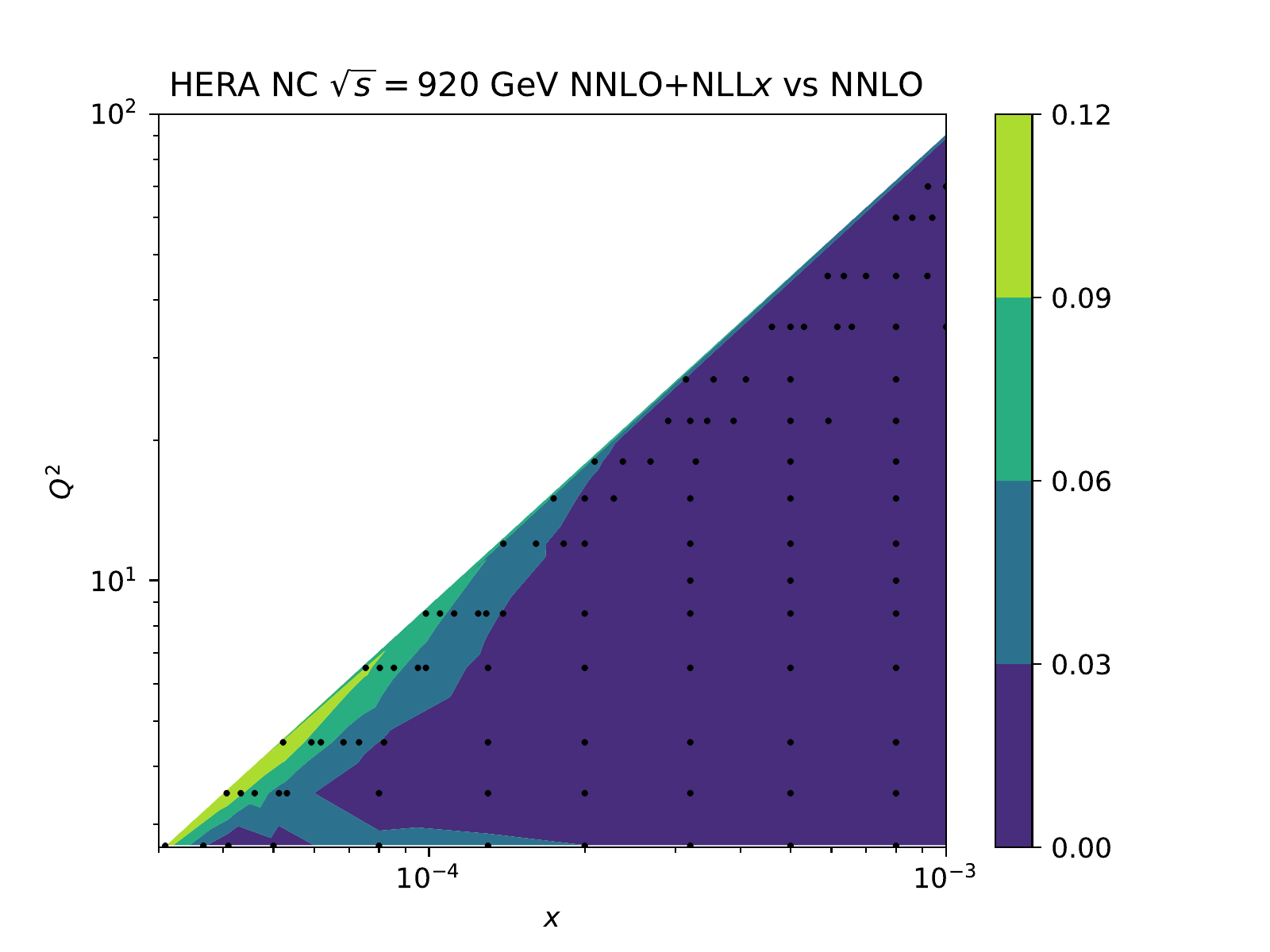}
  \caption{Same as Fig.~\ref{fig:reldiff-data}, now for the relative difference
    in the theoretical predictions of the HERA reduced cross-sections
    between the NNLO and NNLO+NLL$x$ fits, Eq.~(\ref{eq:relativedistances2}).
    Note the different color code and $x$ and
    $Q^2$ ranges with respect to Fig.~\ref{fig:reldiff-data}.}
  \label{fig:reldiff-th}
\end{figure}

In order to further quantify differences and similarities between the NNLO and NNLO+NLL$x$
theoretical predictions,
in Fig.~\ref{fig:reldiff-th} we show
a similar relative pull as in Eq.~(\ref{eq:relativedistances}),
now between the theoretical predictions
for the HERA reduced cross-sections
obtained with the NNLO and the NNLO+NLL$x$ theory and fits, namely
\be
\label{eq:relativedistances2}
\widetilde{P}_i^{\rm rel}(x,Q^2)\equiv \frac{
  \big|\sigma^\text{NNLO+NLL$x$}_{{\rm th},i}
  - \sigma^{\rm NNLO}_{{\rm th},i}  \big|
}{ (\sigma^\text{NNLO+NLL$x$}_{{\rm th},i}
  + \sigma^{\rm NNLO}_{{\rm th},i})/2  } \, .
\ee
Note that in this comparison
both the color code and the $(x,Q^2)$ ranges are different
from those of Fig.~\ref{fig:reldiff-data}.
From the results of Fig.~\ref{fig:reldiff-th}
we see that the differences between the cross-sections
computed with NNLO and NNLO+NLL$x$ theory are between 5\%  and 10\%
for $Q^2 \lesssim 10$ GeV$^2$ and $x\lesssim 2\times 10^{-4}$.
Once we move away from this region, differences become smaller.
For $x \gtrsim 2 \times 10^{-4}$, we find that the differences are always
smaller than $\sim 3\%$, for any value of $Q^2$.
This comparison provides a detailed snapshot of the region
in the $(x,Q^2)$ plane where the impact of resummation
is phenomenologically more relevant,
and is consistent with the results shown in Fig~\ref{fig:datathcomp} and 
the conclusions of the $\chi^2$ profile analysis of Sect.~\ref{sec:chi2}.

\subsubsection{Sensitivity to subleading logarithms}

\begin{figure}[ht]
  \centering
    \includegraphics[width=0.49\textwidth]{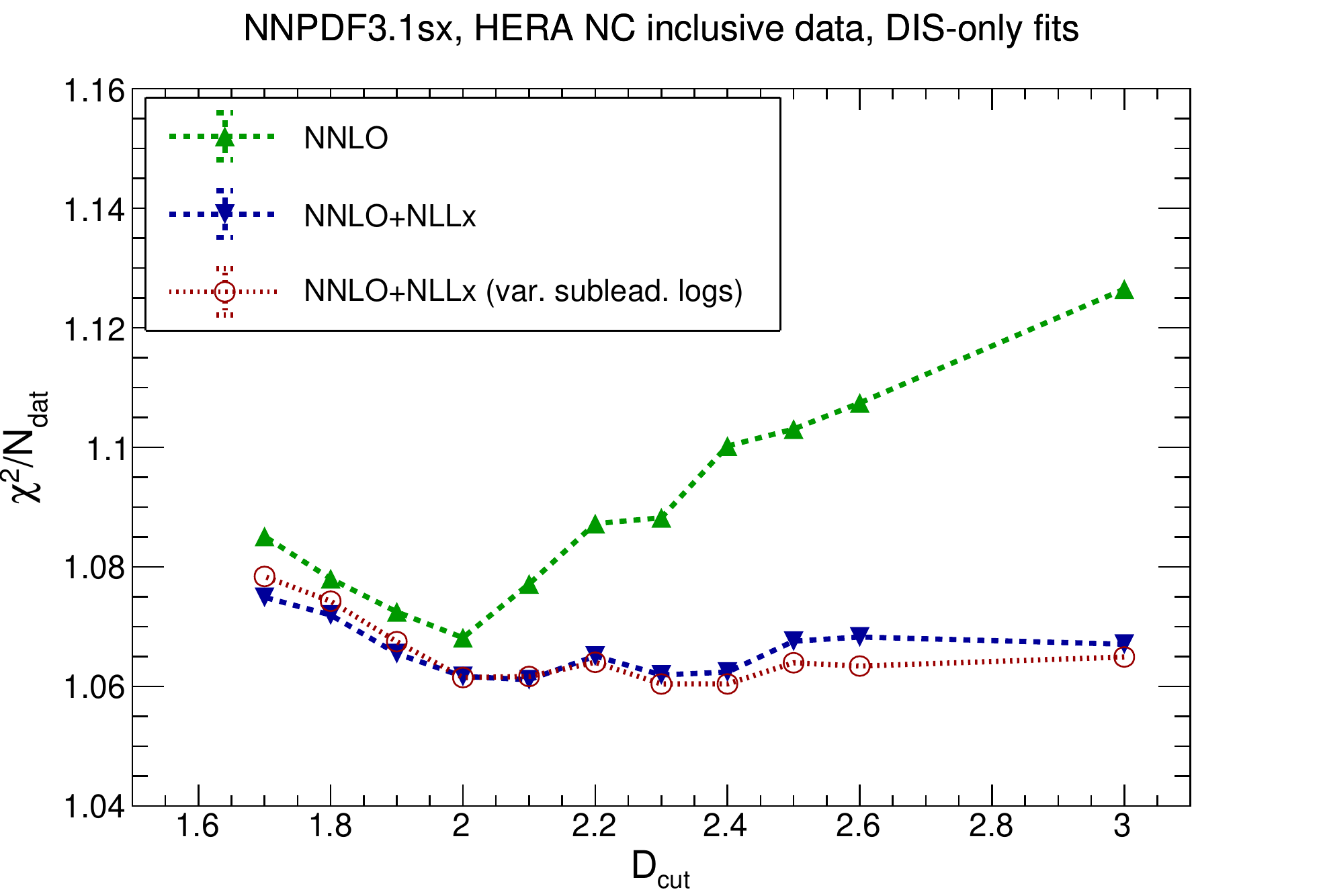}
  \includegraphics[width=0.49\textwidth]{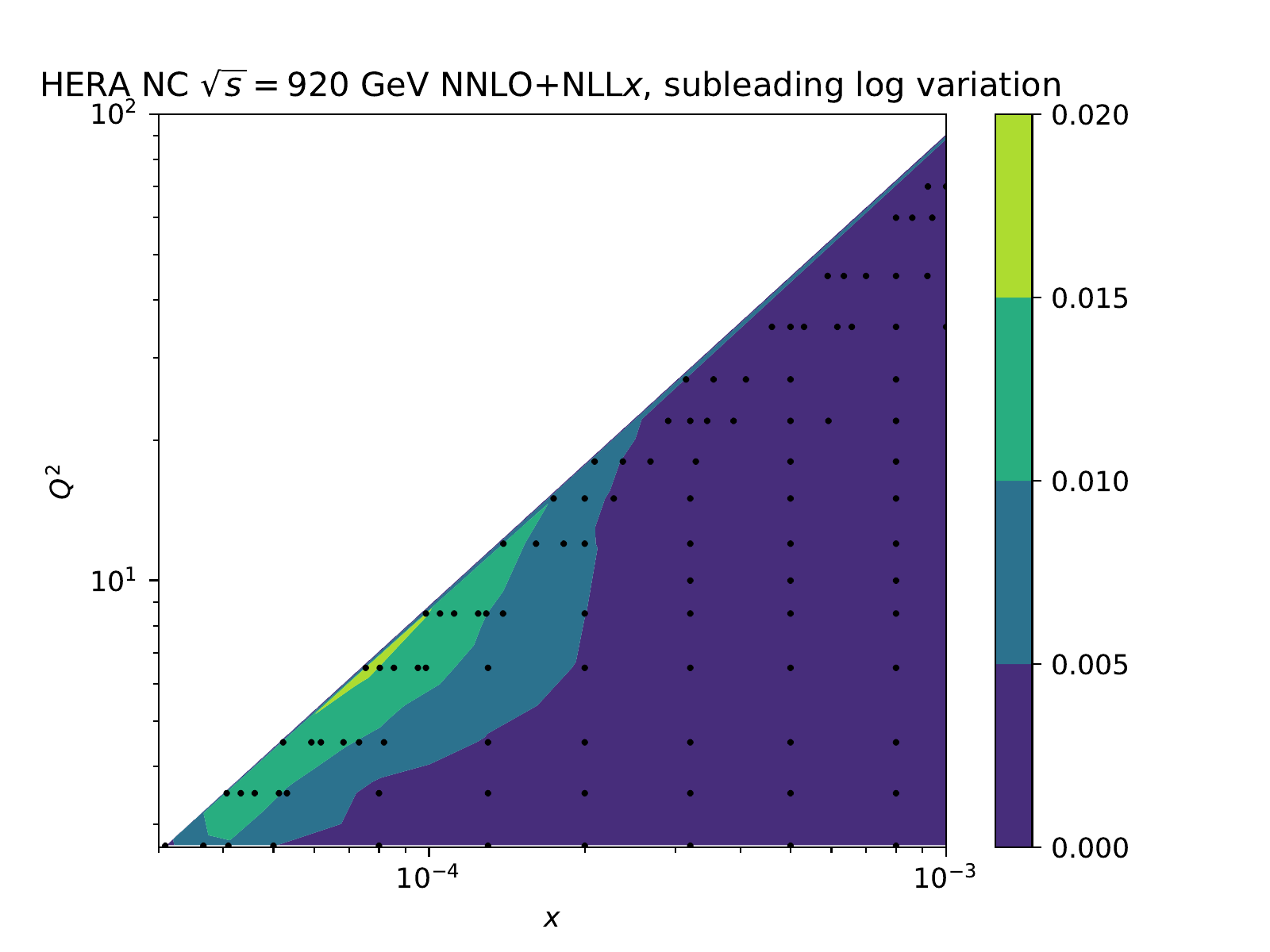}
  \caption{
      Left panel: the values of $\chi^2/N_{\rm dat}$ in the NNPDF3.1sx DIS-only fits
    for the HERA NC inclusive structure function data for different
    values of the cut $D_{\rm cut}$ Eq.~(\ref{eq:dcutdef}), compared to the results of a fit
    where the subleading logarithms are varied. 
      Right panel: same as Fig.~\ref{fig:reldiff-th}, now for the relative difference
    between the NNLO+NLL$x$ DIS-only fit and the NNLO+NLL$x$ DIS-only fit performed with a variation of subleading logarithms.
    Note the different color code with respect to Fig.~\ref{fig:reldiff-th}.
    }
  \label{fig:hellvariation}
\end{figure}

Finally, we can study if our conclusions are stable with respect to variations of unknown subleading logarithms. 
To this end, in the left panel of Fig.~\ref{fig:hellvariation} we show the values of $\chi^2/N_{\rm{dat}}$ for the HERA NC inclusive reduced cross-section as a function of $D_{\rm{cut}}$, now comparing the DIS-only fit at NNLO and NNLO+NLL$x$ to the NNLO+NLL$x$ fit where subleading logarithms are introduced as described in  Section~\ref{sec:disonlyfits}. 
We observe that the two NNLO+NLL$x$ profiles are very similar, with the $\chi^2/N_{\rm{dat}}$ of the alternative fit being slightly lower at larger $D_{\rm{cut}}$. 
To better quantify the differences between the two variants, in the right panel of 
Fig.~\ref{fig:hellvariation} we also show the relative pull Eq.~\eqref{eq:relativedistances2} between the
theoretical predictions for the two NNLO+NLL$x$ fits for the HERA neutral current $\sqrt{s}=920$ GeV dataset.
We observe that the relative difference is at most $2\%$ for the smallest values of $x$ and $Q^2$ probed by the dataset, and is below $0.5\%$ for all $x \gsim 3\times 10^{-4}$, independently of the value of $Q^2$.
This analysis shows that our results are stable with respect to variations of subleading logarithms.

\subsection{Impact of the small-$x$ HERA data on PDFs at medium and large-$x$}
\label{sec:impactlargex}
In the last part of
this section, we present results of additional NNPDF3.1sx NNLO fits where
we have {\it removed} a number of
HERA structure function data points in the small-$x$ and
$Q^2$ region, in order to study how the resulting PDFs are affected
by the use of such reduced dataset.
This exercise allows us to understand to what extent 
existing NNLO global PDF fits might be biased by
fitting low-$x$ data while neglecting the effects of small-$x$ resummation.
Since we have just demonstrated that at small-$x$ 
the HERA structure functions prefer
NNLO+NLL$x$ theory to the NNLO one,
it may be advisable to apply
dedicated kinematic cuts in the small-$x$ and $Q^2$ region in standard
NNLO analyses.
This would ensure on one hand that the fitting 
dataset corresponds to a region where a fixed-order
perturbative expansion is reliable, and on the other hand that
the estimate of the uncertainties at small-$x$ is more reliable.

For this purpose, we have performed variants of the NNPDF3.1sx NNLO global fit
without any cut on the hadronic data (that is, $H_{\rm cut}=\infty$) but where
instead we impose 
the cut Eq.~(\ref{eq:dcutdef}) to the DIS structure function data
{\it before} fitting,
thus reducing the number of data points in the small-$x$ and small-$Q^2$ region. 
Specifically, we have performed NNLO fits with $D_{\rm cut}=1.7,\, 2.0$ and $2.3$,
as a well as a fit without cutting any data ($D_{\rm cut}=\infty$) as a reference.
The motivation for this range of $D_{\rm cut}$ values
is the observation (see Fig.~\ref{fig:chi2-profile-hera-smallx})
that $D_{\rm cut}\simeq 2$ indicates
the region where the effects of
small-$x$ resummation start to become significant.

\begin{figure}[t]
  \centering
  \includegraphics[width=0.49\textwidth]{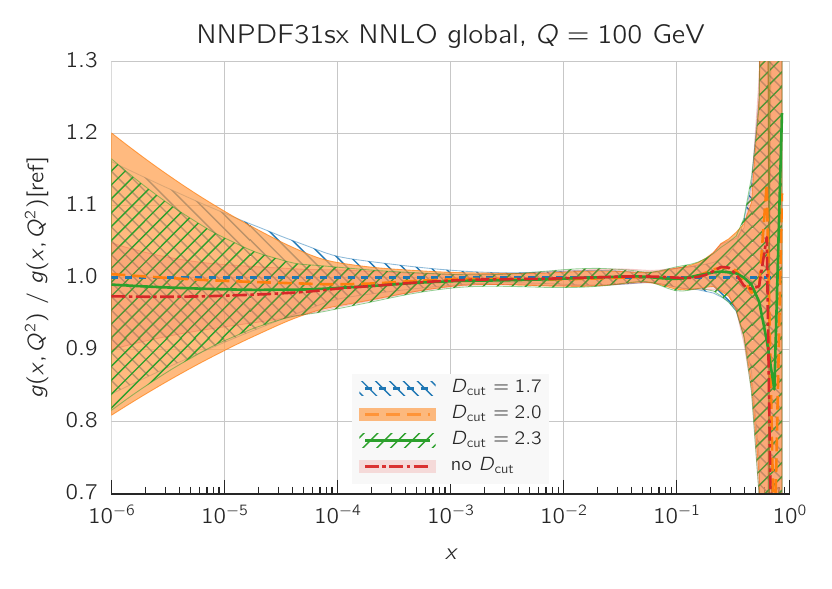}
  \includegraphics[width=0.49\textwidth]{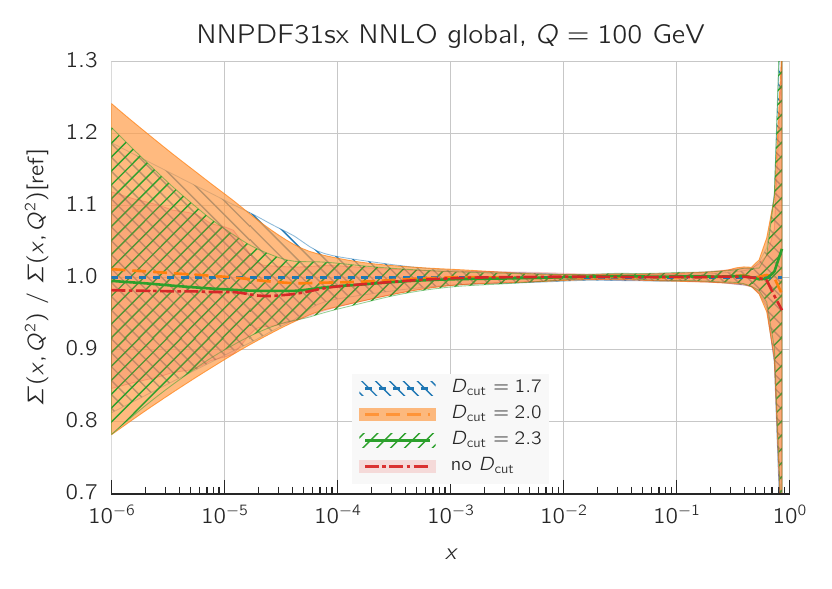}
  \includegraphics[width=0.49\textwidth]{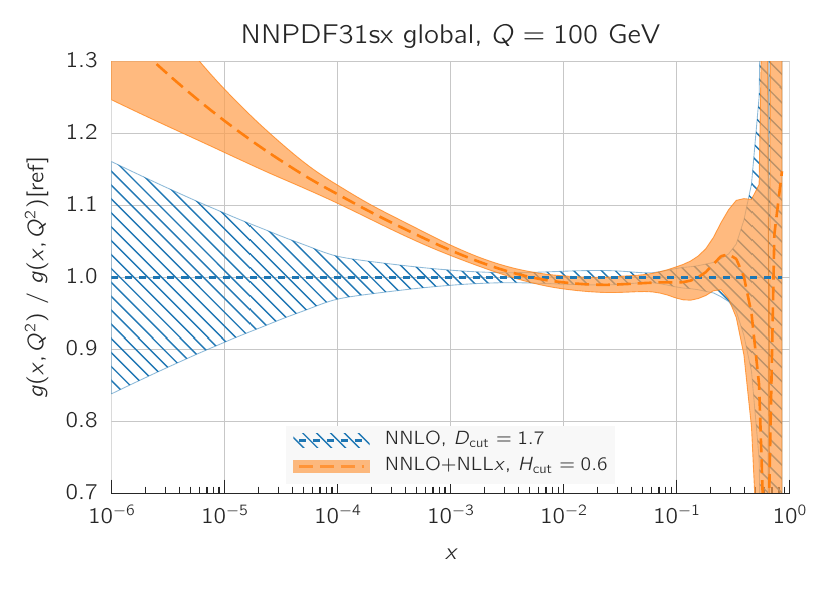}
  \includegraphics[width=0.49\textwidth]{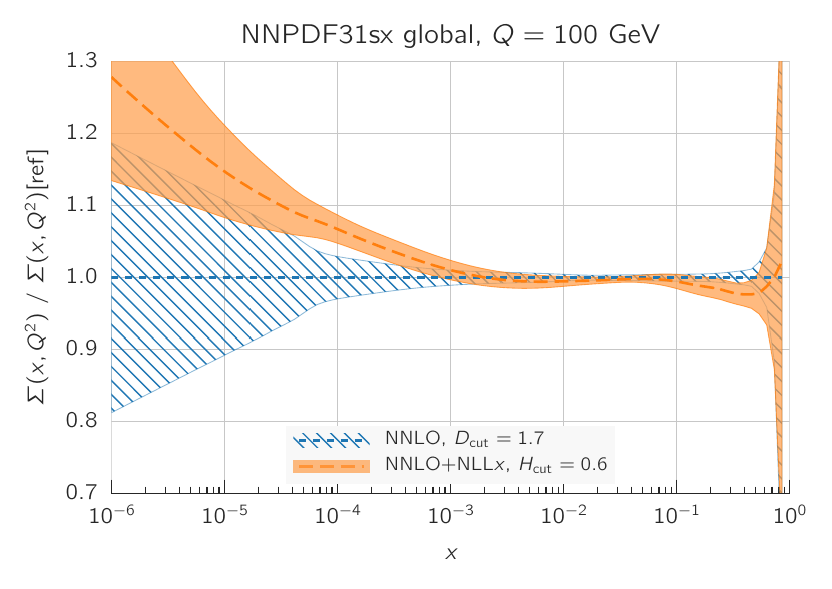}
  \caption{\small Upper plots: comparison between the gluon
    and quark singlet at  $Q=100$ GeV from the NNPDF3.1sx NNLO fits with various values
    of $D_{\rm cut}$ with the corresponding fit without that kinematic cut.
    Bottom plots: comparison between the NNLO fit with $D_{\rm cut}=1.7$
    and the baseline NNLO+NLL$x$ fit (with $H_{\rm cut}=0.6$).
    Both comparisons are shown
    normalized to the central value of the $D_{\rm cut}=1.7$ fit.}
  \label{fig:pdfcomp-Dcut}
\end{figure}

The comparison between the NNPDF3.1sx NNLO fits with different cuts to
the DIS data is shown in the upper plots of Fig.~\ref{fig:pdfcomp-Dcut}.
Specifically, we show the comparison between the gluon
and quark singlet from the fits with various values
of $D_{\rm cut}$ with the corresponding fit without that cut
($D_{\rm cut}=\infty$) at $Q=100$ GeV.
From this comparison we can see that --- as expected --- the higher the value
of $D_{\rm cut}$, the smaller are the PDF uncertainties at small-$x$ due
the increase in kinematic coverage of the fitted HERA data. However, the central values remain very stable, even at the lowest values of $x$.
Additionally, we also see that for $x\gsim 5\times 10^{-4}$
the gluon and quark singlet are extremely stable with respect to
the $D_{\rm cut}$ variations, both in terms of central value and
of PDF uncertainties.
Therefore, we conclude
that current NNLO fits are not biased
in the region relevant for precision LHC phenomenology, even
if the fits include points at small-$x$ while neglecting  
resummation effects.
      
It is also interesting to compare the NNPDF3.1sx NNLO fit with $D_{\rm cut}=1.7$
with our default NNLO+NLL$x$ global fit, namely the one with the cut in hadronic data corresponding to $H_{\rm cut}=0.6$, but $D_{\rm cut}=\infty$.
This comparison allows us to understand if the PDF uncertainties of the conservative
NNLO fits, where data points at small-$x$ and small-$Q^2$ have been removed,
account for
the PDF shift induced by using the more accurate NNLO+NLL$x$ theory.
We show this comparison in the bottom 
plots of Fig.~\ref{fig:pdfcomp-Dcut} at the scale $Q=100$ GeV.
Whereas for the quark singlet the shift between the NNLO+NLL$x$
and NNLO fits is mostly covered by the corresponding PDF uncertainties,
for the gluon the shift in central values is bigger
and is not covered by the PDF uncertainties, despite the 
larger PDF errors of the fit with the conservative dataset.

The results of Fig.~\ref{fig:pdfcomp-Dcut}
suggest that at small-$x$ the theoretical uncertainties associated
with the NNLO gluon are comparable to or larger than the PDF uncertainties.
Moreover, we observe that the shift induced
by NNLO+NLL$x$ theory is covered by the PDF uncertainties only for
values of $x$ larger than $x\simeq 3\times 10^{-3}$.
Therefore, for processes sensitive to the small-$x$ region,
including a number of LHC cross-sections,
current NNLO PDF uncertainties do not fully account for the total
theoretical uncertainty, suggesting that the use of NNLO+NLL$x$ theory would
lead to more reliable theoretical predictions.


\section{\boldmath Small-$x$ phenomenology at the LHC and beyond}
\label{sec:pheno}

In this section we explore some of the phenomenological implications of the
NNPDF3.1sx fits.
First of all, we present a first assessment
of the possible impact of NLL$x$ resummation at the LHC.
We then move to DIS-like processes, for which we can produce
fully consistent NNLO+NLL$x$ predictions. 
In this context, we consider the implications of the NNPDF3.1sx fits
for the ultra-high energy (UHE) neutrino-nucleus cross-sections as well
as for future
high-energy lepton-proton colliders such as the LHeC~\cite{AbelleiraFernandez:2012cc}
and the FCC-eh~\cite{Mangano:2016jyj,Contino:2016spe},
illustrating the key role that small-$x$ resummation could play
in shaping their physics program.

\subsection{Small-$x$ resummation at the LHC}
In this section we perform a first
exploration of the potential effects
of small-$x$ resummation on precision LHC phenomenology.
We start by considering 
parton luminosities and then we
estimate the effects of small-$x$ resummation for electroweak
gauge boson production at the LHC.
The latter study will however
be necessarily only qualitative, since as explained
in Sect.~\ref{sec:th-review} the relevant small-$x$ resummed
partonic cross-sections are not yet implemented in a format amenable
for phenomenological applications.
The studies of this subsection will thus be complementary 
to previous estimates of the effects of small-$x$ resummation on inclusive LHC 
processes in which both evolution and cross-section were resummed, 
but the PDFs used were fixed rather than refitted~\cite{Ball:2007ra,Marzani:2009hu}.

\subsubsection{Parton luminosities}
In order to provide
a first insight on the possible
impact of NLL$x$ resummation effects on hadronic cross-sections,
it is useful to consider its effects on the parton luminosities. 
We consider both the integrated parton luminosity Eq.~(\ref{eq:1D-lumi})
and also luminosities differential in rapidity (see {\it e.g.}~\cite{Campbell:2006wx})
\begin{equation}
\label{eq:lumi2D}
\frac{d \mathcal L_{ij}}{dy} \left(x,\mu^2, y \right) = f_i \left(\sqrt{x} e^y,\mu^2\right) f_j \left( \sqrt{x}e^{-y},\mu^2 \right) \, .
\end{equation}
We assume the production of a hypothetical final state with invariant mass $M_X$,
so that $x=M_X^2/s$ with $\sqrt{s}$ being the LHC center-of-mass energy,
and we take the factorization scale to be $\mu^2=M_X^2$.
Despite offering only a qualitative estimate of the effects of small-$x$ resummation, 
parton luminosities contain the bulk of the information 
about the partonic contributions to a given process.
In particular, rapidity-dependent PDF luminosities 
provide a direct mapping between regions in the
$(x,Q^2) $ plane (PDF sensitivity) and those in the $\lp M_X,y\rp $ plane (kinematic
coverage for collider production), assuming leading order production kinematics.

\begin{figure}[t]
\centering
  \includegraphics[width=0.49\textwidth,page=1]{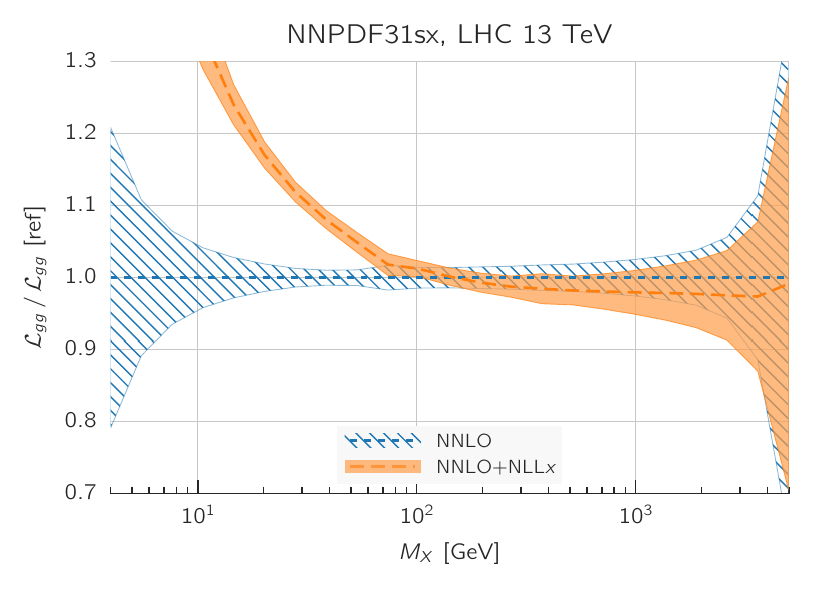}
  \includegraphics[width=0.49\textwidth,page=2]{plots/NNLO_1D.pdf}
  \includegraphics[width=0.49\textwidth,page=3]{plots/NNLO_1D.pdf}
  \includegraphics[width=0.49\textwidth,page=4]{plots/NNLO_1D.pdf}
  \caption{\small The gluon-gluon, quark-gluon, quark-antiquark
    and quark-quark  PDF luminosities, Eq.~(\ref{eq:1D-lumi}), at $\sqrt{s}=13$
    TeV as a function of the final-state invariant mass $M_X$,
    comparing the NNPDF3.1sx  NNLO and NNLO+NLL$x$ global fits.}
  \label{fig:1Dlumi-nnlo}
\end{figure}

Let us start with the integrated parton luminosities, Eq.~(\ref{eq:1D-lumi}).
In Fig.~\ref{fig:1Dlumi-nnlo} we show the  gluon-gluon, quark-gluon, quark-antiquark
    and quark-quark PDF luminosities at $\sqrt{s}=13$
    TeV as a function of the invariant mass $M_X$,
    comparing the NNPDF3.1sx global fits based on
    NNLO and NNLO+NLL$x$ theory respectively.
    For the two gluon-initiated
    luminosities, the effects are very large for $M_X \lesssim 100$ GeV,
    and smaller above that value.
    For the $gg$ luminosity, for instance, the ratio
    between the NNLO and NNLO+NLL$x$ results can be more than
    $\sim 30\%$ for $M_X \simeq 10$ GeV, a region
    relevant for instance for open $B$-meson
    production.
    Even larger effects can be expected for processes at smaller invariant masses,
    such as $D$-meson or $J/\Psi$ production.
   At $M_X \gtrsim 100$~GeV, a region relevant for e.g.\ top quark pair 
   production, the gluon induced luminosities are instead reduced, albeit by only a few per cent.

     In the case of the quark-antiquark and quark-quark luminosities,
     the differences due to resummation are less significant, with the NNLO and
     NNLO+NLL$x$  luminosities in agreement within PDF uncertainties for the entire $M_X$ range.
    However, the effects of small-$x$ resummation are not negligible; for instance,
    they could still be as large as 10\% at $M_X=10$ GeV, a region
    probed by the LHC in processes such as
    low-mass Drell-Yan production.
    At larger invariant masses the differences between NNLO
    and NNLO+NLL$x$ are again down to 1-2\%.
    Thus, in this region the effects are small, but nevertheless of the same order
    as the experimental uncertainties of recent high-precision LHC measurements,
    such as for instance the ATLAS 2011 $W,Z$ rapidity distributions~\cite{Aaboud:2016btc} or
    the CMS $Z$ $p_T$ distributions~\cite{Khachatryan:2015oaa}.
         
    It is important to emphasize here that the luminosity
    comparison in Fig.~\ref{fig:1Dlumi-nnlo} provides 
    only a rough estimate of the actual differences
    between the NNLO and the fully resummed
    NNLO+NLL$x$ cross-sections, since a quantitative assessment requires 
    the resummation of the partonic cross-sections for the relevant 
    processes, and this can be as large as the difference in the 
    luminosities~\cite{Ball:2007ra,Marzani:2009hu}.
    This said, the results of Fig.~\ref{fig:1Dlumi-nnlo} show that the effects
    of small-$x$ resummation are potentially significant for LHC cross-sections,
    in particular for those with large gluon-initiated contributions.

\begin{figure}[t]
\centering
  \includegraphics[width=0.49\textwidth,page=1]{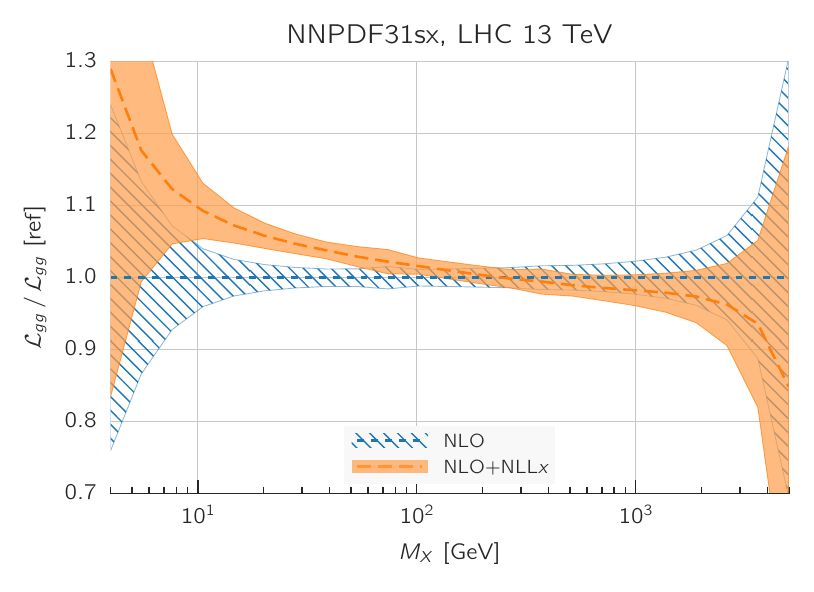}
  \includegraphics[width=0.49\textwidth,page=2]{plots/NLO_1D.pdf}
  \includegraphics[width=0.49\textwidth,page=3]{plots/NLO_1D.pdf}
  \includegraphics[width=0.49\textwidth,page=4]{plots/NLO_1D.pdf}
  \caption{\small Same as Fig.~\ref{fig:1Dlumi-nnlo}, now
    comparing the results of the NLO+NLL$x$ and NLO fits.}
  \label{fig:1Dlumi-nlo}
\end{figure}

Next, in Fig.~\ref{fig:1Dlumi-nlo} we show same comparison but
this time between the NLO and NLO+NLL$x$ fits. As discussed
in Sect.~\ref{sec:results}, we expect the differences to be more moderate
compared to the NNLO fits case.
Indeed, the differences are now much smaller, both for the gluon-initiated
and for the quark-initiated luminosities.
The most significant effect of resummation can again be seen in the 
$gg$ luminosity, but now only at the $10\%$ level at $M_X\lesssim 10$~GeV. The other luminosities all agree within uncertainties.
Henceforth, we will focus on the comparison between the NNLO+NLL$x$ and and the NNLO fits, as in all cases the corresponding
differences between NLO+NLL$x$ and NLO would always be much smaller.

\begin{figure}[p]
\centering
  \includegraphics[width=0.49\textwidth,page=1]{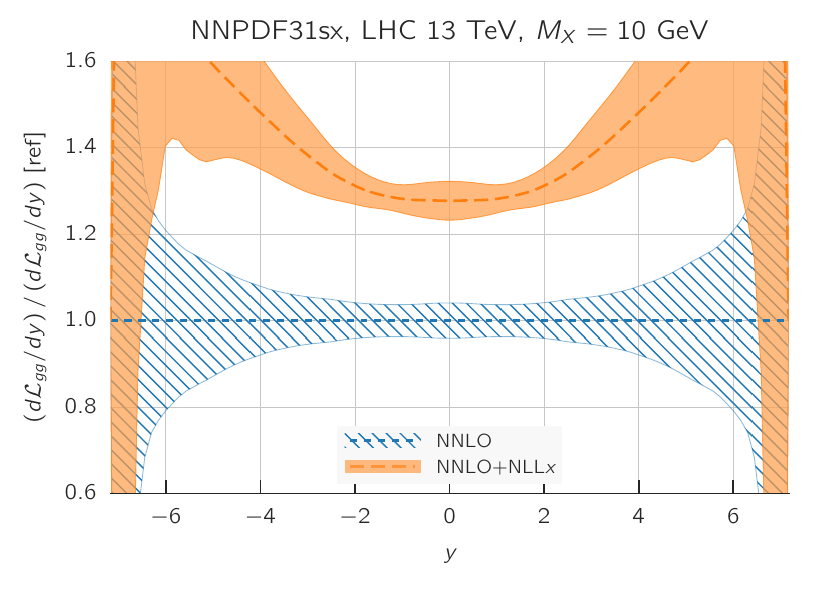}
  \includegraphics[width=0.49\textwidth,page=4]{plots/NNLO_2D.pdf}
  \includegraphics[width=0.49\textwidth,page=2]{plots/NNLO_2D.pdf}
  \includegraphics[width=0.49\textwidth,page=5]{plots/NNLO_2D.pdf}
  \includegraphics[width=0.49\textwidth,page=3]{plots/NNLO_2D.pdf}
  \includegraphics[width=0.49\textwidth,page=6]{plots/NNLO_2D.pdf}
  \caption{\small The double differential PDF luminosities 
    as a function of $\mu=M_X$ and $y$, Eq.~(\ref{eq:lumi2D}), comparing
    the gluon-gluon (left plots) and quark-antiquark (right plots)
    luminosities between the NNLO and NNLO+NLL$x$ fits normalized
    to the central value of the former.
    We show the results as a function of $y$ for $M_X=10,30,100$~GeV (top to bottom).}
  \label{fig:2Dlumi}
\end{figure}

Now we move to compare PDF luminosities which are differential in rapidity, Eq.~(\ref{eq:lumi2D}).
As already mentioned,
these luminosities allow for a more direct mapping between the final
state kinematics and the regions of $x,Q^2$ of the underlying PDFs.
For simplicity, we focus here on the gluon-gluon and quark-antiquark luminosities, as
the behaviour of the gluon-quark and quark-quark is closely related to 
these two.
In Fig.~\ref{fig:2Dlumi} we compare the PDF luminosities of
    the NNLO and NNLO+NLL$x$ fits, normalized
    to the central value of the former.
    We show the results as a function of $y$ for three different values of $M_X$,
    namely 10 GeV, 30 GeV, and 100 GeV.

From the comparisons of Fig.~\ref{fig:2Dlumi} we see that
the impact of small-$x$ resummation depends on
the final state rapidity $y$, and it increases close
to the kinematic endpoints. This is expected as large (or small) values of the rapidity probe small-$x$ values in one of the two partons that make up the parton luminosity, see Eq.~(\ref{eq:lumi2D}).
For instance, in the case of the $gg$ luminosity, for the production of a final state with
invariant mass $M_X=10 \, (30)$ GeV, the ratio between NNLO+NLL$x$ and NNLO
is between 30\% and 50\% (10\% and 20\%), depending on the specific value
of the rapidity.
The differences are smaller in the case of the quark-antiquark
luminosities, though we note that they could become more relevant if the
PDF uncertainties were reduced by
including in the fit data sensitive to the small-$x$ region of the PDFs,
such as the LHCb $W,Z$ forward production cross-sections. 
This would however require the inclusion of
small-$x$ resummation in the partonic cross-sections for the relevant processes.

\subsubsection{Implications for Drell-Yan production}
We now present a first exploration of the possible phenomenological
consequences of small-$x$ resummation for LHC cross-sections, specifically
for the Drell-Yan production process.
We do this by providing estimates for some recent Drell-Yan
cross-section measurements from the LHC,
focusing on those more sensitive to the possible presence of small-$x$ effects,
and comparing the results of the predictions from the NNPDF3.1sx NNLO
and NNLO+NLL$x$ fits, using in both cases the fixed-order NNLO
hard-scattering cross-sections.
These differences likely overestimate the real effect, and in particular 
might be reduced once the resummation in the partonic cross-sections is 
taken into account~\cite{Marzani:2009hu}.
However, we believe 
they provide a 
reliable though conservative estimate of the possible size of the resummation 
effects that can be expected.

\begin{figure}[t]
\centering
  \includegraphics[width=0.49\textwidth,page=1]{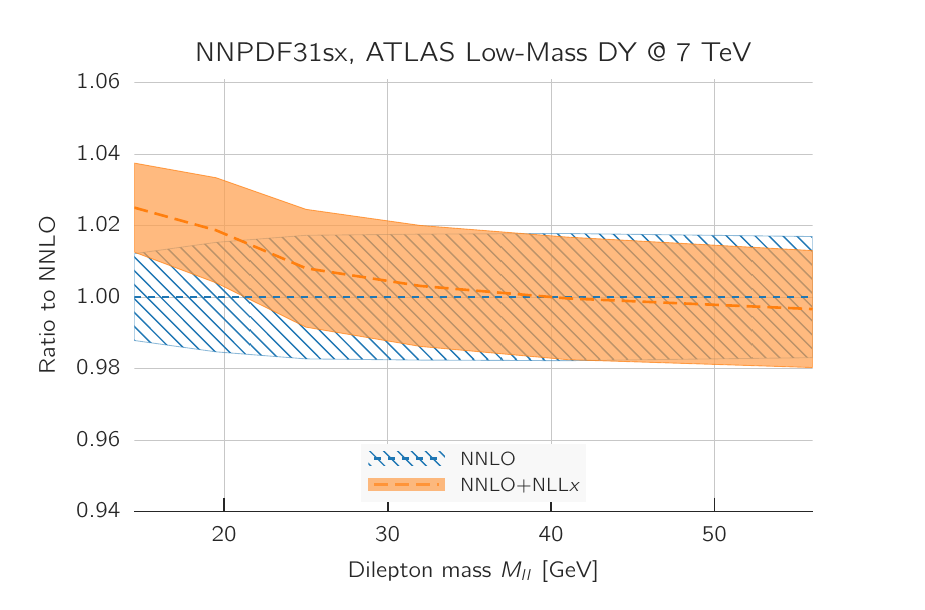}
  \includegraphics[width=0.49\textwidth,page=2]{plots/dyprod.pdf}\\
  \includegraphics[width=0.49\textwidth,page=4]{plots/dyprod.pdf}
  \includegraphics[width=0.49\textwidth,page=5]{plots/dyprod.pdf}
  \caption{\small Comparison between the NNPDF3.1sx NNLO and NNLO+NLL$x$ predictions
    for selected Drell-Yan measurements at the LHC.
    From left to right and from top to bottom, we show the ATLAS low-mass
   measurements at 7 TeV, the CMS low-mass measurements at 8 TeV,
   and the LHCb $W^+$ and $Z$ rapidity distributions at 8 TeV.
   For the  NNLO+NLL$x$ predictions,
   the effects of small-$x$ resummation
   are included in the PDF evolution but not in the partonic cross-sections.}
  \label{fig:LHCpheno}
\end{figure}

Specifically, we show in Fig.~\ref{fig:LHCpheno} the  predictions for the low-mass DY cross-sections
from ATLAS at 7 TeV~\cite{Aad:2014qja}, the lowest invariant mass
bin for the CMS Drell-Yan
cross-sections double-differential in $y$ and $M_{ll}$ at 8 TeV~\cite{CMS:2014jea},
as well as for forward $W^+$ and $Z$ production at 8 TeV from LHCb~\cite{Aaij:2015zlq}.
Note that none of these datasets is included in the NNPDF3.1sx fits,
since as discussed in Sect.~\ref{sec:fitsetup} they are removed by the $H_{\rm cut}$ cut,
Eq.~(\ref{eq:hcutdef}).
We stress once again that we calculate the NNLO+NLL$x$ and NNLO cross-sections 
using the corresponding PDFs from the NNPDF3.1sx fits, but using in both cases the fixed order NNLO
coefficient functions, using the same settings described in Sect.~\ref{sec:fitsetup}.
We do not show the experimental data points in this comparison, as our
aim is to focus on the impact of the resummation rather than a 
comparison with the measured cross-sections.

From the results shown in Fig.~\ref{fig:LHCpheno}, we find that the NNLO and NNLO+NLL$x$ predictions
are consistent within uncertainties in almost all cases.
The differences are more marked for the kinematic regions directly sensitive to
small-$x$, such as small $M_{ll}$ for the ATLAS data and large rapidities in the case
of the LHCb and CMS measurements.
In the latter case, the shift due to NNLO+NLL$x$ theory could be as large as $\sim 5\%$
at the largest rapidities, and the two PDF bands do not overlap for $y > 2$ in the
invariant mass bin considered.

Moreover, since the experimental uncertainties for the cross-sections
shown in Fig.~\ref{fig:LHCpheno}
can be smaller than the corresponding PDF errors (and in
some cases also smaller than the shift between
the NNLO and NNLO+NLL$x$ curves),
we can conclude from this exercise that the inclusion of these data into a fully consistent small-$x$
resummed global PDF fit might provide further evidence for 
BFKL dynamics, this time from high-precision electroweak LHC cross-sections
as opposed to from lepton-proton deep-inelastic scattering.

\subsection{The ultra-high energy neutrino-nucleus cross-section}

We next briefly explore the implication of the NNPDF3.1sx fits
for the calculation of the total neutrino-nucleus cross-sections
at ultra-high energies (UHE).
The interpretation of available
and future UHE data from neutrino telescopes,
such as IceCube~\cite{Aartsen:2014gkd}
and KM3NET~\cite{Adrian-Martinez:2016fdl}, requires
precision predictions for the UHE cross-sections.
With this motivation, a number of phenomenological studies of the UHE
cross-sections and the associated uncertainties has been presented,
both in the framework of collinear
DGLAP
factorization~\cite{Gandhi:1998ri,CooperSarkar:2011pa,
  Anchordoqui:2006ta,Connolly:2011vc,Block:2010ud,CooperSarkar:2007cv,Gluck:1998js}
and beyond it~\cite{Albacete:2015zra,Armesto:2007tg,Arguelles:2015wba,Fiore:2005wf,JalilianMarian:2003wf}, the latter
including for instance the effects of non-linear evolution or saturation.

Here we focus on the charged-current (CC) neutrino-nucleus inclusive cross-sections.
Measuring neutrino-nucleus interactions 
at the highest values of $E_\nu$ accessible at neutrino telescopes 
explores values of $x$ down to $\sim 10^{-9}$ for $Q\sim M_W$, thus representing
a unique testing ground of small-$x$ QCD dynamics.
We have computed the theoretical predictions with {\tt APFEL+HELL} for NNLO and
NNLO+NLL$x$ theory, using the corresponding NNPDF3.1sx fits as input.
Heavy quark mass effects are included using the FONLL scheme,
although these mass corrections are negligible at the
relevant intermediate and high neutrino energies, so the calculation is effectively
a massless one.

\begin{figure}[t]
\centering
  \includegraphics[width=0.49\textwidth]{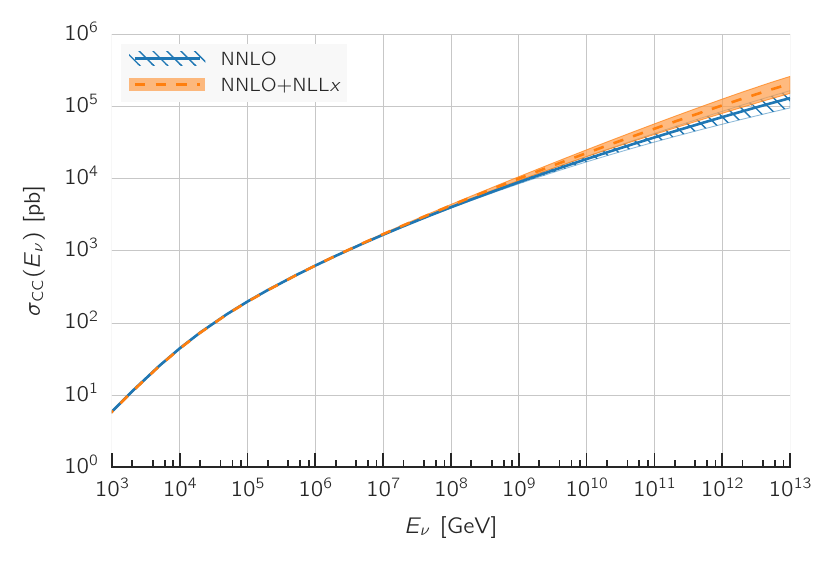}
  \includegraphics[width=0.49\textwidth]{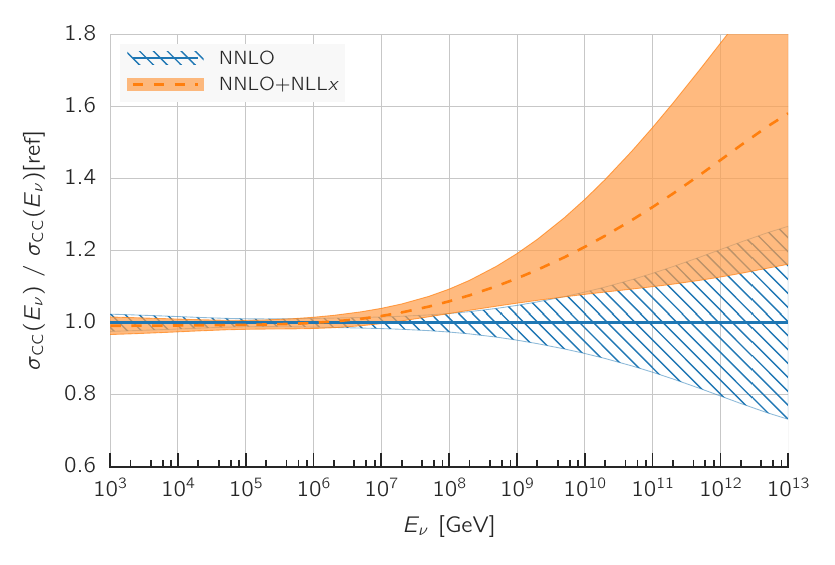}
  \caption{\small The UHE neutrino-nucleus
    charged-current cross-section
    $\sigma_{\rm CC}(E_{\nu})$ as a function of the neutrino energy $E_{\nu}$,
    comparing the results obtained using the NNPDF3.1sx NNLO fits with
    those of the 
    its resummed  NNLO+NLL$x$ counterpart.
  }
  \label{fig:UHExsec}
\end{figure}

In Fig.~\ref{fig:UHExsec} we show the
UHE neutrino-nucleus
    charged-current cross-section
    $\sigma_{\rm CC}(E_{\nu})$ as a function of the neutrino energy $E_{\nu}$ for the fixed-order and for the resummed predictions.
    We show both the absolute cross-sections, and the cross-sections normalized
    to the central value of the NNLO prediction.
    The error bands indicate the one-sigma PDF uncertainties.
    The upper limit in $E_{\nu}$ corresponds to the foreseeable range
    of the current generation of neutrino telescopes.

As we can see from the comparison of Fig.~\ref{fig:UHExsec}, the main effect
of small-$x$ resummation is to increase the cross-section at the highest
energies, by an amount that can be as large as $50\%$ or more.
The PDF errors are however large, and the NNLO and NNLO+NLL$x$ predictions agree at the one-sigma level on the whole range of energy considered.
Given the large PDF uncertainties, it appears difficult to tell apart distinctive BFKL signatures in the total UHE inclusive cross-section.
However, it is interesting to note that the effect of small-$x$ resummation
on the UHE cross-sections is the opposite of that obtained in calculations based on
non-linear QCD dynamics, which instead predict a smaller cross-section at high energy (see e.g.~\cite{Albacete:2015zra}).

A promising strategy towards reducing the large PDF errors that affect $\sigma_{\rm CC}(E_{\nu})$
in Fig.~\ref{fig:UHExsec}
is provided by the inclusion of charm production data from
LHCb~\cite{Aaij:2015bpa,Aaij:2016jht,Aaij:2013mga} in the PDF fits.
As demonstrated in~\cite{Gauld:2016kpd,Gauld:2015yia,Zenaiev:2015rfa},
the inclusion of
LHCb $D$-meson production cross-sections gives a significant reduction in
PDF uncertainties in the small-$x$ region (up to an order of magnitude
at $x\simeq 10^{-6}$), which in turns leads to UHE
cross-sections with few-percent
theory errors up to $E_{\nu}=10^{12}$ GeV~\cite{Gauld:2016kpd}.
In this respect, the combination of NLL$x$ resummation and 
the additional constraints provided by the LHCb charm data 
would make possible a calculation of the UHE cross-sections 
with unprecedented theoretical and experimental uncertainties.

\subsection{Small-$x$ resummation at future electron-hadron colliders}
\label{sec:futurecoll}

Since we have demonstrated the onset of BFKL dynamics in the inclusive HERA structure
function data, it is natural to expect that the effects of small-$x$ resummation
will become even more relevant at the
proposed future high-energy electron-hadron colliders:
the higher their center of mass energy $\sqrt{s}$, the smaller the values of
$x$ kinematically accessible in the perturbative region of $Q^2$.

\begin{figure}[t]
\centering
  \includegraphics[width=0.495\textwidth]{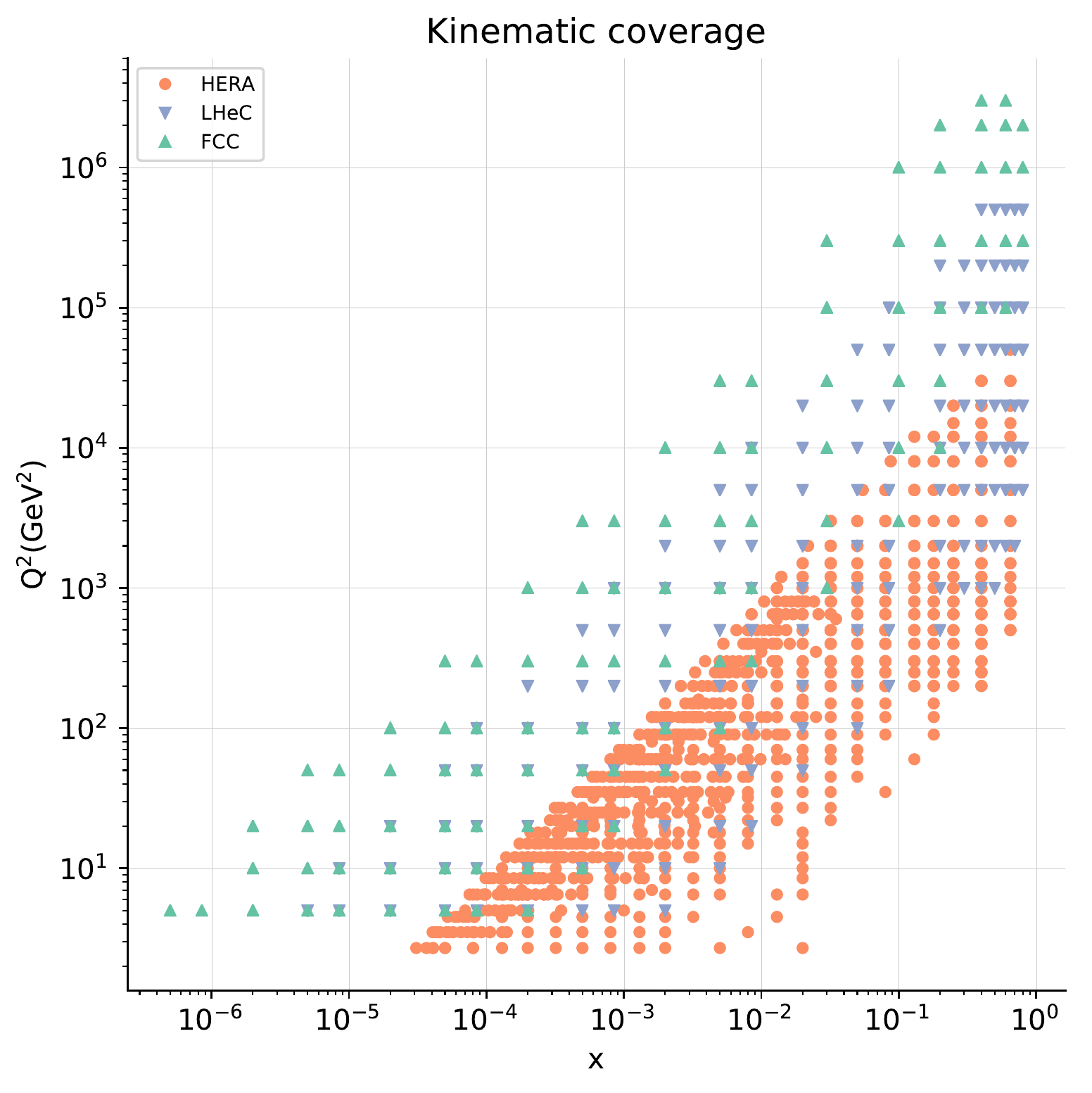}
 \caption{\small Kinematic coverage in the $(x,Q^2)$ plane of the FCC-eh and the LHeC experiments, compared to the kinematic coverage of the HERA structure function data.}
 \label{fig:FCkin}
\end{figure}

One such future $ep$ collider is the Large Hadron-electron Collider
(LHeC)~\cite{AbelleiraFernandez:2012cc,AbelleiraFernandez:2012ty}.
In its latest design, a proton beam from the LHC with $E_{p}=7$ TeV would collide with an electron/positron
beam with $E_e=60$ GeV coming from a new LinAc, thus enabling to access the region down to $x_{\rm min}\simeq 2\cdot 10^{-6}$ at $Q^2=2$ GeV$^2$.
A more extreme incarnation of the same idea corresponds to colliding the same  $E_e=60$ GeV electrons
with the $E_p=50$ TeV beam of the proposed 100 TeV Future Circular Collider (FCC)~\cite{Mangano:2016jyj,Contino:2016spe}.
The resulting collider, dubbed FCC-eh, would be able to reach
 $x_{\rm min}\simeq 2\cdot 10^{-7}$ at
$Q^2=2$ GeV$^2$.
We show in Fig.~\ref{fig:FCkin} the kinematic coverage in the $(x,Q^2)$ plane of the two
machines, compared with that of the HERA structure function data included in the NNPDF3.1sx fits.
It is clear that these two machines would probe into the small-$x$ region much deeper than HERA, thus allowing an unprecedented exploration of new QCD dynamics beyond fixed-order collinear DGLAP framework.

\begin{figure}[t]
\centering
  \includegraphics[width=0.495\textwidth]{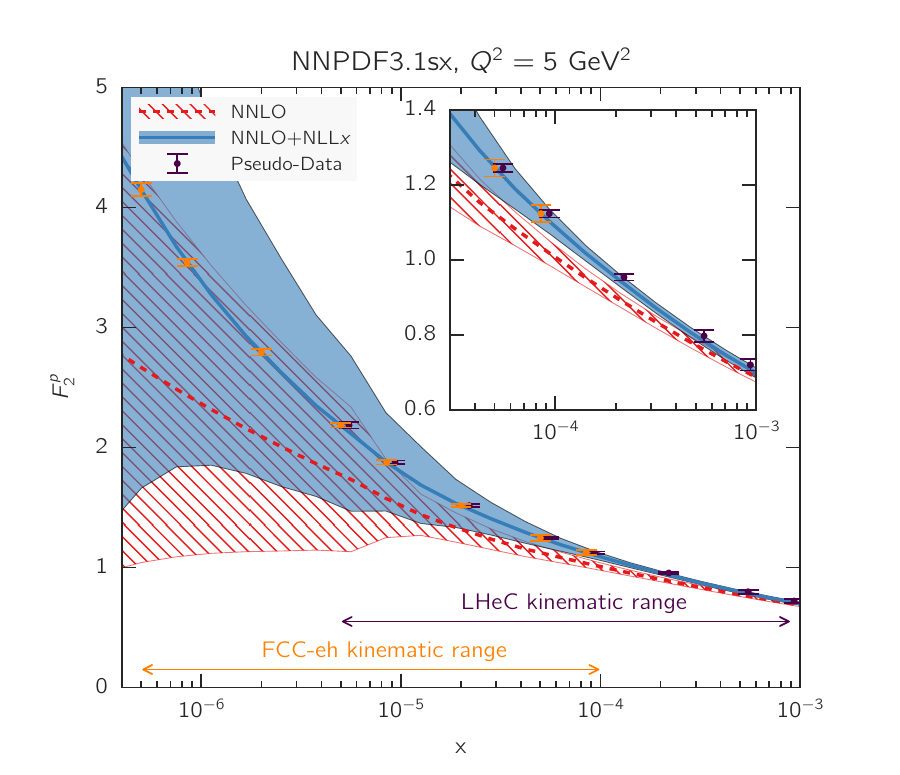}
  \includegraphics[width=0.495\textwidth]{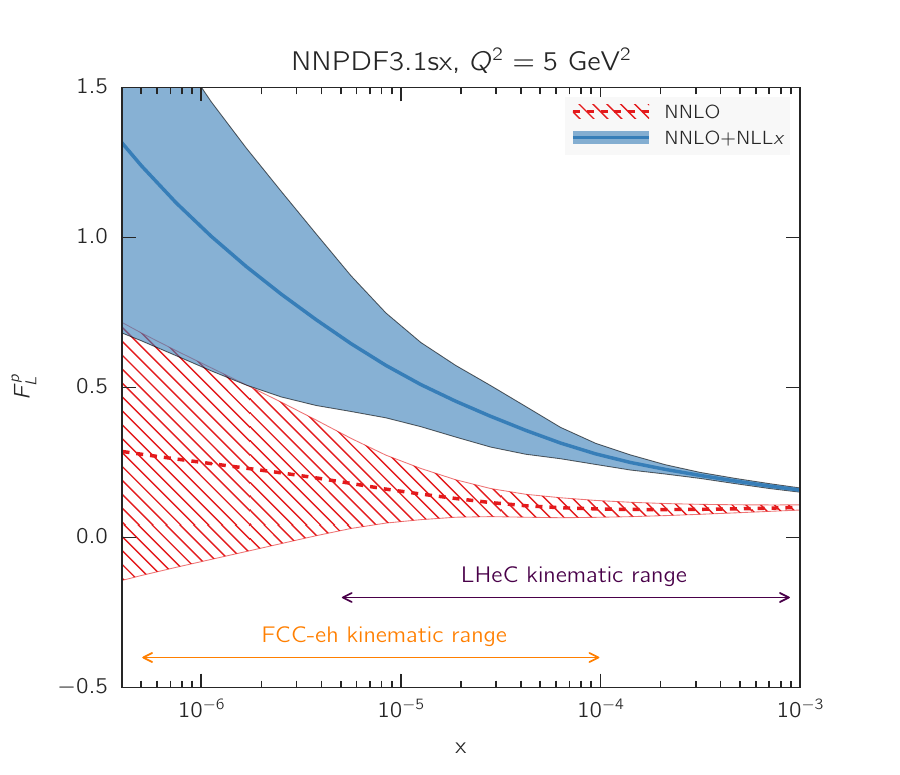}
 \caption{\small Predictions for the $F_2$ and $F_L$ structure functions using the NNPDF3.1sx NNLO
   and NNLO+NLL$x$ fits at $Q^2=5$ GeV$^2$ for the simulated
   kinematics of the LHeC and FCC-eh.
   In the case of $F_2$, we also show the expected total experimental uncertainties
   based on the simulated pseudo-data, assuming the NNLO+NLL$x$ values as central prediction. 
   A small offset has been applied to the LHeC pseudo-data as some of the values of $x$ overlap
   with the FCC-eh pseudo-data points.
   The inset in the left plot shows a magnified view in the kinematic region $x>3\times 10^{-5}$, corresponding to the reach of HERA data.}
 \label{fig:LHeCpredictions}
\end{figure}

In the following, we perform an initial exploration of the potential of the LHeC/FCC-eh for small-$x$ studies.
We use {\tt APFEL} in conjunction with {\tt HELL} to produce NNLO and NNLO+NLL$x$ predictions
for various DIS structure functions, assuming the latest version of the simulated 
LHeC/FCC-eh kinematics.\footnote{We thank Max Klein for providing us with the LHeC/FCC-eh pseudo-data.}
In Fig.~\ref{fig:LHeCpredictions} we show these predictions for the $F_2$ and $F_L$ structure functions using the NNPDF3.1sx NNLO
   and NNLO+NLL$x$ fits at $Q^2=5$ GeV$^2$ for the kinematics of the LHeC
   and the FCC-eh.
    In the case of $F_2$, we also show the expected total experimental uncertainties
    based on the simulated pseudo-data, assuming the NNLO+NLL$x$ curve as central prediction.
    To compare with the kinematic region within the reach of HERA data, we also show in the
    inset of the left plot the values of $F_2$ in a range restricted to $x> 3\times 10^{-5}$.
    The total uncertainties of the simulated pseudo-data are at the few percent
    level at most, hence much smaller than the PDF uncertainties in most of the kinematic range.
    No simulated pseudo-data is currently available for $F_L$ using the latest scenarios for the two colliders, thus in this case we show only the theoretical predictions. 

We now discuss in turn some of the
interesting features in Fig.~\ref{fig:LHeCpredictions}.
First of all, we clearly see how with the FCC-eh one can probe the 
small-$x$ region deeper than the LHeC by about an order of magnitude.
Second, we find that the differences between NNLO and NNLO+NLL$x$ are moderate for $F_2$, especially if we take into account the large PDF uncertainties.
The difference between the central values is in fact at the $15$\% level at $x\simeq 10^{-6}$,
but the current PDF uncertainties are much larger.
However, given the precision that the data could have, measuring $F_2$ (or alternatively the reduced
cross-section $\sigma_{r, \rm NC}$) at the LHeC/FCC-eh would provide discrimination between the two
theoretical scenarios of small-$x$ dynamics.
 Indeed, we see that the differences between the central values of the
 fixed-order and resummed fits in the restricted kinematic region covered by HERA are
 already comparable or larger than the size of the simulated pseudo-data uncertainties.
This suggests that the inclusion of the LHeC/FCC-eh data for $F_2$ into a global fit
would also provide discrimination power between the two theories,
 even if restricted to the HERA kinematic range.
Finally, we see that differences are more marked for $F_L$, with central values
differing by several sigma (in units of the PDF uncertainty) in a good part
of the accessible kinematic range.
This is yet another illustration of the crucial relevance of measurements of
$F_L$ to probe QCD in the small-$x$ region (as highlighted also by Fig.~\ref{fig:FLres}).

The comparisons of Fig.~\ref{fig:LHeCpredictions} do not do justice to the immense
potential of future high-energy lepton-proton colliders to probe 
QCD in a new dynamical regime.
A more detailed analysis, along the lines of Ref.~\cite{Rojo:2009ut}, involves including
various combinations of LHeC/FCC-eh pseudo-data ($\sigma^{\rm red}_{\rm NC}, F_{L}, F_{2}^{c}$, etc.) into the PDF global analysis, allowing one to use the pseudo-data to reduce the PDF uncertainties
and to quantify more precisely the discriminating power for small-$x$ 
resummation effects with various statistical estimators,
generalizing the analysis of the HERA data 
presented in Sect.~\ref{sec:diagnosis}.
Such a program would illustrate the unique role of the LHeC/FCC-eh in
the characterization of small-$x$ QCD dynamics, 
and would provide an important input to strengthen the
physics case of future high-energy lepton-proton colliders.

As a first step in this direction, we have performed variants of the NNPDF3.1sx fits
including various combinations of the LHeC and FCC-eh pseudo-data of $\sigma^{\rm red}_{\rm NC}$.
Specifically, we have used the LHeC (FCC-eh) pseudo-data on $E_p=7$~($50$)~TeV + $E_e=60$ GeV
collisions, where the central value of the pseudo-data has
been assumed to correspond to the NNLO+NLL$x$ prediction computed with the corresponding resummed PDFs.
All experimental uncertainties of the pseudo-data have been added in quadrature.
The fits have been performed at the DIS-only level, since we have demonstrated
in Sect.~\ref{sec:diagnosis} that the small-$x$ results are independent
of the treatment of the hadronic data.
Here we will show results of the fits including both LHeC and FCC-eh pseudo-data,
other combinations lead to similar qualitative results.

\begin{table}[t]
\centering
   \footnotesize
   \renewcommand{\arraystretch}{1.10}
   \begin{tabular}{l C{1.7cm} C{1.7cm}C{1.7cm}C{1.3cm}}
  &  $N_{\rm dat}$  & \multicolumn{2}{c}{$\chi^2/N_{\rm dat}$}  & $\Delta\chi^2$  \\
 & &   NNLO & NNLO+NLL$x$   &  \\
 \toprule
HERA I+II incl. NC    &  922   & 1.22   & 1.07    & -138   \\
\midrule
LHeC incl. NC    &  148    & 1.71    & 1.22    &  -73   \\
\midrule
FCC-eh incl. NC    &  98    &   2.72 &   1.34  & -135   \\
\bottomrule
    {\bf Total}
    &  \bf 1168    &\bf  1.407    & \bf 1.110   & \bf -346        \\
  \end{tabular}
  \vspace{0.5cm}
  \caption{\small Same as Table~\ref{tab:chi2tab_dis} for the
    NNPDF3.1sx NNLO and NNLO+NLL$x$
    fits including both the LHeC and the FCC-eh pseudo-data.
    We show only the $\chi^2/N_{\rm dat}$ values for the HERA inclusive cross-sections
and for the LHeC and FCC-eh pseudo-data, since for all other experiments
the values presented in Table~\ref{tab:chi2tab_dis} are essentially unchanged.
The last row corresponds to the sum of the three experiments listed on the table.
}
\label{tab:chi2tab_dis_lhec}
\end{table}
First of all we discuss the fit results at the $\chi^2/N_{\rm dat}$ level.
For simplicity, we show only the results of the HERA inclusive cross-sections
as well as that of the LHeC and FCC-eh pseudo-data: for all other experiments,
the values presented in Table~\ref{tab:chi2tab_dis} are essentially unchanged.
As shown in Table~\ref{tab:chi2tab_dis_lhec},
it is not possible to find a satisfactory fit to the LHeC/FCC-eh pseudo-data on
inclusive cross-sections using NNLO theory while
assuming that NNLO+NLL$x$ theory is the correct underlying theory,
as we have done here.
As expected, the most marked differences are observed for the FCC-eh pseudo-data.
Note that the last row in Table~\ref{tab:chi2tab_dis_lhec}
corresponds to the sum of the three experiments listed on the table.
By performing the same analysis as in Fig.~\ref{fig:chi2-profile-hera-smallx},
we have verified that the significant improvement in $\chi^2/N_{\rm dat}$
between the NNLO and NNLO+NLL$x$ fits arises from the bins
in the small-$x$ and small-$Q^2$ region.

\begin{figure}[t]
\centering
  \includegraphics[width=0.495\textwidth]{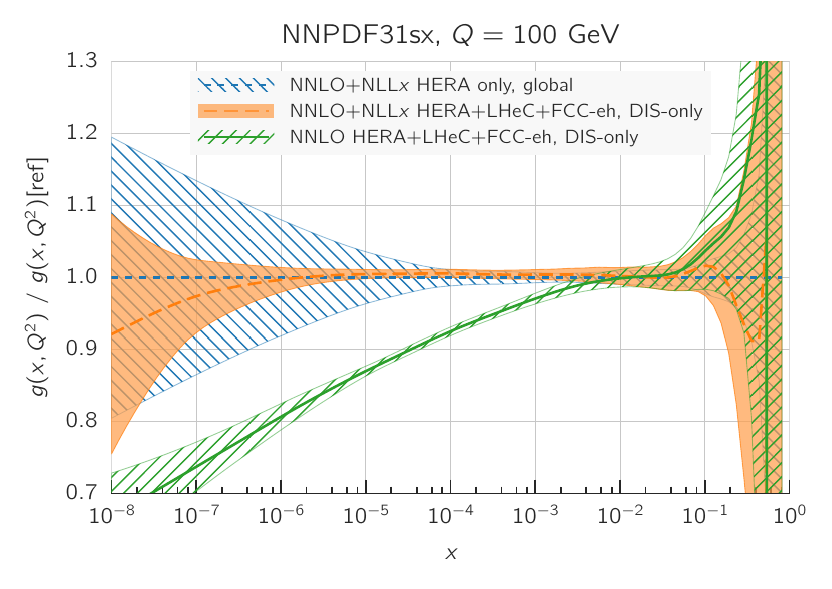}
  \includegraphics[width=0.495\textwidth]{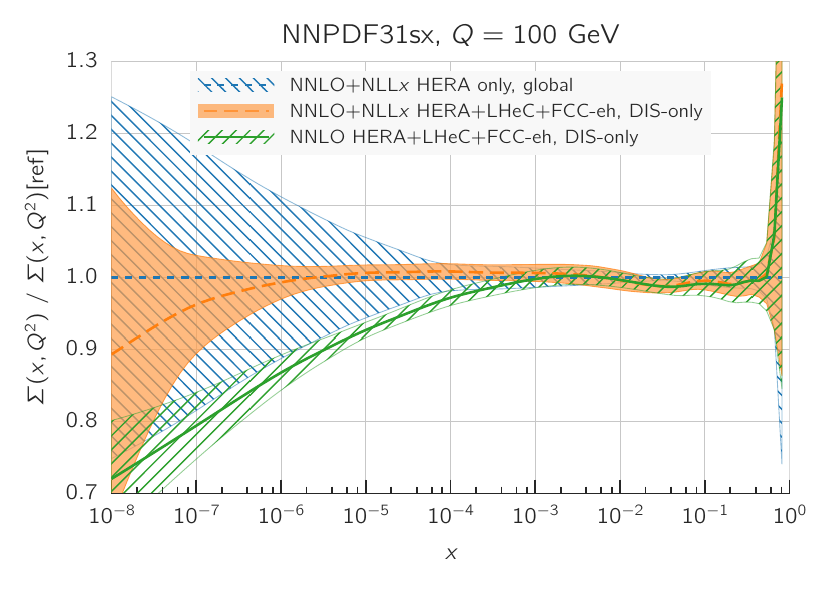}
  \caption{\small Comparison between the gluon (left plot)
    and the singlet (right plot) PDFs in the NNPDF3.1sx NNLO+NNL$x$ fits
    without and with the LHeC+FCC-eh pseudo-data on inclusive structure
    functions.
   For completeness, we also show the results of the corresponding
    NNPDF3.1sx NNLO fit with LHeC+FCC-eh pseudo-data.
  }
 \label{fig:g_sigma_lhecfits}
\end{figure}
Next in Fig.~\ref{fig:g_sigma_lhecfits} we show
the comparison between the gluon 
and the singlet PDFs at $Q=100$~GeV in the NNPDF3.1sx NNLO+NNL$x$ fits
without and with the LHeC+FCC-eh pseudo-data on inclusive structure
functions.
Note that the latter is a DIS-only fit, hence the differences
observed at large-$x$.
For completeness, we also show the results of the corresponding
NNPDF3.1sx NNLO fit with LHeC+FCC-eh pseudo-data.
In the case of the NNLO+NLL$x$, we see that the central values
coincide within uncertainties (as expected by construction)
and there is a significant uncertainty reduction both for
the gluon and for the singlet.
In particular, the LHeC+FCC-eh kinematic coverage ensures that
a precision measurement of the small-$x$ gluon, with few-percent
errors down to $x\simeq 10^{-7}$, would be within reach.

From Fig.~\ref{fig:g_sigma_lhecfits} we also see that for the gluon case,
the NNLO and NNLO+NNL$x$ fits with LHeC+FCC-eh pseudo data are very different
from each other.
For instance, at $x\simeq 10^{-5}$, where we gluon can be pinned down
with 1\% errors, the central values of the two fits differ by $\sim 15\%$.
This comparison highlights that the fixed-order description of the
small-$x$ region at these future high-energy colliders would be
completely unreliable, and that accounting for the effects of resummation
at small $x$ is required for any quantitative prediction.
Indeed, the LHeC and FCC-eh would be truly unique machines in their
potential to unveil the new dynamical regimes of QCD that arise
in the deep small-$x$ region.


\section{Summary and outlook}
\label{sec:summary}

The search for evidence of novel dynamics at small-$x$ beyond the linear fixed-order
DGLAP framework has been an ongoing enterprise ever since the HERA collider started
operations about 25 years ago.
While some tantalizing hints have been reported, until now no 
conclusive evidence
had been found in the HERA inclusive deep-inelastic structure functions.
On the contrary, fixed-order perturbative QCD calculations have been remarkably successful,
leading to good agreement with experimental data even
in kinematic regions where they might naively be expected to fail.

From the theoretical point of view, formalisms for consistently including small-$x$ resummation 
in DGLAP evolution and partonic coefficient functions
were developed more than a decade ago~\cite{Ball:1997vf,Ball:1999sh,Altarelli:1999vw,Altarelli:2000mh,Altarelli:2001ji,Altarelli:2003hk,Altarelli:2005ni,Ball:2007ra,Altarelli:2008xp,Altarelli:2008aj,Salam:1998tj,Ciafaloni:1998iv,Ciafaloni:1999yw,Ciafaloni:1999au,Ciafaloni:2000cb,Ciafaloni:2002xf,Ciafaloni:2003rd,Ciafaloni:2003kd,Ciafaloni:2005cg,Ciafaloni:2006yk,Ciafaloni:2007gf,Thorne:1999sg,Thorne:1999rb,Thorne:2001nr,White:2006yh}.
While these were sufficient to explain the success of fixed order perturbation theory in describing the data, a state-of-the-art global PDF fit including the effects of small-$x$ resummation was never performed.\footnote{A global fit including the effect of small-$x$ resummation was performed in Ref.~\cite{White:2006yh} more than a decade ago. Although the framework for implementing small-$x$ resummation was not the same that we use, and the dataset could not contain LHC data at that time, the results were similar, in particular the fact that including the resummation improves the fit quality.}
It was the main goal of this study to bridge this gap, and to present the 
first genuine attempt at cutting-edge global NLO and NNLO PDF analyses which
include the subtle effects of small-$x$ resummation.

This has been made possible thanks to a number of developments
both from the theory and from the implementation points of view.
These include the consistent matching of NLL$x$ small-$x$ resummation to both
the NLO and NNLO fixed-order results, the resummation of the heavy quark
matching conditions and DIS coefficient functions, as well as
the implementation of these theoretical developments
in the public code {\tt HELL} and its interface with the
{\tt APFEL} program~\cite{Bonvini:2016wki,Bonvini:2017ogt}.
Also crucial to the success of the enterprise was the development of the NNPDF fitting technology, which is sensitive enough to identify small effects without
them being masked by systematic methodological uncertainties from
the fit procedure.

The main result of this work is the demonstration that including small-$x$
resummation stabilizes the perturbative expansion of the DIS structure functions
at small-$x$ and $Q^2$, and thus also of the PDFs extracted from them.
Specifically, the PDFs obtained with small-$x$
resummation using NLO+NLL$x$ and NNLO+NLL$x$ theory
are in much closer agreement with each other at medium and small $x$ than the corresponding fixed-order NLO and NNLO PDFs.
This suggests in turn that the theoretical uncertainty due to missing higher order corrections in a NNLO+NLL$x$ resummed calculation is rather less
at small $x$ than that of the fixed-order NNLO calculation.
This result is reflected in the marked improvement in the quantitative 
description of the HERA inclusive structure function data at small-$x$ when NNLO+NLL$x$ resummed theory is used rather than NNLO: the NNLO+NLL$x$ theory describes the low $Q^2$ and
low $x$ bins of the HERA data just as well as it describes the data at higher $Q^2$ and larger $x$.
This effect is seen in both the inclusive neutral current, and in the 
charm cross-sections, as expected from small-$x$ resummation.
We thus find no need for higher twist contributions at low $x$, as proposed e.g.\ in Refs.~\cite{Abt:2016vjh,Harland-Lang:2016yfn},
at least in the region where the resummed perturbative calculation is valid.

We have also presented here a first exploration of the phenomenological
implications of our results. It has been understood for some time that the 
effect of resummation on the evolution of the PDFs can have a significant 
impact on the shape of parton luminosities and thus of hadronic cross-sections at LHC~\cite{Ball:2007ra}. We have now shown that the further effect on parton luminosities of including small-$x$ resummation in a PDF fit to low-$Q^2$ data at small $x$ also remains sizeable even at higher scales.
We therefore expect that at the LHC small-$x$ resummation might have significant effects, at either low invariant masses or at high rapidities, and thus that 
the accurate description of processes in these kinematic regions will require small-$x$ resummation
Conversely,
present and future LHC measurements might provide further evidence for the onset of BFKL dynamics,
this time in proton-proton collisions.

Small-$x$ resummation also plays a crucial role in shaping the physics case
for future high-energy lepton-proton colliders such as the LHeC and the 
FCC-eh, which would extend the coverage of HERA by up to two orders of 
magnitude into the small-$x$ region.
In this respect, the NNPDF3.1sx fits can be used to improve the accuracy of
existing calculations of deep inelastic scattering processes at these 
new machines.
We have also demonstrated that a clear probe of BFKL dynamics is provided
by the UHE neutrino-nucleus cross-sections, where differences
in event rates could be observed by upcoming measurements
with neutrino telescopes such as IceCube and KM3NET.

The main limitation of the present analysis is the need to impose stringent
cuts to the fitted hadronic data, in particular for Drell-Yan production,
in order to ensure that the contamination from 
unresummed partonic cross-sections is kept to a minimum.
On the one hand, it is well understood
how to combine resummation corrections to partonic cross-sections with 
resummed parton luminosities to obtain fully resummed cross-sections 
even when the coupling runs~\cite{Ball:2007ra}, and small-$x$
resummed partonic cross-sections have been computed for many of the 
relevant collider processes 
~\cite{Catani:1990xk,Catani:1990eg,Collins:1991ty,Ball:2001pq,Marzani:2008uh,Marzani:2010ap,Diana:2009xv,Diana:2010ef,Hautmann:2002tu,Marzani:2008az,Caola:2011wq,Caola:2010kv,Forte:2015gve,Marzani:2015oyb}.
On the other hand, these calculations are still not available in a 
format amenable to systematic phenomenology, and some effort is still
required before they can be used in PDF fits.  
Future work in this direction will allow us to include a wider range of hadron
collider data into a fully consistent small-$x$ resummed global fit by removing
the need for such cuts, and therefore
allow us to achieve the same experimental precision for the resummed PDFs
as is now possible in fixed-order fits.
Moreover, an accurate description of 
processes at high rapidity, such as forward Drell-Yan and
$D$ meson production at LHCb, 
is likely to require the simultaneous resummation of both small-$x$ and 
large-$x$ logarithms, since at high rapidity while one of the partons is 
at very small $x$, the other is at very large $x$.

Finally, we would like to emphasize that the
implications  of our results go beyond what is traditionally thought of
as ``small-$x$ physics''. As LHC data become ever more precise, the 
theoretical challenge is to reduce theoretical uncertainties down to the 1\% level, 
and this will require consistent calculations in perturbative QCD at N$^3$LO.
Recent progress with four-loop splitting 
functions~\cite{Davies:2016jie,Moch:2017uml} suggests that this
may  be possible rather sooner than was previously thought.
However, while at NNLO the most singular term in the gluon splitting 
function is of order $\frac{\alpha_s^3}{x}\ln\frac{1}{x}$ (the term with 
two logarithms being accidentally zero), at 
N$^3$LO the most singular term is or order 
$\frac{\alpha_s^4}{x}\ln^3\frac{1}{x}$. We thus expect the instability in 
fixed order perturbative evolution at small $x$ to be rather worse 
at N$^3$LO than it was at NNLO. Small-$x$ resummation would then be 
mandatory for improved precision, and this would require 
N$^3$LO+NNLL$x$ calculations to properly resum all the small $x$ logarithms. 
While there has been some progress in extending the BFKL 
kernel to NNLL$x$~\cite{Marzani:2007gk,DelDuca:2008jg,Bret:2011xm,DelDuca:2011ae,DelDuca:2014cya,Caron-Huot:2016tzz,Caron-Huot:2017fxr}, much work 
remains to be done.

\bigskip

\noindent
{\large\bf Delivery}\\[2ex]
\noindent
The fits presented in this work are available in the
{\tt LHAPDF6} format~\cite{Buckley:2014ana}
from the webpage of the NNPDF collaboration:
\begin{center}
\url{http://nnpdf.mi.infn.it/nnpdf3-1sx}
  \end{center}
These sets are based on the global dataset and contain $N_{\rm rep}=100$ replicas.
Specifically, the following fits are available:

\begin{itemize}

\item Baseline NLO and NNLO NNPDF3.1sx sets, which are based on the global
dataset with the kinematical cut of $H_{\rm cut}=0.6$ applied to the hadronic data:
\begin{flushleft}
\tt NNPDF31sx\_nlo\_as\_0118 \\
\tt NNPDF31sx\_nnlo\_as\_0118 \\
\end{flushleft}
\item Resummed NLO+NLL$x$ and NNLO+NLL$x$ NNPDF3.1sx sets,
  which are the resummed counterparts of the baseline sets above, based
  on an identical input dataset with the only difference of the theory settings:
  \begin{flushleft}
\tt NNPDF31sx\_nlonllx\_as\_0118 \\
\tt NNPDF31sx\_nnlonllx\_as\_0118 \\
  \end{flushleft}
\end{itemize}
In addition, the other NNPDF3.1sx fits presented in this work, such as the DIS-only fits,
are available upon request from the authors.

The DIS-only fits with various
combinations of the LHeC and FCC-eh pseudo-data, discussed
in Sect.~\ref{sec:futurecoll},
are also available in the same webpage:
\begin{flushleft}
\tt NNPDF31sx\_nnlo\_as\_0118\_DISonly\_LHeC \\
\tt NNPDF31sx\_nnlo\_as\_0118\_DISonly\_FCC \\
\tt NNPDF31sx\_nnlo\_as\_0118\_DISonly\_FCC+LHeC \\
\tt NNPDF31sx\_nnlonllx\_as\_0118\_DISonly\_LHeC \\
\tt NNPDF31sx\_nnlonllx\_as\_0118\_DISonly\_FCC \\
\tt NNPDF31sx\_nnlonllx\_as\_0118\_DISonly\_FCC+LHeC \\
\end{flushleft}

\bigskip

\noindent
{\large\bf Acknowledgments}\\[2ex]
\noindent
We are grateful to our colleagues of the NNPDF collaboration, especially
to Stefano Forte for many useful discussions and suggestions about this
project, and to
Nathan Hartland for his assistance in generating the {\tt FK} tables
and for producing some of the plots in Sect.~5.
We thank Rhorry Gauld for help with the computation of the UHE
neutrino-nucleus cross-sections.
We would also like to thank Mandy Cooper-Sarkar and Robert Thorne
for discussions about the HERA structure functions data, and Max Klein
for providing us with the LHeC and FCC-eh pseudo-data.

\noindent
R.~D.~B.
is supported by the UK STFC grants ST/L000458/1 and ST/P000630/1, while 
V.~B., J.~R. and L.~R. are supported by an European
Research Council Starting Grant ``PDF4BSM''.
The work of J.~R. is also supported by the Dutch
Organization for Scientific Research (NWO),
and part of the work of M.~B. is supported by the Marie Sk\l{}odowska Curie grant HiPPiE@LHC.

\FloatBarrier

\phantomsection
\addcontentsline{toc}{section}{References}
\bibliography{nnpdf31smallx}

\providecommand{\href}[2]{#2}\begingroup\raggedright\begin{thebibliography}{100}

\bibitem{Ball:2012cx}
{\bf NNPDF} Collaboration, R.~D. Ball, V.~Bertone, S.~Carrazza, C.~S. Deans,
  L.~Del~Debbio, et~al., {\it {Parton distributions with LHC data}},  {\em
  Nucl.Phys.} {\bf B867} (2013) 244--289,
  [\href{http://arxiv.org/abs/1207.1303}{{\tt arXiv:1207.1303}}].

\bibitem{Alekhin:2017kpj}
S.~Alekhin, J.~Bl{\"u}mlein, S.~Moch, and R.~Placakyte, {\it {Parton
  distribution functions, $\alpha_s$, and heavy-quark masses for LHC Run II}},
  {\em Phys. Rev.} {\bf D96} (2017), no.~1 014011,
  [\href{http://arxiv.org/abs/1701.05838}{{\tt arXiv:1701.05838}}].

\bibitem{Accardi:2016qay}
A.~Accardi, L.~T. Brady, W.~Melnitchouk, J.~F. Owens, and N.~Sato, {\it
  {Constraints on large-$x$ parton distributions from new weak boson production
  and deep-inelastic scattering data}},  {\em Phys. Rev.} {\bf D93} (2016),
  no.~11 114017, [\href{http://arxiv.org/abs/1602.03154}{{\tt
  arXiv:1602.03154}}].

\bibitem{Dulat:2015mca}
S.~Dulat, T.-J. Hou, J.~Gao, M.~Guzzi, J.~Huston, P.~Nadolsky, J.~Pumplin,
  C.~Schmidt, D.~Stump, and C.~P. Yuan, {\it {New parton distribution functions
  from a global analysis of quantum chromodynamics}},  {\em Phys. Rev.} {\bf
  D93} (2016), no.~3 033006, [\href{http://arxiv.org/abs/1506.07443}{{\tt
  arXiv:1506.07443}}].

\bibitem{Harland-Lang:2014zoa}
L.~A. Harland-Lang, A.~D. Martin, P.~Motylinski, and R.~S. Thorne, {\it {Parton
  distributions in the LHC era: MMHT 2014 PDFs}},  {\em Eur. Phys. J.} {\bf
  C75} (2015) 204, [\href{http://arxiv.org/abs/1412.3989}{{\tt
  arXiv:1412.3989}}].

\bibitem{Jimenez-Delgado:2014twa}
P.~Jimenez-Delgado and E.~Reya, {\it {Delineating parton distributions and the
  strong coupling}},  {\em Phys.Rev.} {\bf D89} (2014), no.~7 074049,
  [\href{http://arxiv.org/abs/1403.1852}{{\tt arXiv:1403.1852}}].

\bibitem{Gao:2017yyd}
J.~Gao, L.~Harland-Lang, and J.~Rojo, {\it {The Structure of the Proton in the
  LHC Precision Era}},  \href{http://arxiv.org/abs/1709.04922}{{\tt
  arXiv:1709.04922}}.

\bibitem{Ball:2015oha}
R.~D. Ball, {\it {Global Parton Distributions for the LHC Run II}},  in {\em
  {29th Rencontres de Physique de La Vallee d'Aoste La Thuile, Aosta, Italy,
  March 1-7, 2015}}, 2015.
\newblock \href{http://arxiv.org/abs/1507.07891}{{\tt arXiv:1507.07891}}.

\bibitem{Forte:2013wc}
S.~Forte and G.~Watt, {\it {Progress in the Determination of the Partonic
  Structure of the Proton}},  {\em Ann.Rev.Nucl.Part.Sci.} {\bf 63} (2013) 291,
  [\href{http://arxiv.org/abs/1301.6754}{{\tt arXiv:1301.6754}}].

\bibitem{Butterworth:2015oua}
J.~Butterworth et~al., {\it {PDF4LHC recommendations for LHC Run II}},  {\em J.
  Phys.} {\bf G43} (2016) 023001, [\href{http://arxiv.org/abs/1510.03865}{{\tt
  arXiv:1510.03865}}].

\bibitem{Rojo:2015acz}
J.~Rojo et~al., {\it {The PDF4LHC report on PDFs and LHC data: Results from Run
  I and preparation for Run II}},  {\em J. Phys.} {\bf G42} (2015) 103103,
  [\href{http://arxiv.org/abs/1507.00556}{{\tt arXiv:1507.00556}}].

\bibitem{Beenakker:2015rna}
W.~Beenakker, C.~Borschensky, M.~KrÃ¤mer, A.~Kulesza, E.~Laenen, S.~Marzani,
  and J.~Rojo, {\it {NLO+NLL squark and gluino production cross-sections with
  threshold-improved parton distributions}},  {\em Eur. Phys. J.} {\bf C76}
  (2016), no.~2 53, [\href{http://arxiv.org/abs/1510.00375}{{\tt
  arXiv:1510.00375}}].

\bibitem{LHCb:2012fja}
{\bf LHCb} Collaboration, J.~Anderson and K.~Mueller, {\it {Inclusive low mass
  Drell-Yan production in the forward region at $\sqrt{s}$ = 7 TeV}}, .

\bibitem{Gauld:2015yia}
R.~Gauld, J.~Rojo, L.~Rottoli, and J.~Talbert, {\it {Charm production in the
  forward region: constraints on the small-x gluon and backgrounds for neutrino
  astronomy}},  {\em JHEP} {\bf 11} (2015) 009,
  [\href{http://arxiv.org/abs/1506.08025}{{\tt arXiv:1506.08025}}].

\bibitem{Bonvini:2015ira}
M.~Bonvini, S.~Marzani, J.~Rojo, L.~Rottoli, M.~Ubiali, R.~D. Ball, V.~Bertone,
  S.~Carrazza, and N.~P. Hartland, {\it {Parton distributions with threshold
  resummation}},  {\em JHEP} {\bf 09} (2015) 191,
  [\href{http://arxiv.org/abs/1507.01006}{{\tt arXiv:1507.01006}}].

\bibitem{Corcella:2005us}
G.~Corcella and L.~Magnea, {\it {Soft-gluon resummation effects on parton
  distributions}},  {\em Phys. Rev.} {\bf D72} (2005) 074017,
  [\href{http://arxiv.org/abs/hep-ph/0506278}{{\tt hep-ph/0506278}}].

\bibitem{Korchemsky:1988si}
G.~P. Korchemsky, {\it {Asymptotics of the Altarelli-Parisi-Lipatov Evolution
  Kernels of Parton Distributions}},  {\em Mod. Phys. Lett.} {\bf A4} (1989)
  1257--1276.

\bibitem{Albino:2000cp}
S.~Albino and R.~D. Ball, {\it {Soft resummation of quark anomalous dimensions
  and coefficient functions in MS-bar factorization}},  {\em Phys. Lett.} {\bf
  B513} (2001) 93--102, [\href{http://arxiv.org/abs/hep-ph/0011133}{{\tt
  hep-ph/0011133}}].

\bibitem{Lipatov:1976zz}
L.~N. Lipatov, {\it {Reggeization of the Vector Meson and the Vacuum
  Singularity in Nonabelian Gauge Theories}},  {\em Sov. J. Nucl. Phys.} {\bf
  23} (1976) 338--345.

\bibitem{Fadin:1975cb}
V.~S. Fadin, E.~Kuraev, and L.~Lipatov, {\it {On the Pomeranchuk Singularity in
  Asymptotically Free Theories}},  {\em Phys.Lett.} {\bf B60} (1975) 50--52.

\bibitem{Kuraev:1976ge}
E.~A. Kuraev, L.~N. Lipatov, and V.~S. Fadin, {\it {Multi - Reggeon Processes
  in the Yang-Mills Theory}},  {\em Sov.Phys.JETP} {\bf 44} (1976) 443--450.

\bibitem{Kuraev:1977fs}
E.~A. Kuraev, L.~N. Lipatov, and V.~S. Fadin, {\it {The Po\-me\-ran\-chuk
  Singularity in Nonabelian Gauge Theories}},  {\em Sov. Phys. JETP} {\bf 45}
  (1977) 199--204.

\bibitem{Balitsky:1978ic}
I.~I. Balitsky and L.~N. Lipatov, {\it {The Pomeranchuk Singularity in Quantum
  Chromodynamics}},  {\em Sov. J. Nucl. Phys.} {\bf 28} (1978) 822--829.

\bibitem{Abt:1993cb}
{\bf H1} Collaboration, I.~Abt et~al., {\it {Measurement of the proton
  structure function F2 (x, Q**2) in the low x region at HERA}},  {\em Nucl.
  Phys.} {\bf B407} (1993) 515--538.

\bibitem{Derrick:1993fta}
{\bf ZEUS} Collaboration, M.~Derrick et~al., {\it {Measurement of the proton
  structure function F2 in e p scattering at HERA}},  {\em Phys. Lett.} {\bf
  B316} (1993) 412--426.

\bibitem{Ball:1994du}
R.~D. Ball and S.~Forte, {\it Double asymptotic scaling at hera},  {\em Phys.
  Lett.} {\bf B335} (1994) 77--86,
  [\href{http://arxiv.org/abs/hep-ph/9405320}{{\tt hep-ph/9405320}}].

\bibitem{Ball:1994kc}
R.~D. Ball and S.~Forte, {\it {A Direct test of perturbative QCD at small x}},
  {\em Phys. Lett.} {\bf B336} (1994) 77--79,
  [\href{http://arxiv.org/abs/hep-ph/9406385}{{\tt hep-ph/9406385}}].

\bibitem{Forte:1995vs}
S.~Forte and R.~D. Ball, {\it Universality and scaling in perturbative qcd at
  small x},  {\em Acta Phys. Polon.} {\bf B26} (1995) 2097--2134,
  [\href{http://arxiv.org/abs/hep-ph/9512208}{{\tt hep-ph/9512208}}].

\bibitem{Gluck:1994uf}
M.~Gluck, E.~Reya, and A.~Vogt, {\it {Dynamical parton distributions of the
  proton and small x physics}},  {\em Z. Phys.} {\bf C67} (1995) 433--448.

\bibitem{Lai:1996mg}
H.~L. Lai, J.~Huston, S.~Kuhlmann, F.~I. Olness, J.~F. Owens, D.~E. Soper,
  W.~K. Tung, and H.~Weerts, {\it {Improved parton distributions from global
  analysis of recent deep inelastic scattering and inclusive jet data}},  {\em
  Phys. Rev.} {\bf D55} (1997) 1280--1296,
  [\href{http://arxiv.org/abs/hep-ph/9606399}{{\tt hep-ph/9606399}}].

\bibitem{Martin:1994kn}
A.~D. Martin, W.~J. Stirling, and R.~G. Roberts, {\it {Parton distributions of
  the proton}},  {\em Phys. Rev.} {\bf D50} (1994) 6734--6752,
  [\href{http://arxiv.org/abs/hep-ph/9406315}{{\tt hep-ph/9406315}}].

\bibitem{Fadin:1996nw}
V.~S. Fadin and L.~N. Lipatov, {\it {Next-to-leading corrections to the BFKL
  equation from the gluon and quark production}},  {\em Nucl. Phys.} {\bf B477}
  (1996) 767--808, [\href{http://arxiv.org/abs/hep-ph/9602287}{{\tt
  hep-ph/9602287}}].

\bibitem{Fadin:1997hr}
V.~S. Fadin, R.~Fiore, A.~Flachi, and M.~I. Kotsky, {\it {Quark - anti-quark
  contribution to the BFKL kernel}},  {\em Phys. Lett.} {\bf B422} (1998)
  287--293, [\href{http://arxiv.org/abs/hep-ph/9711427}{{\tt hep-ph/9711427}}].

\bibitem{Fadin:1997zv}
V.~S. Fadin, M.~I. Kotsky, and L.~N. Lipatov, {\it {One-loop correction to the
  BFKL kernel from two gluon production}},  {\em Phys. Lett.} {\bf B415} (1997)
  97--103.

\bibitem{Camici:1997ij}
G.~Camici and M.~Ciafaloni, {\it {Irreducible part of the next-to-leading BFKL
  kernel}},  {\em Phys. Lett.} {\bf B412} (1997) 396--406,
  [\href{http://arxiv.org/abs/hep-ph/9707390}{{\tt hep-ph/9707390}}]. [Erratum:
  Phys. Lett.B417,390(1998)].

\bibitem{Fadin:1998py}
V.~S. Fadin and L.~Lipatov, {\it {BFKL pomeron in the next-to-leading
  approximation}},  {\em Phys.Lett.} {\bf B429} (1998) 127--134,
  [\href{http://arxiv.org/abs/hep-ph/9802290}{{\tt hep-ph/9802290}}].

\bibitem{Ball:1997vf}
R.~D. Ball and S.~Forte, {\it {Asymptotically free partons at high-energy}},
  {\em Phys.Lett.} {\bf B405} (1997) 317--326,
  [\href{http://arxiv.org/abs/hep-ph/9703417}{{\tt hep-ph/9703417}}].

\bibitem{Ball:1999sh}
R.~D. Ball and S.~Forte, {\it {The Small x behavior of Altarelli-Parisi
  splitting functions}},  {\em Phys. Lett.} {\bf B465} (1999) 271--281,
  [\href{http://arxiv.org/abs/hep-ph/9906222}{{\tt hep-ph/9906222}}].

\bibitem{Altarelli:1999vw}
G.~Altarelli, R.~D. Ball, and S.~Forte, {\it {Resummation of singlet parton
  evolution at small x}},  {\em Nucl. Phys.} {\bf B575} (2000) 313--329,
  [\href{http://arxiv.org/abs/hep-ph/9911273}{{\tt hep-ph/9911273}}].

\bibitem{Altarelli:2000mh}
G.~Altarelli, R.~D. Ball, and S.~Forte, {\it {Small x resummation and HERA
  structure function data}},  {\em Nucl. Phys.} {\bf B599} (2001) 383--423,
  [\href{http://arxiv.org/abs/hep-ph/0011270}{{\tt hep-ph/0011270}}].

\bibitem{Altarelli:2001ji}
G.~Altarelli, R.~D. Ball, and S.~Forte, {\it {Factorization and resummation of
  small x scaling violations with running coupling}},  {\em Nucl.Phys.} {\bf
  B621} (2002) 359--387, [\href{http://arxiv.org/abs/hep-ph/0109178}{{\tt
  hep-ph/0109178}}].

\bibitem{Altarelli:2003hk}
G.~Altarelli, R.~D. Ball, and S.~Forte, {\it {An Anomalous dimension for small
  x evolution}},  {\em Nucl.Phys.} {\bf B674} (2003) 459--483,
  [\href{http://arxiv.org/abs/hep-ph/0306156}{{\tt hep-ph/0306156}}].

\bibitem{Altarelli:2005ni}
G.~Altarelli, R.~D. Ball, and S.~Forte, {\it {Perturbatively stable resummed
  small x evolution kernels}},  {\em Nucl. Phys.} {\bf B742} (2006) 1--40,
  [\href{http://arxiv.org/abs/hep-ph/0512237}{{\tt hep-ph/0512237}}].

\bibitem{Ball:2007ra}
R.~D. Ball, {\it {Resummation of Hadroproduction Cross-sections at High
  Energy}},  {\em Nucl.Phys.} {\bf B796} (2008) 137--183,
  [\href{http://arxiv.org/abs/0708.1277}{{\tt arXiv:0708.1277}}].

\bibitem{Altarelli:2008xp}
G.~Altarelli, R.~D. Ball, and S.~Forte, {\it {Structure Function Resummation in
  small-x QCD}},  {\em PoS} {\bf RADCOR2007} (2007) 028,
  [\href{http://arxiv.org/abs/0802.0968}{{\tt arXiv:0802.0968}}].

\bibitem{Altarelli:2008aj}
G.~Altarelli, R.~D. Ball, and S.~Forte, {\it {Small x Resummation with Quarks:
  Deep-Inelastic Scattering}},  {\em Nucl.Phys.} {\bf B799} (2008) 199--240,
  [\href{http://arxiv.org/abs/0802.0032}{{\tt arXiv:0802.0032}}].

\bibitem{Salam:1998tj}
G.~Salam, {\it {A Resummation of large subleading corrections at small x}},
  {\em JHEP} {\bf 9807} (1998) 019,
  [\href{http://arxiv.org/abs/hep-ph/9806482}{{\tt hep-ph/9806482}}].

\bibitem{Ciafaloni:1998iv}
M.~Ciafaloni and D.~Colferai, {\it {The BFKL equation at next-to-leading level
  and beyond}},  {\em Phys. Lett.} {\bf B452} (1999) 372--378,
  [\href{http://arxiv.org/abs/hep-ph/9812366}{{\tt hep-ph/9812366}}].

\bibitem{Ciafaloni:1999yw}
M.~Ciafaloni, D.~Colferai, and G.~Salam, {\it {Renormalization group improved
  small x equation}},  {\em Phys.Rev.} {\bf D60} (1999) 114036,
  [\href{http://arxiv.org/abs/hep-ph/9905566}{{\tt hep-ph/9905566}}].

\bibitem{Ciafaloni:1999au}
M.~Ciafaloni, D.~Colferai, and G.~P. Salam, {\it {A collinear model for small x
  physics}},  {\em JHEP} {\bf 10} (1999) 017,
  [\href{http://arxiv.org/abs/hep-ph/9907409}{{\tt hep-ph/9907409}}].

\bibitem{Ciafaloni:2000cb}
M.~Ciafaloni, D.~Colferai, and G.~P. Salam, {\it {On factorization at small
  x}},  {\em JHEP} {\bf 07} (2000) 054,
  [\href{http://arxiv.org/abs/hep-ph/0007240}{{\tt hep-ph/0007240}}].

\bibitem{Ciafaloni:2002xf}
M.~Ciafaloni, D.~Colferai, G.~P. Salam, and A.~M. Stasto, {\it {Expanding
  running coupling effects in the hard pomeron}},  {\em Phys. Rev.} {\bf D66}
  (2002) 054014, [\href{http://arxiv.org/abs/hep-ph/0204282}{{\tt
  hep-ph/0204282}}].

\bibitem{Ciafaloni:2003rd}
M.~Ciafaloni, D.~Colferai, G.~P. Salam, and A.~M. Stasto, {\it {Renormalization
  group improved small x Green's function}},  {\em Phys. Rev.} {\bf D68} (2003)
  114003, [\href{http://arxiv.org/abs/hep-ph/0307188}{{\tt hep-ph/0307188}}].

\bibitem{Ciafaloni:2003kd}
M.~Ciafaloni, D.~Colferai, G.~P. Salam, and A.~M. Stasto, {\it {The Gluon
  splitting function at moderately small x}},  {\em Phys. Lett.} {\bf B587}
  (2004) 87--94, [\href{http://arxiv.org/abs/hep-ph/0311325}{{\tt
  hep-ph/0311325}}].

\bibitem{Ciafaloni:2005cg}
M.~Ciafaloni and D.~Colferai, {\it {Dimensional regularisation and
  factorisation schemes in the BFKL equation at subleading level}},  {\em JHEP}
  {\bf 09} (2005) 069, [\href{http://arxiv.org/abs/hep-ph/0507106}{{\tt
  hep-ph/0507106}}].

\bibitem{Ciafaloni:2006yk}
M.~Ciafaloni, D.~Colferai, G.~P. Salam, and A.~M. Stasto, {\it {Minimal
  subtraction vs. physical factorisation schemes in small-x QCD}},  {\em Phys.
  Lett.} {\bf B635} (2006) 320--329,
  [\href{http://arxiv.org/abs/hep-ph/0601200}{{\tt hep-ph/0601200}}].

\bibitem{Ciafaloni:2007gf}
M.~Ciafaloni, D.~Colferai, G.~Salam, and A.~Stasto, {\it {A Matrix formulation
  for small-$x$ singlet evolution}},  {\em JHEP} {\bf 0708} (2007) 046,
  [\href{http://arxiv.org/abs/0707.1453}{{\tt arXiv:0707.1453}}].

\bibitem{Thorne:1999sg}
R.~S. Thorne, {\it {Explicit calculation of the running coupling BFKL anomalous
  dimension}},  {\em Phys. Lett.} {\bf B474} (2000) 372--384,
  [\href{http://arxiv.org/abs/hep-ph/9912284}{{\tt hep-ph/9912284}}].

\bibitem{Thorne:1999rb}
R.~S. Thorne, {\it {NLO BFKL equation, running coupling and renormalization
  scales}},  {\em Phys. Rev.} {\bf D60} (1999) 054031,
  [\href{http://arxiv.org/abs/hep-ph/9901331}{{\tt hep-ph/9901331}}].

\bibitem{Thorne:2001nr}
R.~S. Thorne, {\it {The Running coupling BFKL anomalous dimensions and
  splitting functions}},  {\em Phys. Rev.} {\bf D64} (2001) 074005,
  [\href{http://arxiv.org/abs/hep-ph/0103210}{{\tt hep-ph/0103210}}].

\bibitem{White:2006yh}
C.~D. White and R.~S. Thorne, {\it {A Global Fit to Scattering Data with NLL
  BFKL Resummations}},  {\em Phys. Rev.} {\bf D75} (2007) 034005,
  [\href{http://arxiv.org/abs/hep-ph/0611204}{{\tt hep-ph/0611204}}].

\bibitem{Bonvini:2016wki}
M.~Bonvini, S.~Marzani, and T.~Peraro, {\it {Small-$x$ resummation from HELL}},
   {\em Eur. Phys. J.} {\bf C76} (2016), no.~11 597,
  [\href{http://arxiv.org/abs/1607.02153}{{\tt arXiv:1607.02153}}].

\bibitem{Bonvini:2017ogt}
M.~Bonvini, S.~Marzani, and C.~Muselli, {\it {Towards parton distribution
  functions with small-$x$ resummation: HELL 2.0}},
  \href{http://arxiv.org/abs/1708.07510}{{\tt arXiv:1708.07510}}.

\bibitem{Harland-Lang:2016yfn}
L.~A. Harland-Lang, A.~D. Martin, P.~Motylinski, and R.~S. Thorne, {\it {The
  impact of the final HERA combined data on PDFs obtained from a global fit}},
  {\em Eur. Phys. J.} {\bf C76} (2016), no.~4 186,
  [\href{http://arxiv.org/abs/1601.03413}{{\tt arXiv:1601.03413}}].

\bibitem{Caola:2009iy}
F.~Caola, S.~Forte, and J.~Rojo, {\it {Deviations from NLO QCD evolution in
  inclusive HERA data}},  {\em Phys. Lett.} {\bf B686} (2010) 127--135,
  [\href{http://arxiv.org/abs/0910.3143}{{\tt arXiv:0910.3143}}].

\bibitem{Caola:2010cy}
F.~Caola, S.~Forte, and J.~Rojo, {\it {HERA data and DGLAP evolution: Theory
  and phenomenology}},  {\em Nucl.Phys.} {\bf A854} (2011) 32--44,
  [\href{http://arxiv.org/abs/1007.5405}{{\tt arXiv:1007.5405}}].

\bibitem{Rojo:2015nxa}
J.~Rojo, {\it {Progress in the NNPDF global analysis and the impact of the
  legacy HERA combination}},  in {\em {Proceedings, 2015 European Physical
  Society Conference on High Energy Physics (EPS-HEP 2015)}}, 2015.
\newblock \href{http://arxiv.org/abs/1508.07731}{{\tt arXiv:1508.07731}}.

\bibitem{Abramowicz:2015mha}
{\bf ZEUS, H1} Collaboration, H.~Abramowicz et~al., {\it {Combination of
  measurements of inclusive deep inelastic ${e^{\pm }p}$ scattering cross
  sections and QCD analysis of HERA data}},  {\em Eur. Phys. J.} {\bf C75}
  (2015), no.~12 580, [\href{http://arxiv.org/abs/1506.06042}{{\tt
  arXiv:1506.06042}}].

\bibitem{Abt:2016vjh}
I.~Abt, A.~M. Cooper-Sarkar, B.~Foster, V.~Myronenko, K.~Wichmann, and M.~Wing,
  {\it {Study of HERA ep data at low Q$^2$ and low $x_{Bj}$ and the need for
  higher-twist corrections to standard perturbative QCD fits}},  {\em Phys.
  Rev.} {\bf D94} (2016), no.~3 034032,
  [\href{http://arxiv.org/abs/1604.02299}{{\tt arXiv:1604.02299}}].

\bibitem{Abt:2017nkc}
I.~Abt, A.~M. Cooper-Sarkar, B.~Foster, V.~Myronenko, K.~Wichmann, and M.~Wing,
  {\it {Investigation into the limits of perturbation theory at low $Q^2$ using
  HERA deep inelastic scattering data}},  {\em Phys. Rev.} {\bf D96} (2017),
  no.~1 014001, [\href{http://arxiv.org/abs/1704.03187}{{\tt
  arXiv:1704.03187}}].

\bibitem{DelDebbio:2007ee}
{\bf The NNPDF} Collaboration, L.~Del~Debbio, S.~Forte, J.~I. Latorre,
  A.~Piccione, and J.~Rojo, {\it {Neural network determination of parton
  distributions: The nonsinglet case}},  {\em JHEP} {\bf 03} (2007) 039,
  [\href{http://arxiv.org/abs/hep-ph/0701127}{{\tt hep-ph/0701127}}].

\bibitem{Ball:2008by}
{\bf The NNPDF} Collaboration, R.~D. Ball et~al., {\it {A determination of
  parton distributions with faithful uncertainty estimation}},  {\em Nucl.
  Phys.} {\bf B809} (2009) 1--63, [\href{http://arxiv.org/abs/0808.1231}{{\tt
  arXiv:0808.1231}}].

\bibitem{Ball:2009mk}
{\bf The NNPDF} Collaboration, R.~D. Ball et~al., {\it {Precision determination
  of electroweak parameters and the strange content of the proton from neutrino
  deep-inelastic scattering}},  {\em Nucl. Phys.} {\bf B823} (2009) 195--233,
  [\href{http://arxiv.org/abs/0906.1958}{{\tt arXiv:0906.1958}}].

\bibitem{Ball:2010de}
{\bf {The NNPDF }} Collaboration, R.~D. Ball et~al., {\it {A first unbiased
  global NLO determination of parton distributions and their uncertainties}},
  {\em Nucl. Phys.} {\bf B838} (2010) 136,
  [\href{http://arxiv.org/abs/1002.4407}{{\tt arXiv:1002.4407}}].

\bibitem{Ball:2011uy}
{\bf The NNPDF} Collaboration, R.~D. Ball et~al., {\it {Unbiased global
  determination of parton distributions and their uncertainties at NNLO and at
  LO}},  {\em Nucl.Phys.} {\bf B855} (2012) 153--221,
  [\href{http://arxiv.org/abs/1107.2652}{{\tt arXiv:1107.2652}}].

\bibitem{Ball:2013gsa}
{\bf The NNPDF} Collaboration, R.~D. Ball et~al., {\it {Theoretical issues in
  PDF determination and associated uncertainties}},  {\em Phys.Lett.} {\bf
  B723} (2013) 330, [\href{http://arxiv.org/abs/1303.1189}{{\tt
  arXiv:1303.1189}}].

\bibitem{Ball:2014uwa}
{\bf NNPDF} Collaboration, R.~D. Ball et~al., {\it {Parton distributions for
  the LHC Run II}},  {\em JHEP} {\bf 04} (2015) 040,
  [\href{http://arxiv.org/abs/1410.8849}{{\tt arXiv:1410.8849}}].

\bibitem{Ball:2016neh}
{\bf NNPDF} Collaboration, R.~D. Ball, V.~Bertone, M.~Bonvini, S.~Carrazza,
  S.~Forte, A.~Guffanti, N.~P. Hartland, J.~Rojo, and L.~Rottoli, {\it {A
  Determination of the Charm Content of the Proton}},  {\em Eur. Phys. J.} {\bf
  C76} (2016), no.~11 647, [\href{http://arxiv.org/abs/1605.06515}{{\tt
  arXiv:1605.06515}}].

\bibitem{Ball:2017nwa}
{\bf NNPDF} Collaboration, R.~D. Ball et~al., {\it {Parton distributions from
  high-precision collider data}},  {\em Eur. Phys. J.} {\bf C77} (2017), no.~10
  663, [\href{http://arxiv.org/abs/1706.00428}{{\tt arXiv:1706.00428}}].

\bibitem{Ball:1995vc}
R.~D. Ball and S.~Forte, {\it {Summation of leading logarithms at small x}},
  {\em Phys.Lett.} {\bf B351} (1995) 313--324,
  [\href{http://arxiv.org/abs/hep-ph/9501231}{{\tt hep-ph/9501231}}].

\bibitem{Ball:1995qd}
R.~D. Ball and S.~Forte, {\it {Determination of alpha-s from F2(p) at HERA}},
  {\em Phys. Lett.} {\bf B358} (1995) 365--378,
  [\href{http://arxiv.org/abs/hep-ph/9506233}{{\tt hep-ph/9506233}}].

\bibitem{Ellis:1995gv}
R.~K. Ellis, F.~Hautmann, and B.~R. Webber, {\it {QCD scaling violation at
  small x}},  {\em Phys. Lett.} {\bf B348} (1995) 582--588,
  [\href{http://arxiv.org/abs/hep-ph/9501307}{{\tt hep-ph/9501307}}].

\bibitem{Forshaw:1995ga}
J.~R. Forshaw, R.~G. Roberts, and R.~S. Thorne, {\it {Analytic approach to
  small x structure functions}},  {\em Phys. Lett.} {\bf B356} (1995) 79--88,
  [\href{http://arxiv.org/abs/hep-ph/9504336}{{\tt hep-ph/9504336}}].

\bibitem{Moch:2004pa}
S.~Moch, J.~A.~M. Vermaseren, and A.~Vogt, {\it {The three-loop splitting
  functions in QCD: The non-singlet case}},  {\em Nucl. Phys.} {\bf B688}
  (2004) 101--134, [\href{http://arxiv.org/abs/hep-ph/0403192}{{\tt
  hep-ph/0403192}}].

\bibitem{Vogt:2004mw}
A.~Vogt, S.~Moch, and J.~A.~M. Vermaseren, {\it {The three-loop splitting
  functions in QCD: The singlet case}},  {\em Nucl. Phys.} {\bf B691} (2004)
  129--181, [\href{http://arxiv.org/abs/hep-ph/0404111}{{\tt hep-ph/0404111}}].

\bibitem{Davies:2016jie}
J.~Davies, A.~Vogt, B.~Ruijl, T.~Ueda, and J.~A.~M. Vermaseren, {\it
  {Large-$n_f$ contributions to the four-loop splitting functions in QCD}},
  {\em Nucl. Phys.} {\bf B915} (2017) 335--362,
  [\href{http://arxiv.org/abs/1610.07477}{{\tt arXiv:1610.07477}}].

\bibitem{Moch:2017uml}
S.~Moch, B.~Ruijl, T.~Ueda, J.~A.~M. Vermaseren, and A.~Vogt, {\it {Four-Loop
  Non-Singlet Splitting Functions in the Planar Limit and Beyond}},
  \href{http://arxiv.org/abs/1707.08315}{{\tt arXiv:1707.08315}}.

\bibitem{Marzani:2007gk}
S.~Marzani, R.~D. Ball, P.~Falgari, and S.~Forte, {\it {BFKL at
  next-to-next-to-leading order}},  {\em Nucl. Phys.} {\bf B783} (2007)
  143--175, [\href{http://arxiv.org/abs/0704.2404}{{\tt arXiv:0704.2404}}].

\bibitem{DelDuca:2008jg}
V.~Del~Duca, C.~Duhr, and E.~W.~N. Glover, {\it {Iterated amplitudes in the
  high-energy limit}},  {\em JHEP} {\bf 12} (2008) 097,
  [\href{http://arxiv.org/abs/0809.1822}{{\tt arXiv:0809.1822}}].

\bibitem{Bret:2011xm}
V.~Del~Duca, C.~Duhr, E.~Gardi, L.~Magnea, and C.~D. White, {\it {An infrared
  approach to Reggeization}},  {\em Phys. Rev.} {\bf D85} (2012) 071104,
  [\href{http://arxiv.org/abs/1108.5947}{{\tt arXiv:1108.5947}}].

\bibitem{DelDuca:2011ae}
V.~Del~Duca, C.~Duhr, E.~Gardi, L.~Magnea, and C.~D. White, {\it {The Infrared
  structure of gauge theory amplitudes in the high-energy limit}},  {\em JHEP}
  {\bf 12} (2011) 021, [\href{http://arxiv.org/abs/1109.3581}{{\tt
  arXiv:1109.3581}}].

\bibitem{DelDuca:2014cya}
V.~Del~Duca, G.~Falcioni, L.~Magnea, and L.~Vernazza, {\it {Analyzing
  high-energy factorization beyond next-to-leading logarithmic accuracy}},
  {\em JHEP} {\bf 02} (2015) 029, [\href{http://arxiv.org/abs/1409.8330}{{\tt
  arXiv:1409.8330}}].

\bibitem{Caron-Huot:2016tzz}
S.~Caron-Huot and M.~Herranen, {\it {High-energy evolution to three loops}},
  \href{http://arxiv.org/abs/1604.07417}{{\tt arXiv:1604.07417}}.

\bibitem{Caron-Huot:2017fxr}
S.~Caron-Huot, E.~Gardi, and L.~Vernazza, {\it {Two-parton scattering in the
  high-energy limit}},  {\em JHEP} {\bf 06} (2017) 016,
  [\href{http://arxiv.org/abs/1701.05241}{{\tt arXiv:1701.05241}}].

\bibitem{Dittmar:2005ed}
M.~Dittmar et~al., {\it {Working Group I: Parton distributions: Summary report
  for the HERA LHC Workshop Proceedings}},
  \href{http://arxiv.org/abs/hep-ph/0511119}{{\tt hep-ph/0511119}}.

\bibitem{Forte:2009wh}
S.~Forte, G.~Altarelli, and R.~D. Ball, {\it {Can we trust small x
  resummation?}},  {\em Nucl.Phys.Proc.Suppl.} {\bf 191} (2009) 64--75,
  [\href{http://arxiv.org/abs/0901.1294}{{\tt arXiv:0901.1294}}].

\bibitem{Jaroszewicz:1982gr}
T.~Jaroszewicz, {\it {Gluonic Regge Singularities and Anomalous Dimensions in
  QCD}},  {\em Phys. Lett.} {\bf B116} (1982) 291.

\bibitem{Catani:1989sg}
S.~Catani, F.~Fiorani, and G.~Marchesini, {\it {Small-$x$ behavior of initial
  state radiation in perturbative QCD}},  {\em Nucl. Phys.} {\bf B336} (1990)
  18--85.

\bibitem{Ball:2005mj}
R.~D. Ball and S.~Forte, {\it {All order running coupling BFKL evolution from
  GLAP (and vice-versa)}},  {\em Nucl. Phys.} {\bf B742} (2006) 158--175,
  [\href{http://arxiv.org/abs/hep-ph/0601049}{{\tt hep-ph/0601049}}].

\bibitem{Catani:1990xk}
S.~Catani, M.~Ciafaloni, and F.~Hautmann, {\it {Gluon contributions to
  small-$x$ heavy flavor production}},  {\em Phys.Lett.} {\bf B242} (1990) 97.

\bibitem{Catani:1990eg}
S.~Catani, M.~Ciafaloni, and F.~Hautmann, {\it {High energy factorization and
  small-$x$ heavy flavour production}},  {\em Nucl. Phys.} {\bf B366} (1991)
  135--188.

\bibitem{Catani:1993ww}
S.~Catani, M.~Ciafaloni, and F.~Hautmann, {\it {High-energy factorization in
  QCD and minimal subtraction scheme}},  {\em Phys.Lett.} {\bf B307} (1993)
  147--153.

\bibitem{Catani:1994sq}
S.~Catani and F.~Hautmann, {\it {High-energy factorization and small x deep
  inelastic scattering beyond leading order}},  {\em Nucl. Phys.} {\bf B427}
  (1994) 475--524, [\href{http://arxiv.org/abs/hep-ph/9405388}{{\tt
  hep-ph/9405388}}].

\bibitem{Caola:2010kv}
F.~Caola, S.~Forte, and S.~Marzani, {\it {Small x resummation of rapidity
  distributions: The Case of Higgs production}},  {\em Nucl.Phys.} {\bf B846}
  (2011) 167--211, [\href{http://arxiv.org/abs/1010.2743}{{\tt
  arXiv:1010.2743}}].

\bibitem{Forte:2015gve}
S.~Forte and C.~Muselli, {\it {High energy resummation of transverse momentum
  distributions: Higgs in gluon fusion}},  {\em JHEP} {\bf 03} (2016) 122,
  [\href{http://arxiv.org/abs/1511.05561}{{\tt arXiv:1511.05561}}].

\bibitem{Marzani:2015oyb}
S.~Marzani, {\it {Combining $Q_T$ and small-$x$ resummations}},  {\em Phys.
  Rev.} {\bf D93} (2016), no.~5 054047,
  [\href{http://arxiv.org/abs/1511.06039}{{\tt arXiv:1511.06039}}].

\bibitem{Collins:1991ty}
J.~C. Collins and R.~K. Ellis, {\it {Heavy quark production in very high energy
  hadron collisions}},  {\em Nucl. Phys.} {\bf B360} (1991) 3--30.

\bibitem{Ball:2001pq}
R.~Ball and R.~K. Ellis, {\it {Heavy quark production at high-energy}},  {\em
  JHEP} {\bf 0105} (2001) 053, [\href{http://arxiv.org/abs/hep-ph/0101199}{{\tt
  hep-ph/0101199}}].

\bibitem{Catani:1993rn}
S.~Catani and F.~Hautmann, {\it {Quark anomalous dimensions at small x}},  {\em
  Phys.Lett.} {\bf B315} (1993) 157--163.

\bibitem{Catani:1996sc}
S.~Catani, {\it {Physical anomalous dimensions at small x}},  {\em Z. Phys.}
  {\bf C75} (1997) 665--678, [\href{http://arxiv.org/abs/hep-ph/9609263}{{\tt
  hep-ph/9609263}}].

\bibitem{Marzani:2008uh}
S.~Marzani and R.~D. Ball, {\it {High Energy Resummation of Drell-Yan
  Processes}},  {\em Nucl.Phys.} {\bf B814} (2009) 246--264,
  [\href{http://arxiv.org/abs/0812.3602}{{\tt arXiv:0812.3602}}].

\bibitem{Marzani:2010ap}
S.~Marzani, {\it {High-energy resummation at the LHC: The Case of Drell-Yan
  processes}},  {\em Nucl. Phys. Proc. Suppl.} {\bf 205-206} (2010) 25--30,
  [\href{http://arxiv.org/abs/1006.2314}{{\tt arXiv:1006.2314}}].

\bibitem{Diana:2009xv}
G.~Diana, {\it {High-energy resummation in direct photon production}},  {\em
  Nucl. Phys.} {\bf B824} (2010) 154--167,
  [\href{http://arxiv.org/abs/0906.4159}{{\tt arXiv:0906.4159}}].

\bibitem{Diana:2010ef}
G.~Diana, J.~Rojo, and R.~D. Ball, {\it {High energy resummation of direct
  photon production at hadronic colliders}},  {\em Phys.Lett.} {\bf B693}
  (2010) 430--437, [\href{http://arxiv.org/abs/1006.4250}{{\tt
  arXiv:1006.4250}}].

\bibitem{Hautmann:2002tu}
F.~Hautmann, {\it {Heavy top limit and double logarithmic contributions to
  Higgs production at $m_H^2/s$ much less than 1}},  {\em Phys.Lett.} {\bf
  B535} (2002) 159--162, [\href{http://arxiv.org/abs/hep-ph/0203140}{{\tt
  hep-ph/0203140}}].

\bibitem{Marzani:2008az}
S.~Marzani, R.~D. Ball, V.~Del~Duca, S.~Forte, and A.~Vicini, {\it {Higgs
  production via gluon-gluon fusion with finite top mass beyond next-to-leading
  order}},  {\em Nucl.Phys.} {\bf B800} (2008) 127--145,
  [\href{http://arxiv.org/abs/0801.2544}{{\tt arXiv:0801.2544}}].

\bibitem{Caola:2011wq}
F.~Caola and S.~Marzani, {\it {Finite fermion mass effects in pseudoscalar
  Higgs production via gluon-gluon fusion}},  {\em Phys.Lett.} {\bf B698}
  (2011) 275--283, [\href{http://arxiv.org/abs/1101.3975}{{\tt
  arXiv:1101.3975}}].

\bibitem{Bartels:2002uz}
J.~Bartels, D.~Colferai, S.~Gieseke, and A.~Kyrieleis, {\it {NLO corrections to
  the photon impact factor: Combining real and virtual corrections}},  {\em
  Phys. Rev.} {\bf D66} (2002) 094017,
  [\href{http://arxiv.org/abs/hep-ph/0208130}{{\tt hep-ph/0208130}}].

\bibitem{Bertone:2013vaa}
V.~Bertone, S.~Carrazza, and J.~Rojo, {\it {APFEL: A PDF Evolution Library with
  QED corrections}},  {\em Comput.Phys.Commun.} {\bf 185} (2014) 1647--1668,
  [\href{http://arxiv.org/abs/1310.1394}{{\tt arXiv:1310.1394}}].

\bibitem{Carrazza:2014gfa}
S.~Carrazza, A.~Ferrara, D.~Palazzo, and J.~Rojo, {\it {APFEL Web}},  {\em J.
  Phys.} {\bf G42} (2015), no.~5 057001,
  [\href{http://arxiv.org/abs/1410.5456}{{\tt arXiv:1410.5456}}].

\bibitem{Forte:2010ta}
S.~Forte, E.~Laenen, P.~Nason, and J.~Rojo, {\it {Heavy quarks in
  deep-inelastic scattering}},  {\em Nucl. Phys.} {\bf B834} (2010) 116--162,
  [\href{http://arxiv.org/abs/1001.2312}{{\tt arXiv:1001.2312}}].

\bibitem{Ball:2015dpa}
R.~D. Ball, M.~Bonvini, and L.~Rottoli, {\it {Charm in Deep-Inelastic
  Scattering}},  {\em JHEP} {\bf 11} (2015) 122,
  [\href{http://arxiv.org/abs/1510.02491}{{\tt arXiv:1510.02491}}].

\bibitem{Kretzer:1998ju}
S.~Kretzer and I.~Schienbein, {\it {Heavy quark initiated contributions to deep
  inelastic structure functions}},  {\em Phys.Rev.} {\bf D58} (1998) 094035,
  [\href{http://arxiv.org/abs/hep-ph/9805233}{{\tt hep-ph/9805233}}].

\bibitem{Ball:2015tna}
R.~D. Ball, V.~Bertone, M.~Bonvini, S.~Forte, P.~Groth~Merrild, J.~Rojo, and
  L.~Rottoli, {\it {Intrinsic charm in a matched general-mass scheme}},  {\em
  Phys. Lett.} {\bf B754} (2016) 49--58,
  [\href{http://arxiv.org/abs/1510.00009}{{\tt arXiv:1510.00009}}].

\bibitem{Arneodo:1996kd}
{\bf New Muon} Collaboration, M.~Arneodo et~al., {\it {Accurate measurement of
  $F_2^d/F_2^p$ and $R_d-R_p$}},  {\em Nucl. Phys.} {\bf B487} (1997) 3--26,
  [\href{http://arxiv.org/abs/hep-ex/9611022}{{\tt hep-ex/9611022}}].

\bibitem{Arneodo:1996qe}
{\bf New Muon} Collaboration, M.~Arneodo et~al., {\it {Measurement of the
  proton and deuteron structure functions, $F_2^p$ and $F_2^d$, and of the
  ratio $\sigma_L/\sigma_T$}},  {\em Nucl. Phys.} {\bf B483} (1997) 3--43,
  [\href{http://arxiv.org/abs/hep-ph/9610231}{{\tt hep-ph/9610231}}].

\bibitem{bcdms1}
{\bf BCDMS} Collaboration, A.~C. Benvenuti et~al., {\it A high statistics
  measurement of the proton structure functions $f_2 (x, q^2)$ and $r$ from
  deep inelastic muon scattering at high $q^2$},  {\em Phys. Lett.} {\bf B223}
  (1989) 485.

\bibitem{bcdms2}
{\bf BCDMS} Collaboration, A.~C. Benvenuti et~al., {\it A high statistics
  measurement of the deuteron structure functions $f_2 (x, q^2)$ and $r$ from
  deep inelastic muon scattering at high $q^2$},  {\em Phys. Lett.} {\bf B237}
  (1990) 592.

\bibitem{Whitlow:1991uw}
L.~W. Whitlow, E.~M. Riordan, S.~Dasu, S.~Rock, and A.~Bodek, {\it {Precise
  measurements of the proton and deuteron structure functions from a global
  analysis of the SLAC deep inelastic electron scattering cross-sections}},
  {\em Phys. Lett.} {\bf B282} (1992) 475--482.

\bibitem{Onengut:2005kv}
{\bf CHORUS} Collaboration, G.~Onengut et~al., {\it {Measurement of nucleon
  structure functions in neutrino scattering}},  {\em Phys. Lett.} {\bf B632}
  (2006) 65--75.

\bibitem{Goncharov:2001qe}
{\bf NuTeV} Collaboration, M.~Goncharov et~al., {\it {Precise measurement of
  dimuon production cross-sections in $\nu_{\mu}$Fe and $\bar{\nu}_{\mu}$Fe
  deep inelastic scattering at the Tevatron}},  {\em Phys. Rev.} {\bf D64}
  (2001) 112006, [\href{http://arxiv.org/abs/hep-ex/0102049}{{\tt
  hep-ex/0102049}}].

\bibitem{MasonPhD}
D.~A. Mason, {\it {Measurement of the strange - antistrange asymmetry at NLO in
  QCD from NuTeV dimuon data}}, . FERMILAB-THESIS-2006-01.

\bibitem{Abramowicz:1900rp}
{\bf H1 , ZEUS} Collaboration, H.~Abramowicz et~al., {\it {Combination and QCD
  Analysis of Charm Production Cross Section Measurements in Deep-Inelastic ep
  Scattering at HERA}},  {\em Eur.Phys.J.} {\bf C73} (2013) 2311,
  [\href{http://arxiv.org/abs/1211.1182}{{\tt arXiv:1211.1182}}].

\bibitem{Webb:2003ps}
{\bf NuSea} Collaboration, J.~C. Webb et~al., {\it {Absolute Drell-Yan dimuon
  cross sections in 800-GeV/c p p and p d collisions}},
  \href{http://arxiv.org/abs/hep-ex/0302019}{{\tt hep-ex/0302019}}.

\bibitem{Webb:2003bj}
J.~C. Webb, {\it {Measurement of continuum dimuon production in 800-GeV/c
  proton nucleon collisions}},  \href{http://arxiv.org/abs/hep-ex/0301031}{{\tt
  hep-ex/0301031}}.

\bibitem{Towell:2001nh}
{\bf FNAL E866/NuSea} Collaboration, R.~S. Towell et~al., {\it {Improved
  measurement of the anti-d/anti-u asymmetry in the nucleon sea}},  {\em Phys.
  Rev.} {\bf D64} (2001) 052002,
  [\href{http://arxiv.org/abs/hep-ex/0103030}{{\tt hep-ex/0103030}}].

\bibitem{Moreno:1990sf}
G.~Moreno et~al., {\it {Dimuon production in proton - copper collisions at
  $\sqrt{s}$ = 38.8-GeV}},  {\em Phys. Rev.} {\bf D43} (1991) 2815--2836.

\bibitem{Aaltonen:2010zza}
{\bf CDF} Collaboration, T.~A. Aaltonen et~al., {\it {Measurement of
  $d\sigma/dy$ of Drell-Yan $e^+e^-$ pairs in the $Z$ Mass Region from
  $p\bar{p}$ Collisions at $\sqrt{s}=1.96$ TeV}},  {\em Phys. Lett.} {\bf B692}
  (2010) 232--239, [\href{http://arxiv.org/abs/0908.3914}{{\tt
  arXiv:0908.3914}}].

\bibitem{Abazov:2007jy}
{\bf D0} Collaboration, V.~M. Abazov et~al., {\it {Measurement of the shape of
  the boson rapidity distribution for $p \bar{p} \to Z/\gamma^* \to e^{+}
  e^{-}$ + $X$ events produced at $\sqrt{s}$=1.96-TeV}},  {\em Phys. Rev.} {\bf
  D76} (2007) 012003, [\href{http://arxiv.org/abs/hep-ex/0702025}{{\tt
  hep-ex/0702025}}].

\bibitem{Aaltonen:2008eq}
{\bf CDF} Collaboration, T.~Aaltonen et~al., {\it {Measurement of the Inclusive
  Jet Cross Section at the Fermilab Tevatron p-pbar Collider Using a Cone-Based
  Jet Algorithm}},  {\em Phys. Rev.} {\bf D78} (2008) 052006,
  [\href{http://arxiv.org/abs/0807.2204}{{\tt arXiv:0807.2204}}].

\bibitem{Abazov:2013rja}
{\bf D0} Collaboration, V.~M. Abazov et~al., {\it {Measurement of the muon
  charge asymmetry in $p\bar{p}$ $\to$ W+X $\to$ $\mu$$\nu$ + X events at
  $\sqrt{s}$=1.96 TeV}},  {\em Phys.Rev.} {\bf D88} (2013) 091102,
  [\href{http://arxiv.org/abs/1309.2591}{{\tt arXiv:1309.2591}}].

\bibitem{D0:2014kma}
{\bf D0} Collaboration, V.~M. Abazov et~al., {\it {Measurement of the electron
  charge asymmetry in $\boldsymbol{p\bar{p}\rightarrow W+X \rightarrow e\nu
  +X}$ decays in $\boldsymbol{p\bar{p}}$ collisions at
  $\boldsymbol{\sqrt{s}=1.96}$ TeV}},  {\em Phys. Rev.} {\bf D91} (2015), no.~3
  032007, [\href{http://arxiv.org/abs/1412.2862}{{\tt arXiv:1412.2862}}].
  [Erratum: Phys. Rev.D91,no.7,079901(2015)].

\bibitem{Aad:2011dm}
{\bf ATLAS} Collaboration, G.~Aad et~al., {\it {Measurement of the inclusive
  $W^{\pm}$ and $Z/\gamma^*$ cross sections in the electron and muon decay
  channels in pp collisions at $\sqrt{s}$= 7 TeV with the ATLAS detector}},
  {\em Phys.Rev.} {\bf D85} (2012) 072004,
  [\href{http://arxiv.org/abs/1109.5141}{{\tt arXiv:1109.5141}}].

\bibitem{Aad:2013iua}
{\bf ATLAS} Collaboration, G.~Aad et~al., {\it {Measurement of the high-mass
  Drell--Yan differential cross-section in pp collisions at $\sqrt{s}$=7 TeV
  with the ATLAS detector}},  {\em Phys.Lett.} {\bf B725} (2013) 223,
  [\href{http://arxiv.org/abs/1305.4192}{{\tt arXiv:1305.4192}}].

\bibitem{Aad:2011fp}
{\bf ATLAS} Collaboration, G.~Aad et~al., {\it {Measurement of the Transverse
  Momentum Distribution of $W$ Bosons in $pp$ Collisions at $\sqrt{s}=7$ TeV
  with the ATLAS Detector}},  {\em Phys.Rev.} {\bf D85} (2012) 012005,
  [\href{http://arxiv.org/abs/1108.6308}{{\tt arXiv:1108.6308}}].

\bibitem{Aad:2011fc}
{\bf ATLAS} Collaboration, G.~Aad et~al., {\it {Measurement of inclusive jet
  and dijet production in pp collisions at $\sqrt{s}$ = 7 TeV using the ATLAS
  detector}},  {\em Phys. Rev.} {\bf D86} (2012) 014022,
  [\href{http://arxiv.org/abs/1112.6297}{{\tt arXiv:1112.6297}}].

\bibitem{Aad:2013lpa}
{\bf ATLAS} Collaboration, G.~Aad et~al., {\it {Measurement of the inclusive
  jet cross section in pp collisions at $\sqrt{s}$=2.76 TeV and comparison to
  the inclusive jet cross section at $\sqrt{s}$=7 TeV using the ATLAS
  detector}},  {\em Eur.Phys.J.} {\bf C73} (2013) 2509,
  [\href{http://arxiv.org/abs/1304.4739}{{\tt arXiv:1304.4739}}].

\bibitem{ATLAS:2012aa}
{\bf ATLAS} Collaboration, G.~Aad et~al., {\it {Measurement of the cross
  section for top-quark pair production in $pp$ collisions at $\sqrt{s}=7$ TeV
  with the ATLAS detector using final states with two high-pt leptons}},  {\em
  JHEP} {\bf 1205} (2012) 059, [\href{http://arxiv.org/abs/1202.4892}{{\tt
  arXiv:1202.4892}}].

\bibitem{ATLAS:2011xha}
{\bf ATLAS} Collaboration, G.~Aad et~al., {\it {Measurement of the $t\bar{t}$
  production cross-section in pp collisions at sqrt{s} = 7 TeV using kinematic
  information of lepton+jets events}},
  \href{http://arxiv.org/abs/ATLAS-CONF-2011-121, ATLAS-COM-CONF-2011-132}{{\tt
  ATLAS-CONF-2011-121, ATLAS-COM-CONF-2011-132}}.

\bibitem{TheATLAScollaboration:2013dja}
{\bf ATLAS} Collaboration, G.~Aad et~al., {\it {Measurement of the $t\bar{t}$
  production cross-section in $pp$ collisions at $\sqrt{s}=8$ TeV using $e\mu$
  events with $b$-tagged jets}},
  \href{http://arxiv.org/abs/ATLAS-CONF-2013-097, ATLAS-COM-CONF-2013-112}{{\tt
  ATLAS-CONF-2013-097, ATLAS-COM-CONF-2013-112}}.

\bibitem{Aad:2015auj}
{\bf ATLAS} Collaboration, G.~Aad et~al., {\it {Measurement of the transverse
  momentum and $\phi ^*_{\eta }$ distributions of DrellâYan lepton pairs
  in protonâproton collisions at $\sqrt{s}=8$ TeV with the ATLAS
  detector}},  {\em Eur. Phys. J.} {\bf C76} (2016), no.~5 291,
  [\href{http://arxiv.org/abs/1512.02192}{{\tt arXiv:1512.02192}}].

\bibitem{Aaboud:2016btc}
{\bf ATLAS} Collaboration, M.~Aaboud et~al., {\it {Precision measurement and
  interpretation of inclusive $W^+$ , $W^-$ and $Z/\gamma ^*$ production cross
  sections with the ATLAS detector}},  {\em Eur. Phys. J.} {\bf C77} (2017),
  no.~6 367, [\href{http://arxiv.org/abs/1612.03016}{{\tt arXiv:1612.03016}}].

\bibitem{Aad:2014kva}
{\bf ATLAS} Collaboration, G.~Aad et~al., {\it {Measurement of the $t\bar{t}$
  production cross-section using $e\mu $ events with b-tagged jets in pp
  collisions at $\sqrt{s}$ = 7 and 8 $\,\mathrm{TeV}$ with the ATLAS
  detector}},  {\em Eur. Phys. J.} {\bf C74} (2014), no.~10 3109,
  [\href{http://arxiv.org/abs/1406.5375}{{\tt arXiv:1406.5375}}]. [Addendum:
  Eur. Phys. J.C76,no.11,642(2016)].

\bibitem{Aaboud:2016pbd}
{\bf ATLAS} Collaboration, M.~Aaboud et~al., {\it {Measurement of the
  $t\bar{t}$ production cross-section using $e\mu$ events with b-tagged jets in
  pp collisions at $\sqrt{s}$=13 TeV with the ATLAS detector}},  {\em Phys.
  Lett.} {\bf B761} (2016) 136--157,
  [\href{http://arxiv.org/abs/1606.02699}{{\tt arXiv:1606.02699}}].

\bibitem{Aad:2015mbv}
{\bf ATLAS} Collaboration, G.~Aad et~al., {\it {Measurements of top-quark pair
  differential cross-sections in the lepton+jets channel in $pp$ collisions at
  $\sqrt{s}=8$ TeV using the ATLAS detector}},  {\em Eur. Phys. J.} {\bf C76}
  (2016), no.~10 538, [\href{http://arxiv.org/abs/1511.04716}{{\tt
  arXiv:1511.04716}}].

\bibitem{Aad:2014qja}
{\bf ATLAS} Collaboration, G.~Aad et~al., {\it {Measurement of the low-mass
  Drell-Yan differential cross section at $\sqrt{s}$ = 7 TeV using the ATLAS
  detector}},  {\em JHEP} {\bf 06} (2014) 112,
  [\href{http://arxiv.org/abs/1404.1212}{{\tt arXiv:1404.1212}}].

\bibitem{Aad:2014xaa}
{\bf ATLAS} Collaboration, G.~Aad et~al., {\it {Measurement of the $Z/\gamma^*$
  boson transverse momentum distribution in $pp$ collisions at $\sqrt{s}$ = 7
  TeV with the ATLAS detector}},  {\em JHEP} {\bf 09} (2014) 145,
  [\href{http://arxiv.org/abs/1406.3660}{{\tt arXiv:1406.3660}}].

\bibitem{Chatrchyan:2012xt}
{\bf CMS} Collaboration, S.~Chatrchyan et~al., {\it {Measurement of the
  electron charge asymmetry in inclusive W production in pp collisions at
  $\sqrt{s}$ = 7 TeV}},  {\em Phys.Rev.Lett.} {\bf 109} (2012) 111806,
  [\href{http://arxiv.org/abs/1206.2598}{{\tt arXiv:1206.2598}}].

\bibitem{Chatrchyan:2013mza}
{\bf CMS} Collaboration, S.~Chatrchyan et~al., {\it {Measurement of the muon
  charge asymmetry in inclusive pp to WX production at $\sqrt{s}$ = 7 TeV and
  an improved determination of light parton distribution functions}},  {\em
  Phys.Rev.} {\bf D90} (2014) 032004,
  [\href{http://arxiv.org/abs/1312.6283}{{\tt arXiv:1312.6283}}].

\bibitem{Chatrchyan:2013tia}
{\bf CMS} Collaboration, S.~Chatrchyan et~al., {\it {Measurement of the
  differential and double-differential Drell-Yan cross sections in
  proton-proton collisions at $\sqrt{s} =$ 7 TeV}},  {\em JHEP} {\bf 1312}
  (2013) 030, [\href{http://arxiv.org/abs/1310.7291}{{\tt arXiv:1310.7291}}].

\bibitem{Chatrchyan:2013uja}
{\bf CMS} Collaboration, S.~Chatrchyan et~al., {\it {Measurement of associated
  W + charm production in pp collisions at $\sqrt{s}$ = 7 TeV}},  {\em JHEP}
  {\bf 02} (2014) 013, [\href{http://arxiv.org/abs/1310.1138}{{\tt
  arXiv:1310.1138}}].

\bibitem{Chatrchyan:2013faa}
{\bf CMS} Collaboration, S.~Chatrchyan et~al., {\it {Measurement of the $t
  \bar{t}$ production cross section in the dilepton channel in pp collisions at
  $\sqrt{s}$ = 8 TeV}},  {\em JHEP} {\bf 1402} (2014) 024,
  [\href{http://arxiv.org/abs/1312.7582}{{\tt arXiv:1312.7582}}].

\bibitem{Chatrchyan:2012bra}
{\bf CMS} Collaboration, S.~Chatrchyan et~al., {\it {Measurement of the
  $t\bar{t}$ production cross section in the dilepton channel in $pp$
  collisions at $\sqrt{s}=7$ TeV}},  {\em JHEP} {\bf 1211} (2012) 067,
  [\href{http://arxiv.org/abs/1208.2671}{{\tt arXiv:1208.2671}}].

\bibitem{Chatrchyan:2012ria}
{\bf CMS} Collaboration, S.~Chatrchyan et~al., {\it {Measurement of the
  $t\bar{t}$ production cross section in $pp$ collisions at $\sqrt{s}=7$ TeV
  with lepton + jets final states}},  {\em Phys.Lett.} {\bf B720} (2013)
  83--104, [\href{http://arxiv.org/abs/1212.6682}{{\tt arXiv:1212.6682}}].

\bibitem{Khachatryan:2016pev}
{\bf CMS} Collaboration, V.~Khachatryan et~al., {\it {Measurement of the
  differential cross section and charge asymmetry for inclusive $\mathrm
  {p}\mathrm {p}\rightarrow \mathrm {W}^{\pm }+X$ production at ${\sqrt{s}} =
  8$ TeV}},  {\em Eur. Phys. J.} {\bf C76} (2016), no.~8 469,
  [\href{http://arxiv.org/abs/1603.01803}{{\tt arXiv:1603.01803}}].

\bibitem{Khachatryan:2015luy}
{\bf CMS} Collaboration, V.~Khachatryan et~al., {\it {Measurement of the
  inclusive jet cross section in pp collisions at $\sqrt{s} = 2.76\,\text
  {TeV}$}},  {\em Eur. Phys. J.} {\bf C76} (2016), no.~5 265,
  [\href{http://arxiv.org/abs/1512.06212}{{\tt arXiv:1512.06212}}].

\bibitem{Khachatryan:2016mqs}
{\bf CMS} Collaboration, V.~Khachatryan et~al., {\it {Measurement of the t-tbar
  production cross section in the e-mu channel in proton-proton collisions at
  sqrt(s) = 7 and 8 TeV}},  {\em JHEP} {\bf 08} (2016) 029,
  [\href{http://arxiv.org/abs/1603.02303}{{\tt arXiv:1603.02303}}].

\bibitem{Khachatryan:2015oqa}
{\bf CMS} Collaboration, V.~Khachatryan et~al., {\it {Measurement of the
  differential cross section for top quark pair production in pp collisions at
  $\sqrt{s} = 8\,\text {TeV} $}},  {\em Eur. Phys. J.} {\bf C75} (2015), no.~11
  542, [\href{http://arxiv.org/abs/1505.04480}{{\tt arXiv:1505.04480}}].

\bibitem{Khachatryan:2015oaa}
{\bf CMS} Collaboration, V.~Khachatryan et~al., {\it {Measurement of the Z
  boson differential cross section in transverse momentum and rapidity in
  protonâproton collisions at 8 TeV}},  {\em Phys. Lett.} {\bf B749}
  (2015) 187--209, [\href{http://arxiv.org/abs/1504.03511}{{\tt
  arXiv:1504.03511}}].

\bibitem{Aaij:2012vn}
{\bf LHCb} Collaboration, R.~Aaij et~al., {\it {Inclusive $W$ and $Z$
  production in the forward region at $\sqrt{s} = 7$ TeV}},  {\em JHEP} {\bf
  1206} (2012) 058, [\href{http://arxiv.org/abs/1204.1620}{{\tt
  arXiv:1204.1620}}].

\bibitem{Aaij:2012mda}
{\bf LHCb} Collaboration, R.~Aaij et~al., {\it {Measurement of the
  cross-section for $Z \to e^+e^-$ production in $pp$ collisions at
  $\sqrt{s}=7$ TeV}},  {\em JHEP} {\bf 1302} (2013) 106,
  [\href{http://arxiv.org/abs/1212.4620}{{\tt arXiv:1212.4620}}].

\bibitem{Chatrchyan:2012bja}
{\bf CMS} Collaboration, S.~Chatrchyan et~al., {\it {Measurements of
  differential jet cross sections in proton-proton collisions at $\sqrt{s}=7$
  TeV with the CMS detector}},  {\em Phys.Rev.} {\bf D87} (2013) 112002,
  [\href{http://arxiv.org/abs/1212.6660}{{\tt arXiv:1212.6660}}].

\bibitem{Aaij:2015gna}
{\bf LHCb} Collaboration, R.~Aaij et~al., {\it {Measurement of the forward $Z$
  boson production cross-section in $pp$ collisions at $\sqrt{s}=7$ TeV}},
  {\em JHEP} {\bf 08} (2015) 039, [\href{http://arxiv.org/abs/1505.07024}{{\tt
  arXiv:1505.07024}}].

\bibitem{Aaij:2015zlq}
{\bf LHCb} Collaboration, R.~Aaij et~al., {\it {Measurement of forward W and Z
  boson production in $pp$ collisions at $ \sqrt{s}=8 $ TeV}},  {\em JHEP} {\bf
  01} (2016) 155, [\href{http://arxiv.org/abs/1511.08039}{{\tt
  arXiv:1511.08039}}].

\bibitem{Ball:2016qeg}
R.~D. Ball, {\it {Charm Production: Pole Mass or Running Mass?}},  {\em AIP
  Conf. Proc.} {\bf 1819} (2017), no.~1 030002,
  [\href{http://arxiv.org/abs/1612.03790}{{\tt arXiv:1612.03790}}].

\bibitem{Carli:2010rw}
T.~Carli et~al., {\it {A posteriori inclusion of parton density functions in
  NLO QCD final-state calculations at hadron colliders: The APPLGRID Project}},
   {\em Eur.Phys.J.} {\bf C66} (2010) 503,
  [\href{http://arxiv.org/abs/0911.2985}{{\tt arXiv:0911.2985}}].

\bibitem{Wobisch:2011ij}
{\bf fastNLO} Collaboration, M.~Wobisch, D.~Britzger, T.~Kluge, K.~Rabbertz,
  and F.~Stober, {\it {Theory-Data Comparisons for Jet Measurements in
  Hadron-Induced Processes}},  \href{http://arxiv.org/abs/1109.1310}{{\tt
  arXiv:1109.1310}}.

\bibitem{Bertone:2016lga}
V.~Bertone, S.~Carrazza, and N.~P. Hartland, {\it {APFELgrid: a high
  performance tool for parton density determinations}},  {\em Comput. Phys.
  Commun.} {\bf 212} (2017) 205--209,
  [\href{http://arxiv.org/abs/1605.02070}{{\tt arXiv:1605.02070}}].

\bibitem{Czakon:2016dgf}
M.~Czakon, D.~Heymes, and A.~Mitov, {\it {Dynamical scales for multi-TeV
  top-pair production at the LHC}},  {\em JHEP} {\bf 04} (2017) 071,
  [\href{http://arxiv.org/abs/1606.03350}{{\tt arXiv:1606.03350}}].

\bibitem{Czakon:2015owf}
M.~Czakon, D.~Heymes, and A.~Mitov, {\it {High-precision differential
  predictions for top-quark pairs at the LHC}},  {\em Phys. Rev. Lett.} {\bf
  116} (2016), no.~8 082003, [\href{http://arxiv.org/abs/1511.00549}{{\tt
  arXiv:1511.00549}}].

\bibitem{Czakon:2016olj}
M.~Czakon, N.~P. Hartland, A.~Mitov, E.~R. Nocera, and J.~Rojo, {\it {Pinning
  down the large-x gluon with NNLO top-quark pair differential distributions}},
   {\em JHEP} {\bf 04} (2017) 044, [\href{http://arxiv.org/abs/1611.08609}{{\tt
  arXiv:1611.08609}}].

\bibitem{Boughezal:2017nla}
R.~Boughezal, A.~Guffanti, F.~Petriello, and M.~Ubiali, {\it {The impact of the
  LHC Z-boson transverse momentum data on PDF determinations}},  {\em JHEP}
  {\bf 07} (2017) 130, [\href{http://arxiv.org/abs/1705.00343}{{\tt
  arXiv:1705.00343}}].

\bibitem{Boughezal:2015ded}
R.~Boughezal, J.~M. Campbell, R.~K. Ellis, C.~Focke, W.~T. Giele, X.~Liu, and
  F.~Petriello, {\it {Z-boson production in association with a jet at
  next-to-next-to-leading order in perturbative QCD}},  {\em Phys. Rev. Lett.}
  {\bf 116} (2016), no.~15 152001, [\href{http://arxiv.org/abs/1512.01291}{{\tt
  arXiv:1512.01291}}].

\bibitem{Gavin:2012sy}
R.~Gavin, Y.~Li, F.~Petriello, and S.~Quackenbush, {\it {W Physics at the LHC
  with FEWZ 2.1}},  {\em Comput.Phys.Commun.} {\bf 184} (2013) 208--214,
  [\href{http://arxiv.org/abs/1201.5896}{{\tt arXiv:1201.5896}}].

\bibitem{Marzani:2009hu}
S.~Marzani and R.~D. Ball, {\it {Drell-Yan processes in the high-energy
  limit}},  in {\em {Proceedings, 17th International Workshop on Deep-Inelastic
  Scattering and Related Subjects (DIS 2009): Madrid, Spain, April 26-30,
  2009}}, 2009.
\newblock \href{http://arxiv.org/abs/0906.4729}{{\tt arXiv:0906.4729}}.

\bibitem{Ball:2009qv}
{\bf The NNPDF} Collaboration, R.~D. Ball et~al., {\it {Fitting Parton
  Distribution Data with Multiplicative Normalization Uncertainties}},  {\em
  JHEP} {\bf 05} (2010) 075, [\href{http://arxiv.org/abs/0912.2276}{{\tt
  arXiv:0912.2276}}].

\bibitem{Pumplin:2002vw}
J.~Pumplin et~al., {\it {New generation of parton distributions with
  uncertainties from global QCD analysis}},  {\em JHEP} {\bf 07} (2002) 012,
  [\href{http://arxiv.org/abs/hep-ph/0201195}{{\tt hep-ph/0201195}}].

\bibitem{Andreev:2013vha}
{\bf H1 Collaboration} Collaboration, V.~Andreev et~al., {\it {Measurement of
  inclusive $e p$ cross sections at high $Q^2$ at $\sqrt s =$ 225 and 252 GeV
  and of the longitudinal proton structure function $F_L$ at HERA}},  {\em
  Eur.Phys.J.} {\bf C74} (2014) 2814,
  [\href{http://arxiv.org/abs/1312.4821}{{\tt arXiv:1312.4821}}].

\bibitem{Abramowicz:2014jak}
{\bf ZEUS} Collaboration, H.~Abramowicz et~al., {\it {Deep inelastic
  cross-section measurements at large y with the ZEUS detector at HERA}},  {\em
  Phys. Rev.} {\bf D90} (2014), no.~7 072002,
  [\href{http://arxiv.org/abs/1404.6376}{{\tt arXiv:1404.6376}}].

\bibitem{Harris:1997zq}
B.~W. Harris and J.~Smith, {\it {Charm quark and D*+- cross-sections in deeply
  inelastic scattering at HERA}},  {\em Phys. Rev.} {\bf D57} (1998)
  2806--2812, [\href{http://arxiv.org/abs/hep-ph/9706334}{{\tt
  hep-ph/9706334}}].

\bibitem{AbelleiraFernandez:2012cc}
{\bf LHeC Study Group} Collaboration, J.~Abelleira~Fernandez et~al., {\it {A
  Large Hadron Electron Collider at CERN: Report on the Physics and Design
  Concepts for Machine and Detector}},  {\em J.Phys.} {\bf G39} (2012) 075001,
  [\href{http://arxiv.org/abs/1206.2913}{{\tt arXiv:1206.2913}}].

\bibitem{Mangano:2016jyj}
M.~L. Mangano et~al., {\it {Physics at a 100 TeV pp collider: Standard Model
  processes}},  {\em CERN Yellow Report} (2017), no.~3 1--254,
  [\href{http://arxiv.org/abs/1607.01831}{{\tt arXiv:1607.01831}}].

\bibitem{Contino:2016spe}
R.~Contino et~al., {\it {Physics at a 100 TeV pp collider: Higgs and EW
  symmetry breaking studies}},  {\em CERN Yellow Report} (2017), no.~3
  255--440, [\href{http://arxiv.org/abs/1606.09408}{{\tt arXiv:1606.09408}}].

\bibitem{Campbell:2006wx}
J.~M. Campbell, J.~W. Huston, and W.~J. Stirling, {\it {Hard interactions of
  quarks and gluons: A primer for LHC physics}},  {\em Rept. Prog. Phys.} {\bf
  70} (2007) 89, [\href{http://arxiv.org/abs/hep-ph/0611148}{{\tt
  hep-ph/0611148}}].

\bibitem{CMS:2014jea}
{\bf CMS} Collaboration, V.~Khachatryan et~al., {\it {Measurements of
  differential and double-differential Drell-Yan cross sections in
  proton-proton collisions at 8 TeV}},  {\em Eur. Phys. J.} {\bf C75} (2015),
  no.~4 147, [\href{http://arxiv.org/abs/1412.1115}{{\tt arXiv:1412.1115}}].

\bibitem{Aartsen:2014gkd}
{\bf IceCube} Collaboration, M.~G. Aartsen et~al., {\it {Observation of
  High-Energy Astrophysical Neutrinos in Three Years of IceCube Data}},  {\em
  Phys. Rev. Lett.} {\bf 113} (2014) 101101,
  [\href{http://arxiv.org/abs/1405.5303}{{\tt arXiv:1405.5303}}].

\bibitem{Adrian-Martinez:2016fdl}
{\bf KM3Net} Collaboration, S.~Adrian-Martinez et~al., {\it {Letter of intent
  for KM3NeT 2.0}},  {\em J. Phys.} {\bf G43} (2016), no.~8 084001,
  [\href{http://arxiv.org/abs/1601.07459}{{\tt arXiv:1601.07459}}].

\bibitem{Gandhi:1998ri}
R.~Gandhi, C.~Quigg, M.~H. Reno, and I.~Sarcevic, {\it {Neutrino interactions
  at ultrahigh-energies}},  {\em Phys. Rev.} {\bf D58} (1998) 093009,
  [\href{http://arxiv.org/abs/hep-ph/9807264}{{\tt hep-ph/9807264}}].

\bibitem{CooperSarkar:2011pa}
A.~Cooper-Sarkar, P.~Mertsch, and S.~Sarkar, {\it {The high energy neutrino
  cross-section in the Standard Model and its uncertainty}},  {\em JHEP} {\bf
  08} (2011) 042, [\href{http://arxiv.org/abs/1106.3723}{{\tt
  arXiv:1106.3723}}].

\bibitem{Anchordoqui:2006ta}
L.~A. Anchordoqui, A.~M. Cooper-Sarkar, D.~Hooper, and S.~Sarkar, {\it {Probing
  low-x QCD with cosmic neutrinos at the Pierre Auger Observatory}},  {\em
  Phys. Rev.} {\bf D74} (2006) 043008,
  [\href{http://arxiv.org/abs/hep-ph/0605086}{{\tt hep-ph/0605086}}].

\bibitem{Connolly:2011vc}
A.~Connolly, R.~S. Thorne, and D.~Waters, {\it {Calculation of High Energy
  Neutrino-Nucleon Cross Sections and Uncertainties Using the MSTW Parton
  Distribution Functions and Implications for Future Experiments}},  {\em Phys.
  Rev.} {\bf D83} (2011) 113009, [\href{http://arxiv.org/abs/1102.0691}{{\tt
  arXiv:1102.0691}}].

\bibitem{Block:2010ud}
M.~M. Block, P.~Ha, and D.~W. McKay, {\it {Ultrahigh energy neutrino
  scattering: An Update}},  {\em Phys. Rev.} {\bf D82} (2010) 077302,
  [\href{http://arxiv.org/abs/1008.4555}{{\tt arXiv:1008.4555}}].

\bibitem{CooperSarkar:2007cv}
A.~Cooper-Sarkar and S.~Sarkar, {\it {Predictions for high energy neutrino
  cross-sections from the ZEUS global PDF fits}},  {\em JHEP} {\bf 01} (2008)
  075, [\href{http://arxiv.org/abs/0710.5303}{{\tt arXiv:0710.5303}}].

\bibitem{Gluck:1998js}
M.~Gluck, S.~Kretzer, and E.~Reya, {\it {Dynamical QCD predictions for
  ultrahigh-energy neutrino cross-sections}},  {\em Astropart. Phys.} {\bf 11}
  (1999) 327--334, [\href{http://arxiv.org/abs/astro-ph/9809273}{{\tt
  astro-ph/9809273}}].

\bibitem{Albacete:2015zra}
J.~L. Albacete, J.~I. Illana, and A.~Soto-Ontoso, {\it {Neutrino-nucleon cross
  section at ultrahigh energy and its astrophysical implications}},  {\em Phys.
  Rev.} {\bf D92} (2015), no.~1 014027,
  [\href{http://arxiv.org/abs/1505.06583}{{\tt arXiv:1505.06583}}].

\bibitem{Armesto:2007tg}
N.~Armesto, C.~Merino, G.~Parente, and E.~Zas, {\it {Charged current neutrino
  cross-section and tau energy loss at ultra-high energies}},  {\em Phys. Rev.}
  {\bf D77} (2008) 013001, [\href{http://arxiv.org/abs/0709.4461}{{\tt
  arXiv:0709.4461}}].

\bibitem{Arguelles:2015wba}
C.~A. Argüelles, F.~Halzen, L.~Wille, M.~Kroll, and M.~H. Reno, {\it
  {High-energy behavior of photon, neutrino, and proton cross sections}},  {\em
  Phys. Rev.} {\bf D92} (2015), no.~7 074040,
  [\href{http://arxiv.org/abs/1504.06639}{{\tt arXiv:1504.06639}}].

\bibitem{Fiore:2005wf}
R.~Fiore, L.~L. Jenkovszky, A.~V. Kotikov, F.~Paccanoni, and A.~Papa, {\it
  {Asymptotic neutrino-nucleon cross section and saturation effects}},  {\em
  Phys. Rev.} {\bf D73} (2006) 053012,
  [\href{http://arxiv.org/abs/hep-ph/0512259}{{\tt hep-ph/0512259}}].

\bibitem{JalilianMarian:2003wf}
J.~Jalilian-Marian, {\it {Enhancement and suppression of the neutrino nucleon
  total cross-section at ultrahigh-energies}},  {\em Phys. Rev.} {\bf D68}
  (2003) 054005, [\href{http://arxiv.org/abs/hep-ph/0301238}{{\tt
  hep-ph/0301238}}]. [Erratum: Phys. Rev.D70,079903(2004)].

\bibitem{Aaij:2015bpa}
{\bf LHCb} Collaboration, R.~Aaij et~al., {\it {Measurements of prompt charm
  production cross-sections in $pp$ collisions at $ \sqrt{s}=13 $ TeV}},  {\em
  JHEP} {\bf 03} (2016) 159, [\href{http://arxiv.org/abs/1510.01707}{{\tt
  arXiv:1510.01707}}]. [Erratum: JHEP05,074(2017)].

\bibitem{Aaij:2016jht}
{\bf LHCb} Collaboration, R.~Aaij et~al., {\it {Measurements of prompt charm
  production cross-sections in pp collisions at $ \sqrt{s}=5 $ TeV}},  {\em
  JHEP} {\bf 06} (2017) 147, [\href{http://arxiv.org/abs/1610.02230}{{\tt
  arXiv:1610.02230}}].

\bibitem{Aaij:2013mga}
{\bf LHCb} Collaboration, R.~Aaij et~al., {\it {Prompt charm production in pp
  collisions at sqrt(s)=7 TeV}},  {\em Nucl. Phys.} {\bf B871} (2013) 1--20,
  [\href{http://arxiv.org/abs/1302.2864}{{\tt arXiv:1302.2864}}].

\bibitem{Gauld:2016kpd}
R.~Gauld and J.~Rojo, {\it {Precision determination of the small-$x$ gluon from
  charm production at LHCb}},  {\em Phys. Rev. Lett.} {\bf 118} (2017), no.~7
  072001, [\href{http://arxiv.org/abs/1610.09373}{{\tt arXiv:1610.09373}}].

\bibitem{Zenaiev:2015rfa}
{\bf PROSA} Collaboration, O.~Zenaiev et~al., {\it {Impact of heavy-flavour
  production cross sections measured by the LHCb experiment on parton
  distribution functions at low x}},  {\em Eur. Phys. J.} {\bf C75} (2015),
  no.~8 396, [\href{http://arxiv.org/abs/1503.04581}{{\tt arXiv:1503.04581}}].

\bibitem{AbelleiraFernandez:2012ty}
{\bf LHeC Study Group} Collaboration, J.~L. Abelleira~Fernandez et~al., {\it
  {On the Relation of the LHeC and the LHC}},
  \href{http://arxiv.org/abs/1211.5102}{{\tt arXiv:1211.5102}}.

\bibitem{Rojo:2009ut}
J.~Rojo and F.~Caola, {\it {Parton distributions and small-x QCD at the Large
  Hadron Electron Collider}},  in {\em {Proceedings, 17th International
  Workshop on Deep-Inelastic Scattering and Related Subjects (DIS 2009):
  Madrid, Spain, April 26-30, 2009}}, 2009.
\newblock \href{http://arxiv.org/abs/0906.2079}{{\tt arXiv:0906.2079}}.

\bibitem{Buckley:2014ana}
A.~Buckley, J.~Ferrando, S.~Lloyd, K.~NordstrÃ¶m, B.~Page, et~al., {\it
  {LHAPDF6: parton density access in the LHC precision era}},  {\em
  Eur.Phys.J.} {\bf C75} (2015) 132,
  [\href{http://arxiv.org/abs/1412.7420}{{\tt arXiv:1412.7420}}].

\end{thebibliography}\endgroup

\end{document}